\documentclass[11pt,letterpaper]{amsart}

\numberwithin{equation}{section}

\usepackage{amsmath}
\usepackage{cancel}
\usepackage{ulem}

\usepackage{amsfonts,amssymb,amsmath}
\usepackage{enumerate, xspace}
\usepackage{changebar}
\usepackage[colorlinks]{hyperref}
\usepackage[utf8]{inputenc}

\makeindex
\usepackage{calligra}
\usepackage[T1]{fontenc}

\usepackage{dutchcal}

\newtheorem{theorem}{Theorem}[section]
\newtheorem{lemma}[theorem]{Lemma}
\newtheorem{proposition}[theorem]{Proposition}
\newtheorem{corollary}[theorem]{Corollary}

\newtheorem{remark}[theorem]{Remark}

\theoremstyle{definition}
\newtheorem{definition}[theorem]{Definition}
\newtheorem{assumption}[theorem]{Assumption}

\newcommand{\dd  }{\mathrm{d}}

\newcommand{\mb}[1]{{\mathbf #1}}
\newcommand{\bbR}{\mathbb R}
 \newcommand{\bbP}{\mathbb P}
\newcommand{\bbZ}{\mathbb Z}

\newcommand{\eps}{\varepsilon}

\newcommand{\ga}{\gamma}
\newcommand{\la}{\lambda}
\newcommand{\al}{\alpha}

\newcommand{\bbE}{\mathbb E}

\newcommand{\rv}{\mb r}
\newcommand{\pv}{\mb p}
\newcommand{\Om}{\Omega}
\newcommand{\sD}{ {\scriptstyle\Delta} }

\newcommand{\cal}{\mathcal}

\newcommand\lang{\big\langle\!\big\langle}
\newcommand\rang{\big\rangle\!\big\rangle}

\newcounter{savesection}
\newcounter{apdxsection}

\newcommand\unappendix{\par
  \setcounter{apdxsection}{\value{section}}%
  \setcounter{section}{\value{savesection}}%
  \setcounter{subsection}{0}%
  \gdef\thesection{\large{\bf Supplement}}}

\newcounter{supsection}

\begin{document}

\title[$3$ conservation laws]{Thermal boundary conditions in fractional superdiffusion
    of energy
   }


\author{Tomasz Komorowski}
\address{Tomasz Komorowski\\ IMPAN,  \'{S}niadeckich 8, 00-656, Warsaw, Poland
}
 \email{\tt tkomorowski@impan.pl}

 \author{Stefano Olla}
 \address{Stefano Olla, CEREMADE,
   Universit\'e Paris-Dauphine, PSL Research University \\
 \and Institut Universitaire de France\\ \and
GSSI, L'Aquila}
  \email{\tt olla@ceremade.dauphine.fr}

\begin{abstract}
   {We study heat conduction in a one-dimensional {finite},
    unpinned chain of  atoms
  perturbed by stochastic momentum exchange and coupled to Langevin
  heat baths at {possibly} distinct temperatures placed at the
  endpoints of the chain.
  While infinite systems without boundaries are known to exhibit
  superdiffusive energy transport described by a fractional heat
  equation with the generator $-|\Delta|^{3/4}$,
  the corresponding boundary conditions induced by heat baths
  remain less understood.
  We establish the hydrodynamic limit for a finite chain with $n+1$
  atoms connected to thermostats at the endpoints,
  deriving the macroscopic evolution of the averaged energy profile.
  The limiting equation is governed by a non-local L\'evy-type operator,
  with boundary terms determined by explicit interaction kernels
  that encode absorption, reflection, and transmission of long-wavelength
  phonons at the baths. Our results provide the first rigorous identification
  of boundary conditions for fractional superdiffusion arising directly
  from microscopic dynamics {with local interactions},
  highlighting their distinction from both diffusive and pinned-chain settings.}
   
      \end{abstract}

 \thanks{T.K acknowledges the support of
the NCN grant 2024/53/B/ST1/00286}
\maketitle

\section{Introduction}
\label{sec:intro}


Heat conduction {in dielectric materials} involves the transfer of energy {through}
  the vibrations
of atoms, which  {generate} waves
that   propagate throughout the material.
Heat superdiffusion is generically expected in acoustic
(unpinned) one-dimensional chains of atoms,
where the dispersion relation allows long waves to travel with non-vanishing velocity.
Thermal conductivity is defined by the Green-Kubo formula that involves space-time
correlations of the dynamics. 
Numerical evidence suggests that the thermal conductivity diverges with the system size
{in chains of non-linear oscillators} 
\cite{sll,llp97}.
{Anharmonic deterministic Hamiltonian dynamics are mathematically difficult and at the moment
  there are no rigorous results in this direction. Even the convergence or the divergence of the
Green-Kubo formula is not proven.}
 Rigorous mathematical results were obtained for acoustic {\it harmonic} chains
with a random exchange
of velocities between nearest-neighboring atoms. {Such a stochastic
  mechanism emulates elastic
  collisions between the atoms.
It conserves the total energy
and momentum while}
breaking the complete integrability of the harmonic chain,
{thus induces} scattering of the waves.
The corresponding scattering rates are inversely proportional
to the wave length, which, in turn, induces a macroscopic 
  fractional  L\'evy superdiffusion of energy,
carried predominantely by the long waves.
This  {behaviour contrasts with that of} optical  (pinned)
chains, where long waves {propagate}
slowly and energy diffusivity is finite.
 {The divergence} of thermal conductivity in stochastically perturbed
acoustic chains was proven in \cite{bborev,bbo2}.
A kinetic equation was  {derived} in a low noise limit in
\cite{BOS},  {and} a superdiffusive
scaling limit from the kinetic equation was obtained in
\cite{kjo,babo},  {yielding a heat equation governed by the fractional Laplacian $|\Delta|^{3/4}$.}
A direct space-time scaling limit from the microscopic dynamics
(i.e. the   hydrodynamic limit), without  {relying on} the kinetic equation, was first proven in \cite{JKO15}.
The aforementioned results concern  infinite  {systems}
without boundary conditions; a review can be found in \cite{BBJKO}.


A natural question arises   regarding
the  boundary conditions  {that emerge when}
heat baths are in contact with a chain whose dynamics leads
to a superdiffusive {evolution of the energy}.
In the diffusive case -- e.g., when the chain is pinned or the noise does
not conserve momentum, see \cite{KLO23,KLO23-2} -- a heat bath generates
{a fixed (Dirichlet) boundary condition determined} by its
temperature.
The situation is   more complicated in the case of a fractional diffusion. In fact, due
to the non-locality of the fractional Laplacian operator,
various boundary conditions  can
be defined and it is not apriori
{ evident which one   emerges from}
the underlying microscopic dynamics.

{
According to  heuristic physics literature, see \cite{LM08,LM10,CV17,KB18}, the type of the macroscopic boundary condition
for the energy superdiffusion depends on whether the chain is  microscopically pinned or unpinned
at the boundary.} 
In particular in \cite{LM10,KB18} the authors   characterize (heuristically)
the macroscopic boundary conditions   when the chain is
pinned {at its endpoints} (i.e. the first and the last particle cannot move).
{ In the unpinned case (at the endpoints)   boundary conditions involving boundary layers
are expected to appear,
but their exact formulation {is not clear}, see \cite{LM10,CV17}. In the present paper we determine and mathematically justify  the boundary conditions and the boundary layer
emerging from the unpinned dynamics.
Our method can be  also applied to the   pinned boundary case (in fact
the argument is somewhat simpler then),
providing a mathematical proof of the heuristic results of \cite{LM10,KB18}.

Most of the   numerical simulations of the anharmonic dynamics are performed on
open chains connected to heat baths at different temperatures \cite{sll,llp97}.
We expect that non-linear dynamics will macroscopically lead to a
superdiffusion, with boundary conditions  similar
to those   found here. This highlights the importance of obtaining a
precise formulation of 
such conditions.
Furthermore, according to our knowledge, the particular  boundary conditions for a fractional Laplacian, described  in the present paper,
appear to be new in the existing mathematical literature.}

\medskip
\subsection*{Model.}

A  {standard setup consists in a finite chain of} $n+1$ atoms with
two Langevin  heat baths, at temperatures $T_L$ 
and $T_R$,  attached to the left and right endpoints, respectively.

We consider atoms labeled by $x\in \bbZ_n:=\{0,\ldots,n\}$, with positions
$\mathbf q(t)=\big(q_x(t)\big)_{x\in\bbZ_n}$ and momenta
$\mathbf p(t)=\big(p_x(t)\big)_{x\in\bbZ_n}$,
where $q_x(t),p_x(t)\in\bbR$.
The dynamics is given by:
\begin{equation} \label{eq:bas1}
  \begin{split}
    &\dot q_x(t) = p_x(t), \qquad x\in\bbZ_n, \quad \text{and in the bulk for } x=1,\ldots,n-1, \\
   &\dd p_x(t) = \Delta_{\rm N} q_x(t)\dd t +
   \big[\nabla^\star p_{x+1}(t-)\dd N_{x,x+1}(\gamma t)
     - \nabla^\star p_x(t-)\dd N_{x-1,x}(\gamma t)\big].
\end{split}
\end{equation}
At the boundaries $x=0,n$  the energy is exchanged with two
Langevin heat baths at temperatures $T_L>0$ and $T_R>0$,
respectively:
   \begin{equation} \label{eq:bas1a}
  \begin{split} 
     &\dd p_{0}(t) = \Delta_{\rm N}  q_0(t)\dd t   +\nabla^\star
     p_1(t-) \dd N_{0,1}(\gamma t)  - \tilde\gamma  p_{0}\dd t + \sqrt{2T_L \tilde \gamma} \dd
     w_{L}\\
     & \\
     &  \dd  p_n(t) = \Delta_{\rm N}  q_n(t)\dd t  + \nabla^*  p_n(t-)\dd N_{n-1,n}(\gamma t)
     - \tilde\gamma  p_n(t)\dd t + \sqrt{2T_R \tilde \gamma} \dd w_R(t)
  \end{split}
\end{equation}
{Here $w_L(t)$ and $w_R(t)$ are independent standard Brownian motions, and 
$\{N_{x,x+1}(t), x=0,\dots, n-1\}$ are independent Poisson processes of intensity $1$,
independent of the Brownian motions. We denote $\nabla^\star f_x = f_x - f_{x-1}$.
The Neumann discrete Laplacian, $\Delta_{\rm N}$ is defined as
$\Delta_{\rm N}f_x=f_{x+1}+f_{x-1}-2f_x = \nabla \nabla^\star f_x$,
with boundary conditions 
$f_{n+1}:=f_n$ and $f_{-1}= f_0$.
The parameters $\ga,\tilde
\ga>0$ determine 
the respective rates of the momentum exchange and the strength
of the heat bath.

This dynamics is {\it unpinned} and {consequently}
  invariant under
translations of the positions ($q_x \rightarrow q_x + a$, $a\in\bbR$).
{It is therefore} convenient to work with the
configuration space
\begin{equation}
  \label{eq:1}
  (\mathbf r, \mathbf p) =
  (r_1, \dots, r_n, p_0, \dots, p_n) \in \Om_n:=\bbR^{n}\times\bbR^{n+1},
\end{equation}
where $\mathbf r=(r_1, \dots, r_n)$ 
correspond to the inter-particle stretches $r_x:=q_x-q_{x-1}$,
$x=1,\ldots,n$.
{The momentum exchange mechanism guarantees that 
the energy and momenta and stretches are locally   conserved.}

We assume that the initial data $(\mathbf r(0),\mathbf
p(0))$  is randomly
distributed according to some probability measure $\mu_n$ on $\Om_n$,
{with zero means for both stretches and momenta.
  We further assume certain regularity conditions on $\mu_n$, notably
that the {relative} entropy of $\mu_n$},
with respect to the Gibbs equilibrium measure at some temperature, is
bounded by a constant times the size $n$ of the system.

\medskip
\subsection*{Scaling limit}
We   study the averaged energy profile in the superdiffusive scaling
\begin{equation}
  \label{eq:2}
 \frac 12 \mathbb E_n\left(r^2_{[nu]}(n^{3/2}t) + p^2_{[nu]}(n^{3/2}t)\right), \qquad u\in [0,1], t\ge 0.
\end{equation}
{Here $\mathbb E_n$ denotes the expectation with respect to the
  randomness coming from the initial data, Langevin thermostats and
  momenta exchanges.}
{As shown in Theorem \ref{thm012911-23}, the averaged
energy profiles, viewed as measure-valued functions on $[0,1]$,}
converge weakly, as $n\to\infty$, to the solution $T(t,u)$ of the equation:
\begin{equation}
  \label{eq:3}
  \begin{split}
    \partial_t T(t, u)&= \int_0^1 r(u,u') \left[T(t,u') - T(t,u)\right] du'
    + \sum_{v = 0,1} b(u;v) \left[T_v - T(t,u)\right],
  \end{split}
\end{equation}
{with} $T_0=T_L$ and $T_1:=T_R$.

In \eqref{eq:b-gamma} the coefficients $b(u;v)$, $v=0,1$ represent the rates of absorption and creation at $u\in(0,1)$
due to the heat baths. One can show, see Proposition \ref{lm011210-26}, that 
 $b(u;v) \to +\infty$, for $u\to v$, {where
  $v=0,1$}, see Proposition \ref{lm011210-26},  
 {ensuring}
 the boundary conditions $T(t,v) = T_v$ are satisfied.

{The kernel $r(u,u')$ is symmetric (see \eqref{r-G}) and
determined by the jump rates of the L\'evy-type process whose
generator  is a nonlocal operator with a resolvent-dependent Robin boundary interaction
     on $[0,1]$, 
 see Proposition \ref{lem:positive-effective-kernel}. 
}

The above can be expressed equivalently as:
\begin{equation}
  \begin{split}
      &\partial_t T(t, u)=-c_{\rm bulk}|\Delta_{\rm N}|^{3/4}T(t, u)  \\
    & + c_{\rm
    bd}   \sum_{v=0,1} \int_0^{+\infty}  
   V_{\varrho}(u,v) \int_0^1 V_{\varrho}(u',v)[ T_v -  T(t, u') ]\dd
   u' \frac{F(\tilde\ga,\rho^{1/2}) \dd\varrho}{\varrho^{3/4}} \\
   &
  -  c_{\rm
    bd} \sum_{v=0,1}V_{\varrho}(u,1-v) 
\int_0^1 V_{\varrho}(u',v)[ T_v -  T(t, u') ]\dd
   u' 
    \,\frac{G(\tilde\ga,\rho^{1/2}) d\rho}{\rho^{3/4}},
\end{split}
  \label{eq:eqmacro}
\end{equation}
{where $c_{\rm bulk},c_{\rm bd}>0$ and functions
$F(\cdot,\cdot)$, $G(\cdot,\cdot)$     are given in \eqref{022111-24}
and \eqref{F-G}, res[ectively.} The operator
$-|\Delta_{\rm N}|^{3/4}$ is the
           spectral fractional power of the Neumann Laplacian on $[0,1]$
and
$V_{\varrho}(u',u)=\varrho G_{\varrho}(u',u)$, where
$G_{\varrho}=(\rho-\Delta_{\rm N})^{-1}$ is its Green's function.

Concerning the boundary condition, {we require that for $v=0,1$}
\begin{equation}
          \label{012111-24z}
                   \int_0^{+\infty} \Big\{ \int_0^t\dd s \Big(
  \int_0^1 V_{\varrho}(u',v)  \big(T_v -  T(s,u')\big)  \dd u'\Big)^2 \Big\}\frac{\dd \varrho}{\varrho^{3/4}}<+\infty
\end{equation}
for any $t>0$. The precise notion of a solution of \eqref{eq:eqmacro}
and \eqref{012111-24z} is  {given} in
Definition \ref{df1.5}.
The result informally described above is rigorously formulated in  Theorem
\ref{thm012911-23} below.

\subsection*{{Context}}
{As mentioned in the foregoing,} 
similar dynamics but with pinned boundaries (i.e. with the microscopic
Dirichlet Laplacian in
\eqref{eq:bas1a}, {where} $f_{-1} = 0$, $f_{n+1}=0$)
has been studied heuristically in \cite{LM08,KB18}. 
Because of the pinned microscopic boundary, different macroscopic
boundary conditions {are} expected. {Namely}, the second term on the right
hand side of  \eqref{eq:eqmacro} does not appear.
This could be understood as follows:  the boundary pinning changes
locally the dispersion relation of the chain,
slowing down the long waves when they approach the boundary. 
In that respect the limit obtained in the present paper differs from
the regional fractional Laplacian describing the superdiffusion
of the {\it density} of particles {as formulated} in \cite{BCS-24}.
{Notice that in \cite{BCS-24} particles  perform symmetric random walks
  with   long jumps and subject to the exclusion rule.  {This makes  the dynamics
   non-local  already  at the microscopic level.}
  Our dynamics is completely local on the microscopic level, non-locality emerges for the
  {\it energy} evolution at the macroscopic space-time scale.}

  {Some intuition about the result can be construed from the {\it kinetic limit approach}.
  It is a weak noise (or weak anharmonicity in the non-linear case) limit,
  where the averaged Wigner distribution of the waves of a given frequency (mode or wave number $k$)
  converges
  to a density distribution of {\it phonons}   in space-mode domain,
  that evolves according to  an inhomogeneous, linear
  kinetic equation \cite{BOS}. In the kinetic limit, for an unpinned chain,  phonons of mode
  $k$ move with positive velocity, almost independent of $k$,
  and change the mode  ({\it scatter}) with rate proportional to
  $k^2$. Consequently, most of the energy is carried
  by low mode $k$ phonons (long waves),
  that,  in the proper space-time scaling limit, perform a $3/2$-L\'evy superdiffusion
  \cite{kjo,babo}.}
  
The kinetic limit of the infinite dynamics with  Langevin, or {Poisson type} heat baths
attached at a point was studied in \cite{KORS20,KO20,KO20a}.
A phonon
can get absorbed, reflected, or transmitted when its
  trajectory {intersects} the heat bath,
  {and can also be} created at {explicitly computable} rate
depending on the wave number.
{Then, starting from the kinetic equation, the
superdiffusive hydrodynamic limit has been obtained in
\cite{KOR20}. It is
described by a
 $3/2$-L\'evy superdiffusion with an interface that corresponds to the
 location of the thermostat. The behavior of the process at the interface can be described
  as follows: when the particle    tries to jump over the heat bath,
it is either  absorbed,
transmitted, or reflected with explicitly computable probabilities. In addition,
particles are created at the interface at the rates
depending on the   bath  temperature.}

 {The direct derivation of the hydrodynamic limit
from the microscopic dynamics, in the presence of one or several heat baths at
different temperatures, has remained open.}
{To obtain such a limit is the main goal of the present paper.}
}

\medskip
\subsection*{Outline of the paper.}


{In Section \ref{sec:def} we
present the main results. In particular, Theorem \ref{lm032209-24}
asserts the existence and uniqueness of weak solutions  of equation \eqref{eq:eqmacro}. {Theorem
\ref{thm012611-24} concerns} the regularity of the solution, provided
that the initial data in the equation is sufficiently regular. Our
main result concerning the  
limit of average energy functional is formulated in Theorem \ref{thm012911-23}.   
{A key ingredient in the} 
proof is the fact that we can resolve the covariance matrix of the
stretch/momentum ensemble process $\big(\mathbf r(t), \mathbf
p(t)\big)$, where $\mathbf r(t)=\big( r_x(t)\big)_{x\in\bbZ_n}$, $\mathbf
p(t)=\big( p_x(t)\big)_{x\in\bbZ_n}$, obtained in Section \ref{sec4}.

{Section \ref{Sec3} is devoted to the discussion about the limiting
equation formulated in  \eqref{eq:eqmacro}. In Section \ref{prob-int}
we highlight the probabilistic interpretation of  \eqref{eq:eqmacro}.}

{Preliminary to the proof of the main Theorem \ref{thm012911-23},
  we need to establish an
energy bound, formulated in Theorem \ref{energy-bound}.
A simple argument using an identity satisfied by the covariances of
the  field $\big(\mathbf r(t), \mathbf
p(t)\big)$, see \eqref{energy-bal}, under an additional assumption that
$T_L=T_R$,  is presented in Section \ref{sec4.3}.
We postpone the more technical argument, covering the  case of
  heat baths at different temperatures, till Sections \ref{sec-ent1}
  and \ref{sec13}.}

{The energy bound of} Theorem \ref{energy-bound}. allows us to
conclude compactness of the energy distribution  
in the $\star$-weak topology over $C[0,1]$. The limit identification is
conducted in Sections \ref{lim-id}--\ref{sec:proof-os-time}.
{In particular, the most delicate part of the proof is the
  identification of the boundary layer, see
  Theorem \ref{thm011001-25}, proven in Section \ref{sec7}.}
Section \ref{sec6}
is devoted to proving some technical results used throughout the
argument. In Section \ref{sec2.7} of the Appendix we formulate some basic linear analysis
facts concerning the spectral resolution of the discrete Neumann
Laplacian and the gradient and divergence operators. Section
\ref{secSs} is devoted to finding the solution of equation
\eqref{f-diff1} with the prescribed boundary condition using the orthonormal
base of the Neumann Laplacian {on $[0,1]$} (cosine functions). In particular, 
Section \ref{sec34} {identifies} the unique stationary solution of
the equation, {while Section \ref{sec43} contains the proof of
Theorem \ref{thm012611-24}.}
Section \ref{appA} is devoted to the proofs of the results appearing
in Section \ref{Sec3}.}

\section{Preliminaries}

 \label{sec:def}

\subsection{Dynamics in the stretch/momentum configuration space}

Because of the translation invariance property of the dynamics, we only need
to consider the relative distance between the particles
$r_x:= q_x - q_{x-1} = \nabla^\star q_x$, $x=0,\ldots,n$.
The configuration of particle stretches and momenta are described by
$(\mathbf r, \mathbf p)$ as in \eqref{eq:1}.
The total energy of the chain is defined by the Hamiltonian:
\begin{equation}\label{eq:hn}
  \mathcal{H}_n (\mathbf r, \mathbf p):=
  \sum_{x=0}^n {\cal E}_x (\mathbf r, \mathbf p),
\end{equation}
where the microscopic energy per particle is given by
\begin{equation}
\label{Ex}
{\cal E}_x (\mathbf r, \mathbf p):=  \frac{1}2 (p_x^2+
r_{x}^2),\
\quad x = 0, \dots, n,
\end{equation}
with the convention that $r_0:=0$.

The  microscopic dynamics of
the process $\{(\mathbf r(t), \mathbf p(t))\}_{t\ge0}$
describing the chain is  given   by: 
\begin{equation} 
\label{eq:flip1}
\begin{split}
  &
  \dot   r_x(t) = \nabla^\star p_x(t) , \qquad x\in \{1, \dots, n\},\\
  &\dd   p_x(t) =  \nabla r_x \dd t + \left[\nabla^\star  p_{x+1}(t-)\dd N_{x,x+1}(\gamma t)
      -\nabla^\star   p_x(t-)\dd N_{x-1,x}(\gamma t)\right],\\
    &\qquad  \qquad  \qquad  \qquad  \qquad  \qquad  \qquad  \qquad
    \mbox{for }x=1,\ldots,n-1,
\end{split} \end{equation}
and at the boundaries
   \begin{equation} 
\label{eq:flip1b}
\begin{split} 
     \dd   p_0(t) =    r_1  \dd   t
     + &\nabla^\star  p_{1}(t-) \dd N_{0,1}(\gamma t)   
     -   \tilde \gamma p_0(t) \dd t
                       +\sqrt{2 \tilde  \gamma T_L} \dd   w_L(t),
                       \\
                       \\
                       \dd   p_n(t) = - r_{n}  \dd   t                       
                       &- \nabla^\star  p_{n}(t-) \dd N_{n-1,n}(\gamma t) -
                       \tilde\gamma p_n(t) \dd t
                     +\sqrt{2 \tilde  \gamma T_R} \dd   w_R(t). 
  \end{split} \end{equation}

The generator of the dynamics is given by 
\begin{equation}
  \label{eq:7}
  \mathcal G =  \mathcal A +  \gamma S_{\text{ex}}
  +    \tilde\gamma (S_L+  S_R),
\end{equation}
where, with the convention $r_0=r_{n+1}=0$, its Hamiltonian part equals
\begin{equation}
  \label{eq:8}
  \mathcal A= \sum_{x=1}^n \nabla^\star p_x \partial_{r_x}
  + \sum_{x=0}^n  \nabla r_{x} \partial_{p_x},
\end{equation}
the momentum exchange part is
\begin{equation}
  \label{eq:21}
  \begin{split}
 &   S_{\text{ex}} f (\rv,\pv) =    \sum_{x=0}^{n-1}   \Big( f
 (\rv,\pv^{x,x+1}) - f (\rv,\pv)\Big).
\end{split}
  \end{equation}
 Here $f:\bbR^{2n+1}\to\bbR$ is a bounded and measurable function,
 $\pv^{x,x'}$ is the momentum configuration where the velocities at
 sites {$x\not=x'$} have been exchanged, 
   i.e. $\pv^{x,x'}=(p^{x,x'}_0,\ldots,p_n^{x,x'})$, with $p_y^{x,x'}=p_y$,
 $y\not \in\{x,x'\}$ and  $p_{x'}^{x,x'}=p_x$, $p_{x}^{x,x'}=p_{x'}$. Finally, the
 effects of 
 thermostats correspond to 
 \begin{equation}
   \label{eq:10}
   S_{L} = T_L \partial_{p_0}^2 - p_0 \partial_{p_0}\quad  S_{R} = T_R \partial_{p_n}^2 - p_n \partial_{p_n}.
 \end{equation}

We assume that the initial distribution of stretches and momenta
$(\mathbf r(0),\mathbf p(0))\in \Om_n$ is
random and distributed according to a probability measure
$\mu_n$ defined on the $\sigma$-algebra ${\cal B}_n$ of Borel subsets of
the configuration space. Denote by $\bbP_{n}=\mu_n\otimes \bbP$ and $\bbE_{n}$
the  probability measure on the product space $\big(\Om_n\times
\Sigma,{\cal B}_n\otimes{\cal F}\big)$ and its
corresponding expectation. We decompose the configurations   
\begin{equation}
  \label{011704-24}
   r_x(t):=r_x'(t) + \bar r_x(t) , \qquad p_x(t):=p_x'(t) + \bar p_x(t),
\end{equation}
where the configuration of the means
\begin{equation}
  \label{012911-23} \begin{split}
\bar{\mathbf r}(t)&=(\bar r_1(t),\ldots, \bar r_n(t)):=\bbE_{n}[\mathbf
r(t)] ,\\ \bar{\mathbf p}(t)&=(\bar p_0(t),\ldots, \bar p_n(t)):=\bbE_{n}[\mathbf
p(t)], \end{split}
\end{equation}
while $ {\mathbf r}'(t), {\mathbf p}'(t)$ corresponds to the {\it fluctuating parts} of the dynamics.
It turns out that in the scaling we are concerned with the limiting
behavior of the system is not affected by the dynamics of the
means.  For this reason and also to simplify the presentation we adopt
the following.
 \begin{assumption}\label{ass5a} 
  We assume that  
  \begin{equation}
    \label{eq:sec-mom0}
    \bar{\mathbf r}(0)\equiv 0\quad\mbox{and}\quad \bar{\mathbf p}(0)\equiv 0.
    \end{equation}
\end{assumption}
This assumption obviously implies that  $\bar{\mathbf r}(t)\equiv 0$
and $ \bar{\mathbf p}(t)\equiv 0$ for all $t\ge0$.

\subsection{Fractional diffusion equation with  boundary conditions}
\label{fde}

Let
\begin{align*}
  &
    C^\infty_{\rm N}[0,1]:=\big\{\varphi\in C^\infty[0,1]:\,
    \varphi'(0)=\varphi'(1)=0\big\}\quad\mbox{and}\\
  &
    C^\infty_c(0,1):=\big\{\varphi\in C^\infty[0,1]:\,
     {\rm supp}\,\varphi\in(0,1)\big\}.
\end{align*}
Define the Neumann Laplacian $\Delta_N :C^\infty_{\rm N}[0,1]\to L^2[0,1]$ as the closure of the operator
\begin{equation}
  \label{n-lapl}
  \Delta_N\varphi(u)=\varphi''(u),\quad \varphi\in C^\infty_{\rm N}[0,1],\,u\in[0,1].
\end{equation}
Using the spectral decomposition of the Laplacian in the orthonormal
base given by the cosine functions, see Section \ref{secB.4}, we can
define a self-adjoint operator $|\Delta_N|^{3/4}:{\cal D}(|\Delta_N|^{3/4})\to L^2[0,1]$, see \eqref{frac-lap}, and   the
respective 
Sobolev spaces $H^{3/4}[0,1]$, $H^{3/4}_0[0,1]$,  {see Section \ref{secB.5}}.

For $\varrho>0$   define the resolvent operator
\begin{equation}
  \label{greens}
  G_\varrho[\varphi](u):=   (\varrho-\Delta_N)^{-1}\varphi(u)
  =\int_{0}^1G_{\varrho}(u,v) \varphi(v)\dd v,
\end{equation}
with the Green's function $G_{\varrho}(u,v)$ given by formula
\eqref{G-la}.

Denote
\begin{equation}
  \label{Vla}
  V_{\varrho}(u,v):=\varrho  G_{\varrho}(u,v),\quad\varrho>0,\,u,v\in[0,1].
  \end{equation}
   {By applying \eqref{greens} to $\varphi = 1$ it follows that} 
     \begin{align}
    \label{012412-24a}
    & \int_0^1V_{\varrho}(u,v)\dd u=1,\quad v\in[0,1],
     \end{align}
     and by  \eqref{G-la} that $V_{\varrho}(u,v)\ge0$.
Furthermore, see Lemma \ref{lm021509-24},
for $\varphi\in C^1_c(0,1)$ we have 
\begin{equation}
  \label{W-la1}
\int_0^{+\infty}\Big(
\int_0^1\varphi(u) V_{\varrho}(u,v)\dd u\Big)^2\frac{\dd \varrho}{\varrho^{3/4}}<+\infty,
\quad v=0,1.
\end{equation}
{This bound is equivalent with
  \begin{equation*}
    \int_0^{+\infty}\Big(
\int_0^1\varphi(u) G_{\varrho^{4/9}}(u,v)\dd u\Big)^2 \dd \varrho<+\infty,
\end{equation*}
}

{\begin{definition}
  \label{df1.5}
  Suppose that $c_{\rm bulk},c_{\rm bd}>0$, $T_0,T_1>0$ and $T_{\rm
    ini}\in L^2[0,1]$. Assume furthermore that $\tilde \ga>0$ and 
  \begin{align}
    \label{F-G}
    &
      F(\tilde\ga,\varrho):= \frac{
\sinh \varrho
+2\widetilde\gamma\cosh \varrho
 }{
(1+4\widetilde\gamma^2)\sinh \varrho
+4\widetilde\gamma\cosh \varrho
      }\\
    &
     G(\tilde\ga, \varrho):= \frac{
 2\widetilde\gamma
}{
(1+4\widetilde\gamma^2)\sinh \varrho
+4\widetilde\gamma\cosh \varrho
},\quad \varrho\ge 0.\notag
              \end{align}
  We say that a 
  function $T:[0,+\infty)\to L^2[0,1]$
  is a weak solution of 
\begin{equation}
    \label{f-diff1}
    \begin{split}
      \partial_t T(t, u)&=-c_{\rm bulk}|\Delta|^{3/4}T(t, u)   \\
      &
  +c_{\rm  bd}    \sum_{v=0,1} \int_0^{+\infty}  \Big\{ {V_{\varrho}(u,v)}
 \int_0^1 V_{\varrho}(u',v)[ T_v -  T(t, u') ]\dd
  u'\Big\}\frac{F(\tilde\ga,\varrho^{1/2}) \dd\varrho}{\varrho^{3/4}}
  \\
    &
  -c_{\rm  bd}    \sum_{v=0,1} \int_0^{+\infty}  \Big\{ {V_{\varrho}(u,1-v)}
 \int_0^1 V_{\varrho}(u',v)[ T_v -  T(t, u') ]\dd
  u'\Big\}\frac{G(\tilde\ga,\varrho^{1/2}) \dd\varrho}{\varrho^{3/4}} ,
      \end{split}
    \end{equation}
with  the boundary
      values $T(t,v)=T_v$, $v=0,1$,  
    if the following   hold:
    \begin{itemize}
      \item[i)] $T\in C\Big([0,+\infty); L^2_w[0,1]\Big)$, where
        $L^2_w[0,1]$ is equipped with the weak topology,
      \item[ii)] for any $t>0$ and $v=0,1$ we have
        \begin{equation}
          \label{012111-24}
                      \int_0^t\dd s \int_0^{+\infty} 
 \Big( \int_0^1 V_{\varrho}(u',v)  \big(T_v -  T(s,u')\big)\dd u'
 \Big) ^2 \frac{\dd \varrho}{\varrho^{3/4}}<+\infty,
          \end{equation}
        \item[iii)] for any $\varphi\in C^\infty_c(0,1)$ we have:
     \begin{equation}
    \label{042209-24aa}
    \begin{split}
    &  \langle\varphi,
  T(t )\rangle_{L^2[0,1]} -\langle\varphi,
  T_{\rm ini}\rangle _{L^2[0,1]}
    = - c_{\rm bulk}
 \int_0^t \langle|\Delta|^{3/4}\varphi, T(s )\rangle_{L^2[0,1]}  \dd s\\
     &
      +c_{\rm
    bd} \sum_{v=0,1} \int_0^t\dd s \int_0^{+\infty} \langle V_{\varrho}(\cdot
  ,v), \varphi\rangle_{L^2[0,1]}     
   \langle V_{\varrho}(\cdot
  ,v), T_v - T(s)\rangle_{L^2[0,1]}
  \frac{F(\tilde\ga,\rho^{1/2})\dd\varrho}{\varrho^{3/4}}\\
 & -c_{\rm
    bd}      \sum_{v=0,1} \int_0^t\dd s \int_0^{+\infty}  \Big\{
\langle V_{\varrho}(\cdot
  ,1-v), \varphi\rangle_{L^2[0,1]}     
 \int_0^1 V_{\varrho}(u',v)[ T_v -  T(s, u') ]\dd
 u'\Big\}\frac{G(\tilde\ga,\rho^{1/2})\dd\varrho}{\varrho^{3/4}}.
    \end{split}
  \end{equation}
  \end{itemize}
\end{definition}}

 \begin{theorem}
    \label{lm032209-24}
   Suppose that $T_{\rm ini}\in  L^2[0,1]$. Then,  equation
   \eqref{f-diff1} has a unique solution $T(\cdot,\cdot)$. In addition, the
   solution satisfies  
   \begin{equation}
     \label{boundary}
     \begin{split}
       &\int_0^tT(s,\cdot)\dd s\in C[0,1]\quad\mbox{and}\\
     &\int_0^tT(s,0)\dd s= {T_0} t,\quad  \int_0^tT(s,1)\dd s={T_1} t,\quad
     t\ge0.
     \end{split}
     \end{equation}
        \end{theorem}
The proof of Theorem \ref{lm032209-24} is presented in Section \ref{secB.3} of the Appendix.
In fact the results contained there allow us
to claim some additional regularity of solutions of
\eqref{f-diff1}. For this purpose we consider the fractional Sobolev
space $H^{3/4}[0,1]$ introduced in Section  \ref{secB.4}. In particular,   $H^{3/4}[0,1]\subset
C[0,1]$, see Lemma \ref{lm022009-24}.
\begin{theorem}
  \label{thm012611-24}
  Suppose that $T_{\rm ini}\in H^{3/4}[0,1]$ is such that $T_{\rm
    ini}(v)=T_v$, $v=0,1$. Then:
  \begin{itemize}
    \item[i)] the solution $T(t)$ of
      \eqref{f-diff1} belongs to the space
      $$
      C\Big([0,+\infty); H^{3/4}[0,1]\Big)
      $$, where the target space
 is  considered with the strong topology,
  \item[ii)]  we have
        \begin{equation}
          \label{012111-24a}
          T_0= T(t,0)\quad\mbox{and} \quad T_1= T(t,1),\quad\mbox{for
           all }t\ge0.
        \end{equation}
              \end{itemize}
\end{theorem}
The proof of the result is presented in Section \ref{sec43} of the Appendix.


\subsection{Scaled dynamics of the chain}

{From now on we consider the process in the macroscopic time}, i.e.
$
\big(\mathbf r_n(t), \mathbf p_ n(t))\big) =\big(\mathbf r(n^{3/2}t),
\mathbf p(n^{3/2}t))\big) ,
$ $t\ge0$.
The generator of the dynamics is   given by $n^{3/2}\mathcal G$, where
$\mathcal G $ is defined in \eqref{eq:7}.
Since the time scale is fixed we will drop the index $n$ from the notations
for the configurations at macroscopic time $t$.

Denote by $\mu_n(t)$ the probability measure that is the distribution
of the configuration  $(\rv(t),\pv(t))$  on $\Omega_n$. 
{Recall that thanks to Assumption \ref{ass5a},
    we have only the fluctuation part of the dynamics.}

\subsection{The macroscopic limit of the  energy functional}
\label{sec:thermal-energy}

We assume the existence of the initial  macroscopic energy profile.
 
\begin{assumption}\label{ass4} 
  Assume   that there exists a 
 function (the initial temperature   profile)
  $T_{\rm ini}:[0,1]\to (0,+\infty)$ such that, for any   $\varphi\in
  C[0,1] $,
  \begin{align}
    \label{E00}
  \lim_{n\to+\infty}\frac{1}{n}\sum_{x=0}^{n}
   \varphi\left(\frac x{n}\right) \bbE_{n}\big[{\cal E}_{n,x}(0)\big]
  =\int_0^1 T_{\rm ini} (u)\varphi(u)\dd u.
  \end{align}
  We suppose furthermore that $T_{\rm ini}\in H^{3/4}[0,1]$.
\end{assumption}

Let
\begin{align}
  \label{Hn2}
  {\cal H}_n^{(2)}(t) 
&=\frac 1{2(n+1)} 
\sum_{x,x'=0}^n\bigg\{\mathbb E_{n}\left[ p_{x}(t)
  p_{x'}(t)\right]^2+ \mathbb E_{n}\left[ r_x(t) r_{x'}(t)\right]^2 
+2 \mathbb E_{n}\left[ p_x(t) r_{x'}(t)\right]^2\bigg\}. 
\end{align}

\begin{assumption}\label{ass5} 
  We assume that there exists $C_{2,{\cal H}}>0$ such that
  \begin{equation}
    \label{eq:sec-mom}
    {\cal H}_n^{(2)}(0)  \le C_{2,{\cal H}}.\end{equation}
\end{assumption}
Let
\begin{align}
  \label{H1}
  {\cal H}_n (t) = 
\sum_{x=0}^n\bigg\{\mathbb E_{n}\left[ p_{x}^2(t)
   \right]+ \mathbb E_{n}\left[ r_x^2(t) \right] \bigg\}. 
\end{align}
As a consequence of the above assumptions we conclude 
 the energy bound.
\begin{theorem}
  \label{energy-bound}
 Under Assumptions \ref{ass4} and \ref{ass5},  for any $t_*>0$  there exists $C_{\mathcal{H},t_*}>0$ such that
\begin{equation}\label{eq:energyb}
 \mathbb E_{n}  \big[\mathcal{H}_n (t)\big]
  \le C_{\mathcal{H},t_*} n,\qquad t\in[0,t_*],\,n=1,2,\ldots.
\end{equation}
\end{theorem}
 The proof of this result in the case $T_L=T_R$ is presented in
 Section \ref{sec-ent}. The general case is shown in Section \ref{sec-ent1}.

\begin{theorem}[The limit of thermal energy and equipartition]
  \label{thm012911-23}
  Under the assumptions made in the present section for any continuous test function $\varphi:[0,1]\to\bbR$ and any $t\ge 0$, we have
\begin{align}
   \lim_{n\to+\infty}\frac{1}{n}\sum_{x=0}^{n}  \varphi\left(\frac x{n} \right)
 \bbE_{n}\big[ {\cal E}_{n,x}( t)\big] =\int_0^1 T(t,u)\varphi(u)\dd u,
 \label{eq:conv-temp}
\end{align}
where $T(t,u)$ is the solution of \eqref{f-diff1} with the initial data 
$T(0,u)=T_{\rm ini}(u)$ and the boundary conditions $T(t,0)=T_L$,
$T(t,1)=T_R$. Here  
\begin{align}
    \label{022111-24}
       & c_{\rm bulk} =\frac{1}{(2^3\ga)^{1/2}}
       ,\\
         &{c_{\rm bd} =
           \frac{\widetilde\gamma}{2\pi\gamma^{1/2}}=\frac{\sqrt{2}\widetilde\gamma
          }{
           \pi} c_{\rm bulk}}.\notag
          \end{align}
In addition, for any compactly supported,
continuous function $\Phi:\bbR_+\times [0,1]\to\bbR$
\begin{equation}
  \label{eq:conv-temp2}
  \begin{split}
  \lim_{n\to+\infty}\frac{1}{n}\sum_{x=0}^{n} \int_{0}^{+\infty} \Phi\left(t,\frac x{n} \right)
 \bbE_{n}\big[ p_{x}^2( t)\big] \dd t
  &=
   \lim_{n\to+\infty}\frac{1}{n}\sum_{x=0}^{n} \int_{0}^{+\infty} \Phi\left(t,\frac x{n} \right)
   \bbE_{n}\big[ {\cal E}_{x}( t)\big] \dd t
   \\
   &=  \int_{0}^{+\infty} \dd t\int_0^1 T(t,u) \Phi\left(t, u\right)\dd u.
 \end{split}
\end{equation}
\end{theorem}
 {The proof of Theorem \ref{thm012911-23} is given in Section
   \ref{sec:comp-concl-proof} (the
   convergence part) and in Section \ref{sec:equi} (the equipartition property).}

\subsection{Energy currents}

{Recall that now the generator of the process is given by $n^{3/2} \mathcal G$.}
    Energy currents satisfy
\begin{equation}
\label{eq:current}
  \frac {d}{dt}\bbE_n\left[ {\cal E}_{x}(t)\right]
       =n^{3/2}\mathcal G\mathcal E_{x}(t)   =-n^{3/2}\nabla^\star j_{x,x+1}^{(n)} (t),\quad x=0,\ldots,n,
\end{equation}
   with  
   \begin{align}
     \label{jax}
     &
       j_{x,x+1}(t)= j_{x,x+1}^{(a)}(t)+j_{x,x+1}^{(s)},\quad \mbox{where}\\
  & j_{x,x+1}^{(a)}(t):=- p_{x}(t) r_{x+1}(t) , \quad
    j_{x,x+1}^{(s)}=-\frac{\ga}{2}\big(p_{x+1}^2-p_{x}^2\big),\qquad
    \mbox{for }\quad x =0,...,n,  \notag
\end{align}
and at the boundaries 
\begin{equation} \label{eq:current-bound}
     { j_{-1,0}:= \tilde{ \gamma} \left(T_L - p_{0}^2 \right)},
      \qquad
  j_{n,n+1} :=      \tilde{ \gamma} \left( p_{n}^2-T_R \right).
\end{equation}

By a direct  calculation we obtain
\begin{equation}
\label{energyn}
\frac{\dd}{\dd t}\bbE_n {\cal H}_n(t)= n^{3/2}\tilde\ga
\Big[T_L+T_R- \bbE_n\Big(p_0^2(  t)+  p_n^2(t)\Big)\Big].
\end{equation}
Hence
\begin{equation}
\label{energyn1}
\int_0^t\Big(\bbE_n p_{0}^2(  s)+\bbE_n p_{n}^2(  s)\Big)\dd s\le (T_L+T_R)t
+\frac{1}{\tilde\gamma n^{3/2}}\bbE_n {\cal H}_n(0),
\end{equation}
and, cf \eqref{eq:current-bound},
\begin{equation}
  \label{eq:6}
  \int_0^t \bbE_n\left[ j_{-1,0}(s) - j_{n,n+1}(s)\right] \dd s
  \le \frac{C}{\tilde\gamma\sqrt n}.
\end{equation}
{Concerning the current size estimate we have the following.
  \begin{theorem}
    \label{thm-current}
  Under Assumptions \ref{ass4} and \ref{ass5},  for any $t_*>0$  there exists $C_{\mathcal{J},t_*}>0$
  such that
  \begin{equation}
   \label{052102-24zz}
   \sup_{x=0,\ldots,n+2}|\int_0^t\bbE_n\left[ j_{x-1,x}( s)\right] \dd s|\le
   \frac{C_{\mathcal{J},t_*}}{\sqrt{n}},\quad t\in[0,t_*],\,n=1,2,\ldots.
        \end{equation}
  \end{theorem}
The proof of the theorem is presented in Section \ref{sec-ent11}.}

\section{Some remarks about the limiting equation}
\label{Sec3}

\subsection{Discussion about
  the coefficients of \eqref{f-diff1}} 
  A direct calculation, using formula \eqref{p-w}, shows that
for $\varphi\in C^\infty[0,1]$ we have
\begin{equation}
  \label{011504-25}
  \begin{split}
   -|\Delta|^{3/4}\varphi(u)=
 \int_0^1 q(u',u) [\varphi(u')-\varphi(u)]
 \dd u',
 \end{split}
\end{equation}
 with  
  \begin{align}
    \label{012412-24}
   &q(u,u'):=\frac{ 3}{2^{5/2} {\pi}^{1/2}} \sum_{n\in\bbZ} 
        \Big(\frac{1 }{|u+u'+2n|^{5/2}} +\frac{1}{|u-u'+2n|^{5/2}}
      \Big).
  \end{align}

Let
\[
r:=\frac{1-2 \widetilde\gamma}{1+2 \widetilde\gamma}.
\]
Since $\widetilde\gamma>0$, we have $|r|<1$.
From \eqref{F-G} we can write 
\begin{align}
  \label{eq:F-geometric}
F(\widetilde\gamma, \varrho)=
\frac{1}{1+2 \widetilde\gamma}\cdot
  \frac{1-r e^{-2 \varrho}}{1-r^2e^{-2 \varrho}} 
    =
\frac{1}{1+2 \widetilde\gamma}
\left[
1+
\sum_{m=1}^{+\infty}
\big(r^{2m}-r^{2m-1}\big)e^{-2m \varrho}
\right] 
\end{align}
and 
\begin{align}
  \label{eq:G-geometric}
G(\widetilde\gamma, \varrho)
 =
\frac{4 \widetilde\gamma}{(1+2 \widetilde\gamma)^2}\cdot
                               \frac{e^{-\varrho}}{1-r^2e^{-2
                               \varrho}} 
    =
\frac{4 \widetilde\gamma}{(1+2 \widetilde\gamma)^2}
\sum_{m=0}^{+\infty}
r^{2m}e^{-(2m+1) \varrho}.  
\end{align}
Let
\begin{equation}
  \label{gFgG}
  \begin{split}
 &g_F(u,u';v):=\int_0^{+\infty}   V_{\varrho}(u
  ,v)  V_{\varrho}(u'
    ,v) \frac{F(\tilde
      \gamma,\sqrt{\varrho})\dd\varrho}{\varrho^{3/4}}\quad\mbox{and}\\
    &
    \\
    &
    g_G^{\rm cr}(u,u';v):=\int_0^{+\infty}   V_{\varrho}(u
  ,1-v)  V_{\varrho}(u'
    ,v) \frac{G(\tilde
      \gamma,\sqrt{\varrho})\dd\varrho}{\varrho^{3/4}}
    \end{split}
  \end{equation}
Recall that, for the Neumann resolvent,
\[
V_{\varrho}(u,0)
=
\sqrt{\varrho}\frac{\cosh(\sqrt{\varrho} (1-u))}{\sinh \sqrt{\varrho}},
\qquad
V_{\varrho}(u,1)
=
\sqrt{\varrho}\frac{\cosh(\sqrt{\varrho} u)}{\sinh \sqrt{\varrho}}.
\]
Substituting into the definition of $g_F(u,u';v)$ and subsequently using the
expansion \eqref{eq:F-geometric} we get
\begin{equation}
  \begin{split}
   & g_F(u,u';v)
=
\frac{1}{1+2 \tilde
\gamma}\,g_0(u,u';v)
+
\frac{1}{1+2 \tilde
\gamma}
\sum_{m=1}^{+\infty}
\big(r^{2m}-r^{2m-1}\big)
g_{m}(u,u';v),\\
&
g_G^{\rm cr}(u,u';v)
=
\frac{4 \tilde
      \gamma}{(1+2 \tilde
      \gamma)^2}\Big[g_{0}^{\rm cr}(u,u';v)+
\sum_{m=1}^{+\infty}r^{2m}
g_{m}^{\rm cr}(u,u';v)\Big].
\label{eq:gFG-zero}
\end{split}
\end{equation}
Here  
\begin{equation}
  \label{gm}
  \begin{split}
 &   g_m(u,u';v)
:=
\int_0^{+\infty}
V_{\varrho}(u,v)V_{\varrho}(u',v)
e^{-2m\sqrt{\varrho}}
\frac{\dd\varrho}{\varrho^{3/4}},\quad m=0,1,2,\ldots,\\
&  g_m^{\rm cr}(u,u';v)
:=
\int_0^{+\infty}
V_{\varrho}(u,1-v)V_{\varrho}(u',v)
e^{-(2m+1)\sqrt{\varrho}}
\frac{\dd\varrho}{\varrho^{3/4}},\quad m=0,1,2,\ldots.
\end{split}
\end{equation}
Obviously $g_{m+1}(u,u';v)\le g_m(u,u';v)$ and  $g_{m+1}^{\rm cr}(u,u';v)\le g_m ^{\rm cr} (u,u';v)$, $m=0,1,2,\ldots$
In Appendix  Section  \ref{appA}, we show that
\begin{equation}
\label{eq:gm-double-sum}
g_m(u,u';v)
=
\frac{3\sqrt{\pi}}2
\sum_{n,n'\in\bbZ}
\frac{1}{
\bigl(
2m+|u+v+2n|+|u'+v+2n'|
\bigr)^{5/2}}
\end{equation}
and
\begin{equation}
\label{eq:gG-cross-double-series}
\begin{aligned}
g_m^{\rm cr}(u,u';v)
={}&
\frac{3\sqrt{\pi}}2
\sum_{n,n'\in\bbZ}
\frac{1}{
\bigl(
2m+1
+|u+1-v+2n|
+|u'+v+2n'|
\bigr)^{5/2}}.
\end{aligned}
\end{equation}

In the particular case, $\tilde\ga=1/2$
 from
 \eqref{eq:gFG-zero}   we get
 \[
  \begin{split}
 &g_F(u,u';v):= \frac12g_0(u,u';v)\quad\mbox{and} 
    \quad 
    g_G^{\rm cr}(u,u';v):= \frac12g_0^{\rm cr}(u,u';v).
    \end{split}
  \]

  \subsection{Probabilistic interpretation}
  \label{prob-int}
  Suppose that $c_{\rm bulk}$
 and $c_{\rm bd}$ are as in  \eqref{022111-24}. Define
\begin{equation}
 r_{\widetilde\gamma}(u,u')
 :=
 c_{\rm bulk}q(u,u')
 -
 c_{\rm bd}\sum_{v=0,1}
 h_{\widetilde\gamma}(u,u';v),
 \label{eq:r-gamma}
\end{equation}
where
\begin{align}
 h_{\widetilde\gamma}(u,u';v)
 :=
 g_F(u,u';v)-g_G^{\rm cr}(u,u';v)
   .
 \label{h-ga}
\end{align}
  Equation \eqref{f-diff1} can be rewritten in the form
\begin{equation}
 \partial_tT(t,u)
 =
 \int_0^1
 r_{\widetilde\gamma}(u,u')
 \big[T(t,u')-T(t,u)\big]\,\dd u'
 +
  \sum_{v=0,1}
 b_{\widetilde\gamma}(u;v)
 \big[T_v-T(t,u)\big],
 \label{eq:probabilistic-form}
\end{equation}
where $r_{\widetilde\gamma}(u,u')$ is given by \eqref{eq:r-gamma} and 
\begin{equation}
 b_{\widetilde\gamma}(u;v)
 :=
 c_{\rm bd}\int_0^1
 h_{\widetilde\gamma}(u,u';v)\,\dd u'.
 \label{eq:b-gamma}
\end{equation}
 If $r_{\widetilde\gamma}(u,u')\ge 0$ we can interpret  \eqref{eq:3} as the
       equation describing  the evolution of the density $T(t,u)$ of
       a Markov process
     with creation and annihilation. The dynamics of the
     process can be described as follows:
     {a particle}
     jumps from $u$ to $u'$ with rate $r(u,u')$ (this takes into account the
     jumps with reflection of the fractional Laplacian minus
     {the jumps censored by the boundaries}).
     At time $t$ and position $u$ the particle gets annihilated with rate
     {$\big( b_{\widetilde\gamma}(u,0)+ b_{\widetilde\gamma}(u,1)\big)$}
     and it is created at this site at  
     rate $ b_{\widetilde\gamma}(u,0) T_L +  b_{\widetilde\gamma}(u,1)
     T_R$. The fact that  the effective  jump kernel is positive is
     established in the following. 
\begin{proposition} 
\label{lem:positive-effective-kernel}
If $c_{\rm bulk}$
 and $c_{\rm bd}$ are given by  \eqref{022111-24}, then
\begin{equation}
  \label{r-pos}
 r_{\widetilde\gamma}(u,u')\geq 0,
 \qquad
 u,u'\in(0,1),\quad u\neq u'.
\end{equation}
More precisely,
\begin{equation}
  \label{r-G}
  r_{\widetilde\gamma}(u,u')=
 \frac{1}{\sqrt{2}\,\pi}
 \int_0^\infty
 \rho^{3/4}
 G_\rho^{(\widetilde\gamma)}(u,u')\,d\rho,
\end{equation}
where $G_\rho^{(\widetilde\gamma)}(\cdot,\cdot)$ is the  Green's function of
$    \rho-\frac{d^2}{du^2}$
on $[0,1]$, subject to the Robin boundary conditions
\begin{equation}
\label{eq:Robin-bc-rho}
 f'(0)=2\widetilde\gamma\sqrt{\rho}\,f(0),
 \qquad
 f'(1)=-2\widetilde\gamma\sqrt{\rho}\,f(1).
\end{equation}
\end{proposition}
\begin{remark} More explicit formula for the kernel
  $r_{\widetilde\gamma}(u,u')$ is given in \eqref{eq:jump-kernel-explicit}.
\end{remark}

The boundary behavior of the creation/absorption term is described in
the following.
\begin{proposition}
  \label{lm011210-26}
  We have
  \begin{equation}
\label{011410-26}
 b_{\widetilde\gamma}(u;v)\ge 0,
 \qquad u\in(0,1),\quad v\in\{0,1\}.
\end{equation}
In addition,
\begin{equation}
\label{eq:b-boundary-asymptotics}
 b_{\widetilde\gamma}(u;v)
=
 \frac{\sqrt\pi c_{\rm bd}}{1+2\widetilde\gamma}
 |u-v|^{-3/2}+O(1),
 \qquad u\to v,\quad v\in\{0,1\}.
\end{equation}
The term $O(1)$ is uniformly bounded.
In particular,
\[
 \lim_{u\to v}b_{\widetilde\gamma}(u;v)
 =+\infty.
\]
\end{proposition}
 
The proofs of the above results are presented in Sections \ref{secC3}
and \ref{secC4} of
the Appendix.

 
\section{An identity for covariances and proof of the energy bound in the
  case $T_L=T_R$}

\label{sec-ent}

  \medskip  
  
\subsection{{An identity for some covariances}}

 \label{sec-ent2}

 {After a tedious but direct calculation  
  we obtain the following identity, cf \eqref{Hn2}:}
  \begin{proposition}
    \label{prop011612-24}
  For any $t\ge 0$ and $n=2,3,\ldots$ we have
\begin{align}
  \label{energy-bal}
& {\cal H}_n^{(2)}(t) +
     \frac{2\ga
                       n^{3/2}}{n+1}\sum_{x=1}^{n}\mathop{\sum_{x'=0}^{n}}_{x'\not\in\{x-1,x\}}\int_0^t\left\{\mathbb
                       E_n\left[ \nabla^\star p_x( s)
                 p_{x'}( s)\right] \right\}^2 \dd s  \\
  &
    +\frac{\ga
    n^{3/2}}{n+1}\sum_{x=0}^{n-1}\int_0^t\Big[\nabla \mathbb
    E_n p_{x}^2( s)
                   \Big]^2\dd s 
    + \frac {2 \ga n^{3/2}}{n+1}\sum_{x=1}^{n} \sum_{x'=1}^n
    \int_0^t \left\{\mathbb E_n\left[ \nabla^\star p_x( s)
  r_{x'} ( s)\right] \right\}^2\dd s\notag
\end{align}

 \begin{align*}
&
 +\frac{2\tilde \ga n^{3/2}}{n+1}\int_0^t\Big\{T_L-\mathbb E_n\left[
                p^2_0( s)
                  \right] \Big\}^2\dd s 
                  + \frac{2\tilde\ga n^{3/2}}{n+1}\int_0^t
                  \Big\{ T_R-\mathbb  E_n\left[ p^2_n( s)    \right] \Big\}^2\dd s  \\
  &+ \frac{2\tilde \ga n^{3/2}}{n+1} \sum_{x=1}^n
    \int_0^t\left\{\mathbb E_n\left[ p_0( s)
  p_{x}( s)\right] \right\}^2 \dd s+
  \frac{2\tilde \ga n^{3/2}}{n+1}\sum_{x=0}^{n-1}\int_0^t\left\{\mathbb E_n\left[ p_n(s)
  p_{x}(s)\right] \right\}^2\dd s
 \end{align*}

\begin{align*}
 &
 + \frac {2 \tilde \ga n^{3/2}}{n+1} \sum_{x'=1}^n   \int_0^t\left\{ \mathbb E_n\left[ p_0( s)
  r_{x'} (  s)\right] \right\}^2  \dd s+  \frac {2 \tilde \ga n^{3/2}}{n+1} \sum_{x'=1}^n \int_0^t \left\{  \mathbb E_n\left[ p_n( s)
  r_{x'} ( s)\right] \right\}^2\dd s
\\
&
 ={\cal H}_n^{(2)}(0) +\frac{2\tilde \ga
     n^{3/2}}{n+1}\int_0^t\left[T_L\Big(T_L-\mathbb E_n   p^2_0(  s)
                    \Big) 
    +  T_R\Big( T_R-\mathbb E_n   
                 p_n^2(  s) \Big)\right]\dd s.
\end{align*}

\end{proposition} 
The proof of identity \eqref{energy-bal} can be found in
\cite[Section 1 of the Supplement]{KO-supp}.
We have the following result.

\begin{theorem}\label{cor012102-24}
  {Suppose that $T_L,T_R>0$. Then,  for any $t_*>0$  there exists $C>0$ such that
  \begin{equation}
  \label{eq:9}
  {\cal H}_n^{(2)}(t)    \le C
\end{equation}
for all $t\in[0,t_*]$ and $n=2,3,\ldots$.} Furthermore
  \begin{equation}
    \label{072102-24}
    \begin{split}&
      \sum_{x=1}^{n}\mathop{\sum_{x'=0}^{n}}_{x'\not\in\{x-1,x\}}\int_0^t\left\{\mathbb
                       E_n\left[ \nabla^\star p_x( s)
                 p_{x'}( s)\right] \right\}^2 \dd s +
       \sum_{x=1}^{n}\int_0^t\Big[\nabla^\star \mathbb
    E_n p_{x}^2( s)
                   \Big]^2\dd s 
                   \\
                   &
                   +  \sum_{x=1}^{n} \sum_{x'=1}^n
    \int_0^t \left\{\mathbb E_n\left[ \nabla^\star p_x( s)
        r_{x'} ( s)\right] \right\}^2\dd s 
    + \sum_{z=0,n}\sum_{x=0}^n  \int_0^t
    \mathbb E_n\left[ p_z( s) p_{x}( s)\right]^2 \dd s
  \\
    &+  \sum_{z=0,n}\sum_{x'=1}^n   \int_0^t   \mathbb E_n [p_z( s) r_{x}( s)]^2  \dd s \le
 \frac{C}{n^{1/2}}.
  \end{split}
\end{equation}
\end{theorem}
The proof of the theorem in the case $T_L=T_R$ is presented in Section
\ref{sec4.3}. The general case is treated in Section \ref{sec13.4}.

\subsection{Proof of Theorems \ref{energy-bound} and \ref{cor012102-24} in the case $T_L=T_R$}
\label{sec4.3}
We use \eqref{energy-bal} to prove Theorems \ref{energy-bound}
and \ref{cor012102-24}  in the
case $T_L=T_R=T$. Then, from \eqref{energyn} and \eqref{energy-bal} we
conclude in particular that
\begin{align*}
  & {\cal H}_n^{(2)}(t) +\frac{ T}{n+1}{\cal H}_n (0)\le {\cal
    H}_n^{(2)}(0)+ \frac{ T}{n+1}{\cal H}_n (t).
\end{align*}
By the Cauchy-Schwarz inequality
we have
\begin{align}
  \label{H12}
  \frac{1}{n+1} {\cal H}_n (t) \le  \Big(2{\cal H}_n^{(2)}(t)\Big)^{1/2},
\end{align}
therefore
\begin{align*}
  & {\cal H}_n^{(2)}(t) \le {\cal H}_n^{(2)}(0)+\Big(2 {\cal H}_n^{(2)}(t)\Big)^{1/2}-\frac{n^{3/2}T}{n+1}{\cal H}_n (0),
\end{align*}
which in light of Assumptions \ref{ass4} and \ref{ass5}
implies estimate \eqref{eq:9}. By virtue of  \eqref{H12} we
infer Theorem \ref{energy-bound}. Using again \eqref{energy-bal}
we conclude also estimate \eqref{072102-24}.\qed

\section{Covariance matrix}

\label{sec4}

\subsection{Preliminaries}
The stochastic evolution equation  at  macroscopic time are given by 
\begin{equation} 
\label{eq:fflip-1}
\begin{split}
  \dot { r}_{x}(t) &= n^{3/2}\nabla^\star p_{x}(t) ,\quad\mbox{for }x=1, \dots, n,
   \\
  \dd { p}_{x}(t) &=  n^{3/2}\Big(\nabla r_{x}  +\ga   \Delta_{\rm N} p_{x}(t) \Big)\dd
  t + \Big[\nabla^\star  p_{x+1}(t-)\dd \tilde N^{(n)}_{x,x+1}(\gamma t)\\
  &
        -\nabla^\star   p_{x}(t-)\dd \tilde N^{(n)}_{x-1,x}(\gamma
      t)\Big], \quad\mbox{for }x=1, \dots, n-1.
    \end{split}
 \end{equation}
Here $
\tilde N^{(n)}_{x-1,x}(t):=N^{(n)}_{x-1,x}(t)- n^{3/2}t
$ are independent zero mean martingales.
At the  boundaries we have
\begin{equation} 
\label{eq:flip1bn}
\begin{split}
     \dd   p_{0}(t) = n^{3/2}\Big(  \nabla  r_{0}  &+\ga \Delta_{\rm N} p_{0}(t)\Big) \dd  t
  + \nabla^\star  p_{1}(t-) \dd\tilde N_{0,1}^{(n)}(\gamma t)   \\
        &          
                  -
                       n^{3/2}\tilde \gamma p_{0}(t) \dd t
                       +\sqrt{2 n^{3/2}\tilde  \gamma T_L} \dd   w_L(t),
                       \\
                       \dd   p_{n}(t) =   n^{3/2}\Big(  \nabla r_{n}
                       &+\ga \Delta_{\rm N} p_{n}(t) \Big)\dd  t - \nabla^\star  p_{n}(t-) \dd \tilde N^{(n)}_{n-1,n}(\gamma t)   \\
                       &
                                     -
                                     n^{3/2}\tilde \gamma p_{n}(t) \dd
                                     t  +\sqrt{2 n^{3/2}\tilde  \gamma
                                       T_R} \dd  e w_R(t). 
                                   \end{split} \end{equation}


Let 
\begin{equation}
  \label{Xt}
{\bf X}(t)=\left(\begin{array}{c}
{\bf r}(t)\\
{\bf p}(t)
\end{array}\right).
\end{equation}
The solution of  \eqref{eq:fflip-1}-- 
\eqref{eq:flip1b} satisfies
\begin{equation}
  \label{Xts0}
{\bf X} (t)=e^{-n^{3/2}At}{\bf X}(0)+\int_{0}^t
e^{-n^{3/2}A(t-s)}\Sigma \Big({\bf p} (s-)\Big)\dd M_n(s),\quad t\ge0.
\end{equation}
Here $  A$ is a $2\times 2$ block matrix of the form
\begin{equation}
\label{bAz}
  A=
\left(
  \begin{array}{cc}
    0_n&-\nabla^\star\\
   -\nabla& -\ga \Delta_{\rm N}+\tilde\ga  E
  \end{array}
\right),
\end{equation}
where $E=[\delta_{x,0}\delta_{y,0}+\delta_{x,n}\delta_{y,n}]_{x,y=0,\ldots,n}$
and 
$0_{n,m}$ is the $n\times m$ null matrix. We use the shorthand
notation $0_n=0_{n,n}$. Here also $M_n(t):=\int_0^t\dd M_n(s)$ is 
a $2n+2$-dimensional, zero mean vector martingale, where
$$
\dd M(s) =\left(\begin{array}{c}
                   0_{n,1}\\
                  n^{3/4}\dd   w_L(s)\\
                  \dd \tilde N^{(n)}_{0,1}(\gamma s)\\
                  \vdots\\
                  \dd \tilde N^{(n)}_{n-1,n}(\gamma s)\\
                  n^{3/4}\dd   w_R(s)
\end{array}\right).
$$
Its covariation matrix is
a block matrix of the form $n^{3/2}\Sigma ({\bf p})$, where
\begin{equation}
  \begin{split}
\Sigma ({\bf p}) =\left[\begin{array}{cc}
0_{n,n}&0_{n,n+2}\\
0_{n+1,n}&D({\bf p})
                        \end{array}\right].
      \end{split}
\label{eq:22}
\end{equation}
Here
 $D({\bf p}) $ is an $(n+1)\times (n+2)$-dimensional matrix, given by
\begin{equation}
  \begin{split} D({\bf p}) =
       \begin{bmatrix}
\sqrt{2 \tilde \gamma T_L} & \nabla^\star p_1 & 0 &\dots&0&0&0\\
                     0& -\nabla^\star p_1  &  \nabla^\star p_2  &\dots&0&0&0\\
                     0 & 0 & -\nabla^\star p_2 &\dots&0&0&0\\
                     \vdots & \vdots & \vdots &
                     \vdots&\vdots&\vdots&\vdots\\
                      0& 0 & 0 & \dots &\nabla^\star p_{n-1} &0 &0\\
                     0& 0 & 0 & \dots &-\nabla^\star p_{n-1} &\nabla^\star p_n &0\\
                     0& 0 & 0 & \dots &0&-\nabla^\star p_n &\sqrt{2 \tilde \gamma T_R} 
                        \end{bmatrix}.
 \end{split}
\label{eq:22a}
\end{equation}

Denote  by    $S(t)$ the
  the covariance matrix
\begin{equation}
\label{S1ts}
S(t) =\bbE_{\mu_n}\Big[{\bf X}_n(t)\otimes {\bf X}_n(t)\Big]
=\left[
  \begin{array}{cc}
    {S^{(r)}(t)}&S^{(r,p)}(t)\\
   S^{(p,r)}(t)& S^{(p)}(t)
  \end{array}
\right],
\end{equation}
where
\begin{align}
\label{S1ts1}
&S^{(r)}(t)=\Big[\bbE_{n}[r_{x}( t)r_{y}(t)]\Big]_{x,y=1,\ldots,n},\quad S^{(r,p)}(t)=\Big[\bbE_{n}[r_{x}( t)p_{y}(t)]\Big]_{x=1,\ldots,n,y=0,\ldots,n},\notag\\
&\\
&
S^{(p)}(t)=\Big[\bbE_{n}[p_{x}( t)p_{y}(t)]\Big]_{x,y=0,\ldots,n}\quad \mbox{and}\quad S^{(p,r)}(t)=\Big[S^{(r,p)}(t)\Big]^T. \notag
\end{align}
Furthermore for a vector  ${\bf x}=[x_1,\ldots,x_n]$ we let
\begin{equation}
  \begin{split}
\Sigma_2 ({\bf x}) =\left[\begin{array}{cc}
0_{n,n}&0_{n,n+1}\\
0_{n+1,n}&\ga D_2({\bf x})+2\tilde \ga D_1
                        \end{array}\right].
      \end{split}
\label{eq:11}
\end{equation}                 
Here 
$D_1=[T_L\delta_{x,0}\delta_{0,y}+T_R\delta_{x,n}\delta_{0,n}]_{x,y=0,\ldots,n}$ and
\begin{equation}
  \begin{split}
    &D_2({\bf x}) \\
    &=
       \begin{bmatrix}
                   x_1& - x_1 & 0 &\dots&0&0&0\\
                     -  x_1 & x_1+x_2 & -  x_2  &\dots&0&0&0\\
                     0 & -  x_2 & x_2+x_3 &\dots&0&0&0\\
                     \vdots & \vdots & \vdots &
                     \vdots&\vdots&\vdots&\vdots\\
                      0& 0 & 0 & \dots &x_{n-2}+x_{n-1}&- x_{n-1}  &0\\
                     0& 0 & 0 & \dots &- x_{n-1}  &x_{n-1}+x_n &- x_{n} \\
                     0& 0 & 0 & \dots &0&- x_{n} & x_n 
                        \end{bmatrix}.
 \end{split}
\label{eq:11a}
\end{equation}

From \eqref{Xts0} we obtain
\begin{align*}
&S(t) =\bbE_n\Big[e^{-An^{3/2}t}{\bf X}(0)\otimes {\bf X}(0)
  e^{-A^Tn^{3/2}t}\Big]\\
  &
    +n^{3/2}\int_{0}^t
    e^{-An^{3/2}(t-s)} \Sigma_2 \Big(\overline{\frak {\bf (\nabla p)}^2}(s)\Big)
    e^{-A^Tn^{3/2}(t-s)}\dd s
\end{align*}
where $A$ is given by \eqref{bAz}
{and
  $$
  \overline{\frak {\bf (\nabla p)}^2}(s)=\Big[\bbE_{n}\big(\nabla^\star
  p_1(s)\big)^2,\ldots, \bbE_{n}\big(\nabla^\star
  p_n(s)\big)^2\Big].$$}

Consequently
           \begin{equation}
       \label{eq:23}
       A \lang S \rang_t +\lang S \rang_t A^T -
       \Sigma_2 \Big(\lang \overline{\frak {\bf (\nabla p)}^2}\rang_t \Big) = \frac{1}{n^{3/2}}\delta_{0,t}S,
     \end{equation}
    where for a given stochastic process $\big(f(t)\big)_{t\ge0}$, taking values
    in an appropriate space, we let
    \begin{equation}
      \label{lang-t}
\lang f\rang_t:=\int_0^t\bbE_n f(s)\dd s,\quad \delta_{0,t}f:=\bbE_nf(0)-\bbE_nf(t).
    \end{equation}

    \subsection{Resolution of the covariance matrix}

Equation \eqref{eq:23} leads to the following equations on the blocks 
\begin{align}
  \label{163011-21}
  &\Big[\lang S^{(p,r)}\rang_t\Big]^T=\lang S^{(r,p)}\rang_t,\\
  &\lang S^{(r,p)}\rang_t\nabla
    -\nabla^\star\lang S^{(p,r)}\rang_t=\frac{1}{n^{3/2}}\delta_{0,t}S^{(r)},\notag\\
  &-\nabla\lang S^{(r)}\rang_t- \Big(\gamma\Delta_{\rm N}-
    \tilde \ga E\Big) \lang S^{(p,r)}\rang_t
    + \lang S^{(p)}\rang_t \nabla=\frac{1}{n^{3/2}}\delta_{0,t}S^{(p,r)} , \notag\\
  &\lang S^{(r)}\rang_t \nabla^\star-\lang S^{(r,p)}\rang_t \Big(\gamma\Delta_{\rm N}
    -\tilde \ga E\Big)
    - \nabla^\star\lang S^{(p)}\rang_t=\frac{1}{n^{3/2}}\delta_{0,t}S^{(r,p)},\notag\\
  &-\nabla\lang S^{(r,p)}\rang_t+\lang S^{(p,r)}\rang_t \nabla^\star
    =\ga D_2\Big(\lang \overline{\frak {\bf (\nabla p)}^2}\rang_t
    \Big)+2\tilde \ga D_1t   \label{eq:main}
  \\
  &
    \qquad \qquad \qquad \qquad  \qquad \qquad \qquad 
    +\lang S^{(p)}\rang_t \Big(\gamma\Delta_{\rm N}-\tilde \ga E\Big)  \notag
    \\
  & \qquad \qquad \qquad \qquad  \qquad \qquad \qquad
    +\Big(\gamma\Delta_{\rm N}-\tilde \ga E\Big) \lang S^{(p)}\rang_t
    +\frac{1}{n^{3/2}}\delta_{0,t}S^{(p)}. \notag
\end{align}
To solve the system it is convenient to work with the Fourier
transforms of the matrices. Let $\psi_0(x),\ldots,\psi_n(x)$ and $\phi_1(x),\ldots,\phi_n(x)$ be
the respective  orthonormal bases of the Neumann and Dirichlet
discrete Laplacians defined in \eqref{secA.1}. Define the Fourier
transforms of the stretch and momenta by
\begin{equation}
  \label{021912-24}
  \tilde r_{j}(t):=\sum_{x=1}^n \phi_j(x)r_x(t)\quad \mbox{and}\quad
  \tilde p_{j}(t):=\sum_{x=0}^n \psi_j(x)p_x(t).
  \end{equation}

Denote
\begin{align*}
 &\tilde{ S}^{(r,p)}_{j,j'}=\sum_{x=1}^n\sum_{x'=0}^n \lang
                 S^{(r,p)}_{x,x'}\rang_t\phi_j(x) \psi_{j'}(x')=\lang
                \tilde r_j\tilde p_{j'}\rang_t,\\
 &
     \tilde{ S}^{(p,r)}_{j',j}=\sum_{x=1}^n\sum_{x'=0}^n \lang
                 S^{(p,r)}_{x',x}\rang_t\phi_j(x) \psi_{j'}(x')=\lang
                \tilde r_{j'}\tilde p_{j}\rang_t\quad \mbox{and}
\end{align*}
\begin{align*}
 &\tilde{ S}^{(r)}_{j,j'}=\sum_{x,x'=1}^n \lang
                 S^{(r)}_{x,x'}\rang_t\phi_j(x) \phi_{j'}(x')=\lang
                \tilde r_j\tilde r_{j'}\rang_t, \\
  &   \tilde{ S}^{(p)}_{j,j'}=\sum_{x,x'=0}^n \lang
                 S^{(p)}_{x',x}\rang_t\psi_\ell(x) \psi_{\ell'}(x') =\lang
                \tilde p_j\tilde p_{j'}\rang_t
\end{align*}
for $j,j'=1,\ldots,n$.  
Let
\begin{align*}
                  &\tilde F_{j,j'}:=
                 \ga\sum_{y=0}^n\psi_j(y)\psi_{j'}(y)\Big[\lang
                 (\nabla^\star p_y)^2\rang_t+\lang (\nabla^\star p_{y+1})^2\rang_t\Big]     \\          
              &
                -\ga\sum_{y=1}^n\big[\psi_j(y-1)\psi_{j'}(y)+ \psi_j(y)\psi_{j'}(y-1)\big]\lang
                 (\nabla^\star p_y)^2\rang_t   , \quad j,j'=0,\ldots,n.\notag
                \end{align*}
Due to the convention $p_{-1}=p_0$ and $p_{n+1}=p_n$ we have
$\nabla^\star p_0=\nabla^\star p_{n+1}=0$ and a simple calculation
(see \cite[Section 2  of the Supplement]{KO-supp})
shows that
\begin{align}
   \label{tFjj}
                 \tilde F_{j,j'}
=\ga\ga_j\ga_{j'}\sum_{y=1}^n\phi_j(y) \phi_{j'}(y) \lang
                   (\nabla^\star p_y)^2\rang_t.
\end{align}
Here $\la_j=\ga_j^2$, where $\la_j$, $j=0,\ldots,n$ are the
eigenvalues of   $-\Delta_{\rm N}$, see \eqref{laps}.
We also let
\begin{equation}
  \label{Rj}
  \begin{split}
&    R_{j,j'}^{(\iota)}=\frac{1}{n^{3/2}} \delta_{0,t}{\tilde
      S}_{j,j'}^{(\iota)} ,\quad \iota\in I:=\{p,pr,rp,r\}\quad\mbox{and}\\
  \\
  &
    B_{j,j'}^{(pr)}=\psi_{j}(0) \tilde s^{(p,\tilde
    r)}_{0,j'}+\psi_{j}(n) \tilde s^{(p,\tilde r)}_{n,j'},\quad B_{j,j'}^{(rp)}=B_{j',j}^{(pr)},\\
    &
     B_{j,j'}^{(p)}=2 t
 \Big(T_L\psi_j(0)\psi_{j'}(0)+T_R\psi_j(n)\psi_{j'}(n)\Big) 
  \\
  &
    -\Big(\psi_j(0)\tilde s^{(p)}_{0,j'}+\psi_{j}(n)
      \tilde s^{(p)}_{n,j'}+\psi_{j'}(0) \tilde s^{(p)}_{0,j}+\psi_{j'}(n)
      \tilde s^{(p)}_{n,j}\Big),
    \end{split}
    \end{equation}
  with
  \begin{align}
    \label{sx}
 & \tilde s^{(p,\tilde r)}_{x,j}=\tilde s^{(\tilde r,p)}_{j,x}=\sum_{\ell=0}^n\psi_{\ell}(x)\tilde
  S^{(r,p)}_{j,\ell}=\lang \tilde r_j p_x\rang_t,\\
  &
  \tilde s^{(p)}_{x,j}=  \tilde s^{(p)}_{j,x}=\sum_{\ell=0}^n\psi_{\ell}(x)\tilde
  S^{(p)}_{j,\ell}=\lang \tilde p_j p_x\rang_t.\notag
\end{align}
   With the above notation 
we can rewrite \eqref{163011-21} for   
all $j,j'=0,\ldots,n$  as follows 
\begin{equation}
  \label{163011-21czx}
\begin{split}
  \ga_{j'}\tilde S^{(r,p)}_{j,j'}
   +\ga_{j}\tilde
   S^{(p,r)}_{j,j'}&=R_{j,j'}^{(r)} , \\
 -\ga_j \tilde S^{(r)}_{j,j'}+\ga \la_j \tilde
   S^{(p,r)}_{j,j'}
   + \ga_{j'}\tilde
   S^{(p)}_{j,j'}&=R_{j,j'}^{(pr)} 
 -\tilde \ga B_{j,j'}^{(pr)}
,    \\
  -\ga_{j'} \tilde S^{(r)}_{j,j'}+ \gamma \la_{j'}\tilde S^{(r,p)}_{j,j'}
   +\ga_j \tilde S^{(p)}_{j,j'}& =R_{j,j'}^{(rp)}  
 -\tilde \ga B_{j,j'}^{(rp)},  \\
   -\ga_{j}\tilde S^{(r,p)}_{j,j'}-\ga_{j'}\tilde S^{(p,r)}_{j,j'}+\ga(\la_j+\la_{j'})\tilde  S^{(p)}_{j,j'}&=R_{j,j'}^{(p)}  
  +\tilde \ga {B_{j,j'}^{(p)}}+\tilde F_{j,j'} . 
                                                   \end{split}
                                                   \end{equation}

Solving for $\tilde S^{(\iota)}_{j,j'}$, $\iota=p,r,pr$ (see
\cite[Section 3  of the Supplement]{KO-supp}), we obtain
\begin{align}
  \label{011803-23}
  \tilde S^{(\iota)}_{j,j'} =  \Theta_\iota(\la_j,\la_{j'})\tilde F_{j,j'}
  +\sum_{\iota'\in I}\Pi^{(\iota)}_{\iota'}(\la_j,\la_{j'}) B_{j,j'}^{(\iota')}
   + \sum_{\iota'\in I}\Xi^{(\iota)}_{\iota'}(\la_j,\la_{j'}) R_{j,j'}^{(\iota')} .
\end{align}

  Here 
  \begin{equation}
    \label{Theta}
    \begin{split}
   &\Theta_p(c,c')=\frac {2\ga
     cc'}{\theta(c,c')},\quad\mbox{where}\quad 
     \theta(c,c')= (c-c')^2+ 2\ga^2cc'(c+c') ,\\
     &
     \Theta_r(c,c') =\frac{\ga (c+c')
   \sqrt{c c'}}{\theta(c,c')},\quad  
\Theta_{pr}(c,c')=\frac
                 { (c-c' )\sqrt{c'}}{\theta(c,c')}.
\end{split}
\end{equation}
The coefficients $\Xi^{(\iota)}_{\iota'}(c,c')$ are given by 
   \begin{equation}
    \label{Xi}
    \begin{split}
    &
      \Xi^{(\iota)}_p(c,c')= \Theta_\iota(c,c'),\quad \iota=p,pr,r,\\
      &
      \Xi^{(p)}_{r}(c,c')=-\Xi^{(pr)}_{rp}(c,c') =\Theta_r(c,c'),\\
       &
        \Xi^{(p)}_{pr}(c,c')=-\Theta_{pr}(c,c'),\quad   \Xi^{(p)}_{rp}(c,c')=\Xi^{(p)}_{pr}(c',c),
\\
  &
       \Xi^{(pr)}_{pr}(c,c') =\frac
    {  \ga c'(c+c')}{\theta(c,c')} , \\
                  &
    \Xi^{(p,r)}_{r}(c,c') =\frac
                 { 1}{2 \sqrt{c} }\Big[1+\frac{c^2-(c')^2
         }{ \theta(c,c')}\Big],
  \\
         &
 \Xi^{(r)}_{rp}(c,c')=   \Xi^{(r)}_{pr}(c',c)=-\Xi^{(pr)}_{r}(c,c') ,\\
    &
    \Xi^{(r)}_{r}(c,c')=\ga\frac{
    c^2+(c')^2+\ga^2cc'(c+c')}{\theta(c,c')}.
\end{split}
\end{equation}
Finally $\Pi^{(\iota)}_{\iota'}(c,c')$ are determined from
\begin{equation}
    \label{Pi}
    \begin{split}
      &\Pi^{(\iota)}_p(c,c')=\tilde \ga\Theta_\iota(c,c'),\quad \iota=p,pr,r,\\
      &
      \Pi^{(\iota)}_{\iota'}(c,c')=-\tilde \ga  \Xi^{(\iota)}_{\iota'}(c,c'),\quad
   \iota=p,pr,r,\,\iota'=pr,\,rp,\\
    &   \Pi^{(\iota)}_{r}(c,c')=0,\quad \iota=p,pr,rp,r.
\end{split}
\end{equation}

 \subsection{Further covariance bounds from \eqref{energy-bal}}

 Recall definition \eqref{021912-24} of $\tilde r_j, \tilde p_j$.  
 \begin{corollary}
  \label{cor011912-24}
  For any $t_*>0$ there exists $C>0$ such that
 \begin{equation}
    \label{010901-25}
    \begin{split}
    \sum_{j,j'=0}^n \Big[\big(\bbE_n[\tilde r_j(t)\tilde
   r_{j'}(t)]\big)^2&+\big(\bbE_n[\tilde r_j(t)\tilde
   p_{j'}(t)]\big)^2\\
   &
   +\big(\bbE_n[\tilde p_j(t)\tilde p_{j'}(t)]\big)^2\Big]\le C(n+1),
   \end{split}
 \end{equation}
and
  \begin{equation}
    \label{011912-24}
    \begin{split}
   & \sup_{j=0,\ldots,n}\big(\bbE_n\tilde r_j^2(t)+\bbE_n\tilde
   p_j^2(t)\big)\le C(n+1)^{1/2},\\
   &
   \sup_{j,j'=0,\ldots,n}\big|\bbE_n\big[\tilde r_j(t)\tilde
   p_{j'}(t)\big]\big|\le C(n+1)^{1/2}\quad\mbox{ for $t\in[0,t_*]$, $n=1,2,\ldots$}.
   \end{split}
 \end{equation}
 In addition, for $z=0,n$ we have
 \begin{equation}
    \label{071912-24}
    \begin{split}
   & \sup_{j=1,\ldots,n}\big|\int_0^t\bbE_n[\tilde r_j(s)p_z(s)]\dd s\big| \le
   \frac{C}{(n+1)^{1/4}}\quad \mbox{and}\\
   &
   \sup_{j=0,\ldots,n}\big|\int_0^t\bbE_n[\tilde p_j(s)p_z(s)]\dd s\big| \le
   \frac{C}{(n+1)^{1/4}}\quad\mbox{ for $t\in[0,t_*]$, $n=1,2,\ldots$}.
   \end{split}
 \end{equation}
\end{corollary}
\subsubsection*{Proof of \eqref{010901-25}}
We show that
\begin{equation}
    \label{010901-25a}
     \sum_{j,j'=0}^n \big(\bbE_n[\tilde p_j(t)\tilde p_{j'}(t)]\big)^2 \le C(n+1),
   \end{equation}
   the arguments in the other cases are similar.
The expression on the left hand side equals
\begin{align*}
   &\sum_{j,j'=0}^n
  \sum_{x,x',y,y'=0}^n\psi_j(x)\psi_{j}(x')\psi_{j'}(y)\psi_{j'}(y')
     \bbE_n[  p_x(t) p_{x'}(t)]\bbE_n[  p_y(t) p_{y'}(t)]\\
   &
     = \sum_{x,x'=0}^n 
     \big(\bbE_n[  p_x(t) p_{x'}(t)]\big)^2\le C(n+1),
  \end{align*}
thus \eqref{010901-25a} follows by Theorem \ref{cor012102-24}.

\subsubsection*{Proof of \eqref{011912-24}}
 
Using the Cauchy-Schwarz  inequality 
we can write
\begin{align*}
  \bbE_n\tilde
    p_j^2(t)\le 2\left\{\sum_{x,x'=0}^n
  \big(\bbE_n\big[p_x(t)   p_{x'}(t)\big]\big)^2\right\}^{1/2}
\end{align*}
and the desired estimate is a consequence of Theorem
\ref{cor012102-24}. The
proofs of the other estimates in \eqref{011912-24} are analogous.\qed

\subsubsection*{Proof of \eqref{071912-24}}
Using the Cauchy-Schwarz  inequality 
we get
\begin{align*}
  \big|\int_0^t\bbE_n[\tilde r_j(s)p_z(s)]\big| \dd s
    \le\left\{\sum_{x=1}^n \big\{\int_0^t\bbE_n[ 
    r_x(s)p_z(s)]\dd s\big\}^2 \right\}^{1/2}\le \frac{C}{(n+1)^{1/4}}.
\end{align*}
The desired estimate is a consequence of \eqref{072102-24}. The
proofs of the other estimates in \eqref{071912-24} follow the same argument.
\qed

\medskip

\section{Limit identification. Formulation of the results}

\label{lim-id}

\subsection{Time evolution of the energy density}
\label{sec:time-evol-energy}

Consider   a   test function $\varphi\in C^\infty_c(0,1)$. Define
$\varphi_x = \varphi(u_x)$, where
\begin{equation}
  \label{u-x}
  u_x=\frac{x}{n+1},
  \end{equation} and
\begin{equation}
  \label{Enf}
{\rm E}_n(t; \varphi)=\frac{1}{n+1} \sum_{x=0}^{n}  \varphi_{x} 
  \mathbb E_n\left[{\cal E}_x(t)\right].
  \end{equation}
 We have
\begin{equation}
  \label{eq:12b}
  \begin{split}
    {\rm E}_n (t, \varphi) -  {\rm E}_n (0, \varphi)   
   & =
 - \frac{n^{3/2}}{n+1}\sum_{x=0}^{n}\int_0^{t} \varphi_{x} \mathbb
 E_n\left[ \nabla^\star j_{x,x+1} ( s)\right] \dd s   \\
   &
  = \frac{n^{1/2}}{n+1} \sum_{x=0}^{n-1}  \varphi_{n,x} '
  \int_0^{t} \mathbb E_n\left[ j_{x,x+1} ( s) \right] \dd s.
\end{split}
\end{equation}
Here $ \varphi_{n,x}' :=n(\varphi_{x+1} -\varphi_{x})$. Using the
energy bound \eqref{eq:energyb} we can replace  $ \varphi_{n,x}'$ by
$\varphi'_x:=\varphi'(u_x)$ at the expense of an
error of size $o_n(1)$. 
Separating the parts of the current due to the Hamiltonian and
stochastic parts of the dynamics we get, cf \eqref{jax},
\begin{align}
  & {\rm E}_n  (t;\varphi) -  {\rm E}_n  (0; \varphi)  = J_n(t;\varphi')+
    J_n^{(s)}(t;\varphi')+o_n(1),\quad\mbox{where}\label{022308-25}\\
  &
    J_n(t; \varphi'):=
    \frac 1{\sqrt n} \sum_{x=1}^n \varphi'_{x}\int_0^t\bbE_n\left[ j_{x-1,x}^{(a)}(s)\right] \dd s
    = -\frac 1{\sqrt n} \sum_{x=1}^n \varphi'_{x} \lang S^{(pr)}_{x-1,x} \rang_t
    ,\notag\\
&
       J_n^{(s)}(t;\varphi') :=  - \frac{\ga}{2\sqrt n} \sum_{x=0}^{n-1}  \varphi_{x} '
    \int_0^{t} \mathbb E_n\left[ \nabla p_x^2(s)\right] \dd s \\
  &
  =\frac{\ga }{2n^{3/2}} \sum_{x=0}^{n-1}  \varphi_{x} ''
    \int_0^{t} \mathbb E \left[p_x^2( s) \right] \dd s
    +o_n(1), \quad\mbox{and }\varphi''_x:=\varphi''(u_x). \notag
\end{align}
Because of the energy bound \eqref{eq:energyb}, the contribution of $ J_n^{(s)}$ is
negligible.

Using \eqref{011803-23} for $\iota = pr$, we can write
\begin{equation}
  \label{050603-24}
  \begin{split}
    J_n(t;\varphi')  &= -\frac 1{\sqrt n} \sum_{j,j'} {\tilde S}^{(pr)}_{j',j}
    \sum_{x=1}^n \phi_j(x) \psi_{j'}(x-1) \varphi'_{x} 
    \\
    &= -\frac 1n \theta_{pr}(\varphi';n)
    - \frac 1n
    \sum_{\iota\in I}\xi^{(pr)}_{\iota}(\varphi';n)
    -   \frac 1n \sum_{\iota\in I}\pi^{(pr)}_{\iota}(\varphi';n).
\end{split}
\end{equation}
Here $I=\{p,pr,rp,r\}$ and 
\begin{align}
 & \theta_{pr}(\varphi';n)= \sum_{j,j'=1}^{n} {\cal W}_{j,j'}
   \sqrt{\la_j\la_{j'}}  \Theta_{pr}  (\la_j,\la_{j'})    F_{j,j'}, \label{eq:theta}\\
  &
    \pi^{(pr)}_{\iota} (\varphi';n)
  =     \sum_{j=0}^{n}\sum_{j'=1}^{n} {\cal W}_{j,j'} \Pi^{(pr)}_{\iota} 
    (\la_j,\la_{j'})B_{j,j'}^{(\iota)},\label{eq:pi} \\
     &
  \xi^{(pr)}_{\iota}(\varphi'; n) = \sum_{j=0}^{n}\sum_{j'=1}^{n} {\cal W}_{j,j'}
      \Xi^{(pr)}_\iota(\la_j,\la_{j'})
       R_{j,j'}^{(\iota)}, \label{eq:xi}\\
  &
  {\cal W}_{j,j'}:= \sqrt n \sum_{x=1}^{n} 
    \phi_{j'}(x) \psi_{j}(x-1)  \varphi' (u_{x}) , \label{eq:Wjj}\\
        &
  F_{j,j'} =\ga\sum_{y=1}^n\phi_j(y) \phi_{j'}(y) \lang
    (\nabla^\star p_y)^2\rang_t \label{eq:Fjj}. 
\end{align}

{We refer to $\theta_{pr}(\varphi';n)$, $\pi^{(pr)}_{\iota}
(\varphi';n)$ and  $\xi^{(pr)}_{\iota}(\varphi'; n)$ as the bulk,
boundary and time-coboundary terms respectively.
Before formulating the result for each of them 
  we introduce some notation.
Denote by
\begin{equation}
  \label{eq:trig}
  \begin{split}
   & c_0(u) := 1, \qquad c_{\ell}(u) := \sqrt{2} \cos(\pi\ell u),\\
&
 s_{\ell}(u) := \sqrt{2} \sin(\pi\ell u),   \qquad \ell= 1,2,
 \dots,\,u\in[0,1],
 \end{split}
\end{equation}
the cosine and sine orthonormal bases in $L^2[0,1]$.
Given a function $\varphi\in L^2[0,1]$ we denote
\begin{equation}
  \label{eq:50}
  \hat\varphi_c(\ell) := \int_0^1 \varphi(u)  c_{\ell}(u) du,\quad  \hat\varphi_s(\ell) := \int_0^1 \varphi(u)  s_{\ell}(u) du
\end{equation}
its Fourier coefficients in the respective bases.}

{For $f:[0,1]\to\bbR$ and $j=1,\dots,n$ we define 
\begin{equation}
  \label{eq:18}
  \widehat{f_{n,o}}(j) := \frac{\sqrt 2}{n+1} \sum_{x=1}^n \sin(\pi j u_x) f(u_x),
  \qquad \widehat{f_{n,e}}(j) := \frac{\sqrt 2}{n+1} \sum_{x=0}^n \cos(\pi j u_x) f(u_x),
\end{equation}
and for $j=0$
\begin{equation}
  \label{eq:33}
  \widehat{f_{n,e}}(0) := \frac{1}{n+1} \sum_{x=0}^n f(u_x).
\end{equation}
 Suppose that $f\in C^\infty_c(0,1)$. By \cite[Lemma B.1]{KLO23}, for any $k>0$ we have for some constant $C>0$:
\begin{equation}
  \label{eq:19}
  |\widehat{f_{n,\iota}}(j)| \le  \frac{C}{\chi_n^k(j)},\quad j\in\bbZ,\,n=1,2,\ldots, \qquad \iota = o, e,
\end{equation}
where $\chi_n$ is $2n+2$-periodic extension of the function
$$
\chi_n(j)=(1+j)\wedge (2n+2-j), \quad j=0,\ldots,2n+1.
$$
In addition, if $\kappa\in(0,1)$, then there exists $C>0$ such that
\begin{equation}
  \label{eq:18a}
 \sup_{|j|\le n^{\kappa}}\big(|\widehat{f_{n,o}}(j)-\hat f_s(j)|
 +|\widehat{f_{n,e}}(j)-\hat f_c(j)|\big)\le \frac{C}{n^{1-\kappa}}.
\end{equation}}

{The following results deal with each of the terms appearing on the
right hand side of \eqref{050603-24}. In all of them we shall assume
that the test function $\varphi\in C_c^\infty(0,1)$. In addition, both here and in what follows $o_n(1)$ denotes an arbitarry term that
satisfies
\begin{equation}
  \label{on1}
  \lim_{n\to+\infty}o_n(1)=0.
  \end{equation}}
\begin{proposition}[{Asymptotics} of the bulk term]
  \label{lem-pr}
We have
    \begin{equation}
  \label{051712-24}
  \begin{split}
    \frac{1}{n}  \theta_{pr}(\varphi';n) &=
    {-} \frac{1}{(2^3 \gamma)^{1/2}n} 
    \sum_{y=1}^n  \lang\cal E_y\rang_t
    \sum_{\ell=1}^{+\infty}
      c_\ell (u_y) (\pi\ell)^{1/2}\widehat{(\varphi')}_{s}(\ell)  \\
    &=  \frac{1}{(2^3 \gamma)^{1/2}n}
      \sum_{y=1}^n 
     \lang\cal E_y\rang_t
      \sum_{\ell=1}^{+\infty} (\pi\ell)^{3/2}\hat{\varphi}_c(\ell) c_\ell(u_y) + o_n(1)\\
      & =  \frac{1}{(2^3 \gamma)^{1/2}}
      \int_0^t {\rm E}_{n} \left(s, |\Delta_{\rm N}|^{3/4} \varphi \right) \dd s
      + o_n(1),
    \end{split}
  \end{equation}
  as $n\to+\infty$.
  The operator $|\Delta_{\rm N}|^{3/4}$ is defined in   \eqref{frac-lap}.
\end{proposition}

{\begin{proposition}[{Asymptotics}  of the boundary term]
  \label{lem-pi}
For $n\to+\infty$ we have 
  \begin{equation}
    \label{eq:15}
    \begin{split}
      &\sum_{\iota\in I}\frac{1}{n} \pi^{(pr)}_{\iota}(\varphi';n) =
              - \frac{\widetilde\gamma}{2\pi\gamma^{1/2}} \sum_{v=0,1} \int_0^{+\infty}  
   V_{\varrho}(u,v) \int_0^1 V_{\varrho}(u',v)[ T_v -  T(t, u') ]\dd
   u' \frac{F(\tilde\ga,\rho^{1/2}) \dd\varrho}{\varrho^{3/4}} \\
   &
  +  \frac{\widetilde\gamma}{2\pi\gamma^{1/2}}\sum_{v=0,1}V_{\varrho}(u,1-v) 
\int_0^1 V_{\varrho}(u',v)[ T_v -  T(t, u') ]\dd
   u' 
    \,\frac{G(\tilde\ga,\rho^{1/2}) d\rho}{\rho^{3/4}}+ o_n(1),
\end{split}
\end{equation}
with   $V_{\rho}(\cdot,\cdot)$ defined in \eqref{Vla} and
$F(\cdot,\cdot)$, $G (\cdot,\cdot)$ given by \eqref{F-G}.
\end{proposition}}

\begin{proposition}[Negligible time--boundary terms]
  \label{lem-xi}
  For $n\to+\infty$ we have
  \begin{equation}
    \label{eq:16}
    \sum_{\iota\in I}\frac{1}{n} \xi^{(pr)}_{\iota}(\varphi';n) = o_n(1).
  \end{equation}
\end{proposition}
The proofs of the above results are presented in Sections
\ref{sec:proof-bulk-term} -- \ref{sec:proof-os-time}.

\section{Compactness and conclusion of the proof of Theorem \ref{thm012911-23}}
\label{sec:comp-concl-proof}

\subsection{Compactness}
\label{ssec:compact}

Consider the subset $\mathcal M_{+,E_*}([0,1])$ of  $\mathcal
   M_+([0,1])$ - the space of all positive, finite
  {Borel} measures on  $[0,1]$ -
consisting of measures with total mass less than or equal to $E_*$. It is
compact in the topology of weak convergence of measures. In addition,  the topology is metrizable
when restricted to this set. As a consequence of Corollary   \ref{energy-bound} for any $t_*>0$ the total energy is bounded by
$ C_{\mathcal{H},t_*}$ (see \eqref{eq:energyb}) and we have that $
{\rm E}_n (\cdot)\in  {\cal C}_{\mathcal M}[0,t_*] :=C\left([0,t_*], \mathcal M_{+, C_{\mathcal{H},t_*}
  }([0,1])\right)$, where the latter space 
is endowed with the   topology of the uniform convergence.

  Since $\mathcal M_{+,E_*}([0,1])$ is compact, in order to show that $\big({\rm E}_n (\cdot)\big)_{n\ge1}$
  is compact, 
  we only need to control  
its modulus of continuity in time for any
$\varphi\in C[0,1]$, see e.g. ~\cite[p. 234]{kelley}.
This will be a consequence of the following proposition.

{\begin{proposition}
  \label{mod-cont}
  For any $\varphi\in C[0,1]$ we have 
  \begin{equation}
    \label{eq:14}
    \lim_{\delta \downarrow 0} \limsup_{n\to\infty} 
      \sup_{0\le s ,t\le t_*, |t-s| < \delta} \left| {\rm E}_n ( t, \varphi) -  {\rm E}_n ( s, \varphi)\right| = 0
  \end{equation}
\end{proposition}
\proof From \eqref{eq:12b}  
one can infer  that for any $\varphi\in C_c^\infty(0,1)$ the mean
energy field ${\rm
  E}_n ( t, \varphi)$, see \eqref{eq:12b},  can be decomposed as follows
$$
{\rm
  E}_n ( t, \varphi)={\rm
  E}_n^{\theta} ( t, \varphi) +{\rm
  E}_n^{\pi} ( t, \varphi)+{\rm
  E}_n^{\xi} ( t, \varphi) +o_n(t),
$$ 
where ${\rm
  E}_n^{\theta} ( t, \varphi) $ corresponds to the bulk terms, ${\rm
  E}_n^{\pi} ( t, \varphi) $ - to the boundary terms and ${\rm
  E}_n^{\xi} ( t, \varphi) $ - to the time co-boundary terms. They are considered in Sections \ref{sec:proof-bulk-term},  \ref{sec:proof-boundary-terms}
and \ref{sec:proof-os-time} respectively. The term $o_n(t)$ that
gathers the stochastic current $J_n^{(s)}(t;\varphi')$ and negligible
terms that arise from replacing $\varphi'$ by  $ \varphi_{n,x}' $ (see \eqref{022308-25})
satisfies
\begin{equation}
  \label{on}
\lim_{n\to+\infty}\sup_{t\in[0,t_*]}|o_n(t)|=0,\quad\mbox{for any }t_*>0.
\end{equation}
It follows from the analysis of the bulk and boundary term (see
Sections \ref{sec:proof-bulk-term} and  \ref{sec:proof-boundary-terms})
that for any $t_*>0$
there exists $C>0$
such that
$$
\left| {\rm E}_n^{(\iota)} ( t, \varphi) -  {\rm E}_n ( s, \varphi)\right|\le
C(t-s),\quad \iota\in\{\theta,\pi\},\,  s,t\in [0,t_*],\,n=1,2,\ldots.
$$
As for the time co-boundary term, the argument of Section
\ref{sec:proof-os-time} shows that it is of order $o_n(t)$ satisfying
\eqref{on}. 
This   implies \eqref{eq:14} for any test function $\varphi\in C_c^\infty(0,1)$.
If $\varphi\in C[0,1]$, then we can approximate it by a sequence of
$(\varphi_N)_{N\ge1}\subset C_c^\infty(0,1)$ in the $L^2$ sense, as $N\to+\infty$. Thanks to
\eqref{cor012102-24} we conclude that
$$\lim_{N\to+\infty} 
      \sup_{0\le t\le t_*,n\ge1} \left| {\rm E}_n ( t,
        \varphi_N) -  {\rm E}_n ( t,
        \varphi) \right| = 0
      $$
      and equality \eqref{eq:14} follows as well.
      \qed}

\subsection{Properties of   limiting points of $\left( {\rm E}^{(n)}(\cdot)\right)$}

         As we have already pointed out in   Section \ref{ssec:compact}
the sequence  
$\left( {\rm E}^{(n)}(\cdot)\right)\subset {\cal C}_{\mathcal M}[0,t_*] :=C\left([0,t_*], \mathcal M_{+, C_{\mathcal{H},t_*}
  }([0,1])\right)$ is compact in the uniform convergence topology  for
each $t_*>0$. Any limiting point  ${\rm E} (\cdot)$ is a continuous function ${\rm E}:[0,+\infty)\to \mathcal M_{+
}([0,1]) $. In fact, it follows directly from Theorem
\ref{cor012102-24} that
\begin{equation}
  \label{012308-25}
\frac{1}{n}\sum_{x=0}^{n}
   \big(\bbE_{n}\big[{\cal E}_{n,x}(t)\big]\big)^2\le {\cal
     H}_n^{(2)}(t) \le C,\quad n=1,2,\ldots.
\end{equation}
Therefore  ${\rm E} (\cdot)$ is of the form ${\rm
  E}(t,\varphi)= \int_0^1T(t,u)\varphi(u)\dd u$, where $T(t,\cdot)$ is
a square integrable   function w.r.t. the Lebesgue measure for
any $t>0$. In what follows we shall identify  the measure  valued
function ${\rm E} (\cdot)$ with its density $T(\cdot)$.  Let  $L^2_w[0,1]$
denote the space of all square integrable functions on $[0,1]$, equipped with the
weak topology.  
\begin{theorem}
  \label{thm021401-25}
  Suppose that $T(\cdot)$ is a limiting point of $
\Big({\rm E}_n (\cdot)\Big)_{n\ge1}$.
 Then, under the assumptions made in Section \ref{sec:thermal-energy},
 we have   $
T (\cdot)\in   C\left([0,+\infty), L^2_w[0,1]\right)$. The functions ${\cal b}^{(v)}:[0,+\infty)^2\to\bbR$,
$v=0,\,1$ given by
\begin{align}
  \label{021601-25z}
  {\cal b}^{(v)}(s,\varrho):= T_v-\int_0^1 V_{\rho}(u,v) T(s,u)\dd u
       ,\quad s,\varrho>0,
\end{align}
  satisfy
 \begin{align}
    \label{012212-24az}
  \int_0^t\dd s\int_{0}^{+\infty}\big[{\cal b}^{(v)}(s,\varrho)\big]^2\frac{\dd
   \varrho}{\varrho^{3/4}}<+\infty, \quad t>0,\,v=0,1.
\end{align}
In addition, for each $\varphi\in C_c^\infty(0,1)$  equation
\eqref{042209-24aa} holds.
\end{theorem}
\proof
The fact that $T (t,\cdot)$ satisfies equation \eqref{042209-24aa}
follows directly from equation \eqref{022308-25} and Propositions \ref{lem-pr}--\ref{lem-xi}.
From \eqref{012308-25} it follows that $\sup_{t\ge0}\|T
(t,\cdot)\|_{L^2[0,1]}<+\infty$, which combined with
\eqref{042209-24aa} implies that $
T (\cdot)\in   C\left([0,+\infty), L^2_w[0,1]\right)$. Finally,
\eqref{012212-24az} follows from Corollary \ref{cor012308-25}.
\qed

\subsection{{The end of the proof of convergence of the energy functional}}

\label{sec6.6}

{According to Propositions  \ref{lem-pr}--\ref{lem-xi} and Theorem
  \ref{thm021401-25} any limiting
point for the sequence  $\Big({\rm E}_n (\cdot)\Big)_{n\ge1}\subset
{\cal C}_{\mathcal M}[0,t_*]$  is a weak solution to \eqref{f-diff1} in
the sense of Definition \ref{df1.5}. {To complete the argument for
  the convergence of $\big({\rm E}_n ( \cdot)\big)_{n\ge1}$ it suffices to invoke Theorem \ref{lm032209-24} that
asserts the uniqueness of such solutions.}

\section{The asymptotics of the bulk term. Proof of Proposition \ref{lem-pr}  }
\label{sec:proof-bulk-term}


We first compute ${\cal W}_{j,j'} $ defined in \eqref{eq:Wjj}. Denote
\begin{equation}
  \label{k-j}
  k_j:=\frac{j}{n+1}.
  \end{equation}

A direct calculation (see \cite[Section 4  of the Supplement]{KO-supp}) leads to the following formula.
\begin{lemma} 
  \label{lmW}
  For any function $\varphi:[0,1]\to\bbR$ such that ${\rm
    supp}\,\varphi\subset(0,1)$ we have
  \begin{equation}
    \label{W}
    \begin{split}
        {\cal W}_{j,j'} = & - \Big(\frac{n}{2}\Big)^{1/2} \left(1-\frac{\delta_{0,j}}{2}\right)^{1/2}
      \cos\Big(\frac{\pi k_j}{2}\Big) \Big[ \widehat{(\varphi')_{n,o}}(j-j')
      -\widehat{(\varphi')_{n,o}}(j+j')\Big]
  \\
  &
    -\Big(\frac{n}{2}\Big)^{1/2} 
\sin\Big(\frac{\pi k_j}{2}\Big)    \Big[ \widehat{(\varphi')_{n,e}}(j+j')
-\widehat{(\varphi')_{n,e}}(j-j')  \Big].
\end{split}
\end{equation}
  \end{lemma}

   Substituting from \eqref{W} into \eqref{eq:theta} and recalling \eqref{Theta},
   we have
   \begin{equation}
     \label{eq:20}
      \theta_{pr}(\varphi';n)  =\theta_{pr}^{(o)}(\varphi';n)  +\theta_{pr}^{(e)}(\varphi';n),
    \end{equation}
    with
    \begin{equation}
      \label{eq:24}
      \theta_{pr}^{(o)}(\varphi';n) 
      = -\Big(\frac{n}{2}\Big)^{1/2} \sum_{j,j'=1}^{n}
      \sin(\pi k_j) \Big[ \widehat{(\varphi')_{n,o}}(j-j')
      -\widehat{(\varphi')_{n,o}}(j+j')\Big]
      \frac{\la_{j'}  (\la_j - \la_{j'}) }{\theta (\la_j,\la_{j'})}  F_{j,j'},
       \end{equation}
    and
    \begin{equation}
      \label{eq:25}
      \begin{split}
        \theta_{pr}^{(e)}(\varphi';n) =
        -\Big(\frac{n}{2^3}\Big)^{1/2}\sum_{j,j'=1}^{n} 
        \Big[ \widehat{(\varphi')_{n,e}}(j+j') -\widehat{(\varphi')_{n,e}}(j-j')  \Big]
         \frac{ \la_j\la_{j'}  (\la_j - \la_{j'}) }{\theta (\la_j,\la_{j'})} F_{j,j'}\\
      \end{split}
    \end{equation}
   By a symmetry argument,  interchanging
the roles of indices $j$, $j'$ in \eqref{eq:25}, we have $
\theta_{pr}^{(e)}(\varphi';n) =0$.
   Using the above and   the parity $F_{-j,j'} =F_{j,-j'}=-F_{j,j'}$
we conclude that (see \cite[Section 5  of the Supplement]{KO-supp} for detailed calculation)
  \begin{equation}
      \label{eq:27}
    \theta_{pr}(\varphi';n)  =\theta_{pr}^{(o)}(\varphi';n)  = - \left(\frac{n}{2^3}\right)^{1/2} \sum_{j,j'=-n-1}^{n}   
      \sin(\pi k_j) \widehat{(\varphi')_{n,o}}(j-j')
      \frac{\la_{j'}  (\la_j - \la_{j'}) }{\theta (\la_j,\la_{j'})}  F_{j,j'},
    \end{equation}

    Recalling \eqref{tFjj} we have
    \begin{equation}
      \label{eq:28}
      \begin{split}
         \theta_{pr}(\varphi';n) =&  - \gamma \Big(\frac{n}{2^3}\Big)^{1/2} \sum_{j,j'=-n-1}^{n}
      \sin(\pi k_j) \widehat{(\varphi')_{n,o}}(j-j')\\
      &\times \frac{\gamma_{j'}^2  (\gamma_j^2 - \gamma_{j'}^2) }
      {\theta (\gamma_j^2,\gamma_{j'}^2)}
      \sum_{y=1}^n \phi_j(y) \phi_{j'}(y)
      \lang \left(\nabla^* p_y\right)^2\rang_t\\
      =& - \gamma \Big(\frac{n}{2^3 }\Big)^{1/2} \sum_{y=1}^n
      \lang \left(\nabla^* p_y\right)^2\rang_t
      \sum_{j,j'=-n-1}^{n}\sin(\pi k_j)\frac{\gamma_{j'}^2  (\gamma_j^2 - \gamma_{j'}^2) }
      {\theta (\gamma_j^2,\gamma_{j'}^2)}
    \\
      &\times   \widehat{(\varphi')_{n,o}}(j-j')\frac{\cos(\pi u_y (j-j')) - \cos(\pi u_y(j+j'))}{n+1}
      \\
      =: &  \theta_{pr,-}(\varphi';n) -  \theta_{pr,+}(\varphi';n)
      \end{split}
    \end{equation}
Using elementary  trigonometric formulas we conclude that
    \begin{equation}
      \label{eq:34}
      \begin{split}
         \theta_{pr,-}(\varphi';n) =      -\frac{\gamma n^{1/2}}{2^{3/2} (n+1)} 
     & \sum_{j,j' =-n-1}^{n}  \frac{\sin(\pi k_j) \sin^2\left(\frac{\pi k_{j'}}{2}\right)
            \sin\left(\frac{\pi (k_j + k_{j'})}{2}\right) \sin\left(\frac{\pi (k_j - k_{j'})}{2}\right)}
       {\sin^2\left(\frac{\pi (k_j + k_{j'})}{2}\right)
         \sin^2\left(\frac{\pi (k_j - k_{j'})}{2}\right) + 2^3 \gamma^2\Gamma(k_j,k_{j'})}
    \\
      &\times  \sum_{y=1}^n
      \lang \left(\nabla^* p_y\right)^2\rang_t \widehat{(\varphi')_{n,o}}(j-j') \cos(\pi u_y (j-j'))
      \\
    =   - \frac{\gamma n^{1/2}}{2^{3/2} (n+1)} 
      & \sum_{\ell,j'=-n-1 }^{n } 
      \frac{\sin(\pi k_{\ell + j'}) \sin^2\left(\frac{\pi k_{j'}}{2}\right)
            \sin\left(\frac{\pi k_{\ell +2j'}}{2}\right) \sin\left(\frac{\pi k_\ell}{2}\right)}
       {\sin^2\left(\frac{\pi k_{\ell +2j'}}{2}\right)
         \sin^2\left(\frac{\pi k_\ell}{2}\right) + 2^3 \gamma^2\Gamma(k_{\ell+j'},k_{j'})}
    \\
      &\times  \sum_{y=1}^n
      \lang \left(\nabla^* p_y\right)^2\rang_t \widehat{(\varphi')_{n,o}}(\ell) \cos(\pi u_y \ell),
      \end{split}
    \end{equation}
    where
   \begin{equation}
      \label{Gjj} \Gamma(k_j,k_{j'}) = \sin^2\left(\frac{\pi k_j}{2}\right) \sin^2\left(\frac{\pi k_{j'}}{2}\right) 
           \left(\sin^2\left(\frac{\pi k_j}{2}\right)
             +\sin^2\left(\frac{\pi k_{j'}}{2}\right) \right).
         \end{equation}
         Choose $\kappa\in(0,1)$. We further adjust the parameter later on.
    Thanks to \eqref{eq:19} we can consider only the terms $|\ell| \le
    n^{\kappa}$  
     and we have 
    \begin{equation}
      \label{eq:35}
      \begin{split}
        \theta_{pr,-}(\varphi';n) =  - \frac{\gamma n^{1/2}}{2^{3/2} (n+1)} 
      & \sum_{|\ell|\le n^\kappa} \sum_{j'=-n-1}^{n}
      \frac{\sin(\pi k_{\ell + j'}) \sin^2\left(\frac{\pi k_{j'}}{2}\right)
            \sin\left(\frac{\pi k_{\ell +2j'}}{2}\right) \sin\left(\frac{\pi k_\ell}{2}\right)}
       {\sin^2\left(\frac{\pi k_{\ell +2j'}}{2}\right)
         \sin^2\left(\frac{\pi k_\ell}{2}\right) +  2^3 \gamma^2\Gamma(k_{\ell+j'},k_{j'})}
    \\
      &\times  \sum_{y=1}^n
      \lang \left(\nabla^* p_y\right)^2\rang_t \widehat{(\varphi')_{n,o}}(\ell) \cos(\pi u_y \ell)
      + o_n(1).
    \end{split}
    \end{equation}
 Since $|k_\ell|=|\ell|/(n+1)\le (n+1)^{\kappa-1}$ we can use approximate
  equalities
  \begin{align}
    \label{sinl}
&\sin\Big(\frac{\pi k_\ell}{2}\Big)\approx
\frac{\pi \ell}{2(n+1)}\quad \mbox{and}\quad 2k_{j'}+k_{\ell}
                   \approx 2k_{j'}.
  \end{align}
  Then we have
  \begin{equation}
    \label{eq:65}
    \begin{split}
      \Gamma(k_{j'+\ell},k_{j'}) 
          \approx  2 \sin^6\left(\frac{\pi k_{j'}}{2}\right) 
        \end{split}
  \end{equation}
  and, as a result, obtain
  \begin{equation}
    \label{eq:36}
    \begin{split}
          \frac{1}{n}\theta_{pr,-}(\varphi';n) &=- \frac{\gamma  }{2^{3/2} (n+1)^{3/2}} 
      \sum_{|\ell|\le n^\kappa} \sum_{j'=-n-1}^{n}
      \frac{\sin(\pi k_{ j'}) \sin^2\left(\frac{\pi k_{j'}}{2}\right)
            \sin\left(\frac{\pi k_{2j'}}{2}\right)}
       {\sin^2\left(\frac{\pi k_{2j'}}{2}\right)
         \left(\frac{\pi\ell}{2n}\right)^2 + 2^4 \gamma^2 \sin^6\left(\frac{\pi k_{j'}}{2}\right)}
    \\
      &\times \frac{\pi k_\ell}{2} \sum_{y=1}^n
      \lang \left(\nabla^* p_y\right)^2\rang_t \widehat{(\varphi')_{n,o}}(\ell) \cos(\pi u_y \ell)
      + o_n(1)\\
                &=  - \frac{\gamma  }{2^2 (n+1)^{3/2}} 
                               \sum_{|\ell|\le n^\kappa}  \pi \ell\widehat{(\varphi')_{n,o}}(\ell) \sum_{j'=-n-1}^{n}
      \frac{\sin^2\left(\pi k_{j'}\right)}
       {\cos^2\left(\frac{\pi k_{j'}}{2}\right)
         \left(\frac{\pi\ell}{n}\right)^2 + 2^4\gamma^2 \sin^4\left(\frac{\pi k_{j'}}{2}\right)}
    \\
      &\times   \frac{1 }{ 2(n+1) } \sum_{y=1}^n
      \lang \left(\nabla^* p_y\right)^2\rang_t 2^{1/2}\cos(\pi u_y \ell)
      + o_n(1)\\
       \end{split}
     \end{equation}
  Choose   $\delta\in(0,1)$. Observe that if $|j'|\ge \delta n$ the denominator in the
  last expression is larger than $c\gamma^2 {\delta^4}$ for some $c>0$.
  Because of the factor $n^{-3/2}$ in front and the energy bound, this implies that when the sum is restricted
 to  $|j'|\ge \delta n$, the respective  expression is of order $o_n(1)$. So we can write \eqref{eq:36} as
 \begin{equation}
   \label{eq:37}
   \begin{split}
      \frac{1}{n}\theta_{pr,-}(\varphi';n) &=-\frac{\gamma}{2^2 (n+1)^{3/2}}
        \sum_{|\ell|\le n^\kappa} \pi \ell\widehat{(\varphi')_{n,o}}(\ell)\sum_{|j'|\le \delta n }
      \frac{\sin^2\left(\pi k_{j'}\right)}
       { \left(\frac{\pi\ell}{n}\right)^2 \cos^2\left(\frac{\pi k_{j'}}{2}\right)
        + 2^4 \gamma^2 \sin^4\left(\frac{\pi k_{j'}}{2}\right)}
    \\
      &\times \frac{1}{2( n+1)} \sum_{y=1}^n
      \lang \left(\nabla^* p_y\right)^2\rang_t  \sqrt 2
      \cos(\pi u_y \ell) + o_n(1).
    \end{split}
  \end{equation}
  Using the  approximations $\sin\Big(\frac{\pi k_{j'}}{2}\Big)\approx
\frac{\pi k_{j'}}{2}$ and $\cos\Big(\frac{\pi k_{j'}}{2}\Big)\approx
1$, valid for a sufficiently small $\delta$, we can rewrite it as
  \begin{equation}
    \label{eq:38}
    \begin{split}
        \frac{1}{n}\theta_{pr,-}(\varphi';n) &= - \frac{\gamma}{2^2 n^{3/2}}
       \sum_{|\ell|\le n^\kappa}  \pi \ell\widehat{(\varphi')_{n,o}}(\ell)\sum_{j'=-\delta n}^{\delta n}
      \frac{\left(\pi k_{j'}\right)^2}{ \left(\frac{\pi\ell}{n}\right)^2 +  \gamma^2 (\pi k_{j'})^4}
    \\
      &\times \frac{1}{2(n+1)} \sum_{y=1}^n
      \lang \left(\nabla^* p_y\right)^2\rang_t  \sqrt 2
      \cos(\pi u_y \ell) + o_n(1)\\
       =   -\frac{\gamma}{2^2 n^{1/2}}
      & \sum_{|\ell|\le n^\kappa} \pi \ell\widehat{(\varphi')_{n,o}}(\ell)\int_{-\delta}^{\delta}
      \frac{(\pi u)^2 \dd u}{ \left(\frac{\pi\ell}{n}\right)^2 +  \gamma^2 (\pi u)^4}
    \\
      &\times \frac{1}{2( n+1)} \sum_{y=1}^n
      \lang \left(\nabla^* p_y\right)^2\rang_t  \sqrt 2
      \cos(\pi u_y \ell) + o_n(1).
       \end{split}
   \end{equation}
Changing variables $v=\left(\frac{\pi \ga n}{\ell}\right)^{1/2}u$ we
conclude that
 \begin{equation}
    \label{eq:38a}
    \begin{split}
      \frac{1}{n}\theta_{pr,-}(\varphi';n) &
      =  - \frac{1}{2^2 \pi\gamma^{1/2}}
      \sum_{|\ell|\le n^\kappa} (\pi |\ell|)^{1/2}\widehat{(\varphi')_{n,o}}(\ell)
            \int_{\mathbb R}  \frac{v^2 \dd v}{ 1 + v^4}
    \\
      &\times \frac{1 }{2(n+1)} \sum_{y=1}^n
      \lang \left(\nabla^* p_y\right)^2\rang_t  \sqrt 2
      \cos(\pi u_y \ell) + o_n(1).
       \end{split}
   \end{equation}


  Using the residue theorem one can calculate
  \begin{equation}
    \label{eq:39}
       \frac 1{2\pi}  \int_{\mathbb R} \frac{v^2 \dd v}{ 1 + v^4} = \frac 1{2^{3/2}}.
  \end{equation}

  In Section \ref{sec:equi} we prove the following
  \begin{proposition}
    For any $\varphi \in C_c(0,1)$, 
    \label{prop:energy}
    \begin{equation}
   \lim_{n\to\infty} \frac{1}{n} \sum_{y=1}^n
    \int_0^t \varphi_y \mathbb E_n
    \left(\left(\nabla^* p_y(s)\right)^2 - 2 \mathcal E_x(s)\right)  \; \dd s = 0.
    \label{eq:40}
  \end{equation}
  \end{proposition}
  Then from \eqref{eq:39} and  \eqref{eq:40} we conclude that
  \begin{equation}
    \label{eq:32}
    \begin{split}
        \frac{1}{n}\theta_{pr,-}(\varphi';n) 
    &=  -\frac{1}{(2^5 \gamma)^{1/2}} \int_0^t \dd s
    \frac{1}{n} \sum_{y=1}^n  \mathbb E_n\left(\mathcal E_y(s)\right)
    \sum_{|\ell|\le n^{\kappa} } c_\ell (u_y)   (\pi|\ell|)^{1/2}\widehat{(\varphi')_{n,o}}(\ell) +o_n(1)\\
    &=
   - \frac{1}{(2^3 \gamma)^{1/2}} \int_0^t \dd s
    \frac{1}{n} \sum_{y=1}^n  \mathbb E_n\left(\mathcal E_y(s)\right)
    \sum_{\ell=1}^{+\infty} c_\ell (u_y) (\pi\ell)^{1/2}\widehat{(\varphi')}_s(\ell) +o_n(1).
  \end{split}
\end{equation}
To obtain the last equality we use \eqref{eq:18a} and choose $\kappa\in(0,2/3)$.
Since $\widehat{(\varphi')}_s(\ell)
=-\pi\ell\hat{\varphi}_c(\ell)  $ the formula \eqref{051712-24}
 follows, provided we prove that
 \begin{equation}
   \label{eq:42}
    \lim_{n\to\infty}   \theta_{pr,+}(\varphi';n) = 0.
  \end{equation}

  

  In order to show \eqref{eq:42} we need the following bound  that will be proven
  in Section \ref{sec6.1}.  Define
  \begin{equation}
    \label{eq:43}
    \widehat{\frak E}_n(t,\ell) :=
    \frac 1{2n} \sum_{y=1}^n \lang\left(\nabla^* p_y\right)^2\rang_t c_\ell(u_y)
  \end{equation}
  \begin{lemma}
    \label{lm011712-25}
   For any $t>0$ there exists $C>0$ such that
    \begin{equation}
      \label{eq:44}
      \left|\widehat{\frak E}_n(t,\ell) \right| \le C \left( 1 \wedge \frac{n^{1/4}}{\ell}\right).
    \end{equation}
    and
    \begin{equation}
      \label{eq:45}
      \sum_{\ell =1}^n\frac 1\ell \left|\widehat{\frak E}_n(t,\ell) \right| \le C.
    \end{equation}
  \end{lemma}

  As for \eqref{eq:37}, arguing as in the case of
  $\theta_{pr,-}(\varphi';n)$, cf. \eqref{eq:38},  we conclude that
  \begin{equation}
   \label{eq:37+}
   \begin{split}
    \theta_{pr,+}(\varphi';n) &=  \frac{\gamma}{2^{2}n^{3/2}}
       \sum_{|\ell|\le n^\kappa} |\pi \ell| \widehat{(\varphi')_{n,o}}(\ell) \sum_{j' =-\delta n}^{\delta n}
      \frac{\left(\pi k_{j'}\right)^2}{ \left(\frac{\pi\ell}{n}\right)^2 +  \gamma^2 (\pi k_{j'})^4}
      \widehat{\frak E}_n(t,2j'+\ell)
      + o_n(1).
    \end{split}
  \end{equation}
  We split the summation over $j'$ in the above expression into the case $|j'|> n^{1/3}$ and
  $|j'|\le n^{1/3}$ and choose $\kappa\in(0, 1/3)$.
  
  When $|j'|\ge n^{1/3}$ we have 
  $ \left|\widehat{\frak E}_n(t,2j'+\ell) \right| \le C n^{-1/12}$, by \eqref{eq:44},
  and using the same calculation as in \eqref{eq:38} the corresponding term
  will be also bounded by  $C n^{-1/12}$ and it  vanishes, as  $n\to\infty$.

  In the case when $|j'| \le n^{1/3}$, the respective term  is estimated by
       \begin{align*}
         &   \frac{C}{ n^{3/2}} \sum_{\ell=1}^{n^{\kappa}}
           (\pi\ell)  |\widehat{(\varphi')_o}(\ell)|
           \sum_{| j'| \le n^{1/3}} \frac{ (\pi k_{j'})^2}{ \Big(\frac{\ell\pi}{n}\Big)^2
           + \ga^2 (\pi k_{j'})^4} 
   \le      \frac{ C}{n^{1/2}} \int_0^{n^{-2/3}}    
           \frac{\varrho^2\dd \varrho}{\Big(\frac{\pi}{n}\Big)^2  +\ga^2 \varrho^4},
                  \end{align*}           
                  that again tends to $0$, as $n\to+\infty$.
 This ends the proof of  \eqref{051712-24}.\qed

                  \section{Asymptoptics of the boundary terms. Proof
                    of Proposition \ref{lem-pi}}
\label{sec:proof-boundary-terms}

Recall that $\pi^{(pr)}_{\iota} (\varphi';n)$  is given by formulas
\eqref{eq:pi} and \eqref{Pi}.
Then,
\begin{equation}
  \label{012208-25}
  \begin{split}
    &\sum_{\iota \in I} \frac 1n   \pi^{(pr)}_{\iota}  
    =\bar \pi^{(pr)}_p (n)
    +\bar \pi^{(pr)}_{pr} (n) ,\quad \mbox{where}\\
    &
   \bar \pi^{(pr)} (n) := \frac {\tilde \ga }{n} \sum_{j=0}^{n}\sum_{j'=1}^{n} {\cal W}_{j,j'} 
   \Theta_{pr} (\la_j,\la_{j'})B_{j,j'}^{(p)}\quad\mbox{and}\\
   &
  \bar \pi^{(pr)}_{pr} (n):=  \frac {1 }{n}
  (\pi^{(pr)}_{pr}+\pi^{(pr)}_{rp})   .
  \end{split}
\end{equation}
Here  $I=\{p,pr,rp,r\}$ and $B_{j,j'}^{(\iota)}$, $\iota\in I$
are given by   \eqref{Rj}. The  term   $\bar \pi^{(pr)}_{pr} (n)$ is negligible, as  we shall see in
Section \ref{sec8.1} below. We deal first with $\bar \pi^{(pr)}_p
( n) $

\subsection{Asymptotics of $\bar \pi^{(pr)}_p(n)
 $}
\label{sec4.1.1a}
 
By using equation \eqref{W} we have
\begin{equation}
  \label{eq:47}
    \bar\pi^{(pr)} _p(n)=   \bar\pi_{p,o}^{(pr)}  (n)+ \bar \pi_{p,e}^{(pr)} (n).
  \end{equation}
By the same symmetry argument as used in \eqref{eq:25}:
\begin{equation}
  \label{eq:48}
  \begin{split}
    \bar\pi_{p,e}^{(pr)} (n):=   -\tilde\ga\Big(\frac{1}{2n}\Big)^{1/2}
    \sum_{j,j'=1}^{n} 
     \sin\Big(\frac{\pi k_j}{2}\Big)
     \Big[ \widehat{(\varphi')_{n,e}}(j+j') -\widehat{(\varphi')_{n,e}}(j-j')  \Big]
     \Theta_{pr} (\la_j,\la_{j'})B_{j,j'}^{(p)}\\
     =  -\tilde\ga\Big(\frac{n}{2}\Big)^{1/2}
    \sum_{j=1}^{n}\sum_{j'=1}^{n}
     \Big[ \widehat{(\varphi')_{n,e}}(j+j') -\widehat{(\varphi')_{n,e}}(j-j')  \Big]
      \frac{\sqrt{\lambda_j\lambda_{j'}}(\lambda_j - \lambda_{j'})}{\theta (\la_j,\la_{j'})}
     B_{j,j'}^{(p)} = 0.
  \end{split}
\end{equation}
Furthermore, we define  $\bar \pi_{p,o}^{(pr)} =  \bar\pi_{p,o}^{(pr,0)} +
\bar\pi_{p,o}^{(pr,n)} $, where we  separate the contributions
coming  from the left and right endpoints of the chain, writing
\begin{equation}
  \label{eq:49}
  \begin{split}
    \bar\pi_{p,o}^{(pr,z)} (n)&= - \tilde\ga  \Big(\frac{1}{2n}\Big)^{1/2}\sum_{j=0}^{n}\sum_{j'=1}^{n}
    \left(1-\frac{\delta_{0,j}}{2}\right)^{1/2}
      \cos\Big(\frac{\pi k_j}{2}\Big) \\
      &\times\Big[ \widehat{(\varphi')_{n,o}}(j-j')
      -\widehat{(\varphi')_{n,o}}(j+j')\Big]\frac{\sqrt{\lambda_{j'}}(\lambda_j - \lambda_{j'})}{\theta (\la_j,\la_{j'})} B_{j,j'}^{(p,z)}
  \end{split}
\end{equation}
for $z=0,n$.
 Here for $v=0,1$ and $T_0=T_L$, $T_1=T_R$:
\begin{equation}
  \label{Rj1}
  \begin{split}
    &
     B_{j,j'}^{(p)} =  B_{j,j'}^{(p,0)} +
     B_{j,j'}^{(p,n)},\quad\mbox{where}\\
&
    B_{j,j'}^{(p,vn)} = \sum_{x=0}^n \left(\psi_j(vn)\psi_{j'}(x) + \psi_j(x)\psi_{j'}(vn)\right)
    \lang b^{(p)}_{nv,x}
      \rang_t \quad\mbox{and}\\
    &
      b^{(p)}_{nv,nv}(s)= T_v-\bbE_n 
      p_{nv}^2(s)  , \quad
    b^{(p)}_{nv,x}(s)=- \bbE_n\big[
      p_{nv}( s) p_x( s)\big]  ,\quad x\not\in\{0,n\}.
  \end{split}
  \end{equation}
 Consider     $\bar\pi_{p,o}^{(pr,0)}$. Using the fact that
 $B_{j,-j'}^{(p,0)}= B_{j,j'}^{(p,0)}$ and   the definitions
 of $\psi_j(x)$  we have
 \begin{equation}
   \label{barp}
   \begin{split}
   & \bar\pi_{p,o}^{(pr,0)} (n)= -  \frac{2\tilde\ga}{n+1}  \Big(\frac{2}{n}\Big)^{1/2}\sum_{j=0}^{n}\sum_{j'=-n-1}^{n}\sum_{x=0}^n
    \left(1-\frac{\delta_{0,j}}{2}\right)   \sin\Big(\frac{\pi k_{j'}}{2}\Big) 
    \cos\Big(\frac{\pi k_j}{2}\Big) 
     \frac{\lambda_j - \lambda_{j'}}{\theta
       (\la_j,\la_{j'})}\\
     &
     \times \widehat{(\varphi')_{n,o}}(j-j')\left[\cos\Big(\frac{\pi k_j}{2}\Big) \cos\Big(\frac{\pi j'(2x+1)}{2(n+1)}\Big) + \cos\Big(\frac{\pi j(2x+1)}{2(n+1)}\Big)  \cos\Big(\frac{\pi k_{j'}}{2}\Big)\right]
    \lang b^{(p)}_{0,x}\rang_t  \\
     &
     =-  \tilde \ga     \Big(\frac{1}{2n}\Big)^{1/2} 
                    \sum_{j,j'=-n-1}^{n}
                    \Big(1-\frac{\delta_{0,j}}{2}\Big)^{-1/2}\Big(1-\frac{\delta_{0,j'}}{2}\Big)^{-1/2}\psi_j(0) \widehat{(\varphi
                      ')_o}(j-j') \\
                    &
                    \times\sin\Big(\frac{\pi (j'+j)}{2(n+1)}\Big)\frac{ \la_j-\la_{j'} }{\theta(\la_j,\la_{j'})}
                  \lang   \tilde b_{0,j'}^{(p)}   \rang_t
                  ,\quad\mbox{where} \quad
     \tilde b_{z,j}^{(p)}(t) = \sum_{x=0}^{n} \psi_j(x)
     b^{(p)}_{z,x} (t),\,z=0,n. 
     \end{split}
\end{equation}
Thanks to estimate \eqref{072102-24} and the Plancherel identity we conclude that for any $t>0$
there exists $C>0$ such that  
\begin{align}
  \label{bx0}
                                                                         (n+1)^{1/2}\sum_{j=0}^n\int_0^t
  \Big(\tilde b_{0,j}^{(p)}(s) \Big)^2\dd s =  (n+1)^{1/2}\sum_{x=0}^n \int_0^t \Big({
                 b}^{(p)}_{0,x}( s)\Big)^2\dd s\le C
                                             \end{align}
for $n=1,2,\ldots$.
Let
$\varrho_j:= \frac{j\pi}{(n+1)^{1/2}}$ and
$$
\tilde b_{z,j}^{(p)}(t) = \sum_{x=0}^{n} \psi_j(x)
b^{(p)}_{z,x} (t),\,z=0,n.
$$
Define   sequences of functions
$$
{\cal b}^{(p,v)}_{n } :[0,+\infty)^2\to\bbR,\quad v=0,1,\,n=1,2,\ldots,
$$
as follows: for {$\varrho > \varrho_ {(n+1)^{2/3}}$
  and $t\ge0$ we let}                      ${\cal b}^{(p,v)}_{n }  (t, \varrho)=0$.
 For $ 0\le j\le  (n+1)^{2/3}$, $  \varrho\in
 \big[\varrho_j, \varrho_{j+1}\big)$ we  let
 \begin{align}
   \label{tbj}
    &  {\cal b}^{(p,v)}_{n } (t, \varrho)=
     (n+1)^{1/2}\tilde b_{nv,j}^{(p)}(t),\quad v=0,1.
 \end{align}
 Thanks to \eqref{bx0} for any $t>0$ there exists $C>0$ such that
  \begin{align}
    \label{012212-24}
  \int_0^t\dd s\int_{0}^{+\infty}\big[{\cal
    b}^{(p,v)}_{n}(s, \varrho )\big]^2\dd \varrho\le C, \quad \,v=0,1,
    \, n=1,2,\ldots.
\end{align}
Invoking the definitions of $\theta
       (\la_j,\la_{j'})$, see \eqref{Theta}, and   $\Gamma(k_j,k_{j'})$, see \eqref{Gjj},    we can further write
\begin{align*}
&
    \bar\pi^{(pr,0)}_{p,o}( n)  
    = - \frac{\tilde \ga     }{ 2^2 (n+1)n^{1/2}}    
                    \sum_{j,j'=-n-1}^{n}\int_0^t{\cal
                 b}^{(p,0)}_{n}(s, \varrho_{j'})\dd s
                \Big(1-\frac{\delta_{0,j'}}{2}\Big)^{-1/2}\cos\Big(\frac{\pi \varrho_j}{2(n+1)^{1/2}}\Big)
    \\
    &
    \times \sin^2\Big(\frac{\pi (\varrho_j+\varrho_{j'})}{2(n+1)^{1/2}}\Big)\widehat{(\varphi
      ')_o}(j-j') 
      \sin\Big(\frac{\pi ( j- j')}{2(n+1) }\Big)
   \\
&
    \times\Big\{  \sin^2\Big(\frac{\pi ( j- j')}{2(n+1) }\Big)\sin^2\Big(\frac{\pi
                      (\varrho_j+\varrho_{j'})}{2(n+1)^{1/2}}\Big) 
    +2^3\ga^2
     \Gamma\Big(\frac{\varrho_j }{(n+1)^{1/2}}, \frac{\varrho_{j'} }{(n+1)^{1/2}}\Big)\Big\}^{-1} 
     \end{align*}

      Let $\kappa\in(0,1/2)$. Denoting $\ell=j-j'$ and using the
      approximations
      \begin{align*}
&\sin \Big(\frac{\pi     (\varrho_j+\varrho_{j'})}{2(n+1)^{1/2}}\Big)
                    \approx \frac{\pi
                      \varrho_{j'}}{(n+1)^{1/2}},\quad
                       \sin\Big(\frac{\pi ( j- j')}{2(n+1) }\Big)
                       \approx \frac{\pi ( j- j')}{2(n+1) },\\
        &
    \cos\Big(\frac{\pi \varrho_j}{2(n+1)^{1/2}}\Big)\approx 1,      
     \end{align*}                
                    valid for $|j'-j|\le n^\kappa$,
      we can write that
      \begin{align}
        \label{022208-25}
&
                 \bar\pi^{(pr,0)}_{p,o}( n)  
                  =  - \frac{\tilde \ga   }{2(n+1)^{1/2}}    
                    \sum_{|\ell|\le n^{\kappa}}\pi\ell\widehat{(\varphi
      ')_o}(\ell) \sum_{|j'|\le
                       (n+1)^{2/3}} 
                \int_0^t  \frac{{\cal b}^{(p,0)}_{n}(s, \varrho_{j'})
                            \dd s
      }{  (\pi\ell)^2   + \ga^2
      (\pi     \varrho_{j'} )^4  } +o_n(1)\notag
      \\
            &=
              \frac{\tilde \ga   }{2(n+1)^{1/2}}    
                    \sum_{|\ell|\le n^{\kappa}}\hat{\varphi
      }_c(\ell)(\pi \ell)^2 \sum_{|j'|\le
          (n+1)^{2/3}} 
                  \int_0^t\frac{{\cal b}^{(p,0)}_{n}(s, \varrho_{j'})
              \dd s
        }{
         ( \pi\ell )^2  
    +  \ga^2 
          \varrho_{j'}^4  }  +o_n(1) \notag
    \\
             &
        = \frac{2^{1/2}\tilde \ga   }{ \pi}    
                    \sum_{\ell=1}^{+\infty}c_\ell(0)(\pi \ell)^2\hat{\varphi
      }_c(\ell) \int_0^t\dd s\int_0^{+\infty}
                  \frac{{\cal b}^{(p,0)}_{n}(s, \varrho) \dd\varrho 
        }{
         ( \pi\ell )^2  
    +  \ga^2 
           \varrho ^4  }  +o_n(1).
      \end{align}

      We have the following result.
\begin{theorem}
 \label{thm011001-25}
   For any test function
       $f\in    L^2[0,+\infty)$, $t>0$
 and $v=0,1$      we have  
\begin{align}
  \label{061001-25L}
   & \int_0^t\dd s \int_0^{+\infty}{\cal b}^{(p,nv)}_{n}(s,\varrho)f(\varrho)\dd \varrho   
 \\
       &
         =  \sqrt{2}\int_{0}^{+\infty}\Bigg(t T_v- \sum_{\ell=0}^{+\infty} 
   \frac{\ga^2  \varrho^4 c_\ell(v) \widehat{\frak E}_n(t,\ell)}{ (\ell\pi)^2 
    +\ga^2 \varrho^4    }
         \Bigg)  
F(\tilde\ga, \ga\varrho^2)f(\rho) \dd \varrho \notag\\
   &
     -    \sqrt{2}\int_{0}^{+\infty}\Bigg(t T_{1-v}- \sum_{\ell=0}^{+\infty} 
   \frac{\ga^2  \varrho^4 c_\ell(1-v) \widehat{\frak E}_n(t,\ell)}{ (\ell\pi)^2 
     +\ga^2 \varrho^4    } \Bigg) G(\tilde\ga, \ga\varrho^2)f(\rho) \dd \varrho  +o_n(1)   .\notag
\end{align}
where $\widehat{\frak E}_n(t,\ell)$ is defined by \eqref{eq:43} and
$F(\cdot,\cdot)$, $G(\cdot,\cdot)$ are defined in \eqref{F-G}.
  \end{theorem}
The proof Theorem \ref{thm011001-25} is presented in Section
\ref{sec7}. As an immediate conclusion of the theorem and estimate \eqref{012212-24} we formulate the following.
\begin{corollary}
  \label{cor012308-25}
  Suppose that $T(\cdot)$ is a limiting point of $
  \Big({\rm E}_n (\cdot)\Big)_{n\ge1}$ and ${\cal b}^{(v)}(s,\varrho)$
  $v=0,\,1$ are defined in \eqref{021601-25z}.
 Then, \eqref{012212-24az} holds.
  \end{corollary}

Here 
  we apply it to the function
  $f(\varrho) =\left[ (\pi\ell)^2 + \gamma^2 \varrho^4\right]^{-1}$.
  Then by using asymptotics \eqref{022208-25} and Proposition
  \ref{prop:energy} we obtain
  \begin{equation}
    \label{eq:52}
    \begin{split}
    \bar\pi^{(pr,0)}_{p,o}( n)  &=  \frac{2^{1/2}\tilde \ga   }{ \pi}    
       \sum_{\ell = 1}^{+\infty} \int_0^{\infty}  \frac{c_\ell(0)(\pi \ell)^2
      \widehat{\varphi}_{c}(\ell)F(\tilde\ga, \ga\varrho^2) }
    {(\pi\ell)^2 +\ga^2 \varrho^4}  
       \Bigg(tT_0- \sum_{\ell'=0}^{+\infty} 
     \frac{\ga^2  \varrho^4 c_{\ell'}(0) \widehat{\frak E}_n(t,\ell')}{ (\ell'\pi)^2 +\ga^2  \varrho^4}
     \Bigg) \dd\varrho \\
     &
     -\frac{2^{1/2}\tilde \ga   }{ \pi}  \sum_{\ell = 1}^{+\infty}   \int_{0}^{+\infty} \frac{c_\ell(0)(\pi \ell)^2
      \widehat{\varphi}_{c}(\ell)G(\tilde\ga, \ga\varrho^2)}  {(\pi\ell)^2 +\ga^2 \varrho^4}
     \Bigg(t T_{1}- \sum_{\ell=0}^{+\infty} 
   \frac{\ga^2  \varrho^4 c_\ell(1) \widehat{\frak E}_n(t,\ell)}{ (\ell\pi)^2 
     +\ga^2 \varrho^4    } \Bigg) \dd\varrho+ o_n(1).
 \end{split}
   \end{equation}
Observe that, since ${\rm supp}\,\varphi\subset(0,1)$, we have
  \begin{equation}
    \label{eq:59}
    \begin{split}
   & 0=\varphi(0)=  \sum_{\ell = 1}^\infty  \frac{ (\pi \ell)^2  \widehat{\varphi}_{c}(\ell)c_\ell(0)}
      {(\pi\ell)^2 +\ga^2    \varrho^4} +\sum_{\ell = 0}^\infty  \frac{ \varrho^4 \widehat{\varphi}_{c}(\ell) c_\ell(0)}
      {(\pi\ell)^2 +\ga^2    \varrho^4} \\
      &
      =\sum_{\ell = 1}^\infty  \frac{ (\pi \ell)^2  \widehat{\varphi}_{c}(\ell) c_\ell(0)}
      {(\pi\ell)^2 + \ga^2    \varrho^4} +\int_0^1V_{\ga \varrho^4}(u,0)
      \varphi(u)\dd u.
    \end{split}
  \end{equation}
   By virtue of \eqref{012412-24a} we also have
  \begin{equation}
    \label{eq:60}
    \begin{split}
     tT_0 =tT_0\int_0^1V_{ \ga^2\varrho^4}(u,0) \dd u = \Big(\frac {1}{n}
     \sum_{y=0}^n  V_{\ga^2  \varrho^4}(0,u_y)\Big)T_0t+o_n(1).
    \end{split}
  \end{equation}
Using \eqref{eq:59} and \eqref{eq:60} in the expression on the utmost
right hand side of \eqref{eq:52} and changing variables
$\varrho':=\ga^2\varrho^4$ we conclude \eqref{eq:15}.
  The argument for the other boundary $x=n$ {(and $v=1$)} is analogous.
  The   things that yet need to be done to   conclude the proof of
  Proposition \ref{lem-pr} are the proofs of  Theorem
  \ref{thm011001-25} and negligibility of $ \bar \pi^{(pr)}_{pr} (n) $ appearing in \eqref{012208-25}.

  \subsection{Estimates of $\bar \pi^{(pr)}_{pr}( n)$}
  \label{sec8.1}

  Our goal in the present section is to show the following.
\begin{lemma}
  \label{lm012208-25}
  There exists a constant $C>0$ such that 
  \begin{equation}
  |\bar \pi^{(pr)}_{pr}( n)|\le \frac{C}{n^{1/2}},\quad n=1,2,\ldots.
  \label{eq:46}
  \end{equation}
\end{lemma}
\proof
Let
\begin{equation}
  \label{brx}
  \begin{split}
    b^{(pr)}_{z,x}(s)= \bbE_n[
    p_z( s) r_x( s)]  ,\quad \quad x=1,\ldots,n,\, \,z=0,n.
       \end{split}\end{equation}
After a straightforward calculation using \eqref{Pi} and the parities
of  $\widehat{(\varphi ')_o}(j)$ and $\widehat{(\varphi ')_e}(j)$  we conclude that
\begin{align*}
  &
     \bar \pi^{(pr)}_{pr}( n)=\sum_{z=0,n} \bar \pi^{(pr)}_{z}(n)
   ,\quad\mbox{where}\\
&  \bar \pi^{(pr)}_{z}(n):=  \frac{ \tilde \ga\ga    
    }{ 2^{3/2}n^{1/2}}  \sum_{x=1}^n\sum_{j,j'=-n-1}^{n}  \widehat{(\varphi ')_o}(j-j')\sin
    \Big(\frac{\pi(k_j+k_{j'})}{2}\Big) \sin
    \Big(\frac{\pi k_{j'}}{2}\Big) \\
  &
    \times
   \Delta^{-1}(k_j,k_{j'}) 
    \Big[\sin^2 \Big(\frac{\pi k_{j}}{2}\Big)+
    \sin^2 \Big(\frac{\pi k_{j'}}{2}\Big)\Big]\psi_j(z)\phi_{j'}(x)\lang b^{(pr)}_{z,x}\rang_t
 .
\end{align*}
Here
\begin{equation}
  \label{k-j1}
  \begin{split}
    &
                  \Delta(k,k'):=       \sin^2\Big(\frac{ \pi(k-k')}{2}\Big)\sin^2\Big(\frac{ \pi(k+k')}{2}\Big)\\
  &
    +2^3\ga^2 \sin^2\Big(\frac{ \pi k}{2}\Big) \sin^2\Big(\frac{ \pi k'}{2}\Big)
    \left(\sin^2\Big(\frac{ \pi k}{2}\Big) +\sin^2\Big(\frac{ \pi k'}{2}\Big)\right).
\end{split}
\end{equation}
This expression can be further rewritten in the form
\begin{align}
  \label{011307-24a}     \bar\pi^{(pr)}_{z}(n)
    =\frac{\tilde \ga\ga   }{ n^{1/2}}
                                                   \sum_{\ell=-n-1}^n  \widehat{(\varphi ')_o}(\ell) \sum_{x=1}^n
    \lang b^{(pr)}_{z,x}\rang_t{\frak i}_{x,z}^{(pr)}(\ell)
\end{align}
 and 
 \begin{align}
   \label{033012-24}
  &
    {\frak i}_{x,z}^{(pr)}(\ell) =\frac{1}{2^{9/2}}\sum_{j'=1}^{n}\phi_{j'}(x) \psi_{j'+\ell}(z) \sin
    \Big(\frac{\pi(2k_{j'}+k_{\ell})}{2}\Big) \sin
    \Big(\frac{\pi k_{j'}}{2}\Big)
      \\
  &
 \times  \Big[\sin^2 \Big(\frac{\pi(k_{j'}+k_{\ell})}{2}\Big) +
    \sin^2 \Big(\frac{\pi k_{j'}}{2}\Big)\Big] \Delta^{-1}(k_{j'+\ell},k_{j'}).\notag
\end{align}
By the Cauchy-Schwarz inequality
\begin{align}
  \label{021612-24a}
&\Big|\sum_{x=1}^n
    \lang b^{(pr)}_{z,x}\rang_t{\frak i}_{x,z}^{(pr)}(\ell)\Big|\le \Big( {\cal B}^{(pr)}_z\Big)^{1/2}\Big( {\cal I}^{(pr)}_z(\ell)\Big)^{1/2},\quad\mbox{where}\\
    &
    {\cal B}^{(pr)}_z := \sum_{x=0}^n  \lang b^{(pr)}_{z,x}\rang_t^2,\qquad
    {\cal I}^{(pr)}_z(\ell):=\sum_{x=0}^n\big({\frak
                 i}_{x,z}^{(p)}(\ell)\big)^2.\notag
\end{align}
From estimate \eqref{072102-24} we conclude that 
for each $t_*>0$  there exists $C>0$ such that
\begin{equation}
    \label{012106-24}
    \sum_{z=0,n}  {\cal B}^{(pr)}_z 
      \le \frac{C}{n^{1/2}} 
      ,\quad t\in[0,t_*],\,n=1,2,\ldots.
 \end{equation}

 By the Plancherel identity we have 
\begin{align*}
  &
    \sum_{x=1}^n
    ({\frak i}_{x,z}^{(pr)}(\ell))^2=\frac{2}{n+1}\sum_{j'=1}^{n} \sin^2
    \Big(\frac{\pi (2k_{j'}+k_\ell)}{2}\Big) \sin^2
    \Big(\frac{\pi k_{j'}}{2}\Big)    \cos^2\Big(\frac{\pi (k_{j'}+k_{\ell})}{2}\Big)\\
  &
    \times
    \Big[\sin^2\Big(\frac{\pi (k_{j'}+k_{\ell})}{2}\Big)+
    \sin^2  \Big(\frac{\pi k_{j'}}{2}\Big)  \Big]^2
   \Delta^{-2}(k_{j'+\ell},k_{j'}).
\end{align*}
As in Section \ref{sec4.1.1a} we can restrict ourselves to the case
$|\ell|\le n^{\kappa}$ for some $\kappa\in(0,1)$.  This allows us to estimate
\begin{align}
  \label{032708-25}
  &
   \sum_{x=1}^n ({\frak i}_{x,z}^{(p,r)}(\ell))^2    \le
   \frac{ C}{n+1}\sum_{j'=1}^{n}   \frac{  \sin^6\Big(\frac{\pi k_{j'}}{2}\Big)}{
  \Big[(\frac{\ell}{n+1}\Big)^2
    + 
    \sin^4\Big(\frac{\pi k_{j'}}{2}\Big)\Big]^{2}}\\
  &
    \le C\int_0^1
    \frac{  u^6\dd u}{
  \Big[(\frac{\ell}{n+1}\Big)^2
    + 
    u^4\Big]^{2}}
  \le \frac{Cn^{1/2}}{\ell^{1/2}},\quad 1\le \ell\le n^{\kappa}.\notag
\end{align}
Combining this estimate with \eqref{011307-24a},  \eqref{021612-24a}
we conclude  \eqref{eq:46}.\qed


  \section{Proof of Theorem \ref{thm011001-25}}

  \label{sec7}

  We denote  ${\cal b}^{(p,v)}_{n,+}(t,\varrho):={\cal
  b}^{(p,v)}_{n}(t,\varrho)$, where the functions on the right  are defined in  \eqref{tbj}.
Define  also sequences of functions
\begin{align}
  \label{tbpm} 
            {\cal b}^{(p,v)}_{n,-}(t,\varrho):= \tilde
           b_{v,j}^{(p,-)}(t), \quad\mbox{where}\quad \tilde
           b_{v,j}^{(p,-)}(t):=\sum_{x=0}^n\psi_j(n-x)b_{v,x}^{(p)}(t),
\end{align}
and
\begin{align}
    \label{tbr}  {\cal b}^{(pr,v)}_{n }(t, \varrho)=
     (n+1)^{1/2}\tilde b_{nv,j}^{(pr)}(t),\quad \mbox{where} \quad
      \tilde
      b_{z,j}^{(pr)}(t):=\sum_{x=1}^n\phi_j(x)b_{z,x}^{(pr)}(t),\quad z=0,n.
\end{align}
 for $ 0\le j\le  (n+1)^{2/3}$, $t\ge0,\, \varrho\in
  \big[\varrho_j, \varrho_{j+1}\big)$, see \eqref{Rj1} and
  \eqref{brx}.
For $\varrho > \varrho_ {(n+1)^{2/3}}$ and all $t\ge0$ we let ${\cal
  b}^{(p,v)}_{n,\epsilon}(t,\varrho)\equiv0$, $v\in\{0,1\}$, $\eps=\pm$.

By the Plancherel identity  and \eqref{072102-24}
for $\iota=p,pr$ we have
\begin{align*}
   & 
\int_0^t\dd s\int_{0}^{+\infty}\big[{\cal
  b}^{(\iota,v)}_{n,\eps }(s, \varrho)\big]^2\dd
     \varrho\le    (n+1)^{1/2}
     \sum_{x=0}^n \int_0^t\big( b_{nv,x}^{(\iota)}(s)\big)^2\dd s\le C.
\end{align*}

The proof of Theorem  \ref{thm011001-25} is a consequence of the following two propositions.
 \begin{proposition}
    \label{prop010901-25}
    For any test function
       $f\in    L^2[0,+\infty)$, $t>0$  
       we have
\begin{align}
  \label{030901-25a}
         \int_0^t\dd s\int_0^{+\infty}{\cal b}^{(p,v)}_{n,+}(s,\varrho)f(\varrho)\dd \varrho   
     = & \int_{0}^{+\infty}\Bigg(t T_v- \sum_{\ell=0}^{+\infty} 
   \frac{\ga^2  \varrho^4 c_\ell(v) \widehat{\frak E}_n(t,\ell)}{ (\ell\pi)^2 
    +\ga^2 \varrho^4    }
         \Bigg)\frac{\sqrt{2} f(\varrho)}{1+2\tilde\ga\coth(\ga \varrho^2)}\dd \varrho \\
   &
     -   \int_{0}^{+\infty}
  {\cal b}^{(p,1-v)}_{n,-}( \varrho) \frac{2 \tilde\ga f(\varrho)}{\sinh(\ga \varrho^2)+2 \tilde\ga\cosh(\ga \varrho^2)}\dd \varrho +o_n(1) .\notag
   \end{align}
 \end{proposition}

  \begin{proposition}
    \label{prop011001-25}
    For any test function
       $f\in    L^2[0,+\infty) $, $t>0$ and $v=0,1$
       we have
\begin{equation}
  \label{030901-25}
  \begin{split}
  & \int_0^t\dd s  \int_{0}^{\infty} {\cal
     b}^{(p,1-v)}_{n,-}(s,\varrho)   f(\varrho)\dd
     \varrho
     =   \int_{0}^{+\infty}\Bigg(t T_{1-v}- \sum_{\ell=0}^{+\infty} 
   \frac{\ga^2  \varrho^4 c_\ell(1-v) \widehat{\frak E}_n(t,\ell)}{ (\ell\pi)^2 
     +\ga^2 \varrho^4    } \Bigg)\frac{ \sqrt{2}f(\varrho)}{1+2\tilde \ga \coth(\ga\varrho^2)}\dd \varrho \\
   &
     - 2\tilde \ga \int_0^t\dd s  \int_{0}^{\infty} {\cal
       b}^{(p,v)}_{n,+}(s,\varrho) \frac{f(\varrho)}{\sinh(\ga \varrho^2)+2 \tilde\ga\cosh(\ga \varrho^2)}\dd \varrho +o_n(1). 
     \end{split}
\end{equation}
\end{proposition}

\subsection{Asymptotics of $\int_0^t\cal{b}^{(p,0)}_{n}(s,\varrho)\dd s$:
  proof of Proposition \ref{prop010901-25}}

\label{sec10.1}

We prove the result for $v=0$. The argument for $v=1$ is analogous.

\subsubsection{Preliminaries}


Suppose that $f\in    C_c^\infty(0,+\infty)$ is a test function
such that ${\rm supp}f\subset [\delta,M]$ for some
$0<\delta<M<+\infty$.
Summing  the expression  in \eqref{011803-23} corresponding to $\iota=p$,
    over $j'$, 
we can write
\begin{equation}
  \label{011007-24a}
  \begin{split} 
   \int_0^t{\cal b}^{(p,0)}_{n}(s,\varrho) \dd s =   t T_L(\varrho)
   +{\cal T}^{(p)}_{n}(\varrho) 
  +{\cal  P}^{(p)}_{n,p}( \varrho) 
  + {\cal P}^{(p)}_{n,pr}( \varrho)
   +\sum_{\iota'\in I} {\cal X}_{n,\iota'}^{(p)}( \varrho).
  \end{split}
   \end{equation}
   Here $I=\{p,pr,r\}$ and for $\varrho\in
   \big[\varrho_j,\varrho_{j+1}\big)$ (cf \eqref{Theta}, \eqref{Xi}
   and \eqref{Pi})  we let
    \begin{align*}
     &
       T_L(\varrho):= \Big(2-\delta_{j,0}\Big)^{1/2}T_L\cos\Big(\frac{\varrho_j  }{2(n+1)^{1/2}}\Big),\\
 &
{\cal T}_{n}^{(p)}(\varrho)
   =-(n+1)^{1/2}\sum_{j'=0}^n\Theta_p(\la_j,\la_{j'})
   \la_j^{1/2}\la_{j'}^{1/2} 
   F_{j,j'}\psi_{j'}(0),  
  \\
  &
  {\cal X}_{n,\iota'}^{(p)}(\varrho)=  -(n+1)^{1/2}\sum_{j'=0}^n\Xi^{(p)}_{\iota'}(\la_j,\la_{j'})
                 R_{j,j'}^{(\iota')}\psi_{j'}(0),\quad
    \iota'\in\{p,pr,rp,r\},
\end{align*}

\begin{align*}
   {\cal P}^{(p)}_{n,\iota}(\varrho)=&
    - (n+1)^{1/2}\sum_{z=0,n}\sum_{j'=0}^n\Pi^{(p)}_{\iota}(\la_j,\la_{j'})
             \psi_{j}(z) \psi_{j'}(0) \lang\tilde b_{z,j'}^{(\iota)}\rang_t
  \\
  &
    -(n+1)^{1/2}\sum_{z=0,n}\sum_{j'=0}^n\Pi^{(p)}_{\iota}(\la_{j'}, \la_j)
             \psi_{j'}(0) \psi_{j'}(z)  \lang\tilde b_{z,j}^{(\iota)}\rang_t,   \quad\iota\in\{p,pr\}.
\end{align*}

Clearly
\begin{align}\label{eq:730}
  t\int_{0}^{+\infty}T_L(\varrho)f(\varrho)\dd \varrho =
  \sqrt{2} tT_0\int_0^{+\infty}f(\varrho)\dd \varrho+o_n(1).
\end{align}

In the following we prove that
\begin{align}
      &\int_0^\infty {\cal T}_{n}^{(p)}(\varrho) f(\varrho) \; \dd \varrho 
       =- \sum_{\ell=-\infty}^{+\infty}  (1+ \delta_{0,\ell} )^{1/2}
         \widehat{\frak E}_n(t,\ell) 
        \int_{0}^{+\infty}\frac{\ga^2  \varrho^4f(\varrho)\dd  \varrho  }{ (\ell\pi)^2 
    +\ga^2  \varrho^4    }+o_n(1), \label{eq:73}\\
   &
  \label{eq:74}
    \int_0^\infty {\cal  P}^{(p)}_{n,p}(\varrho) f(\varrho) \; \dd \varrho =- 2\tilde \ga \int_0^t\dd s  \int_{0}^{\infty} {\cal
     b}^{(p,0)}_{n,+}(s,\varrho) \coth(\ga\varrho^2) f(\varrho)\dd
     \varrho
    \\
   &
     - 2\tilde \ga \int_0^t\dd s  \int_{0}^{\infty} {\cal b}^{(p,n)}_{n,-}(s,\varrho) {\rm csch}\,(\ga\varrho^2) f(\varrho)\dd \varrho +o_n(1),\notag
\\
&
  \label{eq:75}
        \int_0^\infty {\cal P}^{(p)}_{n,pr}( \varrho) f(\varrho) \; \dd \varrho 
       = o_n(1),
  \\
  &
  \label{eq:76}
   \int_0^\infty \sum_{\iota'\in I} {\cal X}_{n,\iota'}^{(p)}( \varrho) f(\varrho) \; \dd \varrho 
       = o_n(1).
     \end{align}
Here  ${\rm csch}\,(u)=\frac{1}{\sinh u}$ is  the hyperbolic cosecant.
     Adding up \eqref{eq:730}, \eqref{eq:73}, \eqref{eq:74}, \eqref{eq:75}, and \eqref{eq:76} 
     we conclude Proposition \ref{prop010901-25}.
  
  \subsubsection{Proof of \eqref{eq:73}}

We have (cf \eqref{k-j}  and \eqref{eq:43})
\begin{align}
  \label{013112-24}
      &\int_0^{+\infty} {\cal T}_{n}^{(p)}(\varrho) f(\varrho)\dd \varrho
    = {\cal T}_{n,+}^{(p)} - {\cal T}_{n,-}^{(p)} ,\quad\mbox{where}\\
      &{\cal T}_{n,\pm}^{(p)}
        :=2^4\ga^2\sum_{1\le j\le M(n+1)^{1/2}}
        \int_{\varrho_j}^{\varrho_{j+1}}f(\varrho)\dd
        \varrho 
        \sum_{j'=1}^{n}   (1+ \delta_{0,j\pm j'} )^{1/2} 
       \cos\Big(\frac{\pi k_{j'}}{2}\Big)\notag\\
  &\qquad
    \times \sin^3\Big(\frac{\pi k_{j}}{2}\Big)\sin^3\Big(\frac{\pi k_{j'}}{2}\Big)
    \Delta^{-1}(k_j,k_{j'})  \widehat{\frak E}_n(t,j\pm j') ,\notag
\end{align}
with $\Delta(k_j,k_{j'})$  defined in \eqref{k-j1}.  Since $ \pi k_j =
      \frac{\varrho_j}{(n+1)^{1/2} } $ and ${\rm supp}f\subset [\delta,M]$ for some
$0<\delta<M<+\infty$,
       the summation in \eqref{013112-24} has been restricted to $1\le j'
      \le 100M(n+1)^{1/2}$.
       
Using  Lemma \ref{lm011712-25} and repeating the calculations leading
to the estimate of
 $\bar\theta^{(o)}_{pr,+}(\varphi';n)$ in Section \ref{sec:proof-bulk-term}, we
 conclude that ${\cal T}_{n,+}^{(p)} =o_n(1)$.
 Consider now ${\cal  T}_{n,-}^{(p)}$.  We have $\sin^3\Big(\frac{\pi
        k_{j}}{2}\Big)\approx \frac{\varrho^3}{2^3(n+1)^{3/2} }$ for $\varrho\in[\varrho_j, \varrho_{j+1})$. Denoting $\ell:=j-j'$ we can write
      \begin{align}
  \label{013112-24a}
      & {\cal T}_{n,-}^{(p)}(t )
       =  -4\ga^2  \sum_{1\le j\le M(n+1)^{1/2}} \sum_{1-j\le
        \ell \le 100M(n+1)^{1/2}-j}     (1+ \delta_{0,\ell} )^{1/2}  \notag\\
  &
    \times  
    \frac{\varrho_{j+\ell}^3}{\Delta''_\ell(\varrho_{j}, \varrho_{j+\ell})}
    \int_{\varrho_j}^{\varrho_{j+1}}\varrho^3f(\varrho)\dd \varrho 
   \; \hat{\frak E}_n(t,\ell)  +o_n(1),\qquad\mbox{with}\\
        &
      \Delta''_\ell(\varrho , \varrho ')  :=   (\varrho +\varrho ')^2 (\ell \pi)^2 +2\ga^2 
    \varrho^2 (\varrho ')^2( \varrho+\varrho ')^2. \notag
      \end{align}
The last expression can be rewritten as 
                      ${\cal T}_{n,\le}^{(p)}+{\cal T}_{n,>}^{(p)}$,
where ${\cal T}_{n,\le}^{(p)}$, ${\cal T}_{n,>}^{(p)}$ correspond to the summation over
$|\ell|\le n^{1/4}$ and $|\ell|> n^{1/4}$, respectively.
We have
$$
\frac{\varrho_{j+\ell}^3}{(\varrho_{j}+\varrho_{j+\ell})^2}\le C,\quad
\mbox{for }
1\le j,\, j+ \ell \le 100M(n+1)^{1/2},
$$
therefore 
\begin{align*}
                        |{\cal T}_{n,>}^{(p)}| \le     C \sum_{1\le j\le M(n+1)^{1/2}} 
                      \int_{\varrho_j}^{\varrho_{j+1}}\varrho^3|f(\varrho)|\dd
    \varrho  \sum_{|\ell|>n^{1/4} } \frac{1}{\ell^2 } \to0,
\end{align*}
as $n\to+\infty$.
As a result
  \begin{align}
  \label{013112-24b}
      & {\cal T}_{n,-}^{(p)}
        =- \sum_{\ell=-\infty}^{+\infty}  (1+ \delta_{0,\ell} )^{1/2}  \widehat{\frak E}_n(\ell)
        \int_{0}^{+\infty}\frac{\ga^2  \varrho^4f(\varrho)\dd  \varrho  }{ (\ell\pi)^2 
    +\ga^2  \varrho^4    }+o_n(1). 
      \end{align}

        \subsubsection{Proof of \eqref{eq:74}}
\label{sec5.4.2}

       We have
\begin{align*}
  & \int_0^{+\infty} {\cal    P}^{(p)}_{n,p}(\varrho) f(\varrho)\dd \varrho
    =\sum_{z=0,n}\big({\rm I}_{n,z}+{\rm II}_{n,z}\big),
\\
    &
    {\rm I}_{n,z}=  -2\ga\tilde \ga  
      \sum_{0\le j\le M(n+1)^{1/2}}  \sum_{j'=1}^n \int_0^t  {\cal b}^{(p,z)}_{n,+}(s,\varrho_{j'})\dd s \int_{\varrho_j}^{\varrho_{j+1}}f(\varrho)
      \theta_{j,j'}\psi_{j}(z) \psi_{j'}(0)
     \dd \varrho
   \\
  &
    {\rm II}_{n,z}=-2\ga\tilde \ga  
     \sum_{0\le j\le M(n+1)^{1/2}} \sum_{j'=0}^n \int_0^t\dd s \psi_{j'}(z) \psi_{j'}(0) \theta_{j,j'} 
     \int_{\varrho_j}^{\varrho_{j+1}}f(\varrho) {\cal b}^{(p,z)}_{n,+}( s,\varrho)\dd \varrho,\\
    &
\mbox{where}  \qquad \theta_{j,j'}   = 
     \frac{  \la_{j}
    \la_{j'} }{(\la_j-\la_{j'})^2+2\ga^2
    \la_j\la_{j'}(\la_j+\la_{j'})}  .
\end{align*}

\subsubsection*{Calculation of  ${\rm II}_{n,0}$}

For $\ell=j'-j$ fixed we can write
\begin{align*}
 & \theta_{j,j+\ell}  \psi_{j+\ell}^2(0) =\frac{2 }{n+1 } \cdot
    \frac{\varrho_j^2\varrho_{j'}^2}
    {(  \varrho_{j}-\varrho_{j'})^2 (  \varrho_{j}+\varrho_{j'})^2 
    +\frac{2\ga^2}{(n+1)}
     \varrho_{j} ^2  \varrho_{j'}^2
   ( \varrho_{j} ^2  +\varrho_{j'}^2)}\\
  &
    = 
    \frac{\varrho^2_j }
    {  2[  (\pi \ell)^2 
     + \ga^2 \varrho^4_j] }+o_n(1)
\end{align*}
and, as can be easily seen for  fixed $0<\delta<M$ we have
$$
0\le \theta_{j,j+\ell}  \psi_{j+\ell}^2(0) \le\frac{C}{1+\ell^2},\quad \ell\in\bbZ, 
$$
with the constant $C>0$ independent of $n$. 
Therefore, we can
     write
   \begin{align*}  &
    {\rm II}_{n,0}=-\ga\tilde \ga  
     \sum_{0\le j\le M(n+1)^{1/2}} 
     \int_0^t\dd s \int_{\varrho_j}^{\varrho_{j+1}}f(\varrho) {\cal b}^{(p,0)}_{n,+}( s,\varrho)\dd \varrho\Big(\sum_{\ell= -j}^{M(n+1)^{1/2}-\ell} 
    \frac{ \varrho^2 }
    {  (\pi \ell)^2 
     + \ga^2 \varrho^4 }\Big)+o_n(1)
 \end{align*}
 \begin{align*}
   &
     = 
       -\ga\tilde \ga    \sum_{0\le j\le M(n+1)^{1/2}} 
     \int_0^t\dd s \int_{\varrho_j}^{\varrho_{j+1}}f(\varrho) {\cal b}^{(p,0)}_{n,+} ( s,\varrho)\dd \varrho\Big(\sum_{\ell\in\bbZ} 
    \frac{ \varrho^2 }
    {  (\pi \ell)^2 
     + \ga^2 \varrho^4 }\Big)+o_n(1).
\end{align*}
Using the well known formula
$$
\sum_{\ell\in\bbZ} 
    \frac{ A }
    {  (\pi \ell)^2 
     + A^2 }=\coth(A).
$$
We conclude that
\begin{align}
  \label{IIn0}
    {\rm II}_{n,0}=- \tilde \ga  
          \int_0^t\dd s \int_{0}^{+\infty}f(\varrho) \coth(\ga\varrho^2) {\cal b}^{(p,z)}_{n,+}( s,\varrho)\dd \varrho.
 \end{align}

 \subsubsection*{Calculation of  ${\rm II}_{n,n}$}

For $j'=j+\ell$ and $\ell$ fixed
   we can write
\begin{align*}
  &
   \theta_{j,j+\ell}  \psi_{j+\ell} (0) \psi_{j+\ell} (n)=
    \theta_{j,j+\ell}  \psi_{j+\ell}^2(0) (-1)^{j+\ell}\\
  &
    =\frac{2 }{n+1 } \cdot
           \frac{(-1)^{j+\ell}\varrho_j^2\varrho_{j'}^2}
    {(  \varrho_{j}-\varrho_{j'})^2 (  \varrho_{j}+\varrho_{j'})^2 
    +\frac{2\ga^2}{(n+1)}
     \varrho_{j} ^2  \varrho_{j'}^2
    ( \varrho_{j} ^2  +\varrho_{j'}^2)} +o_n(1)\\
  &
    =    
           \frac{(-1)^{j+\ell}\varrho_j^2}
    {2[(\pi \ell)^2  +
     \ga^2 
     \varrho_{j} ^4 ] } +o_n(1)
\end{align*}
As a result, repeating the calculations made for ${\rm II}_{n,0}$ and
using the formula
$$
\sum_{\ell\in\bbZ} 
    \frac{ A(-1)^{\ell} }
    {  (\pi \ell)^2 
     + A^2 }={\rm csch}\,(A),
$$
we get
\begin{align*}
&
  {\rm II}_{n,n} 
     =- \tilde \ga \int_0^t\dd s 
     \int_{0}^{+\infty}{\rm csch}\,(\ga\varrho^2) f(\varrho) {\cal b}^{(p,n)}_{n,-}( s,\varrho)\dd \varrho+o_n(1),
 \end{align*}
   where ${\rm csch}\,(u)=\frac{1}{\sinh u}$ is  the hyperbolic cosecant.

\subsubsection*{Calculation of  ${\rm I}_{n,0}$}

For  $j'=j+\ell$ and a fixed
$\ell$ we can write
\begin{align*}
  &   
    \psi_{j'}^2(0) \theta_{j'-\ell,j'}  =  \frac{\varrho^2_{j'} }
    {  2[  (\pi \ell)^2 
     + \ga^2 \varrho^4_{j'}] }+o_n(1).
\end{align*}
Hence  
\begin{align*}
  &
      {\rm I}_{n,0}
   =- \ga\tilde \ga    \sum_{j'=0}^{ 2M(n+1)^{1/2}}     \sum_{\ell=j'-2M(n+1)^{1/2}}^{j'}
    \frac{ \varrho_{j'}^2}
    {( \pi\ell)^2 
    + \ga^2  \varrho_{j'}^4}  \int_0^t  \int_{\varrho_j'}^{\varrho_{j+1}'} {\cal b}^{(p,0)}_{n,+}(s,\varrho_{j'}) \dd s
     f(\varrho')\dd
    \varrho' +o_n(1)
\end{align*}
As a result we have 
\begin{align*}
  &
      {\rm I}_{n,0}
   =- \ga\tilde \ga    \sum_{j'=0}^{ 2M(n+1)^{1/2}}     \sum_{\ell\in\bbZ}
     \int_0^t  \int_{\varrho_j'}^{\varrho_{j+1}'} \frac{ (\varrho')^2}
    {( \pi\ell)^2 
    + \ga^2  (\varrho')^4}{\cal b}^{(p,0)}_{n,+}(s,\varrho') \dd s
     f(\varrho')\dd
    \varrho' +o_n(1)
\\
&=- \tilde \ga  
          \int_0^t\dd s \int_{0}^{+\infty}f(\varrho) \coth(\ga\varrho^2) {\cal b}^{(p,0)}_{n,+}( s,\varrho)\dd \varrho.
 \end{align*}

Conducting a similar calculation for ${\rm I}_{n,n}$ we obtain that,
due to the cancelation, appearing in the same way as in the case of
${\rm II}_{n,n}$ that
\begin{align*}
  &
      {\rm I}_{n,n}
 =- \tilde \ga  
          \int_0^t\dd s \int_{0}^{+\infty}f(\varrho) {\rm csch}\,(\ga\varrho^2) {\cal b}^{(p,n)}_{n,-}( s,\varrho)\dd \varrho.
 \end{align*}

Summarizing,   we have shown   that
 \begin{align}
    \label{010701-25}
   & \int_0^{+\infty} {\cal    P}^{(p)}_{n,p}(\varrho) f(\varrho)\dd \varrho
     =- 2\tilde \ga \int_0^t\dd s  \int_{0}^{\infty} {\cal
     b}^{(p,0)}_{n,+}(s,\varrho) \coth(\ga\varrho^2) f(\varrho)\dd
     \varrho
   \notag\\
   &
     - 2\tilde \ga \int_0^t\dd s  \int_{0}^{\infty} {\cal b}^{(p,n)}_{n,-}(s,\varrho) {\rm csch}\,(\ga\varrho^2) f(\varrho)\dd \varrho +o_n(1)
 \end{align}
and \eqref{eq:74} follows.

     \subsubsection{Calculation of $ {\cal P}^{(p)}_{n,pr}(\varrho)$}

\label{sec5.4.3}

   We have (cf \eqref{Xi}, \eqref{Pi})
\begin{align*}
  & \int_0^{+\infty} {\cal    P}^{(p)}_{n,pr}(\varrho) f(\varrho)\dd \varrho
    =\sum_{z=0,n}\big({\rm I}_{n,z}+{\rm II}_{n,z}\big)    ,\quad\mbox{where}\\
  &
    {\rm I}_{n,z}=  \tilde \ga 
    \sum_{j=0}^{M(n+1)^{1/2}}  \sum_{j'=1}^n \psi_{j}(z) \psi_{j'}(0)
    \theta^{(p,pr)}_{j,j'}  \int_0^t  {\cal b}^{(pr,z)}_{n}(s,
    \varrho_{j'}) \dd s
    \int_{\varrho_{j}}^{\varrho_{j+1}}f(\varrho)\dd
    \varrho,  
\end{align*}
\begin{align}
  \label{thjj}
  &
    {\rm II}_{n,z}=\tilde \ga \sum_{j=0}^{M(n+1)^{1/2}}  \sum_{j'=0}^n \psi_{j'}(0) \psi_{j'}(z)
    \theta^{(p,pr)}_{j',j}
  \int_0^t\dd s   \int_{\varrho_{j}}^{\varrho_{j+1}}  {\cal
    b}^{(pr,z)}_{n}(s,\varrho)  f(\varrho)\dd    \varrho \notag\\
  &
\mbox{and}  \qquad  \theta^{(p,pr)}_{j,j'}:= \frac{ \la_{j'}^{1/2}(\la_{j'}
    -\la_j)  }{(\la_j-\la_{j'})^2+2\ga^2
    \la_j\la_{j'}(\la_j+\la_{j'})}.
\end{align}

\subsubsection*{Calculation of  ${\rm I}_{n,0}+{\rm II}_{n,0}$}

For $j'=j+\ell$, $\ell$ fixed
and $\kappa\in(0,1/100)$  we can write
\begin{align*}
 & \psi_{j+\ell}^2(0)  
    \theta^{(p,pr)}_{j,j+\ell} =\frac{2}{n+1}\cdot\frac{\varrho_{j'} (\varrho_j-\varrho_{j+\ell})
    (\varrho_j+\varrho_{j+\ell}) } 
      {(  \varrho_{j}-\varrho_{j+\ell})^2 (  \varrho_{j}+\varrho_{j+\ell})^2 
    +\frac{2\ga^2}{(n+1)}
     \varrho_{j} ^2  \varrho_{j+\ell}^2
   ( \varrho_{j} ^2  +\varrho_{j+\ell}^2)}   \\
  &
    = \frac{-\pi\ell
      } 
      {(n+1)^{1/2}[(  \pi\ell)^2  
    + \ga^2
     \varrho_{j} ^4] }+o_n(1)
\end{align*}
and
$$
\psi_{j+\ell}^2(0)  
|\theta^{(p,pr)}_{j,j+\ell}|\le \frac{C|\ell|}{(n+1)^{1/2}(1+\ell^2)}
$$
Taking this into account we can write
  \begin{align*}
  &
    {\rm II}_{n,0}= -\frac{ \tilde \ga}{(n+1)^{1/2}}   
              \sum_{j=1}^{ M(n+1)^{1/2+\kappa }}\sum_{\ell=-j}^{ M(n+1)^{1/2+\kappa }-j} \frac{\pi\ell
      } 
      {(  \pi\ell)^2  
    + \ga^2
     \varrho_{j} ^4 } \\
    &
      \times  \varrho_{j}\int_0^t\dd s \int_{\varrho_{j}}^{\varrho_{j+1}}{\cal
    b}^{(pr,z)}_{n}(s,\varrho) f(\varrho)\dd    \varrho +o_n(1).
  \end{align*}
  Since ${\rm supp}\,f\subset[\delta,M]$ for some $0<\delta<M$, the summation over $j$ extends
  only from $0$ to $M(n+1)^{1/2+\kappa }/2$.  Performing the summation
  over $\ell$ yields
  \begin{align*}
  &
    {\rm II}_{n,0}= -\frac{ \tilde \ga}{(n+1)^{1/2}}   
              \sum_{j=1}^{ M(n+1)^{1/2+\kappa }}\sum_{\ell=j+1}^{ M(n+1)^{1/2+\kappa }-j} \frac{\pi\ell
      } 
      {(  \pi\ell)^2  
    + \ga^2
     \varrho_{j} ^4 } \\
    &
      \times  \varrho_{j}\int_0^t\dd s \int_{\varrho_{j}}^{\varrho_{j+1}}{\cal
    b}^{(pr,z)}_{n}(s,\varrho) f(\varrho)\dd    \varrho +o_n(1).
  \end{align*}
  Hence
  \begin{align*}
  &
   | {\rm II}_{n,0}| \le \frac{ C}{(n+1)^{1/2}}   
              \sum_{j=1}^{ M(n+1)^{1/2+\kappa }}\sum_{\ell=j+1}^{ M(n+1)^{1/2+\kappa }-j} \frac{1
      } 
      {  \pi\ell 
    } \\
    &
      \times  |\varrho_{j}|\int_0^t\dd s \int_{\varrho_{j}}^{\varrho_{j+1}}|{\cal
    b}^{(pr,z)}_{n}(s,\varrho)| |f(\varrho)|\dd    \varrho +o_n(1).
  \end{align*}
  \begin{align*}
  &
   \le
              \sum_{j=1}^{ M(n+1)^{1/2+\kappa }}\frac{ \log
    j}{(n+1)^{1/2}} s \int_0^t\dd s \int_{\varrho_{j}}^{\varrho_{j+1}}|{\cal
      b}^{(pr,z)}_{n}(s,\varrho)| |f(\varrho)|\dd    \varrho +o_n(1)\\
    &
      \le \frac{C\log n}{n} \int_0^t\dd s \int_{0}^{+\infty}|{\cal
      b}^{(pr,z)}_{n}(s,\varrho)| |f(\varrho)|\dd    \varrho +o_n(1),
  \end{align*}
  which tends to $0$, as $n\to+\infty$.

 On the other hand
\begin{align*}&
{\rm I}_{n,0} =   \frac{2\tilde\ga}{\pi}
   \sum_{j,j'=1}^{M(n+1)^{1/2+\kappa}}
  \varrho_{j'}\int_0^t\dd s \int_{\varrho_{j'}}^{\varrho_{j'+1}}
   {\cal b}^{(pr,0)}_{n}( s,\varrho') \dd \varrho'   \int_{\varrho_{j}}^{\varrho_{j+1}}f(\varrho') \dd
    \varrho' \\
     &
    \times  
    \frac{(\varrho_{j'}-\varrho_{j})
    (\varrho_j+\varrho_{j'}) } 
      {(  \varrho_{j}-\varrho_{j'})^2 (  \varrho_{j}+\varrho_{j'})^2 
    +\frac{2\ga^2}{(n+1)}
     \varrho_{j} ^2  \varrho_{j'}^2
    ( \varrho_{j} ^2  +\varrho_{j'}^2)} +o_n(1).
\end{align*}
Fixing $\ell=j'-j$ we get
\begin{align*}
   & \Big(\int_{\varrho_{j}}^{\varrho_{j+1}}f(\varrho') \dd
    \varrho'  \Big) \frac{\varrho_{j'} (\varrho_{j'}-\varrho_{j})
    (\varrho_j+\varrho_{j'}) } 
      {(  \varrho_{j}-\varrho_{j'})^2 (  \varrho_{j}+\varrho_{j'})^2 
    +\frac{2\ga^2}{(n+1)}
     \varrho_{j} ^2  \varrho_{j'}^2
     ( \varrho_{j} ^2  +\varrho_{j'}^2)}
  \\
   &
     = \frac{\pi\ell
      f(\varrho_{j'})} 
      {(n+1)[(  \pi\ell)^2  
    + \ga^2
     \varrho_{j'} ^4] }+o_n(1)
\end{align*}
and, as an be easily seen for $\varrho_j\in[\delta,M]$ we have
\begin{align*}
   & \Big|\int_{\varrho_{j}}^{\varrho_{j+1}}f(\varrho') \dd
    \varrho'  \Big| \frac{\varrho_{j'} |\varrho_{j'}-\varrho_{j}|
    (\varrho_j+\varrho_{j'}) } 
      {(  \varrho_{j}-\varrho_{j'})^2 (  \varrho_{j}+\varrho_{j'})^2 
    +\frac{2\ga^2}{(n+1)}
     \varrho_{j} ^2  \varrho_{j'}^2
     ( \varrho_{j} ^2  +\varrho_{j'}^2)}
  \\
   &
     \le  \frac{C |\ell|
       } 
      {(n+1)[ (\ell)^2  
    +1] }.
\end{align*}
The same argument as in the case of ${\rm II}_{n,0}$ allows us to
conclude that ${\rm I}_{n,0}\to0$, as $n\to+\infty$.

      In the case of  ${\rm I}_{n,n}      +  {\rm II}_{n,n}  $, the presence
of   factors $\psi_{j}(n)$ and  $\psi_{j'}(n)$ does not influence the
situation. As a result we get ${\rm I}_{n,n}      +  {\rm II}_{n,n}
\to0 $. Thus \eqref{eq:75} follows.

\subsubsection{Calculation of $ {\cal X}_{n,\iota}^{(p)}(\varrho)$,
  $\iota=p,pr,rp,r$}

\label{sec6.1.4}

Recall that ${\rm supp}f\subset [\delta,M]$ for some
$0<\delta<M<+\infty$.  We have
   \begin{align*}
    & \int_0^{+\infty}{\cal X}_{n,p}^{(p)}(\varrho) f(\varrho)\dd \varrho\\
     &
       =
     -\frac{(n+1)^{1/2}}{n^{3/2}}\sum_{0\le j\le M(n+1)^{1/2}} \sum_{j'=0}^n\Xi^{(p)}_{p}(\la_j,\la_{j'})
       \psi_{j'}(0) \delta_{0,t}{\tilde S}_{j,j'}^{(p)}
       \int_{\frac{j\pi}{(n+1)^{1/2}}}^{\frac{(j+1)\pi}{(n+1)^{1/2}}}f(\varrho)\dd
    \varrho .
\end{align*}
Substituting for $\Xi^{(p)}_{p}(\la_j,\la_{j'})$ from \eqref{Xi},  we can
write that the right hand side equals  ${\rm I}_{n}+o_n(1)$, where 
\begin{align*}
       &
    {\rm I}_{n}   = -\frac{2\ga}{n(n+1)^{1/2}} \sum_{0\le j, j'\le
         100M(n+1)^{1/2}}   \sin^2\Big(\frac{k_j\pi}{2 }\Big)
    \sin^2\Big(\frac{k_{j'}\pi}{2 }\Big)  \cos
   \Big(\frac{k_{j'}\pi}{2 }\Big)  \delta_{0,t}{\tilde S}_{j,j'}^{(p)}
    \\
  &
    \times\Delta^{-1}(k_j,k_{j'}) \int_{\frac{j\pi}{(n+1)^{1/2}}}^{\frac{(j+1)\pi}{(n+1)^{1/2}}}f(\varrho)\dd
    \varrho     .
\end{align*}

 Hence,
   \begin{align}
     \label{020901-25}
       &
   | {\rm I}_{n}|\le    \frac{C }{n^{3/2} }\sum_{0\le j\le
         M(n+1)^{1/2}} \sum_{0\le j'\le 100M(n+1)^{1/2}}
         \big| \delta_{0,t}{\tilde S}_{j,j'}^{(p)}
         \big|\int_{\frac{j\pi}{(n+1)^{1/2}}}^{\frac{(j+1)\pi}{(n+1)^{1/2}}}|f(\varrho)|\dd
     \varrho \notag
    \\
  &
    \times\frac{\varrho_j^2 \varrho_{j'}^2}{(\varrho_j-\varrho_{j'})^2 (\varrho_j+\varrho_{j'})^2 +\frac{1}{n+1}(\varrho_j\varrho_{j'})^2 (\varrho_j ^2+\varrho_{j'}^2) }
\\
       &
         \le   \frac{C }{n^{2} }  \sum_{0\le j, j'\le 100M(n+1)^{1/2}}
         \big| \delta_{0,t}{\tilde S}_{j,j'}^{(p)}
         \big|      
         ={\rm I}_{n,\le}^{(1)}+{\rm I}_{n,\le}^{(2)}+{\rm I}_{n,\le}^{(3)},\notag
   \end{align}
   where the terms of the summation on the utmost right hand side correspond to the cases $n^{1/4}<|j-j'|$,
   $1\le |j-j'|\le n^{1/4}$ and  $j=j'$.

The term ${\rm I}_{n,\le}^{(1)}$ can be estimated using the
Cauchy-Schwarz inequality and \eqref{eq:9} as follows
  \begin{align*}
       &
         {\rm I}_{n,\le}^{(1)} \le   \frac{C }{n^{3/2} } (n+1)^{1/2} \sum_{0\le j, j'\le 100M(n+1)^{1/2}}
         \big| \delta_{0,t}{\tilde S}_{j,j'}^{(p)}
         \big|  \int_{\frac{j\pi}{(n+1)^{1/2}}}^{\frac{(j+1)\pi}{(n+1)^{1/2}}}|f(\varrho)|\dd
     \varrho \\
       &
         \le       \frac{C }{n }\Big(\sum_{0\le j, j'\le 100M(n+1)^{1/2}}      \big[
         \delta_{0,t}{\tilde S}_{j,j'}^{(p)}
         \big]^2 \Big)^{1/2}\le \frac{C}{n^{1/2}}.
  \end{align*}

  Concerning ${\rm I}_{n,\le}^{(3)}$ we have 
  \begin{align}
    \label{011301-25}
       &
 {\rm I}_{n,\le}^{(3)} \le   \frac{C }{n^{2} } \sum_{\delta(n+1)^{1/2}\le j\le 100M(n+1)^{1/2}}   \varrho_j^4
      \bbE_n\big[\tilde p_j^2(t) + \tilde p_j^2(0)  \big]   \Big( \frac{ \varrho_j^6 }{n+1}
        \Big)^{-1}\notag\\
       &
         \le  C \sum_{\delta(n+1)^{1/2}\le j\le 100M(n+1)^{1/2}}    \frac{
      \bbE_n\big[\tilde p_j^2(t) + \tilde p_j^2(0) \big]  }{j^2}.
         \end{align}
From Corollary \ref{cor011912-24} it follows that
    for any $t_*>0$ there exists $C>0$ such that
  \begin{equation}
    \label{011201-25}
    \frac{1}{n}
    \sum_{j=0}^{ n}
           \Big[ \Big(\bbE_n[\tilde
    p_{j }^2(t)]\Big)^2+\Big(\bbE_n[\tilde
    r_{j }^2(t)]\Big)^2\Big]\le C,\quad n=1,2,\ldots,\,t\in[0,t_*].
    \end{equation}
  Using  \eqref{011201-25} and the Cauchy-Schwarz
  inequality we can write
  \begin{align*}
  &
       {\rm I}_{n,\le}^{(3)} 
         \le C \left\{\sum_{0\le j\le 100M(n+1)^{1/2}}    \frac{
      1  }{j^4} \right\}^{1/2}\left\{\sum_{\delta(n+1)^{1/2}\le j\le 100M(n+1)^{1/2}}    
      \big(\bbE_n\big[\tilde p_j^2(t) 
         \big]\big)^2 \right\}^{1/2}\\
       &
         \le \frac{C}{n^{3/4}} \left\{\sum_{0\le j\le n}    
      \big(\bbE_n\big[\tilde p_j^2(t) 
         \big]\big)^2 \right\}^{1/2}\le \frac{C}{n^{3/4}}
   (n+1)^{1/2}\to0,\quad n\to+\infty.
\end{align*}
Finally,
 \begin{align*}
       &
 {\rm I}_{n,\le}^{(2)} \le   \frac{C }{n^{2} }\sum_{\delta(n+1)^{1/2}\le j\le
         100 M(n+1)^{1/2}} \mathop{\sum_{|\ell|\le (n+1)^{1/4 }} }_{0\le j+\ell\le
         M(n+1)^{1/2}}  \frac{|j|^2|j+\ell|^2}{|\ell|^2 |2j+\ell|^2}
    \big|\bbE_n\big[\tilde p_j(t)\tilde
   p_{j+\ell}(t)\big]\big|
 \end{align*}
 \begin{align*}
       &
 \le    C\sum_{\delta(n+1)^{1/2}\le j\le
         100 M(n+1)^{1/2}} \mathop{\sum_{1\le|\ell|\le (n+1)^{1/4 }} }_{0\le j+\ell\le
         M(n+1)^{1/2}}  \frac{\big|\bbE_n\big[\tilde
         p_j^2(t)\big]\big|+\big|\bbE_n\big[\tilde
         p_{j+\ell}^2(t)\big]\big|}{|\ell|^2 |2j+\ell|^2}\\
       &
         \le    C\Big(\sum_{1\le |\ell|\le (n+1)^{1/4 }} \frac{1}{|\ell|^2}\Big)\sum_{\delta(n+1)^{1/2}\le j\le
         100 M(n+1)^{1/2}} \frac{\big|\bbE_n\big[\tilde
         p_j^2(t)\big]\big|}{ j^2}
         \le \frac{C}{n^{1/4}}. 
   \end{align*}

 In a similar fashion one can also show that
$\int_0^{+\infty}{\cal X}_{r,n}^{(p)}(\varrho) f(\varrho)\dd \varrho\to0$,
as $n\to+\infty$.

On the other hand,
   \begin{align*}
     &  \int_0^{+\infty}\big({\cal X}_{pr,n}^{(p)}(\varrho) +{\cal X}_{rp,n}^{(p)}(u)\big)
       f(\varrho)\dd \varrho =
       -\frac{(n+1)^{1/2}}{n^{3/2}}\sum_{0\le j\le M(n+1)^{1/2}}
       \int_{\frac{j\pi}{(n+1)^{1/2}}}^{\frac{(j+1)\pi}{(n+1)^{1/2}}}f(\varrho)\dd
     \varrho\\
     &
     \times  \sum_{j'=0}^n\Big(\Xi^{(p)}_{pr}(\la_j,\la_{j'})
             \delta_{0,t}{\tilde
     S}_{j,j'}^{(pr)} +\Xi^{(p)}_{rp}(\la_j,\la_{j'}) \delta_{0,t}{\tilde
    S}_{j,j'}^{(rp)}\Big)    \psi_{j'}(0).
   \end{align*}
Choosing $\kappa>0$, to be determined later on, and using the fact
that $\Xi^{(p)}_{pr}(\la_j,\la_{j'})=\Xi^{(p)}_{rp}(\la_{j'}, \la_j) $
we conclude that the left hand side equals   $  {\rm J}_{n}^{(1)}+
{\rm J}_{n}^{(2)} +o_n(1)$, where 
 \begin{align*}
               &
                 {\rm J}_{n}^{(1)}   =-\frac{2^{1/2}}{n(n+1)^{1/2}}\sum_{0\le j,j'\le
       M(n+1)^{1/2+\kappa}}  \Xi^{(p)}_{pr}(\la_j,\la_{j'})
             \delta_{0,t}{\tilde S}_{j,j'}^{(pr)}\int_{\frac{j\pi}{(n+1)^{1/2}}}^{\frac{(j+1)\pi}{(n+1)^{1/2}}}f(\varrho)\dd
       \varrho, \\
   &
    {\rm J}_{n}^{(2)}:=   -\frac{2^{1/2}}{n(n+1)^{1/2}}\sum_{0\le j,j'\le
       M(n+1)^{1/2+\kappa}}  \Xi^{(p)}_{rp}(\la_{j},\la_{j'}) \delta_{0,t}{\tilde
    S}_{j,j'}^{(rp)}\int_{\frac{j\pi}{(n+1)^{1/2}}}^{\frac{(j+1)\pi}{(n+1)^{1/2}}}f(\varrho)\dd
       \varrho .
 \end{align*}
Using the definitions of  $\Xi^{(p)}_{pr}(\la_j,\la_{j'})$ and
$\Xi^{(p)}_{rp}(\la_j,\la_{j'})$ (see \eqref{Xi}) we conclude that
   \begin{align*}
     &
      |{\rm J}_{n}^{(1)}|\le  \frac{C}{n^2}\sup_{t\in[0,t_*]}\sum_{0\le j,j'\le
       M(n+1)^{1/2+\kappa}} \Big|\sin\Big(\frac{k_{j'}\pi}{2 }\Big) \Big| \\
         &
           \times\Big|\sin\Big(\frac{(k_j-k_{j'})\pi}{2 }\Big)
           \sin\Big(\frac{(k_j+k_{j'})\pi}{2 }\Big)\Big|\frac{\big|\bbE_n \big[\tilde r_{j'}( t) \tilde
       p_j( t)\Big]\big|}{\Delta(k_j,k_{j'})}
            \end{align*}

    \begin{align*}
          &
       \le \frac{C}{n}\sup_{t\in[0,t_*]}\sum_{0\le j,j'\le
       M(n+1)^{1/2+\kappa}}
            \frac{j'|j-j'|(j+j')}{(j-j')^2(j+j')^2}\big|\bbE_n
            \big[\tilde r_{j'}( t) \tilde
            p_j( t)\Big]\big|\\
          &
            \le \frac{C}{n}\sup_{t\in[0,t_*]}\sum_{0\le j,j'\le
       M(n+1)^{1/2+\kappa}} \frac{\big|\bbE_n \big[\tilde r_{j'}( t) \tilde
            p_j( t)\Big]\big|}{|j-j'|+1}.
    \end{align*}
    We can estimate the last expression using the Cauchy-Schwarz
    inequality and estimate \eqref{010901-25}. Hence,
    \begin{align*}
                    &
            |{\rm J}_{n}^{(1)}| \le \frac{C}{n}\sup_{t\in[0,t_*]}\left\{\sum_{0\le j,j'\le
       M(n+1)^{1/2+\kappa}}  \big|\bbE_n \big[\tilde r_{j'}(n^{3/2}t) \tilde
                      p_j(n^{3/2}t)\Big]\big|^2 \right\}^{1/2}\\
                    &
                      \times \left\{\sum_{0\le j,j'\le
                      M(n+1)^{1/2+\kappa}}
                      \frac{1}{(|j-j'|+1)^2}\right\}^{1/2} \le \frac{C}{n^{1/4-\kappa/2}}\to0,
    \end{align*}
    provided $\kappa<1/2$.  Similarly, we have 
   $
      {\rm J}_{n}^{(2)}\to0 $, as $n\to+\infty$.
   This ends the proof of Proposition  \ref{prop010901-25}.\qed

   \subsubsection{Summary}

   We have shown the following.
  \begin{align}
    \label{010110-26}
   & \int_0^t\dd s  \int_{0}^{\infty} {\cal
     b}^{(p,0)}_{n,+}(s,\varrho)  [1+2\tilde \ga \coth(\ga\varrho^2)] f(\varrho)\dd
     \varrho\\
    &
     = \sqrt{2} \int_{0}^{+\infty}\Bigg(t T_0- \sum_{\ell=0}^{+\infty} 
   \frac{\ga^2  \varrho^4 c_\ell(v) \widehat{\frak E}_n(t,\ell)}{ (\ell\pi)^2 
     +\ga^2 \varrho^4    } \Bigg)f(\varrho)\dd \varrho \notag\\
   &
     - 2\tilde \ga \int_0^t\dd s  \int_{0}^{\infty} {\cal b}^{(p,n)}_{n,-}(s,\varrho) {\rm csch}\,(\ga\varrho^2) f(\varrho)\dd \varrho +o_n(1).\notag
 \end{align} Substituing $[1+2\tilde \ga \coth(\ga\varrho^2)] f(\varrho)$
 in place $f(\varrho)$ we conclude Proposition \ref{prop010901-25}.\qed

   \subsection{Asymptotics of $\int_0^t\dd s  \int_{0}^{\infty} {\cal
     b}^{(p,1-v)}_{n,-}(s,\varrho) $
     : proof of Proposition \ref{prop011001-25}}

We could repeat the argument made in Section \ref{sec10.1}, but
instead note that using the transformation $x\mapsto n-x$ we can
invert the roles of the left and right endpoints of the chain. With
this observation we can conclude \eqref{030901-25} directly from \eqref{030901-25a}.\qed

\subsection{Proof of Theorem \ref{thm011001-25}}
\label{sec9.3}

We consider only the case $v=0$, as the case $v=1$ can be dealt with
analogously. Combining the results of Propositions \ref{prop010901-25} and
\ref{prop011001-25} we obtain
       $f\in    C_c(0,+\infty)$, $t>0$  
       we   get
\begin{align*}
       &  \int_0^t\dd s\int_0^{+\infty}{\cal b}^{(p,0)}_{n,+}(s,\varrho)f(\varrho)\Bigg[1-\Big(\frac{2 \tilde\ga  }{\sinh(\ga \varrho^2)+2 \tilde\ga\cosh(\ga \varrho^2)}\Big)^2\Bigg]\dd \varrho   
  \notag
  \\
       &
         =  \int_{0}^{+\infty}\Bigg(t T_L- \sum_{\ell=0}^{+\infty} 
   \frac{\ga^2  \varrho^4 c_\ell(0) \widehat{\frak E}_n(t,\ell)}{ (\ell\pi)^2 
    +\ga^2 \varrho^4    }
         \Bigg)\frac{\sqrt{2} f(\varrho)}{1+2\tilde\ga\coth(\ga \varrho^2)}\dd \varrho \\
   &
     -    \int_{0}^{+\infty}\Bigg(t T_R- \sum_{\ell=0}^{+\infty} 
   \frac{\ga^2  \varrho^4 c_\ell(1) \widehat{\frak E}_n(t,\ell)}{ (\ell\pi)^2 
     +\ga^2 \varrho^4    } \Bigg)\frac{ \sqrt{2} }{1+2\tilde \ga \coth(\ga\varrho^2)}\cdot\frac{2 \tilde\ga f(\varrho)}{\sinh(\ga \varrho^2)+2 \tilde\ga\cosh(\ga \varrho^2)}\dd
     \varrho  +o_n(1)   .\notag
\end{align*}
Using the identity
\begin{align*}
&1-\left(
\frac{2\widetilde\gamma}
{\sinh(\gamma\varrho^2)
 +2\widetilde\gamma\cosh(\gamma\varrho^2)}
\right)^2\\
  &
=
\frac{
\sinh(\gamma\varrho^2)
\left[
(1+4\widetilde\gamma^2)\sinh(\gamma\varrho^2)
+4\widetilde\gamma\cosh(\gamma\varrho^2)
\right]
}{
\left[
\sinh(\gamma\varrho^2)
+2\widetilde\gamma\cosh(\gamma\varrho^2)
\right]^2
}.
\end{align*}
we get
\begin{align}
  \label{020310-26}
       &  \int_0^t\dd s\int_0^{+\infty}{\cal b}^{(p,0)}_{n,+}(s,\varrho)f(\varrho) \frac{
\sinh(\gamma\varrho^2)
\left[
(1+4\widetilde\gamma^2)\sinh(\gamma\varrho^2)
+4\widetilde\gamma\cosh(\gamma\varrho^2)
\right]
}{
\left[
\sinh(\gamma\varrho^2)
+2\widetilde\gamma\cosh(\gamma\varrho^2)
\right]^2}\dd \varrho   
  \notag
  \\
       &
         =  \int_{0}^{+\infty}\Bigg(t T_L- \sum_{\ell=0}^{+\infty} 
   \frac{\ga^2  \varrho^4 c_\ell(0) \widehat{\frak E}_n(t,\ell)}{ (\ell\pi)^2 
    +\ga^2 \varrho^4    }
         \Bigg)\frac{\sqrt{2} f(\varrho)}{1+2\tilde\ga\coth(\ga \varrho^2)}\dd \varrho \\
   &
     -    \int_{0}^{+\infty}\Bigg(t T_R- \sum_{\ell=0}^{+\infty} 
   \frac{\ga^2  \varrho^4 c_\ell(1) \widehat{\frak E}_n(t,\ell)}{ (\ell\pi)^2 
     +\ga^2 \varrho^4    } \Bigg)\frac{ \sqrt{2} }{1+2\tilde \ga \coth(\ga\varrho^2)}\cdot\frac{2 \tilde\ga f(\varrho)}{\sinh(\ga \varrho^2)+2 \tilde\ga\cosh(\ga \varrho^2)}\dd
     \varrho  +o_n(1)   .\notag
\end{align}
 Thus \eqref{061001-25L} follows.\qed

\section{The time-coboundary terms}
\label{sec:proof-os-time}

We prove here Proposition \ref{lem-xi}.
Suppose that $\kappa\in(0,1)$.
Using the rapid decay of $| \widehat{(\varphi ')_o}(\ell)|$ and \eqref{eq:20},  we can write 
\begin{align}
  \label{022312-24}
  &
    \frac{1}{n+1}\xi^{(pr)}_{\iota}(\varphi';n) =\bar
    \xi^{(pr)}_{\iota}( n)
    +o_n(1),\quad\iota=p,r,\quad\mbox{and}\\
  &
    \frac{1}{n}\big(\xi^{(pr)}_{rp}(\varphi';n)  +\xi^{(pr)}_{pr}(\varphi';n)\big)  =
                       \bar \xi^{(pr)}_{pr}(n)+o_n(1),\notag
\end{align}
where
\begin{equation}
  \label{022312-24a}
  \bar \xi^{(pr)}_{\iota}( n) =\sum_{|\ell|\le n^{\kappa}}\widehat{(\varphi ')_o}(\ell)
  \pi\ell \bar \xi^{(pr)}_{\iota}(\ell;n)
\end{equation}
and
\begin{equation}
  \label{041312-24a}
  \begin{split}
           &\bar \xi^{(pr)}_{p}( \ell;n) 
   =  \frac{ 1}{ 2n(n+1)^2  }  \sum_{j'=-n-1}^{n}  \frac{ 
  \cos^2\Big(\frac{\pi k_{j'}}{2}\Big) \delta_{0,t}\tilde
            S_{j'+\ell,j'}^{(p)} }{\Delta'(\ell,k_{j'})}\\
                          &
                \bar \xi^{(pr)}_{pr}( \ell;n)  = \frac{\ga}{    2^{1/2} n(n+1) } \sum_{j'=-n-1}^{n}  
    \frac{      \sin\Big(\frac{\pi k_{j'}}{2}\Big) \sin\Big(\frac{\pi k_{j'}}{2}\Big) \delta_{0,t}\tilde       S_{j'+\ell,j'}^{(p,r)}}{\Delta'(\ell,k_{j'})}
                    ,\\
                &
 \bar \xi^{(pr)}_{r}( \ell;n) =  \frac{1  }{2^{5/2} (n+1)^3
 } \mathop{\sum_{j'=-n-1}^{n}}_{j'\not=0,-\ell}   \frac{  \cos
   \Big(\frac{\pi (k_{j'}+k_\ell)}{2}\Big) \sin
   \Big(\frac{\pi (2k_{j'}+k_\ell)}{2}\Big) \delta_{0,t}\tilde S_{j'+\ell,j'}^{(r) } }{\sin
   \Big(\frac{\pi (k_{j'}+k_\ell)}{2}\Big)  \Delta'(\ell,k_{j'})} , \\
                       &
                        \mbox{and}\quad  \Delta'(\ell,k):= \Big(\frac{\ell\pi}{n+1}\Big)^2 \cos^2
                         \Big(\frac{\pi k}{2}\Big)  
                         +2^4\ga^2  \sin^4
                         \Big(\frac{\pi k}{2}\Big)   .
                \end{split}
\end{equation}
We claim that for each $\iota=p,pr,r$ we have
\begin{equation}
  \label{022112-24a}
  \bar \xi^{(pr)}_{\iota}( n) =o_n(1),\quad\mbox{as}\quad n\to+\infty.
\end{equation}
We show \eqref{022112-24a} for $\iota=p$. The
arguments in the remaining cases are similar.
It suffices only to prove that
\begin{equation}
  \label{022112-24}
  \bar \xi^{(pr)}_{p}(\ell;n) =o_n(1),\quad\mbox{as}\quad n\to+\infty.
\end{equation}
for each $\ell\not=0$. We can write
$\bar \xi^{(pr)}_{p}(\ell;n)=\bar
\xi^{(pr)}_{p,\le}( \ell;n) +\bar \xi^{(pr)}_{p,>}(\ell;n)$, where
the terms on the right correspond to  the summation over $|j'|\le
(n+1)^{3/5}$ and  $(n+1)^{3/5}<|j'|\le
(n+1)$, respectively. The term $\bar \xi^{(pr)}_{p,>}(\ell;n)$  can
be estimated by
\begin{align*}
           |\bar \xi^{(pr)}_{p,>}( \ell;n)|
                    \le   Cn \sum_{(n+1)^{3/5}<|j'|\le
n+1}
                \frac{ |\delta_{0,t}\tilde S_{j'+\ell,j'}^{(p)}|  }{   
          (j')^4 } .
           \end{align*}
           Using   estimate \eqref{011912-24} we get
           \begin{align}
             \label{012708-25}
          &  |\bar \xi^{(pr)}_{p,>}( \ell,n)| \le   Cn^{3/2}  \sum_{(n+1)^{3/5}<|j'|  \le n+1}
                \frac{  1
              }{   
            (j')^4 } 
            \le \frac{C}{n^{3/10}} \to0,
            \end{align}      
as $n\to+\infty$.

   Concerning $\bar \xi^{(pr)}_{p,\le}( \ell,n)$,  we write $|\bar
   \xi^{(pr)}_{p,\le}( \ell,n)|\le
         I_n(t)+I_n(0)$, where
\begin{align*}
                 &
         I_n(t):=  \frac{ C }{ n   }   \sum_{|j'|\le (n+1)^{3/5}}  
 \frac{ \bbE_n\tilde p_{j'}^2(t)}{
              (\ell\pi)^2
    +\varrho_{j'}^4 }. 
\end{align*}

 We can write
\begin{align*}
                  &
         I_n(t )=  \frac{ C }{ n   }   \sum_{|j'|\le (n+1)^{3/5}}  \sum_{y,y'=0}^{n} \bbE_n\big[p_y(t)p_{y'}(t)\big]
 \frac{ \psi_{j'}(y) \psi_{j'}(y')}{              (\ell\pi)^2
    +\varrho_{j'}^4}
                    \le  I_{n,+}(t)+
                      I_{n,-}(t)
\end{align*}
where
\begin{align*}
                  &
                   I_{n,\pm}(t):= \frac{ C }{  (n+1)^2   }  \Big|  \sum_{y,y'=0}^{n}
\bbE_n\big[p_y(t)p_{y'}(t)\big]\sum_{|j'|\le (n+1)^{3/5}}  \frac{\exp\{i\pi j'(u_y\pm u_{y'})\}}{
              (\ell\pi)^2
                    + \varrho_{j'}^4}\Big|,
\end{align*}
with $\varrho_{j'}=j'/(n+1)^{1/2}$.
Furthermore,  $I_{n,-}(t )= I_{n,-,\le}(t)+
                    I_{n,-,>}(t)$, where the terms correspond to the
                    summation over $|y-y'|\le (n+1)^{1/4}$ and
                    $|y-y'|>(n+1)^{1/4}$, respectively. 
Then,
\begin{align*}
&
| I_{n,-,\le}(t)  |        \le   \frac{ C }{  (n+1)^2   }
                 \mathop{\sum_{y,y'=0}^n}_{|y-y'|\le
                 (n+1)^{1/4}} \Big(\bbE_n\big[p_y^2(t)\big]+\bbE_n\big[p_{y'}^2(t)\big]\Big)\\
  &
    \times \sum_{|j'|\le
                 (n+1)^{3/5}} \frac{1}{
              (\ell\pi)^2
                    +  \varrho_{j'}^4}    \le   \frac{ C }{  (n+1)^{3/2}   }\int_{\bbR}\frac{\dd \varrho}{
              (\ell\pi)^2
                 + \varrho^4}
\\
&
      \times              \mathop{\sum_{y,y'=0}^n}_{|y-y'|\le
                 (n+1)^{1/4}} \Big(\bbE_n\big[p_y^2(t)\big]+\bbE_n\big[p_{y'}^2(t)\big]\Big)
          \le   \frac{ C }{  n^{1/4}   }
                 \to0,\quad\mbox{as $n\to+\infty$.}
\end{align*}
In the penultimate estimate we have used bound \eqref{eq:9}.

On the other hand
\begin{align*}
                          &
               I_{n,-,>}(t)=     \frac{ C }{  (n+1)^2   }   \Big|\mathop{\sum_{y,y'=0}^n}_{|y-y'|>
                 (n+1)^{1/4}}
                  \frac{\bbE_n\big[p_y(t)p_{y'}(t)\big]}{e^{i\pi
                            (u_y- u_{y'})}-1} \sum_{|j'|\le (n+1)^{3/5}}  \frac{\nabla_{j'}\exp\{i\pi {j'}(u_y- u_{y'})\}}{
              (\ell\pi)^2
                    +  \varrho_{j'}^4}\Big|.
\end{align*}
Denote by $m_n$ ($M_n$) the smallest (resp. largest) integer larger
(resp.  smaller) than $-(n+1)^{3/5}$ (resp. $(n+1)^{3/5}$). Summing by
parts in $j'$ we can write
\begin{align*}
                  &
                   I_{n,-,>}(t)
                    =
                  |  I_{n,-,>}^{(b)}(t )+I_{n,-,>}^{(a)} (t; M_n)-I_{n,-,>}^{(a)}  (t;
                    m_n)|,\quad\mbox{where}
  \\
  &
                             I_{n,-,>}^{(a)}  (t;m):=  \frac{ C }{  (n+1)^2   }   \mathop{\sum_{y,y'=0}^n}_{|y-y'|>
                 (n+1)^{1/4}}
 \frac{\bbE_n\big[p_y(t)p_{y'}(t)\big]e^{i\pi m(u_y-u_{y'})}}{\Big[
              (\ell\pi)^2
                    +
                    \varrho_{m}^4\Big]\Big[e^{i\pi(u_y-u_{y'})}
                    -1\Big]}\\
  &     I_{n,-,>}^{(b)}(t):=  -\frac{ C }{  (n+1)^2   }   \mathop{\sum_{y,y'=0}^n}_{|y-y'|>
                 (n+1)^{1/4}}
  \sum_{j'=m_n+1}^{M_n}  \frac{\bbE_n\big[p_y(t)p_{y'}(t)\big]e^{i\pi j'(u_y-u_{y'})}}{e^{i\pi(u_y-u_{y'})}
                    -1}\\
                  &
                    \qquad \qquad \qquad \qquad \times\nabla_{j'}^\star \Bigg(\frac{1}{
              (\ell\pi)^2+
    \varrho_{j'}^4}\Bigg).
  \end{align*}
  By the Cauchy-Schwarz inequality and the estimates in Corollary \ref{cor011912-24}
  we get
\begin{align}
  \label{032112-24}
   \mathop{\sum_{y,y'=0}^n}_{|y-y'|> (n+1)^{1/4}}
  \frac{|\bbE_n\big[p_y(t)p_{y'}(t)\big]|}{|y-y'|}& \le
  \left\{\mathop{\sum_{y,y'=0}^n}_{|y-y'|> (n+1)^{1/4}}
    \Big[\bbE_n\big[p_y(t)p_{y'}(t)\big]\Big]^2\right\}^{1/2}\notag
\\
  &\times \left\{ \mathop{\sum_{y,y'=0}^n}_{|y-y'|>  (n+1)^{1/4}}
    \frac{1}{(|y-y'|+1)^{2}}\right\}^{1/2}\le Cn^{7/8}.
  \end{align}
Hence,
$$
|I_{n,-,>}^{(a)}  (t;m)|  \le
C\frac{n^{7/8}}{n }=\frac{C}{n^{1/8}}\to0,\quad \mbox{as $n\to+\infty$. }
$$
Using an estimate
$$
\Bigg|\nabla_{j'}^\star \Bigg(\frac{1}{
              (\ell\pi)^2+
   \varrho_{j'}^4}\Bigg)\Bigg|\le \frac{C}{(n+1)^{1/2}}\cdot\frac{|\varrho_{j'}|^3}{[
              (\ell\pi)^2+
    \varrho_{j'}^4]^2}
$$
we can write
\begin{align*}
|I_{n,-,\ge}^{(b)}(t)|  \le  \frac{ C }{  n+1  } \int_{\bbR}\frac{|\varrho|^3\dd \varrho}{[
              (\ell\pi)^2+
    \varrho^4]^2}  \mathop{\sum_{y,y'=0}^n}_{|y-y'|>
                 (n+1)^{1/4}}
    \frac{|\bbE_n\big[p_y(t)p_{y'}(t)\big]|}{|y-y'|}.
\end{align*}
Invoking \eqref{032112-24} we conclude that
\begin{align*}
|I_{n,-,\ge}^{(b)}(t)|  \le  \frac{ Cn^{7/8} }{  n}=\frac{C}{n^{1/8}}\to0.
\end{align*}
as $n\to+\infty$. We have shown therefore that $I_{n,-}(t)=o_n(1)$.
Likewise we can show that
$
I_{n,+}(t)=o_n(1).
$
These facts together imply that
$
I_{n}(t;\ell)\to0.
$
Hence,
$
\bar \xi^{(pr)}_{p,\le}( n)=o_n(1)
$ and in consequence \eqref{022112-24} follows.\qed

                                         \section{Proofs of some
                                           technical results}
                                         \label{sec6}



\subsection{Equivalence of some kinetic energy functionals}

\begin{proposition}
  \label{prop031301-25}
   For any $t>0$  there exists a constant $C>0$ such that
          \begin{equation}
            \label{011811-22}
            \frac{1}{n}\sum_{x=0}^{n-1}\int_0^t\big(\mathbb E_n\left[
                p_{x+1}(s)p_{x}(s)\right]\big)^2\dd s\le
            \frac{C}{n^{1/2}}.
            \end{equation}
  \end{proposition}

   \proof
        From the Cauchy-Schwarz inequality we have
         \begin{align*}
          \left(\mathbb E_n\left[
           p_{x+1}(s)\big(p_{x}(s)-         p_{0}(s)\big)\right]\right)^2
           \le n\sum_{x'=1}^{x}\left(\mathbb E_n\left[
         p_{x+1}(s)\nabla^\star p_{x'}(s)\right]\right)^2.
         \end{align*}
         We also have
         \begin{align*}
         & \sum_{x=0}^{n-1}\int_0^t\left(\mathbb E_n\left[  p_{x+1}(s)p_{x}(s)\right]\right)^2\dd s\\
          & \le 2\sum_{x=0}^{n-1}\int_0^t
             \left(\mathbb E_n\left[ p_{x+1}(s)\big(p_{x}(s)- p_{0}(s)\big)\right]\right)^2\dd s
           +2\sum_{x=0}^{n-1}\int_0^t\left(\mathbb E_n\left[ p_{x+1}(s) p_{0}(s)\right]\right)^2 \dd s\\
           &
             \le 2 n \sum_{x=0}^{n-1}\sum_{x'=1}^{x}\int_0^t\left(\mathbb E_n\left[
         p_{x+1}(s)\nabla^\star p_{x'}(s)\right]\right)^2\dd s
             +2\sum_{x=0}^{n-1}\int_0^t\left(\mathbb E_n\left[ p_{x+1}(s) p_{0}(s)\right]\right)^2 \dd s.
         \end{align*}
        From \eqref{072102-24} we conclude that
        the right hand side can be estimated by
        $C n^{1/2}$.
        \qed

        \medskip

        As a direct application of Proposition  \ref{prop031301-25}
        and estimate \eqref{energy-bal} we conclude the following.
        \begin{corollary}
          \label{cor011401-25}
          For any $t>0$  there exists a constant $C>0$ such that
          \begin{equation}
            \label{021401-25}
            \frac{1}{n+1}\sum_{x=0}^{n}\int_0^t\big\{\mathbb E_n\left[
              (\nabla p_{x}(s))^2\right]-2 \mathbb E_n\left[p^2_{x}(s)\right] \big\}^2
            \dd s\le  \frac{C}{n^{1/2}}. 
            \end{equation}
          \end{corollary}

  \subsection{Equipartition property}
          \label{sec:equi}

\begin{theorem}[Equipartition property]
  \label{thm011401-25}
  For any compactly supported,
continuous function $\Phi: [0,+\infty)\times [0,1]\to\bbR$
\begin{equation}
  \lim_{n\to+\infty}\frac{1}{n}\sum_{x=0}^{n} \int_{0}^{+\infty} \Phi\left(t,\frac x{n} \right)
 \Big\{\bbE_{n}\big[ p_{x}^2( t)\big] -\bbE_{n}\big[ r_{x}^2( t)\big]\Big\}\dd t=0.
 \label{eq:conv-temp1a}
\end{equation}
\end{theorem}
\proof By an approximation it suffices to show that for any
$\varphi\in C^1_c(0,1)$ and $t>0$:
\begin{equation} \lim_{n\to+\infty}\frac{1}{n}\sum_{x=0}^{n} \int_{0}^{t} \varphi_x
 \Big\{\bbE_{n}\big[ p_{x}^2( s)\big] -\bbE_{n}\big[ r_{x}^2( s)\big]\Big\}\dd s=0,
 \label{eq:conv-temp1b}
\end{equation}
where $\varphi_x=\varphi\left(\frac{x}{n}\right)$. 
Define the position functional by letting
  \begin{equation}
    \label{position}
    q_x = \sum_{y=1}^x r_y,\quad x=1,\ldots n, \qquad \text{and} \qquad q_0=0.
    \end{equation} 
Then, cf \eqref{eq:7}, remembering that $\varphi_0=\varphi_n=0$, we get
\begin{equation*}
  \begin{split}
  & \mathcal G \bigg(\frac 1{n} \sum_{x=0}^n \varphi_x  p_x q_x
     \bigg) = \frac 1{n} \sum_{x=0}^{n} \varphi_x
     \Big(  p_x^2 -p_xp_0+ q_x \nabla r_x +\ga q_x\Delta p_x\Big).
     \end{split}
   \end{equation*}
 We now use the identity $\nabla(q_xr_x)=r_{x+1}^2+ q_x\nabla r_x$,
 valid for $x=1,\dots,n-1$.
 Summing by parts we obtain
\begin{equation*}
  \begin{split}
   & \frac 1 n \sum_{x=0}^n \varphi_x \int_0^t\left[\bbE_n p_x^2(s) -
      \bbE_n r_{x+1}^2(s) \right] \dd s={\rm I}_n+{\rm II}_n+{\rm
      III}_n,\quad \mbox{where}\\
&   {\rm I}_n =\frac 1n \sum_{x=0}^n (\nabla^*\varphi_x) \int_0^t\bbE_n
    \big[q_x(s)  r_x(s) \big]\dd s\\
    &
    {\rm II}_n= \frac 1n \sum_{x=0}^n \varphi_x \int_0^t\dd s\Big\{\bbE_n \big[p_x(s)p_0(s) \big]  -\ga\bbE_n \big[q_x(s) \Delta p_x(s)\big]\Big\}\\
    &
    {\rm III}_n=\frac 1{n^{5/2}} \sum_{x=0}^n\varphi_x\Big\{\bbE_{n} \big[p_x(t)q_x(t)\big]
    -\bbE_{n} \big[p_x(0)q_x(0)\big]\Big\}.
  \end{split}
\end{equation*}
By the Cauchy-Schwarz inequality and \eqref{eq:energyb} we have that
\begin{equation*}
   {\rm III}_n \le \frac C{n^{3/2}}\left(\mathcal H_n(t) +  \mathcal H_n(0)\right) \le \frac{C}{n^{1/2}}.
\end{equation*}

We have by \eqref{072102-24}
\begin{equation*}
  \begin{split}
       |{\rm I}_n| &\le  \frac{\|\varphi'\|_\infty}{n^2}\sum_{x,y=0}^n \Big|\int_0^t\bbE_n
    \big[r_x(s)  r_y(s) \big]\dd s \Big|\\
    &
    \le \frac{C}{n}\Big\{\sum_{x,y=0}^n \int_0^t \Big[\bbE_n
    \big[r_x(s)  r_y(s) \big]\Big]^2 \dd s\Big\}^{1/2}\le \frac{C}{n^{1/2}}.
  \end{split}
\end{equation*}
Finally, again from \eqref{072102-24}, we conclude that
\begin{align*}
  \frac 1n \sum_{x=0}^n  \Big|\int_0^t\bbE_n
  \big[p_x(s)  p_0(s) \big]\dd s\Big|
    \le  \frac 1{n^{1/2}} \left\{\sum_{x=0}^n  \Big[\int_0^t\bbE_n
    \big[p_x(s)  p_0(s) \big]\dd s\Big]^2\right\}^{1/2}\le \frac{C}{n^{3/4}}
\end{align*}
and 
\begin{align*}
  &\frac 1n \sum_{x=0}^n  \Big|\int_0^t\bbE_n \big[q_x(s) \Delta p_x(s)\big]\dd s\Big|
    \le  \frac{C}{n} \sum_{x,x'=0}^n  \Big|\int_0^t\bbE_n
    \big[r_{x'}(s) \nabla p_x(s)\big]\dd s\Big|\\
  &
    \le C\left\{\sum_{x,x'=0}^n  \Big[\int_0^t\bbE_n
    \big[r_{x'}(s) \nabla p_x(s)\big]\dd s\Big]^2\right\}^{1/2}\le\frac{C}{n^{1/4}}.
\end{align*}
In conclusion $|{\rm II}_n|\le C/n^{1/4}$ and the theorem has been
proved.\qed

\medskip

Define 
\begin{equation}
  \label{051401-25}
\hat{\cal E}_n(t,\ell) :=\frac{1}{n+1}\sum_{x=0}^n 
                     \bbE_n\left[{\cal E}_x(t)\right]c_\ell(u_x),\quad\ell=0,1,\ldots  .
                     \end{equation}
Combining the results of Corollary \ref{cor011401-25} and Theorem
\ref{thm011401-25} we conclude the following.
 \begin{corollary}
          \label{cor021401-25}
          For any $t>0$ and $\varphi\in C^\infty_c(0,1)$ we have 
          \begin{equation}
            \label{041401-25}
           \lim_{n\to+\infty} \frac{1}{n+1}\sum_{x=0}^{n}\varphi\Big(\frac{x}{n}\Big)\int_0^t\Big\{\mathbb E_n\left[
                (\nabla p_{x}(s))^2\right]-2\mathbb E_n
                {\cal E}_{x}(s)\Big\}\dd s=0
              \end{equation}
            and (cf \eqref{eq:43}) in consequence
              \begin{equation}
                \label{051401-25a}
           \lim_{n\to\infty}
           \sum_{\ell=0}^{+\infty} \hat\varphi_c(\ell) \left[\int_0^t
           \hat{\cal E}_n(s,\ell) \dd s-
           \hat{\frak E}_n(t,\ell)\right]=0.
            \end{equation}
          \end{corollary}

                                          \subsection{Estimates of the
                                            gradient of the kinetic energy}

                                         \begin{proposition}
    \label{corPP}
    For any $t_*>0$ there exists $C>0$s such that 
 \begin{equation} 
    \label{010503-24y}
    \frac{1}{n}\sum_{x=1}^n\int_0^t\Big\{\bbE_n\big[\nabla^\star \big(
    p_x
      p_{x+1}\big)(  s)\big]\Big\}^2\dd s\le
      \frac{C}{n^{3/2}},\quad n=1,2,\ldots,\quad t\in[0,t_*].
    \end{equation}
    \end{proposition}
    \proof
    Since 
   $   \nabla^\star \big(
    p_x  p_{x+1}\big)=  p_{x+1}   \nabla^\star  p_{x} +  p_{x-1}\nabla p_{x+1}$
       the left hand side of \eqref{010503-24y} can be estimated by
    \begin{align*}
         &                \frac{ 2}{n}\sum_{x=1}^{n}\int_0^t\left\{\mathbb
       E_n\left[ \big(  \nabla^\star p_x( s) \big)
           p_{x+1}(s)\right] \right\}^2 \dd s 
       + \frac{ 2}{n}\sum_{x=1}^{n}\int_0^t\left\{\mathbb
       E_n\left[ \big(  \nabla p_x( s)\big)
        p_{x-1}( s)\right] \right\}^2 \dd s\le \frac{C}{n^{3/2}},
      \end{align*}
  by virtue of  \eqref{072102-24}.
   \qed

\subsection{Proof of Lemma \ref{lm011712-25}}
\label{sec6.1}

\subsubsection{Proof of \eqref{eq:44}}
We write
\begin{align}
  \label{042406-24}
  &\hat{\frak E}_n(t,\ell)  =2  {\frak E}_{\rm kin}(t,\ell)  +
  {\frak E}_{\rm cor}(t,\ell)  + \widehat{\cal R}_n(t,\ell)\quad \mbox{where}\notag\\
                       &
                         {\frak E}_{\rm kin}(t,\ell) :=
                         \frac{1}{2 (n+1) }\sum_{y=0}^nc_\ell(u_y)  
                         \lang p_y^2\rang_t, \notag\\
                       &
                       {\frak E}_{\rm cor}(t,\ell):=  -\frac{1}{n+1  }\sum_{y=1}^nc_\ell(u_y)  
                   \lang p_yp_{y-1}\rang_t \\
                       &
                       \widehat{\cal R}_n(t,\ell):=  -\frac{1}{2(n+1) }
                  \Big(\lang p_0^2\rang_t +c_\ell(u_y) \lang p_n^2\rang_t\Big). \notag
\end{align}

Thanks to Proposition \ref{prop011612-24} we have $|\widehat{\cal
  R}_n(t,\ell)|\le C/n$.
By a direct calculation we conclude the following.
\begin{lemma}For any sequence $(a_y)$ of real numbers and
  $\ell\in\bbZ$  we have
                      \begin{equation}
                       \label{032406-24}
                       \begin{split}
                         &2\sum_{y=0}^n a_y \cos(\pi \ell u_y)    = a_0
                         + (-1)^{\ell+1}a_n                                            -\sum_{y=1}^{n}\frac{\sin\big( \pi k_\ell  (y-1/2)
                         \big)} { \sin\big(\frac{\pi k_\ell
                         }{2 }\big)
                         }\nabla^\star a_y  .
                         \end{split}
                      \end{equation}
                     \end{lemma}

                     Using formula \eqref{032406-24} we can write
                     \begin{align*}
                         2{\cal E}_{\rm kin}(t,\ell)= \bar{\rm  E}_n^{(0)}(t,\ell)
                         +{\rm r}_n^{(0)}(t,\ell),\quad {\cal E}_{\rm cor}(t,\ell)
                         = \bar{\rm E}_n^{(1)}(t,\ell)+{\rm  r}_n^{(1)}(t,\ell),
\end{align*}
                      where
                      \begin{align*} &
                       \bar{\rm   E}_n^{(0)}(t,\ell):=  -\frac{1}{ 2(n+1)  }
                                       \sum_{y=1}^{n}
                \frac{\sin\big( \pi k_\ell  (y-1/2) \big)} { \sin\big(\frac{\pi k_\ell }{2 }\big) }
                         \nabla^\star \lang p_y^2\rang_t,\\
                       &
                        {\rm r}_n^{(0)}(t,\ell):= \frac{1}{2(n+1) }\Big( \lang p_0^2\rang_t + (-1)^{\ell+1} 
                         \lang p_n^2\rang_t\Big),
                     \end{align*}
                     \begin{align*}
                       &
                         \bar{\rm E}_n^{(1)}(t,\ell):=  -\frac{1}{ 2(n+1)   }\sum_{y=2}^{n}
                         \frac{\sin\big( \pi k_\ell  (y-1/2) \big)} { \sin\big(\frac{\pi k_\ell}{2 }\big) }
                   \nabla^\star\lang p_yp_{y-1}\rang_t ,\\
                       &
                         {\rm  r}_n^{(1)}(t,\ell):= -\frac{1}{2(n+1) }
                         \Big(\lang p_1p_{0}\rang_t  +{(-1)^{\ell}}  \lang p_np_{n-1}\rang_t \Big).
                     \end{align*}
                     From Proposition \ref{prop011612-24} we conclude
                     \begin{align*}
                        \Big|{\rm r}_n^{(m)}(t,\ell)\Big|\le
                       \frac{C}{n},\quad  |\ell|\le n+1,\,m=0,1.
                     \end{align*}
                     Using trigonometric identities we can verify that
                     \begin{align*}
                                                    2\sum_{y=0}^n
                       \sin\big( \pi k_\ell(y+1/2)
                          \big) \sin\big( \pi k_{\ell'}(y+1/2)
                          \big)=(n+1)\delta_{\ell,\ell'}-(n+1)\delta_{\ell,0}\delta_{\ell',0}.
                          \end{align*}
                          In consequence
                          \begin{align*}
                                                    \sum_{y=0}^n
                            \chi_\ell(y)
                            \chi_{\ell'}(y)=\delta_{\ell,\ell'},\quad \ell,\ell'=1,\ldots,n,
                          \end{align*}
                       where
                       $
                       \chi_\ell(y):=\left(\frac{2}{n+1}\right)^{1/2}\sin\big( \pi k_\ell(y+1/2)
                          \big).
                       $                        
                          We can write therefore 
                        \begin{align*}   
                       \bar{\rm  E}_n^{(0)}(t,\ell)=  -\frac{1}{ 2^{3/2}(n+1)^{1/2} \sin\Big(\frac{\pi k_\ell
                         }{2 }\Big)} \sum_{y=0}^{n}\chi_\ell(y)
                         \nabla^\star \lang p_{y+1}^2\rang_t,
                        \end{align*}
                        where, by a convention, $\nabla^\star \bbE_n[
                        p_{n+1}^2(s)]=0$.
                        By virtue of \eqref{072102-24} we have
                        \begin{align*}   
                       &|\bar{\rm E}_n^{(0)}(t,\ell)|\le  \frac{C(n+1)^{1/2}}{ |\ell|} \left\{\sum_{y=0}^{n}\Big[
                         \nabla^\star \lang p_{y+1}^2\rang_t\Big]^2\right\}^{1/2} \le C\frac{(n+1)^{1/4}}{|\ell|}.
                        \end{align*}
                        Using the same argument and estimate
                        \eqref{010503-24y}  we conclude also that
                        \begin{align*}   
                          &|\bar{\rm E}_n^{(1)}(t,\ell)|\le  \frac{C(n+1)^{1/2}}{ |\ell|} \left\{\sum_{y=2}^{n}
                            \Big[
                         \nabla^\star \lang p_{y} p_{y-1}\rang_t\Big]^2\right\}^{1/2}
                          \le  C\frac{(n+1)^{1/4}}{|\ell|}.
                        \end{align*}
                        This concludes the proof of
                        \eqref{eq:44}.

\subsubsection{Proof of \eqref{eq:45}}
We prove that 
there exists $C>0$
\begin{equation}
                       \label{060901-25a}
                      \sum_{\ell=-n}^n\frac{1}{|\ell|+1}\Big|\hat{\frak E}_n(t,\ell) \Big|\le
                      C\,
                     \end{equation}
                    for $n=1,2,\ldots.$ Using the Cauchy-Schwarz
                    inequality  
                    we can estimate the left hand side of
                    \eqref{060901-25a} by 
\begin{equation}
  \begin{split}
    \label{070901-25}
    &\frac{C}{n+1}\left( \sum_{\ell=-n}^n\frac{1}{(|\ell|+1)^2}\right)^{1/2}\left\{ 
                   \sum_{y,y'=0}^n  \lang p_y^2 \rang_t
                     \lang p_{y'}^2 \rang_t 
                     \sum_{\ell=-n}^n\cos\left(\pi\ell u_y \right) \cos\left(\pi\ell u_{y'}\right)
                     \right\}^{1/2}.
                     \end{split}
                   \end{equation}
                   Recalling an elementary trigonometric identity
                   \begin{align}
                     \label{trig-id}
                     &\sum_{\ell=-n}^n\cos\left(\pi\ell u_y \right) \cos\left(\pi\ell u_{y'}\right)=(2n+1)\Big[\cos\left(n\pi(u_y+u_{y'})\right)1_{ \bbZ}\Big(\frac{u_y+u_y'}{2}\Big) \notag\\
                     &
                       +\cos\left(n\pi(u_y-u_{y'})\right) 1_{ \bbZ}\Big(\frac{u_y-u_y'}{2}\Big) \Big]
                   \end{align}
                   we conclude that the expression in
                   \eqref{070901-25} can be estimated by
                   \begin{equation}
                   \begin{split}
    \label{070901-25a}
    &\frac{C}{(n+1)^{1/2}} \Big\{ \sum_{y=0}^n 
                     \Big(\lang p_y^2 \rang_t\Big)^2 \Big\}^{1/2}\le C,
                     \end{split}
                   \end{equation}
                   by virtue of \eqref{072102-24}. This ends the proof of \eqref{060901-25a}.
                     \qed

   \section{Proof of energy bounds  for arbitrary  $T_L,T_R$}

\label{sec-ent1}

The main purpose of the present section is to provide the
proof of  energy bounds given in Section \ref{sec:thermal-energy}
without the assumption that the temperatures $T_L$, $T_R$ of both heat baths at the
end of the chain are equal, used in Section \ref{sec-ent}. We shall also show the estimate of the
current given in Theorem \ref{thm-current}.

\subsection{Proof of Theorem \ref{energy-bound}}

To indicate the dependence of   the solution \eqref{Xt} of
\eqref{Xts0} on the temperature of the heat baths we denote it  in the present
section  by  
$$
{\bf X}(t;T_L,T_R)=\left(\begin{array}{c}
                           {\bf r}(t;T_L,T_R)\\
                           \\
{\bf p}(t;T_L,T_R)
\end{array}\right).
$$
Likewise we denote
   by    $S(t;T_L,T_R)$ the
  the respective covariance matrix, see \eqref{S1ts}, 
\begin{equation}
\label{051010-26}
S(t;T_L,T_R) =\bbE_{\mu_n}\Big[{\bf X}_n(t;T_L,T_R)\otimes {\bf X}_n(t;T_L,T_R)\Big]
=\left[
  \begin{array}{cc}
    {S^{(r)}(t;T_L,T_R)}&S^{(r,p)}(t;T_L,T_R)\\
   S^{(p,r)}(t;T_L,T_R)& S^{(p)}(t;T_L,T_R)
  \end{array}
\right],
\end{equation}
 From \eqref{eq:23} we conclude that  
 \begin{equation}
             \label{071010-26}
             \begin{split}
      & A \lang S(\cdot;T_L,T_R) \rang_t +\lang S(\cdot;T_L,T_R) \rang_t A^T  +
      \ga \Sigma_2' \Big(\lang \overline{\frak {\bf (\nabla
          p)}^2}(\cdot;T_L,T_R)\rang_t \Big) \\
      &
     +
      2\tilde \ga \Sigma_1' (T_L,T_R)=
      \frac{1}{n^{3/2}}\delta_{0,t}S(\cdot;T_L,T_R),
      \end{split}
    \end{equation}
    where  $A$ is given by \eqref{bAz},
\begin{equation}
  \begin{split}
\Sigma_1'(T_L;T_R) =\left[\begin{array}{cc}
0_{n,n}&0_{n,n+1}\\
0_{n+1,n}&  D_1 (T_L;T_R) 
                        \end{array}\right],\quad \Sigma_2 ( {\bf
x}                        ) =\left[\begin{array}{cc}
0_{n,n}&0_{n,n+1}\\
0_{n+1,n}&  D_2 ( {\bf x}) 
                        \end{array}\right],
      \end{split}
\label{021410-26}
\end{equation}   
$D_1 (T_L;T_R)
=[T_L\delta_{x,0}\delta_{0,y}+T_R\delta_{x,n}\delta_{0,n}]_{x,y=0,\ldots,n}$,
 $D_2 ( {\bf x}) $ is defined in \eqref{eq:11a} 
and
  $$
  \overline{\frak {\bf (\nabla p)}^2}(s;T_L,T_R)=\Big[\bbE_{n}\big(\nabla^\star
  p_1(s;T_L,T_R)\big)^2,\ldots, \bbE_{n}\big(\nabla^\star
  p_n(s;T_L,T_R)\big)^2\Big].$$ Note that
    $$
  \bbE_n{\bf (\nabla p)}^2(t;T_L,T_R)=\Gamma S(t;T_L,T_R)
$$
for some matrix $\Gamma$.
     Differentiating \eqref{071010-26} in $t$ we conclude that $S(t ;T_L,T_R)$ satisfies the
     following  inhomogeneous linear system of  differential equations
\begin{equation}
  \label{071010-26a}
  \begin{split}
   &\frac{\dd S(t ;T_L,T_R)}{\dd t}  =-n^{3/2}\big(  A S (t ;T_L,T_R)
   + S (t ;T_L,T_R) A^T \\
   &
   +
     \ga \Sigma_2' \Gamma S(t ;T_L,T_R) +
     2\tilde \ga \Sigma_1' (T_L,T_R) ,\\
     &
     S(0 ;T_L,T_R)=\bbE_n\Big[ {\bf X}(0 ;T_L,T_R)\otimes {\bf X}(0 ;T_L,T_R)
   \Big]
     .
     \end{split}
     \end{equation}
     The following result holds.
     \begin{proposition}
       \label{prop011210-26}
       Suppose that $C$ is a covariance matrix of a certain zero mean
       random vector ${\bf X}$ and ${\bf X}(t)$ is the solution of
       \eqref{Xts0}. Then its covariance matrix $C(t)$, given by
       \eqref{051010-26}, satisfies \eqref{071010-26a}, with $C(0)=C$.

       Conversely, suppose that $C(t)$ satisfies \eqref{071010-26a}
       with $C(0)$ that is a second moment of a certain zero mean random vector
       ${\bf X}$. Then, $C(t)$ is   the covariance matrix of ${\bf X}(t)$ that is the solution of
       \eqref{Xts0} with the law of ${\bf X}(0)$ coinciding with
       ${\rm X}$.
       \end{proposition}

  Consider two systems with the dynamics described by
  \eqref{eq:fflip-1} and \eqref{eq:flip1bn}: one corresponding to equal
       temperatures $T$ at both endpoint thermostats and the second
       corresponding to $T_L$, $T_R$ and such that
       $\max\{T_L,T_R\}<T$. Suppose furthermore that they have the
       same initial data.
Denote by ${\cal H}_n(t;T,T) $ and $ {\cal H}_n(t;T_L,T_R)$ their
respective energy functionals defined in \eqref{H1}.

       Let $\delta S(t):=S(t;T,T)-S(t;T_L,T_R)$. Then
       $\delta S(t)$ satisfies equation \eqref{071010-26a} with $\Sigma_1'
       (T_L,T_R)$ replaced by $\Sigma_1' (T-T_L,T-T_R)$. According to 
Proposition \ref{prop011210-26} the solution $\delta S(t)$ is the matrix of
covariances of the solution of the system
\eqref{eq:fflip-1} and \eqref{eq:flip1bn}, where the temperatures at the left
and right endpoints equal $T-T_L$ and $T-T_R$, respectively. Since the
diagonal elements of $\delta S(t)$ are all nonnegative, we conclude that
${\cal H}_n(t;T,T) \ge {\cal H}_n(t;T_L,T_R)$. According to the
already proven energy estimate
 in the case when the heat baths are at the same
temperature, see  Section \ref{sec4.3}, for any $t_*>0$ there exists $C>0$ such that 
\begin{equation}
  \label{011210-26}
  \sup_{t\in[0,t_*]}{\cal H}_n(t;T,T) \le Cn,\quad n=1,2,\ldots
  \end{equation}
  thus the conclusion of  Theorem \ref{energy-bound} holds.\qed


\subsection{Estimates of  the energy current}

Given a function $\beta:[0,1]\to (0,+\infty)$  define
\begin{equation}
  \label{Jb}
J_n(t;\beta):=\sum_{x=1}^n\beta_x\int_0^t\bbE_n j_{x-1,x}^{(a)}(s)\dd s,
\end{equation}
where $\beta_x=\beta\big(\frac{x}{n+1}\big)$. 
We have the following.
   \begin{proposition}
     \label{prop022002-24}
   Suppose that
   $\beta:[0,1]\to [0,+\infty)$ is a $C^1$ class function. Then, for
   $z=0$ and $z=n+1$ we have
   \begin{equation}
   \label{eq:7bZ}
   \begin{split}
     &|\int_0^t\bbE_n j_{z-1,z}( s)\dd s|\le
     \Big(\sum_{x=1}^n\beta_x\Big)^{-1}\Bigg\{
   |J_n(t;\beta)|+ 
       \|\beta\|_\infty (T_L+T_R)t \\
     &
    +\frac{(n+1) \|\beta\|_\infty}{n^{3/2}}\big(\bbE_n{\cal
       H}_n(t)+\bbE_n{\cal
       H}_n(0)\big) +
         \frac{\|\beta\|_\infty }{n^{3/2}}\bbE_n{\cal
       H}_n(0)+\frac{\ga \|\beta\|_\infty }{2n} \int_0^t\bbE_n{\cal
       H}_n(s)\dd
       s    
        \Bigg\}
     \end{split}
        \end{equation}
        for all $n=1,2,\ldots$ and $t\ge0$. In addition,
         \begin{equation}
   \label{052102-24a}
   \begin{split}
     &\sup_{x=0,\ldots,n+1}|\int_0^t\bbE_n j_{x-1,x}( s)\dd s|\le \Big(\sum_{x=1}^n\beta_x\Big)^{-1}\Bigg\{| J_n(t;\beta)|+\|\beta\|_\infty  
        (T_L+T_R)t \\
        &
         +\frac{1}{n^{3/2}}\Big[(n+1) \|\beta\|_\infty +\sum_{x=1}^n\beta_x\Big]\big(\bbE_n{\cal
         H}_n(t)+\bbE_n{\cal
         H}_n(0)\big)
        \\
             &
    +\frac{\|\beta\|_\infty  }{n^{3/2}}\bbE{\cal
         H}_n(0)   +\frac{\ga\|\beta'\|_\infty}{2n}\left|\int_0^t{\cal
         H}_n(s)\dd
       s\right|   \Bigg\}.
     \end{split}
        \end{equation}
   \end{proposition}
   \proof
We can obviously write
\begin{align}
  \label{052102-24}
 & \Big(\sum_{x=1}^n\beta_x\Big)\left|\int_0^t\bbE_n j_{-1,0}( 
   s)\dd s\right|\le I_1+I_2,\quad\mbox{where}\notag\\
  &
    I_1:=\left|\int_0^t\sum_{x=1}^{n}\beta_x\bbE_nj_{x-1,x}(  s)\dd
   s\right|\\
  &
    I_2:= \left|\int_0^t\sum_{x=1}^{n}\beta_x\bbE_n\big[j_{x-1,x}(
    s)-j_{-1,0}(  s)\big]\dd
    s\right|.\notag
\end{align}
    We have
    \begin{align*}
      & I_1\le    |J_n(t;\beta)|+ R_{1},\quad\mbox{where}\\
             & R_{1}  :=\frac{\ga}{2}\left|\int_0^t\sum_{x=0}^{n-1}\beta_{x+1}\bbE_n\nabla
        p_{x}^2(  s)\dd
      s\right|.
      \end{align*}
      Summing by parts and using  the fact that $\beta$ is of $C^1$
      class we obtain
\begin{align*}
      & R_1\le \frac{\ga}{2}\left|\int_0^t\sum_{x=1}^{n-1}\nabla^\star\beta_{x+1}\bbE_n
        p_{x}^2(  s)\dd
      s\right|+\beta_1 \int_0^t\bbE_n
        p_{0}^2(s)
   \dd   s+\beta_n \int_0^t\bbE_n
        p_{n}^2(  s)\dd
        s\\
      &
        \le \frac{\ga\|\beta'\|_\infty}{n}\int_0^t\bbE_n{\cal H}_n(s)\dd
        s+\|\beta\|_\infty\left( \int_0^t\bbE_n
        p_{0}^2(  s)\dd
      s+ \int_0^t\bbE_n
        p_{n}^2(  s)\dd
        s\right).
\end{align*}
 Applying estimate \eqref{energyn1}
 we end up with 
 \begin{align}
   \label{032102-24}
       R_{1}  
        \le \frac{\ga\|\beta'\|_\infty}{n}\int_0^t\bbE_n{\cal H}_n(s)\dd
        s
        +\|\beta\|_\infty \Big[(T_L+T_R)t+\frac{1}{n^{3/2}}\bbE_n{\cal
   H}_n(0)\Big]. 
\end{align}
In consequence
\begin{align}
  \label{042102-24}
      & I_{1}  
        \le |J_n(t;\beta)|+\frac{\ga\|\beta'\|_\infty}{n}\int_0^t\bbE_n{\cal H}_n(s)\dd
        s\\
  &
        +\|\beta\|_\infty \Big[(T_L+T_R)t+\frac{1}{n^{3/2}}\bbE_n{\cal
   H}_n(0)\Big].\notag
\end{align}

Concerning $I_2$, using \eqref{eq:current}, we write
\begin{align}
  \label{042102-24a}
  &
   I_2=\left|\int_0^t\sum_{x=1}^{n}\beta_x\sum_{y=0}^{x-1}\bbE_n\nabla^\star
    j_{y,y+1}(
    s) \dd
    s\right|
    =
     \frac{1}{n^{3/2}}\left|\int_0^t\sum_{x=1}^{n-1}\beta_x\sum_{y=0}^{x-1}\bbE_n{\cal
    G}{\cal E}_y(  s)\dd
    s\right|\\
  &
    =\frac{1}{n^{3/2}}\left|\sum_{x=1}^{n-1}\beta_x\sum_{y=0}^{x-1}\big[\bbE_n{\cal
    E}_y(  t) -\bbE_n{\cal E}_y(0)\big]\dd
    s\right|
    \le (n+1)\|\beta\|_\infty\frac{\bbE_n {\cal H}_n(  t)+\bbE_n {\cal H}_n(  0) }{n^{3/2}}.\notag
\end{align}
Combining \eqref{042102-24} with \eqref{042102-24a} we get
\eqref{eq:7bZ} for $z=0$. The proof for $z=n+1$ is analogous.

Using \eqref{eq:current} we get
\begin{equation}
  \label{062102-24}
  \begin{split}
 &\int_0^t \bbE_nj_{x-1,x}
  ( s)\dd s=\sum_{y=0}^{x-1} \int_0^t\bbE_n[\nabla^\star j_{y,y+1}( s)]\dd s+\int_0^t\bbE_n j_{-1,0}
  ( s)\dd s\\
  &
  =\frac{1}{n^{3/2}}\sum_{y=0}^{x-1} \bbE_n\mathcal E_y (0)
  -\frac{1}{n^{3/2}}\sum_{y=0}^{x-1} \bbE_n\mathcal E_y ( t)
  +\int_0^t\bbE_n j_{-1,0}(s)\dd s
  ,\quad x=0,\ldots,n.
  \end{split}
\end{equation}
Combining with \eqref{eq:7bZ} we obtain
  \eqref{052102-24a}  as well.
\qed

\subsection{Proof of Theorem \ref{thm-current}}
 
               We shall prove in Section \ref{sec13} the following estimate: there exists
$C>0$ such that
\begin{equation}
  \label{010403-24}
  \begin{split}
    & n^{1/2} | J_n(t;\beta')|\le 
   C\Big[n 
   +n^{3/4}|J_n(t,\beta')|^{1/2}+ \bbE_n {\cal
  H}_n(0) +  \bbE_n{\cal
  H}_n(t) \\
   &
 +\int_0^t \bbE_n {\cal
  H}_n(s) \dd s + \Big(n \bbE_n {\cal
  H}_n(0)   \Big)^{1/2} + \Big(n \bbE_n {\cal
  H}_n(t)   \Big)^{1/2} + \Big(n\int_0^t \bbE_n{\cal
  H}_n(s) \dd s \Big)^{1/2} \Big].
  \end{split}
\end{equation}
Using Young's inequality
\begin{equation}
  \label{YI}
ab\le \frac{a^2}{2\ga}+\frac{\ga b^2}{2},\quad a,b,\ga>0,
\end{equation}
this leads to the estimate
\begin{equation}
  \label{010403-24a}
  n^{1/2} | J_n(t;\beta')|\le C\Big(n+ \int_0^t \bbE_n{\cal
  H}_n(s) \dd s
     +   \bbE_n{\cal
  H}_n(t)  +  \bbE_n {\cal
  H}_n(0) \Big).
   \end{equation}

\label{sec-ent11}
From estimate \eqref{010403-24a} and Corollary \ref{energy-bound},
shown modulo estimate
\eqref{010403-24}, we conclude that for any $t_*>0$ there exists $C>0$ such that
\begin{equation}\label{052708-25}
| J_n(t;\beta')|\le  C\sqrt{n},\quad t\in[0,t_*],\,n=1,2,\ldots.
\end{equation}
Then, estimate \eqref{052102-24zz} is a conclusion of \eqref{052708-25}
and \eqref{052102-24a}.\qed

\section{Proof of estimate (\ref{010403-24})}

\label{sec13}

 \subsection{Preliminaries}
 \label{sec8}

We consider $\beta:[0,1]\to(0,+\infty)$ that is $C^\infty$ smooth and such
that ${\rm supp}\,\beta'\subset(0,1)$. Recall the defintion of
$J_n(t;\beta')$ given in   \eqref{022308-25}
As in \eqref{050603-24} we can write
\begin{align}
  \label{050603-24z}
   n^{1/2} J_n(t;\beta')=-\theta_{pr}(\beta';n)-\sum_{\iota\in I}\xi^{(pr)}_{\iota}(\beta';n)-\sum_{\iota\in I}\pi^{(pr)}_{\iota}(\beta';n).
\end{align}
Here $I=\{p,pr,rp,r\}$ and the terms on the right hand side has been
defined in \eqref{eq:theta}--\eqref{eq:Fjj}.
Denote also
\begin{equation}\label{011712-24}
  {\frak G}_n(t):=\frac{|J_n(t,\beta)|+1}{n^{1/2}}+\frac{1}{n}\big(\bbE_n{\cal H}_n (0)  +\bbE_n{\cal H}_n (t) \big)+\frac{1}{n^{3/2}} \int_0^t \bbE_n{\cal H}_n (s)\dd
  s.
\end{equation}
Combining \eqref{eq:7bZ} and \eqref{energy-bal} we conclude in
particular the following:
\begin{corollary}
\label{cor012102-24z}  Suppose that $\beta:[0,1]\to[0,+\infty)$ is a function satisfying  the
assumptions of Propostion \ref{prop022002-24} such that
$\sum_{x=0}^n\beta_x\sim n$. 
 Then, there exists a constant
  $C>0$ such that
  \begin{equation}
    \label{010503-24z}
          {\cal H}_n^{(2)}(t)  \le {\cal H}_n^{(2)}(0) 
      +C  {\frak G}_n(t)
      \end{equation}
  and
  \begin{equation}
    \label{072102-24z}
    \begin{split}&
      \sum_{x=1}^{n}\mathop{\sum_{x'=0}^{n}}_{x'\not\in\{x-1,x\}}\int_0^t\left\{\mathbb
                       E_n\left[ \nabla^\star p_x( s)
                 p_{x'}( s)\right] \right\}^2 \dd s +
       \sum_{x=1}^{n}\int_0^t\Big[\nabla^\star \mathbb
    E_n p_{x}^2( s)
                   \Big]^2\dd s 
                   \\
                   &
                   +  \sum_{x=1}^{n} \sum_{x'=1}^n
    \int_0^t \left\{\mathbb E_n\left[ \nabla^\star p_x( s)
        r_{x'} ( s)\right] \right\}^2\dd s 
     + \sum_{z=0,n}\sum_{x=0}^n \int_0^t[b_{z,x}^{(p)}(s)]^2\dd s
  \\
    &+  \sum_{z=0,n}\sum_{x'=1}^n   \int_0^t[b_{z,x}^{(pr)}(s)]^2  \dd s \le
 \frac{1}{n^{1/2}}\Big({\cal H}_n^{(2)}(0) +C  {\frak G}_n(t)\Big)
  \end{split}
\end{equation}
 for all $n=1,2,\ldots$. 
\end{corollary}

\subsection{Estimates of $\theta_{pr}(\beta';n)   $}

Using formula \eqref{eq:27} we can write that  $\theta_{pr}(\beta';n) =\theta_{pr,-}(\beta';n) -\theta_{pr,+}(\beta';n) $
where $\theta_{pr,\pm}(\beta';n)$ are defined in  \eqref{eq:28}, with
$\varphi$ replaced by $\beta$. Thanks to \eqref{eq:38a} we conclude
that for any $t>0$ there exists a constant $C>0$ such that
the estimate
\begin{equation}
  \label{011806-24}
 |\theta_{pr,-} (\beta';n) |\le  
 C\int_0^t\bbE_n{\cal H}_n(s)\dd s,\quad n=1,2,\ldots.
\end{equation}
 Concerning $ \theta_{pr,+}^{(o)}(\beta';n) $, after similar
 calculations to those performed in the case of $
 \theta_{pr,-}^{(o)}(\beta';n) $ in Section \ref{sec:proof-bulk-term}, we conclude that
\begin{align}
  \label{012506-24}
   &
   \theta_{pr,+} (\beta';n) =(n+1)  \bar
    \theta_{pr,+} (\beta';n)(1+o_n(1)) ,\quad \mbox{where}  \\
  &
   \bar\theta_{pr,+} (\beta';n)= - \frac{ \ga
    }{2^2n^{3/2}}\sum_{\ell=-n^{\kappa}}^{n^{\kappa}} \widehat{(\beta
    ')_o}(\ell)  (\pi\ell)  \sum_{j'=-\delta n}^{\delta n}    \frac{(\pi k_{j'})^2\hat{\frak E}_n(t, 2j'+\ell)}{
      \Big(\frac{\ell\pi}{n+1}\Big)^2 
    +\ga^2 (\pi k_{j'})^4} \notag
\end{align}
for some $\delta,\kappa\in(0,1)$. Arguing as in \eqref{eq:38} we get
\begin{equation}
  \label{021806-24}
  |\theta_{pr,+} (\beta';n) |\le C 
     \Big( \int_0^t\bbE_n{\cal H}_n(s)\dd s+o_n(1)\Big).
\end{equation}

Summarizing we have shown the following.
\begin{lemma}
  \label{lm011212-24}
 Suppose that $\beta\in C^\infty[0,1]$ is such that ${\rm
   supp}\,\beta'\subset(0,1)$. Then, there exists $C>0$ such that
  \begin{equation}
  \label{041212-24}
  |\theta_{pr} (\beta';n) |\le C 
     \Big( \int_0^t\bbE_n{\cal H}_n(s)\dd s+o_n(1)\Big),\quad n=1,2,\ldots.
\end{equation}
  \end{lemma}

\subsection{Estimates of $\xi^{(pr)}_{\iota}(\beta';n)   $}

\subsubsection{Estimates of $\xi^{(pr)}_{p}(\beta';n)   $}
\label{sec4.3.1}

Thanks to  \eqref{022312-24} and  \eqref{041312-24a} (replacing
$\varphi$ by $\beta$ in \eqref{022312-24a}) we can write
\begin{align}
  \label{041312-24x}
          &  \xi^{(pr)}_{p}(\beta';n) =
            n\bar\xi^{(pr)}_{p}( n)+ 
            o_n(1)\big(\bbE_n{\cal H}_n(0)+\bbE_n{\cal H}_n(t)\big) .  
\end{align}
   Going back to the defintion of $\bar\xi^{(pr)}_{p}(
n)$ in \eqref{041312-24a}   there exists $c>0$ such that  
\begin{equation}
  \label{022708-25}
  (n+1)^2\Delta'(\ell,k_{j'})\ge c>0,\quad |j'|\ge
  n/100,\,n=1,2,\dots.
\end{equation}
Therefore that part of the sum can
be   estimated by $C\big(\bbE_n{\cal H}_n(0) +\bbE_n{\cal H}_n(t)\big)$. As
a result we can write
\begin{align*}  
  & | \xi^{(pr)}_{p}( n)|\le    C
      \mathop{ \sum_{|\ell|\le n^{\kappa},|j'|\le   n/100}}_{j'\not=-2\ell,\ell\not=0}  \frac{|\widehat{(\beta ')_o}(\ell) |}{|\ell|}\cdot\frac{|j'|}{|2j'+\ell|}
    |\delta_{0,t}\tilde S_{j'+\ell,j'}^{(p)}|+C\big(\bbE_n{\cal H}_n(0) +\bbE_n{\cal H}_n(t)\big)\\
  &
    \le    C
      \mathop{ \sum_{|\ell|\le n^{\kappa},|j'|\le n/100}}_{j'\not=-2\ell,\ell\not=0}  \frac{|\widehat{(\beta ')_o}(\ell) |( 1+|\ell|)}{2|\ell|}
   \Big(\bbE_n[\tilde
    p_{j'+\ell}^2(0)]+\bbE_n[\tilde p_{j' }^2(0)]\\
  &
 +\bbE_n[\tilde
    p_{j'+\ell}^2( t)]+\bbE_n[\tilde p_{j' }^2(t)]\Big)    + C\big(\bbE_n{\cal H}_n(0)+\bbE_n{\cal H}_n(t)\big)
    \le C'\big({\cal H}_n(0)+{\cal H}_n(t)\big)
\end{align*}
for some constant $C'>C$ independent of $n$.
Summarizing, we have shown that there exists $C>0$ such that
\begin{equation}
  \label{011312-24}
|\xi^{(pr)}_{p}(\beta';n)|\le  C\big(\bbE_n{\cal H}_n(0)+\bbE_n{\cal H}_n(t)\big).
\end{equation}

\subsubsection{Estimates of $\xi^{(pr)}_{pr}(\beta';n) +\xi^{(pr)}_{rp}(\beta';n)   $}

We have ,  see \eqref{041312-24a},
$$
\xi^{(pr)}_{pr}(\beta';n) +\xi^{(pr)}_{rp}(\beta';n)  =\bar
\xi^{(pr)}_{rp}(n)+o_n(1) \big(\bbE_n{\cal H}_n(0)+\bbE_n{\cal H}_n(t)\big).
$$
After a direct calculation we obtain from \eqref{Xi}
   \begin{align}
                     \label{061912-24}
                     &
   \xi^{(pr)}_{rp}(\beta';n)  +\xi^{(pr)}_{pr}(\beta';n)  
                                              = \frac{i\ga}{ n }
                       \sum_{|\ell|\le n^{\kappa}    }\sum_{j=-n-1}^{n}  
    \widehat{(\beta ')_o}(\ell) \xi_{pr}'(j'+\ell,j') \delta_{0,t}\tilde
                       S_{j'+\ell,j'}^{(pr)}, \end{align}
                     where
                     \begin{align*}  
                     &
    \xi_{pr}'(j,j'):=   \frac{\sin
    \Big(\frac{\pi(k_j+k_{j'})}{2}\Big) }{2^{3/2}\Delta(k_j,k_{j'})} \left(\sin^2\Big(\frac{\pi
                       k_j}{2}\Big)+\sin^2\Big(\frac{\pi k_{j'}}{2}\Big)\right)\sin
    \Big(\frac{\pi k_{j'}}{2}\Big).                      
                     \end{align*}
Following the same procedure as in Section \ref{sec4.3.1} we get that
for $\kappa\in(0,1)$
\begin{align*}  
  & | \xi^{(pr)}_{pr}(\beta';n)+\xi^{(pr)}_{rp}(\beta';n)|\le     C
      \mathop{ \sum_{|\ell|\le n^{\kappa},|j'|\le
    n/100}}_{2j'\not=-\ell,\ell\not=0}  \frac{|\widehat{(\beta
    ')_o}(\ell) |}{|\ell|}|\delta_{0,t}\tilde S_{j'+\ell,j'}^{(p)}| \\
  &
    \times \frac{ (j'+\ell)^2+(j')^2 }{n|2j'+\ell|} 
    +C\big(\bbE_n{\cal H}_n(0)+\bbE_n{\cal H}_n(t)\big) 
    \le C'\big(\bbE_n{\cal H}_n(0)+\bbE_n{\cal H}_n(t)\big).
\end{align*}
      Summarizing, we have shown that there exists $C>0$ such that
\begin{equation}
  \label{041312-24}
|\xi^{(pr)}_{pr}(\beta';n)+\xi^{(pr)}_{rp}(\beta';n)|\le  C\big(\bbE_n{\cal H}_n(0) \bbE_n+{\cal H}_n(t)\big).
\end{equation}

\subsubsection{Estimates of $\xi^{(pr)}_{r}(\beta';n)  $}

Using formula  \eqref{041312-24a} and estimate \eqref{022708-25} we conclude that
\begin{align}
  \label{071912-24a} 
 &|\xi^{(pr)}_{r}(\beta';n)|\le
   |\hat\xi^{(pr)}_{r}(\beta';n)|+C\big({\cal H}_n(0)+{\cal
   H}_n(t)\big),\quad\mbox{with}\\
 & \bar \xi^{(pr)}_{r}( \ell;n) =  \frac{C }{ (n+1)^2
 } \mathop{ \sum_{|\ell|\le n^{\kappa},|j'|\le
    n/100}}_{2j'\not=-\ell,\ell\not=0}    \frac{  \cos
   \Big(\frac{\pi (k_{j'}+k_\ell)}{2}\Big) \sin
   \Big(\frac{\pi (2k_{j'}+k_\ell)}{2}\Big) \delta_{0,t}\tilde S_{j'+\ell,j'}^{(r) } }{\sin
   \Big(\frac{\pi (k_{j'}+k_\ell)}{2}\Big)  \Delta'(\ell,k_{j'})}.\notag
   \end{align}
   Following the argument used in Section \ref{sec4.3.1} we infer that
   \begin{equation}
     \label{011412-24}
     \begin{split}
    &   | \xi^{(pr)}_{r}(\beta';n)|  \le C\big(\bbE_n{\cal
   H}_n(0)+\bbE_n{\cal H}_n(t)\big)\ .
 \end{split}
   \end{equation}
    Summarizing, from \eqref{011312-24}, \eqref{041312-24} and
    \eqref{011412-24} we obtain that there exists $C>0$ such that
\begin{equation}
  \label{021412-24}
\sum_{\iota\in I}|\xi^{(pr)}_{\iota}(\beta';n)|\le  C\big(\bbE_n{\cal
  H}_n(0)+\bbE_n{\cal H}_n(t)\big),\quad t>0,\,n=1,2,\ldots.
\end{equation}

\subsection{Estimates of $\pi^{(pr)}_{\iota}(\beta';n)   $}

\subsubsection{Estimates of $\pi^{(pr)}_{p}(\beta';n)   $}
\label{sec4.1.1aa}

In the present section we show that for each $t_*>0$ there exists $C>0$ such that
\begin{align}
  \label{031612-24}
 &  | \pi^{(pr)}_{p}(\beta';n) |\le   C\Big(n 
   +n^{3/4}|J_n(t,\beta')|^{1/2}+ \bbE_n{\cal H}_n(0)+\bbE_n {\cal H}_n(t)
  \\
  &
 \qquad    \qquad  \qquad     +  \frac{1}{n^{1/2}} \int_0^t \bbE_n {\cal H}_n(s)\dd
       s\Big),\quad t\in[0,t_*],\,n=1,2,\ldots.\notag
\end{align}

Suppose
that $\kappa\in(0,1/2)$.
According to the calculation performed in Section \ref{sec4.1.1a}
we have
$$
\pi^{(pr)}_{p}(\beta';n) =\sum_{z=0,n}\hat\pi^{(pr,z)}_{p}(\beta';n)+ \frac{o_n(1)}{n^{1/2}}\int_0^t \bbE_n {\cal H}_n(s)\dd
       s,
$$
where (cf. \eqref{Rj1})  
 \begin{align}
   \label{042312-24}
 &
   \hat \pi^{(pr,z)}_{p}(\beta';n) := \tilde \ga    
                       n\sum_{\ell=-n^{\kappa}}^{n^{\kappa}}
                       \widehat{(\beta ')_o}(\ell)\sin\Big(\frac{\pi k_\ell}{2}\Big) \sum_{x=0}^n
                    \lang b^{(p)}_{z,x} \rang_t {\frak
                      i}_{x,z}^{(p)}(\ell)\quad\mbox{and}\notag\\
&
       {\frak
                 i}_{x,z}^{(p)}  (\ell):=          \frac{  n^{1/2}
    }{2^{5/2}(n+1)}  \sum_{j'=-n-1}^{n} \frac{\sin^2\Big(\frac{\pi (2k_{j'}+k_\ell)}{2}\Big) }{\Delta(k_{j'+\ell},k_{j'})
    } \Big(1-\frac{\delta_{0,j'+\ell}}{2}\Big)^{-1/2}\Big(1-\frac{\delta_{0,j'}}{2}\Big)^{-1/2}\psi_{j'+\ell}(z) \psi_{j'}(x).
          \end{align}
We
show how to estimate $\bar\pi^{(pr,0)}_{p,o}(\beta';n)$, as the term
corresponding to $z=n$ can be dealt with in a similar manner. 
By the Cauchy-Schwarz inequality applied to \eqref{022208-25} we conclude that
\begin{align}
&  |\hat  \pi^{(pr,0)}_{p}(\beta';n) |\le \left( {\cal B}^{(p)}_0 \right)^{1/2}\sum_{\ell=1}^{n^{\kappa}} |\widehat{(\beta ')_o}(\ell)||\ell| 
  \left({\cal I}^{(p)}_0(\ell)\right)^{1/2},\quad \mbox{where}\label{021612-24}\\
  &
    {\cal B}^{(p)}_z := \sum_{x=0}^n \lang b^{(p)}_{z,x} \rang_t^2,\qquad
    {\cal I}^{(p)}_z (\ell):=\sum_{x=0}^n\big({\frak
                 i}_{x,z}^{(p)}(\ell)\big)^2. \label{021612-24-1}
\end{align}
Recall that $ {\frak G}_n(t)$ is given by \eqref{011712-24}.
\begin{lemma}
  \label{lm011612-24}
 For each $t_*>0$  there exists $C>0$ such that
\begin{equation}
    \label{072102-24x}
    \begin{split}
      & \sum_{z=0,n}  {\cal B}^{(p)}_z
      \le
       \frac{C}{n^{1/2}} 
      \Big(1+ {\frak G}_n(t)\Big),\quad t\in[0,t_*],\,n=1,2,\ldots.
  \end{split}
\end{equation}
\end{lemma}
The proof of the lemma is presented in Section \ref{sec4.4}. We apply
it first to finish the estimate of $|\pi^{(pr)}_{p}(\beta';n)|$.

By  the Plancherel identity and  the fact that $|\ell|\ll n$ we
conclude that for 
\begin{align*}
&
 \sum_{x=0}^n [{\frak
                 i}_{x,z}^{(p)}(\ell)]^2\le \frac{ C}{ (n+1)^2}\sum_{j'=0}^{n}  \frac{ \cos^2
   \Big(\frac{\pi k_{j'}}{2}\Big)  }{ \Big[\Big(\frac{\ell\pi}{2(n+1)}\Big)^2 
                 +2^4\ga^2 \sin^4 \Big(\frac{\pi k_{j'}}{2}\Big)\Big]^2}.
\end{align*}
Choosing any $\delta\in(0,1)$ we can find $C>0$ such that the last
expression can be estimated by
 \begin{align*}
  &
   \frac{C
                 }{(n+1)^2}   \sum_{j'=0}^{\delta n}    
          \Big\{\Big(\frac{\ell}{n+1}\Big)^2 
                 + (\pi k_{j'})^4
    \Big\}^{-2}+\frac{C}{n+1}\\
   &
     \le \frac{C}{n+1}\int_0^{\delta} \frac{  \dd u }{
          \Big[\Big(\frac{\ell}{n+1}\Big)^2
    +u^4\Big]^{2}}+\frac{C}{n+1}\\
  &
   = \frac{C(n+1)^{5/2}}{\ell^{7/2}}\int_0^{ \delta[(n+1)/\ell]^{1/2}}
                 \frac{   \dd u }{
   (1
                 +u^4)^{2}} +\frac{C}{n+1}\le \frac{C(n+1)^{5/2}}{\ell^{7/2}}.
 \end{align*}
 Hence
 \begin{equation}
   \label{031712-24}
  \sum_{x=0}^n [{\frak
    i}_{x,z}^{(p)}(\ell)]^2\le \frac{C(n+1)^{5/2}}{\ell^{7/2}},\quad
  1\le \ell\le  n^{\kappa}.
  \end{equation}
Combining with    \eqref{021612-24} and \eqref{072102-24x}, and
estimating analogously  $|\hat  \pi^{(pr,n)}_{p}(\beta';n) |$, we conclude that
\begin{align*}
 &  | \pi^{(pr)}_{p}(\beta';n) |\le   C\Big[n 
   +n^{3/4}|J_n(t,\beta')|^{1/2}+n^{1/2}\Big(\big(\bbE_n {\cal H}_n(0)\big)^{1/2}+\big(\bbE_n {\cal H}_n(t)\big)^{1/2}\Big)
  \\
  &
    \qquad \qquad \qquad \qquad  +\frac{o_n(1)}{n^{1/2}}\int_0^t \bbE_n {\cal H}_n(s)\dd
       s+  n^{1/4}\Big(\int_0^t \bbE_n {\cal H}_n(s)\dd
       s\Big)^{1/2}\Big].
\end{align*}
Using the Young's inequality \eqref{YI} with
suitably chosen $a,b,\ga>0$ we conclude \eqref{031612-24}.

\subsection{Estimate of $\pi^{(pr)}_{pr}(\beta';n) +\pi^{(pr)}_{rp}(\beta';n) $}

\label{sec4.5}

Using \eqref{011307-24a} (with $\beta$ replacing $\varphi$) we can write
\begin{align}
  \label{011307-24aa}
  \pi^{(pr)}_{pr}(\beta';n) +\pi^{(pr)}_{rp}(\beta';n)  =\tilde
  \ga\ga n^{1/2}
                                                   \sum_{\ell=-n-1}^n  \widehat{(\beta ')_o}(\ell)  \sum_{z=0,n} \sum_{x=1}^n
    \lang b^{(pr)}_{z,x}\rang_t{\frak i}_{x,z}^{(pr)}(\ell),
\end{align}
with $b^{(pr)}_{z,x}$ and ${\frak i}_{x,z}^{(pr)}(\ell)$ given by
\eqref{brx} and  \eqref{033012-24}, respectively. 
By the
Cauchy-Schwarz inequality, as in \eqref{021612-24a}, we obtain
\begin{align}
  \label{021612-24aB}
&\Big|\sum_{x=1}^n
    \lang b^{(pr)}_{z,x}\rang_t{\frak i}_{x,z}^{(pr)}(\ell)\Big|\le \Big( {\cal B}^{(pr)}_z\Big)^{1/2}\Big( {\cal I}^{(pr)}_z(\ell)\Big)^{1/2},\quad\mbox{where}\\
    &
    {\cal B}^{(pr)}_z := \sum_{x=0}^n  \lang b^{(pr)}_{z,x}\rang_t^2,\qquad
    {\cal I}^{(pr)}_z(\ell):=\sum_{x=0}^n\big({\frak
                 i}_{x,z}^{(p)}(\ell)\big)^2.\notag
\end{align}
We have the following
\begin{lemma}
  \label{lm011712-24a}
 For each $t_*>0$  there exists $C>0$ such that
\begin{equation}
    \label{012106-24B}
    \sum_{z=0,n}  {\cal B}^{(pr)}_z 
      \le \frac{C}{n^{1/2}} 
      \Big(1+ {\frak G}_n(t)\Big),\quad t\in[0,t_*],\,n=1,2,\ldots.
 \end{equation}
\end{lemma}
The proof of the lemma is presented in  Section \ref{sec4.4}.

 Using   rapid decay of $ \widehat{(\beta ')_o}(\ell)$ and
to estimate the right hand side of  \eqref{011307-24aa} we can restrict ourselves to the case
$|\ell|\le n^{\kappa}$ for some $\kappa\in(0,1)$. Combining \eqref{032708-25}
  with     \eqref{021612-24aB} we conclude that
\begin{align}
  \label{041712-24}
 &  | \pi^{(pr)}_{pr}(\beta';n) +\pi^{(pr)}_{rp}(\beta';n) |\le   Cn^{1/2}
   \Big(1+ {\frak G}_n(t)\Big)^{1/2} + \frac{o_n(1)}{n^{1/2}}\int_0^t \bbE_n{\cal H}_n (s)\dd
  s\\
  &
    \le C\Big\{n^{1/2}+ n^{1/4}|J_n(t,\beta')|^{1/2} +o_n(1)\int_0^t \bbE_n{\cal H}_n (s)\dd
  s\notag
   \\
  &
  +\frac{1}{n^{1/2}}\big[\big(\bbE_n{\cal H}_n (0)\big)^{1/2} +\big(\bbE_n{\cal H}_n (t)\big)^{1/2} \big]
  +  \frac{1}{n^{1/4}}\Big(\int_0^t \bbE_n{\cal H}_n (s)\dd
  s\Big)^{1/2}\Big\}.\notag
\end{align}
Estimate \eqref{010403-24} is then a straightforward consequence of the
equality \eqref{050603-24} and estimates \eqref{041212-24},
\eqref{011312-24}, \eqref{041312-24}, \eqref{021412-24}, 
\eqref{031612-24} and  \eqref{041712-24}.

\subsection{Proofs of Lemmas \ref{lm011612-24} and  \ref{lm011712-24a}}

\label{sec4.4}

Suppose that $\beta:[0,1]\to[0,+\infty)$ is a function satisfying  the
assumptions of Propostion \ref{prop022002-24} such that
$\sum_{x=0}^n\beta_x\sim n$. 
 Then, using \eqref{eq:7bZ} to estimate the right hand side of \eqref{energy-bal}
 we conclude that there exists a constant
  $C>0$ such that
  \begin{equation}
    \label{010503-24}
          {\cal H}_n^{(2)}(t)  \le {\cal H}_n^{(2)}(0) 
      +C  {\frak G}_n(t)
      \end{equation}
Using the definition \eqref{021612-24-1} and  the Cauchy-Schwarz inequality in the $t$ variable we get also
\begin{equation}
    \label{072102-24xx}
    \begin{split}
      &   {\cal B}^{(p)}_0=\Big[\int_0^t\Big(T_0-\mathbb E_np^2_0(s)  \Big)
      \dd s\Big]^2+\sum_{x=1}^n \Big[\int_0^t \mathbb E_n[p_x(s) p_0(s)] 
      \dd s\Big]^2\\
      & 
      \le t \Bigg\{\int_0^t\Big(T_0-\mathbb E_np^2_0(s)  \Big)^2
      \dd s +\sum_{x=1}^n  \int_0^t\big\{ \mathbb E_n[p_x(s) p_0(s)] \big\}^2  \dd s\Bigg\}
     \\
     &
     \le\frac{t}{n^{1/2}}\Big({\cal H}_n^{(2)}(0) +C  {\frak G}_n(t)\Big),
  \end{split}
\end{equation}
by virtue of \eqref{072102-24z}. This combined with Assumption \ref{ass5} 
yields  \eqref{072102-24x}. The proofs in the case $z=n$ and for
${\cal B}^{(pr)}_z$, $z=0,n$ are  analogous.\qed

{This ends the proof of \eqref{010403-24}, thus finishing the proof
Theorem  \ref{thm-current}. }

{
\subsection{Proof of Theorem  \ref{cor012102-24} in the general case}

\label{sec13.4}

The proof of Theorem  \ref{cor012102-24} follows  from the already proved estimate 
\eqref{052708-25} and   Proposition \ref{prop011612-24} .\qed}

   \subsection*{Acknowledgements}

   T. Komorowski wishes to express  thanks to   A. Bobrowski, K. Bogdan, T. Klimsiak, J. Ma\l
   ecki and A. Rozkosz for enlightening discussions concerning the
   subject of the paper. Both authors express  profound gratitude to
   an anonymous referee for pointing out errors in the original
   version of the paper.

\section{Compliance with Ethical Standards}

   Ethics declaration: not applicable.

 \appendix

\section{Discrete lattice gradient and Laplacian}

\label{sec2.7}


\subsection{Finite lattice gradient and divergence operators}

\label{secA.1}

Let $\bbZ_n:=\{0,\ldots,n\}$ and suppose that $f:
\bbZ_n\to\bbR$. It can be represented as a vector in finite
dimensional space
$
f =(f_0,\ldots,    f_n)
     $.
Its divergence $\nabla^\star:\bbR^{n+1}\to\bbR^n$ is given by
$\nabla^\star f_x=f_{x}-f_{x-1}$, $x=1,\ldots,n$.
  The gradient operator $\nabla:\bbR^{n}\to\bbR^{n+1}$ assigns to each
  $
f =\left(\begin{array}{c}f_1\\
    \vdots
    \\
    f_n
    \end{array}\right)
  $
  a vector $(\nabla f)_x=f_{x+1}-f_{x}$, $x=0,\ldots,n$, with the
  convention $f_0=f_{n+1}=0$.  
We have
$
\nabla^T=-\nabla^\star
$
and 
\[
  \sum_{x=0}^n \nabla f_x\;  g_x = -\sum_{x=1}^n f_x \nabla^\star
  g_x,\quad f\in \bbR^n,\,g\in\bbR^{n+1}.
\]

\subsection{Discrete Neumann Laplacian $-\Delta_{\rm N}$}
 The discrete Neumann Laplacian is defined as an operator on $\bbR^{n+1}$ given by 
the formula
$$
\Delta_{\rm N}f_x:=f_{x+1}+f_{x-1}-2f_x,\quad x=0,\ldots,n,
$$
with the boundary condition $f_{-1}:=f_0$ and $f_{n+1}:=f_n$. 
Let  $\la_{j,n}$ and $\psi_j$, $j=0,\ldots,n$ be the eigenvalues
and the respective  eigenfunctions of $-\Delta_{\rm N}$. 
They are given by
\begin{align}
\label{laps}
 &\la_j
                 =\ga_j^2,\quad
\psi_j(x)=\left(\frac{2-\delta_{0,j}}{n+1}\right)^{1/2}\cos\left(\frac{\pi
  j(2x+1)}{2(n+1)}\right),
 \quad\mbox{with}\notag\\
  &
 \ga_j=   2\sin\left(\frac{j\pi}{2(n+1)}\right)
\end{align}
for $ x,j=0,\ldots,n.$
We have
$$
\sum_{j={0}}^n\psi_j(x) \psi_j(x')=\delta_{x,x'},\quad\mbox{
and}\quad  
\sum_{x=0}^n\psi_j(x) \psi_{j'}(x)=\delta_{j,j'},\quad x,x',j,j'=0,\ldots,n.
$$


\subsection{Dirichlet Laplacian}

It is defined as an operator on $\bbR^n$ that is given by 
the formula
$$
\Delta_{\rm D}f_x:=f_{x+1}+f_{x-1}-2f_x,\quad x=1,\ldots,n
$$
with the boundary condition $f_0=f_{n+1}:=0$. Its eigenvalues equal $\la_j$ and  the
respective eigenvectors are given by
\begin{align}
  \label{lapsD}
\phi_j(x)=\left(\frac{2}{n+1}\right)^{1/2}\sin\left(\frac{jx\pi}{n+1}\right),\quad\mbox{with
                 } x,j=1,\ldots,n.
\end{align}

We have the orthogonality relations
$$
\sum_{j=1}^n\phi_j(x) \phi_j(x')=\delta_{x,x'},\quad  \mbox{
and}\quad  
\sum_{x=1}^n\phi_j(x) \phi_{j'}(x')=\delta_{j,j'},\quad x,x' ,j,j'=1,\ldots,n.
$$
Note that
\begin{align}
\label{gaps}
&\nabla^\star \psi_j=-\ga_j \phi_j\quad\mbox{and}\quad\nabla \phi_j=\ga_j \psi_j,\quad j=0,\ldots,n.
\end{align}
In addition,
\begin{align*}
&\nabla \nabla^\star f=\Delta_{\rm N}f,\quad
                 f\in\bbR^{n+1}\quad\mbox{and} \quad
\nabla^\star \nabla f=\Delta_{\rm D}f,\quad  f\in\bbR^{n}.
\end{align*}


 
 \section{Proofs of the results from Section \ref{Sec3}}

 \subsection{Green's function of the Neumann Laplacian}
 \label{appC1}

The Neumann Laplacian $\Delta_{\rm N}$ is the generator of the
reflected Brownian motion  $\Big(\sqrt{2} w_t^{({\rm
    N})}\Big)_{t\ge0}$, where
$w_t^{({\rm N})}=\chi\big( w_{t}\big)$, $t\ge0$,
and  $\chi:\bbR\to\bbR$ is the $2$-periodic extension of the function
$\chi(u)=|u|$, $u\in[-1,1]$ and $(w_t)_{t\ge0}$ is the standard Brownian
motion.
Its transition probabilities are given by
\begin{equation}
  \label{p-w}
  \begin{split}
  &
    p_t(u,v)
    =\sum_{n=-\infty}^{+\infty}\Big[p_t (u-v+2n)+p_t
    (2n+u+v)\Big],\quad\mbox{where}\\
  &
  p_t(u):=\frac{1}{\sqrt{4\pi t }}e^{-u^2/(4t)},\quad u,v\in[0,1].
  \end{split}
\end{equation}
The Green's function kernel corresponding to  the operator
$(\la-\Delta_{\rm N})^{-1}$, see \eqref{greens}, is then given by
\begin{align}
  \label{G-la}
&G_\la(u,v)=
                 \sum_{n=0}^{+\infty}\frac{c_n(u)
                 c_n(v)}{\la+(n\pi)^2}\\
  &
    = \sum_{n=-\infty}^{+\infty}\Big[g_\la(u-v+2n)+g_\la(2n+u+v)\Big],\quad\mbox{where}\notag\\
  &
    g_\la(u)=\int_0^{+\infty}e^{-\la
    t}p_t(u)\dd t.\notag
\end{align}
 Using the classical identity
\[
\int_0^\infty t^{-1/2}e^{-at-b/t}\,dt
=\sqrt{\frac{\pi}{a}}e^{-2\sqrt{ab}},
\qquad a>0,\quad b\geq 0,
\]
with $a=\lambda$ and $b=u^2/4$, we obtain
\begin{equation}
\label{eq:g-lambda}
g_\lambda(u)
=\frac{1}{2\sqrt{\lambda}}
e^{-\sqrt{\lambda}|u|}.
\end{equation}
In particular, for $u\in[0,1]$,
\[
g_\lambda(u)
=\frac{1}{2\sqrt{\lambda}}
e^{-\sqrt{\lambda}u}.
\]

    \subsection{Representation of  $g_m(u,u';v)$, $v=0,1$}
\label{appA}

Recall that
\(
V_\varrho(u,u')=\varrho G_\varrho(u,u'),
\)
where, see Section \ref{appC1},
\begin{equation}
\label{eq:G-image}
G_{\varrho }(u,v)
=
\frac{1}{\sqrt{\varrho}}
    \sum_{n\in\mathbb Z}
e^{-\sqrt{\varrho} |u+v+2n|},
\qquad v\in\{0,1\}.
\end{equation} is the Green function of the
Neumann Laplacian on $[0,1]$. In particular we have
\begin{equation}
  \label{011310-26}
G_\varrho(u,0) =\frac{\cosh(\sqrt{\varrho} (1-u))}{\sqrt{\varrho}\sinh \sqrt{\varrho}},
\qquad 
G_\varrho(u,1) =\frac{\cosh(\sqrt{\varrho} u)}{\sqrt{\varrho}\sinh \sqrt{\varrho}}.
\end{equation}
and, as a result,
\begin{equation}
  \label{031310-26}
V_\varrho(u,0) =\sqrt{\varrho}\frac{\cosh(\sqrt{\varrho} (1-u))}{\sinh \sqrt{\varrho}},
\qquad 
V_\varrho(u,1) =\sqrt{\varrho}\frac{\cosh(\sqrt{\varrho} u)}{\sinh \sqrt{\varrho}}.
\end{equation}
Using the above formulas, \eqref{gm} and substitution
$k:=\sqrt{\varrho}$, we get, for $m=0,1,2,\ldots$
and $v\in\{0,1\}$,  
\begin{align*}
&g_m(u,u';v)=
2\sum_{n,n'\in\bbZ}
\int_0^\infty
k^{3/2}
e^{-k(2m+|u+v+2n|+|u'+v+2n'|)}
\,\dd k\\
&=
2\Gamma\left(\frac52\right)
\sum_{n,n'\in\bbZ}
\frac{1}{
\bigl(
2m+|u+v+2n|+|u'+v+2n'|
\bigr)^{5/2}}.
\end{align*}
Since
\(
2\Gamma\left(\frac52\right)
=
\frac{3\sqrt{\pi}}2,
\) formula \eqref{eq:gm-double-sum} follows. Performing analogous
calculations, using the definition of $g_m^{\rm cr}(u,u';v)$, see
\eqref{gm}, we get \eqref{eq:gG-cross-double-series}.

\subsection{Green's function for the Robin problem}
\label{secC2}
Define
\begin{align*}
F_\varrho(u)
&=
\cosh(\sqrt{\varrho }u)+2\widetilde\gamma\sinh(\sqrt{\varrho }u),\\
H_\varrho(u)
&=
\cosh(\sqrt{\varrho }(1-u))
+2\widetilde\gamma\sinh(\sqrt{\varrho }(1-u)).
\end{align*}
These functions solve
\(
-f''+\rho f=0
\)
and satisfy the left and right Robin
boundary conditions, respectively:
\[
F_\varrho '(0)=2\widetilde\gamma \sqrt{\varrho }F_\varrho(0),
\qquad
H_\varrho'(1)=-2\widetilde\gamma \sqrt{\varrho }H_\varrho(1).
\]
Their Wronskian is constant and equals
\begin{align*}
F_\varrho'(u)H_\varrho(u)-F_\varrho(u)H_\varrho'(u)
=
\sqrt{\varrho}\bigl[
&(1+4\widetilde\gamma^2)\sinh \sqrt{\varrho} +4\widetilde\gamma\cosh \sqrt{\varrho}
\bigr].
\end{align*}
Consequently,
\begin{equation}
\label{eq:Robin-Green-explicit}
G_\rho^{(\widetilde\gamma)}(u,u')
=
\frac{
F_\varrho(u\wedge u')H_\varrho(u\vee u')
}{
\sqrt{\varrho}\bigl[
(1+4\widetilde\gamma^2)\sinh \sqrt{\varrho}
+4\widetilde\gamma\cosh \sqrt{\varrho}
\bigr]
} .
\end{equation}
In particular we can see from the above formula that
\begin{equation}
\label{eq:Robin-positive-rho}
 G_\rho^{(\widetilde\gamma)}(u,u')\geq0.
\end{equation}

\subsection{Representation of the Robin Green's function using
  perturbative formula for the Neumann  Green's function} 

   Recall that $G_\rho(u,u')$ denotes the Green's function of the
Neumann resolvent
\(
    (\rho-\Delta_N)^{-1},
\)
and that
\(
    V_\rho(u,v)=\rho G_\rho(u,v).
\)
Introduce the vector
\[
 {\bf G}_\rho(u)
 :=
 \begin{pmatrix}
 G_\rho(u,0)\\
 G_\rho(u,1)
 \end{pmatrix}.
\]
The restriction of the Neumann Green's function to the two
endpoints is (see \eqref{011310-26})
\begin{equation}
\label{eq:B-rho}
 B_\rho
 :=
 \begin{pmatrix}
 G_\rho(0,0)&G_\rho(0,1)\\
 G_\rho(1,0)&G_\rho(1,1)
 \end{pmatrix}
 =
 \frac{1}{\sqrt{\rho}}
 \begin{pmatrix}
 \coth\sqrt{\rho}&(\sinh\sqrt{\rho})^{-1}\\
 (\sinh\sqrt{\rho})^{-1}&\coth\sqrt{\rho}
 \end{pmatrix}.
\end{equation}

Consider also the matrix
\begin{equation}
\label{eq:M-gamma-rho}
 M_{\widetilde\gamma}(\rho)
 :=
 \begin{pmatrix}
 F(\widetilde\gamma,\sqrt{\rho})
 &
 -G(\widetilde\gamma,\sqrt{\rho})
 \\
 &\\
 -G(\widetilde\gamma,\sqrt{\rho})
 &
 F(\widetilde\gamma,\sqrt{\rho})
 \end{pmatrix},
\end{equation}
where $F$ and $G$ are given by \eqref{F-G}.
A direct calculation gives
\begin{equation}
\label{eq:F2-G2-rho}
 F(\widetilde\gamma,\sqrt{\rho})^2
 -
 G(\widetilde\gamma,\sqrt{\rho})^2
 =
 \frac{\sinh\sqrt{\rho}}
 {(1+4\widetilde\gamma^2)\sinh\sqrt{\rho}
 +4\widetilde\gamma\cosh\sqrt{\rho}}.
\end{equation}
Consequently, the inverse matrix  to $M_{\widetilde\gamma}(\rho)$ is
given by 
\begin{align}
 M_{\widetilde\gamma}(\rho)^{-1}
 &= 
 \begin{pmatrix}
 1+2\widetilde\gamma\coth\sqrt{\rho}
 &
 2\widetilde\gamma(\sinh\sqrt{\rho})^{-1}
 \\
 & \\
 2\widetilde\gamma(\sinh\sqrt{\rho})^{-1}
 &
 1+2\widetilde\gamma\coth\sqrt{\rho}
 \end{pmatrix}=
 I+2\widetilde\gamma\sqrt{\rho}\,B_\rho.
 \label{eq:M-inverse-rho}
\end{align}

To verify this, fix $u'\in(0,1)$.  Since $G_\rho$ is the
Neumann Green's function,
\begin{equation}
\label{eq:Neumann-bc-G}
 \partial_uG_\rho(0,u')
 =
 \partial_uG_\rho(1,u')
 =
 0.
\end{equation}
Using \eqref{011310-26} we conclude that
\begin{equation}
\label{eq:boundary-derivatives-G}
 \partial_uG_\rho(0,0)=-1,
 \qquad
 \partial_uG_\rho(1,0)=0,
 \qquad
 \partial_uG_\rho(0,1)=0,
 \qquad
 \partial_uG_\rho(1,1)=1.
\end{equation}
Furthermore, both $G_\rho(\cdot,0)$ and
$G_\rho(\cdot,1)$ satisfy
\[
 \left(\rho-\frac{d^2}{du^2}\right)f=0
 \qquad\text{in }(0,1).
\]

For this fixed $u'$, let
\[
 {\bf c}_\rho(u')
 =
 \begin{pmatrix}
 c_{\rho,0}(u')\\
 c_{\rho,1}(u')
 \end{pmatrix},
\]
with the components, yet to be determined,   such  that 
\begin{equation}
\label{eq:H-rho-def}
 G_\rho^{(\widetilde\gamma)}(u,u')
 =
 G_\rho(u,u')
 -
 c_{\rho,0}(u')G_\rho(u,0)
 -
 c_{\rho,1}(u')G_\rho(u,1)
\end{equation}
would satisfy
\[
 \left(\rho-\frac{d^2}{du^2}\right)
 G_\rho^{(\widetilde\gamma)}(u,u')
 =
 \delta_{u'}(u), \quad u,u'\in(0,1),
\]
with the  Robin boundary conditions \eqref{eq:Robin-bc-rho}.

Using \eqref{eq:Neumann-bc-G} and
\eqref{eq:boundary-derivatives-G}, we have
\[
 \partial_u G_\rho^{(\widetilde\gamma)}(u,u')=c_{\rho,0}(u'),
 \qquad
 \partial_u G_\rho^{(\widetilde\gamma)}(u,u')=-c_{\rho,1}(u').
\]
At the left endpoint,
\[
 G_\rho^{(\widetilde\gamma)}(u,u')
 =
 G_\rho(0,u')
 -
 c_{\rho,0}(u')G_\rho(0,0)
 -
 c_{\rho,1}(u')G_\rho(0,1),
\]
whereas at the right endpoint,
\[
G_\rho^{(\widetilde\gamma)}(u,u')
 =
 G_\rho(1,u')
 -
 c_{\rho,0}(u')G_\rho(1,0)
 -
 c_{\rho,1}(u')G_\rho(1,1).
\]
Therefore the two Robin conditions in
\eqref{eq:Robin-bc-rho} are equivalent to
\begin{align}
 c_{\rho,0}(u')
 &=
 2\widetilde\gamma\sqrt{\rho}
 \Big[
 G_\rho(0,u')
 -
 c_{\rho,0}(u')G_\rho(0,0)
 -
 c_{\rho,1}(u')G_\rho(0,1)
 \Big],
 \label{eq:c0-rho}
 \\
 c_{\rho,1}(u')
 &=
 2\widetilde\gamma\sqrt{\rho}
 \Big[
 G_\rho(1,u')
 -
 c_{\rho,0}(u')G_\rho(1,0)
 -
 c_{\rho,1}(u')G_\rho(1,1)
 \Big].
 \label{eq:c1-rho}
\end{align}
In matrix notation, \eqref{eq:c0-rho}--\eqref{eq:c1-rho}
read
\begin{equation}
\label{eq:c-system-rho}
 \left(
 I+2\widetilde\gamma\sqrt{\rho}\,B_\rho
 \right)
 {\bf c}_\rho(u')
 =
 2\widetilde\gamma\sqrt{\rho}
 \begin{pmatrix}
 G_\rho(0,u')\\
 G_\rho(1,u')
 \end{pmatrix}.
\end{equation}
By symmetry of the Neumann Green's function,
\[
 \begin{pmatrix}
 G_\rho(0,u')\\
 G_\rho(1,u')
 \end{pmatrix}
 =
 {\bf G}_\rho(u').
\]
Using \eqref{eq:M-inverse-rho} in
\eqref{eq:c-system-rho}, we therefore obtain
\begin{equation}
\label{eq:c-solution-rho}
 {\bf c}_\rho(u')
 =
 2\widetilde\gamma\sqrt{\rho}\,
 M_{\widetilde\gamma}(\rho)
 {\bf G}_\rho(u').
\end{equation}
Substituting \eqref{eq:c-solution-rho} into
\eqref{eq:H-rho-def} proves that
\begin{equation}
\label{eq:Robin-resolvent-rho}
 G_\rho^{(\widetilde\gamma)}(u,u')
 =
 G_\rho(u,u')
 -
 2\widetilde\gamma\sqrt{\rho}\,
 {\bf G}_\rho(u)^{\mathsf T}
 M_{\widetilde\gamma}(\rho)
 {\bf G}_\rho(u').
\end{equation}


\subsection{Proof of Proposition \ref{lem:positive-effective-kernel}}

  \label{secC3}
We express 
$r_{\widetilde\gamma}$ in terms of  the kernels that appear in formula \eqref{eq:r-gamma}.
Recall that  $q(u,u')$ is given in \eqref{012412-24}.
Using \eqref{eq:G-image} we obtain
\[
 2\sqrt{\rho}\,G_\rho(u,u')
 =
 \sum_{n\in\mathbb Z}
 \left\{
 e^{-\sqrt{\rho}|u-u'+2n|}
 +
 e^{-\sqrt{\rho}|u+u'+2n|}
 \right\},
\]
therefore
\begin{equation}
\label{eq:q-resolvent-rho}
 q(u,u')
 =
 \frac{1}{\sqrt{2}\,\pi}
 \int_0^\infty
 \rho^{3/4}G_\rho(u,u')\,d\rho.
\end{equation}

On the other hand, from the definition of
$h_{\widetilde\gamma}$ and the identity
$V_\rho(u,v)=\rho G_\rho(u,v)$, we obtain
\begin{align*}
 &\sum_{v=0,1}
 h_{\widetilde\gamma}(u,u';v)
 \\
 &\quad =
 \int_0^\infty \rho^{5/4}
 \Big\{
 F(\widetilde\gamma,\sqrt{\rho})
 \big[
 G_\rho(u,0)G_\rho(u',0)
 +
 G_\rho(u,1)G_\rho(u',1)
 \big]
 \\
 &\hspace{42mm}
 -
 G(\widetilde\gamma,\sqrt{\rho})
 \big[
 G_\rho(u,1)G_\rho(u',0)
 +
 G_\rho(u,0)G_\rho(u',1)
 \big]
 \Big\}\,d\rho.
\end{align*}
By  \eqref{eq:M-gamma-rho}, we can write
\begin{equation}
\label{eq:h-resolvent-rho}
 \sum_{v=0,1}
 h_{\widetilde\gamma}(u,u';v)
 =
 \int_0^\infty
 \rho^{5/4}
 {\bf G}_\rho(u)^{\mathsf T}
 M_{\widetilde\gamma}(\rho)
 {\bf G}_\rho(u')\,d\rho.
\end{equation}
We also have, see \eqref{022111-24},
\begin{equation}
\label{eq:constants-ratio-rho}
 \frac{c_{\rm bd}}{c_{\rm bulk}}
 =
 \frac{\sqrt{2}\,\widetilde\gamma}{\pi}.
\end{equation}
Combining
\eqref{eq:q-resolvent-rho},
\eqref{eq:h-resolvent-rho}, and
\eqref{eq:constants-ratio-rho}, we get
\begin{align}
 &q(u,u')
 -
 \frac{\sqrt{2}\,\widetilde\gamma}{\pi}
 \sum_{v=0,1}
 h_{\widetilde\gamma}(u,u';v)
 \notag\\
 &\quad =
 \frac{1}{\sqrt{2}\,\pi}
 \int_0^\infty
 \rho^{3/4}
 \Big[
 G_\rho(u,u')
 -
 2\widetilde\gamma\sqrt{\rho}\,
 {\bf G}_\rho(u)^{\mathsf T}
 M_{\widetilde\gamma}(\rho)
 {\bf G}_\rho(u')
 \Big]\,d\rho
 \notag\\
 &\quad =
 \frac{1}{\sqrt{2}\,\pi}
 \int_0^\infty
 \rho^{3/4}
 G_\rho^{(\widetilde\gamma)}(u,u')\,d\rho.
 \label{eq:effective-Robin-rho}
\end{align}
The last expression follows from \eqref{eq:Robin-resolvent-rho}.
By \eqref{eq:Robin-positive-rho}, the last expression is
nonnegative, which ends the proof of positivity of
$r_{\widetilde\gamma}(u,u')$ together with formula \eqref{r-G}. \qed

\begin{remark}
Substituting \eqref{eq:Robin-Green-explicit}
into \eqref{r-G}
we obtain
\begin{equation}
\label{eq:jump-kernel-explicit}
\begin{aligned}
r_{\widetilde\gamma}(u,u')
={}&
\frac{1}{\sqrt{2}\pi}
\int_0^\infty \varrho^{1/4} 
\frac{
F_\varrho(u\wedge u')H_\varrho(u\vee u')
}{
(1+4\widetilde\gamma^2)\sinh \sqrt{\varrho}
+4\widetilde\gamma\cosh \sqrt{\varrho}
}
\,\dd \varrho.
\end{aligned}
\end{equation}
The representation
\eqref{eq:jump-kernel-explicit}
immediately implies that
\[
r_{\widetilde\gamma}(u,u')
=
r_{\widetilde\gamma}(u',u)>0,
\qquad u\ne u'.
\]
\end{remark}

 \subsection{Proof of Proposition \ref{lm011210-26}}
 \label{secC4}

From formulas \eqref{gFgG} and \eqref{h-ga} we obtain
\begin{align*}
 b_{\widetilde\gamma}(u;v)
 =c_{\rm bd}\int_0^\infty \big[&F(\widetilde\gamma,\sqrt\varrho)
 V_\varrho(u,v)-G(\widetilde\gamma,\sqrt\varrho)
 V_\varrho(u,1-v)\big]\frac{d\varrho}{\varrho^{3/4}}.
\end{align*}
Since, see \eqref{031310-26},
\[
 V_\varrho(u,1-v)\leq
 \cosh(\sqrt\varrho) V_\varrho(u,v),\qquad v\in\{0,1\}
\]
and
\[
 F(\widetilde\gamma, \sqrt\varrho)-G(\widetilde\gamma, \sqrt\varrho)\cosh \sqrt\varrho
 =\frac{\sinh \sqrt\varrho}
 {(1+4\widetilde\gamma^2)\sinh \sqrt\varrho
 +4\widetilde\gamma\cosh \sqrt\varrho}>0,
\]
we conclude that
\begin{align*}
 &F(\widetilde\gamma,\sqrt\varrho)V_\varrho(u,v)
 -G(\widetilde\gamma,\sqrt\varrho)V_\varrho(u,1-v)\\
 &\qquad\geq
 \big[F(\widetilde\gamma,\sqrt\varrho)
 -G(\widetilde\gamma,\sqrt\varrho)\cosh(\sqrt\varrho)\big]
 V_\varrho(u,v)>0.
\end{align*}
Thus \eqref{011410-26} follows.
 
Concerning \eqref{eq:b-boundary-asymptotics}, by the reflection symmetry, it suffices to consider
$v=0$ and $u\downarrow0$.

Using \eqref{031310-26} and changing variables  $k=\sqrt\rho$ in the
formula \eqref{h-ga}
we obtain
\begin{align}
\label{eq:b-k-representation}
 b_{\widetilde\gamma}(u;0)
 =
 2 c_{\rm bd}\int_0^\infty k^{1/2}
 \Bigg[
 \frac{\cosh(k(1-u))}{\sinh k}
 F(\widetilde\gamma,k)
  -
 \frac{\cosh(ku)}{\sinh k}
 G(\widetilde\gamma,k)
 \Bigg]\,dk.
\end{align}
From \eqref{F-G} we conclude that 
for $k\to+\infty$:
\[
 F(\widetilde\gamma,k)
 =
 \frac{1}{1+2\widetilde\gamma}
 +O(e^{-2k}),
 \qquad
 G(\widetilde\gamma,k)=O(e^{-k}).
\]
Moreover,   for $0<u\leq 1/2$,
\[
 \frac{\cosh(k(1-u))}{\sinh k}
 =
 e^{-ku}+O(e^{-k}),
 \qquad \mbox{uniformly in   }k\geq1,
\]
and
\[
 \frac{\cosh(ku)}{\sinh k}
 =O(e^{-k/2}),
 \qquad \mbox{uniformly in   }k\geq1 .
\]
It follows that the contribution to
\eqref{eq:b-k-representation} from $k\geq1$
equals
\[
  b_{\widetilde\gamma}(u;0)
 = \frac{2 c_{\rm bd}}{1+2\widetilde\gamma}
 \int_1^\infty k^{1/2}e^{-ku}\,dk
 +O(1),
\]
where the error is uniform as $u\downarrow0$.

The integral over $0<k<1$ is   uniformly
bounded,
thus,
\[
 b_{\widetilde\gamma}(u;0)
 =
 \frac{2 c_{\rm bd}}{1+2\widetilde\gamma}
 \int_0^\infty k^{1/2}e^{-ku}\,dk
 +O(1).
\]
By the Gamma integral,
\[
 \int_0^\infty k^{1/2}e^{-ku}\,dk
 =
 \Gamma(3/2)u^{-3/2}
 =
 \frac{\sqrt\pi}{2}u^{-3/2}.
\]
Consequently,
\[
 b_{\widetilde\gamma}(u;0)
 =
 \frac{\sqrt\pi c_{\rm bd}}{1+2\widetilde\gamma}
 u^{-3/2}+O(1),
 \qquad u\downarrow0.
\]
This proves \eqref{eq:b-boundary-asymptotics}.\qed

\section{Proofs of Theorems \ref{lm032209-24} and \ref{thm012611-24}}
\label{secSs}



\subsection{Spectral  fractional power of the Neumann Laplacian}

\label{secB.4}
Given a function $\varphi\in L^2[0,1]$ we denote by $\hat
\varphi_c(\ell)$ the respective Fourier coefficients.
Suppose that $\alpha\in(0,1]$. We define the operator
\begin{align}
  \label{frac-lap}
&|\Delta_N|^{\al}\varphi(u)=\sum_{n=0}^{+\infty}(n\pi)^{2\al}\hat\varphi_c(n)c_n(u),\quad\mbox{with}\notag\\
&
{\cal D}(|\Delta_N|^{\al})=\Big[\varphi\in L^2[0,1]:\, \sum_{n=0}^{+\infty}(n\pi)^{4\al}\hat\varphi_c^2(n)<+\infty\Big].
\end{align}

\subsubsection{Fractional Sobolev spaces}

\label{secB.5}

Suppose that $\al>0$. Define $H^{\alpha}[0,1]$ as the completion of
    $C^\infty[0,1]$ - $C^\infty$ smooth functions -  under that norm
    \begin{equation}
      \label{Hal}
      \begin{split}
       & \|\varphi\|_{\alpha}:=\left(\|\varphi\|^2_{L^2[0,1]}+\|\varphi\|_{\alpha,0}^2\right)^{1/2}\quad\mbox{where}\\
       &
       \|\varphi\|_{L^2[0,1]}=\left(\int_0^1\varphi^2(u)\dd u\right)^{1/2}=\left(\sum_{\ell=0}^{+\infty}
       \hat\varphi_c(\ell) ^2\right)^{1/2},\\
       &
       \|\varphi\|_{\alpha,0}:=\Big(\sum_{\ell=1}^{+\infty}(\pi
       \ell)^{2\alpha} \hat\varphi_c^2(\ell) \Big)^{1/2}.
\end{split}
\end{equation}
Let $P[0,1]$ be the space of all finite linear
 combinations made of the cosine basis.  By $H^{\alpha}_0[0,1]$ we denote the subspace of $H^{\alpha}[0,1]$
being the closure of  $C^\infty_c(0,1)$ -- the set of  $C^\infty$ smooth functions,
compactly supported in $(0,1)$  -- under the  norm
$\|\cdot\|_{\alpha,0}$ in  \eqref{Hal}. The
spaces  $H^{\alpha}[0,1]$ and  $H^{\alpha}_0[0,1]$   are Hilbert and we shall denote by
$\langle\cdot ,\cdot\rangle_{\al}$  and $\langle\cdot
,\cdot\rangle_{\al,0}$    the respective scalar products. The
scalar product $\langle\cdot
,\cdot\rangle_{\al,0}$  obviously extends to a bounded
bilinear form on $H^{\alpha}[0,1]$.

\begin{lemma}
  \label{lm022009-24} Suppose that $\al>1/2$. Then,
  $H^{\alpha}[0,1]\subset C[0,1]$ and {if $\varphi\in H^{\alpha}[0,1]$}
  \begin{equation}
    \label{042009-24}
\varphi(u)=\sum_{\ell=0}^{+\infty}\hat\varphi_c(\ell)c_\ell(u),\quad u\in[0,1]\quad\mbox{pointwise}.
\end{equation}
\end{lemma}
\proof
By the  Cauchy-Schwarz inequality,  we have for  $\alpha >1/2$ 
    \begin{align*}
      \sum_{\ell=0}^{+\infty}|\hat\varphi_c(\ell)|\le \left(\sum_{\ell=0}^{+\infty}(\hat\varphi_c(\ell))^2\ell^{2\al}\right)^{1/2}\left(\sum_{\ell=1}^{+\infty}\frac{1}{\ell^{2\al}}\right)^{1/2}<+\infty
            \end{align*}
and the  conclusion of  Lemma \ref{lm022009-24}    follows.
\qed

\medskip

From the lemma we  conclude also the following.
\begin{corollary}
  \label{lm012011-24} Under the assumption of Lemma \ref{lm022009-24}
  we have 
  \begin{equation}
    \label{010404-25} 
     H^{\alpha}_0[0,1]=\Big[\varphi\in
H^{\alpha}[0,1]:\,\varphi(0)=\varphi(1)=0\Big].
\end{equation}
In addition,
the norms $\|\cdot\|_{\alpha}$ and $\|\cdot\|_{\alpha,0}$ defined in
  \eqref{Hal} are equivalent on the space $H^{\alpha}_0[0,1]$.
\end{corollary}
\proof Denote the space on the right hand side of  \eqref{010404-25} as
${\cal H}_0^{\al}$. From the definition of $H^{\alpha}_0[0,1]$ it  is easy to see that $ {\cal H}_0^{\al}\subset
H^{\alpha}_0[0,1]$.
We show that  $ H^{\alpha}_0[0,1]\subset{\cal H}_0^{\al} $.
For $\alpha=1$, suppose that $\varphi\in H^{1}[0,1]$ and $\varphi(0)=\varphi(1) =0$.
Then $\varphi'\in L^2[0,1]$ and can be approximated by functions
$\chi_n\in C_c^\infty(0,1)$. Consider
$
\psi_n(u):=\int_0^u \chi_n(v)\dd v.
$
We have
$
\lim_{n\to+\infty}\|\psi_n-\varphi\|_{1}=0.
$
Therefore
$
\lim_{n\to+\infty}\sup_{u\in[0,1]}|\psi_n(u)-\varphi(u)|=0.
$
Let us fix a function $\chi'\in C^\infty_c(0,1)$ such that $\chi(0)=0$
and $\chi(1)=1$. Define
$
\varphi_n(u)=\psi_n(u)-\chi(u)\psi_n(1).
$
Then, $\varphi_n\in C^\infty_c(0,1)$  and 
$
\lim_{n\to+\infty}\|\varphi_n-\varphi\|_{1}=0
$
and the conclusion of the lemma follows for $\al=1$.

Suppose now that $\al\in(1/2,1)$ and
$\varphi=\sum_{\ell=0}^{+\infty}\hat\varphi_c(\ell)c_\ell(u)$ belongs
to 
$H^{\alpha}[0,1]$ and satisfies $\varphi(0)= \varphi(1)=0.$ Then, consider  fixed
functions $\chi_j\in C^\infty[0,1]$, $j=1,2$ such that
\begin{equation}
  \label{062009-24}
  \begin{split}
    \chi_1(0)=0,\quad \chi_1(1)=1\quad\mbox{and}
    \quad \chi_2(0)=1,\quad \chi_2(1)=0.
\end{split}
\end{equation}
Define
$
r_n(u):=\sum_{\ell=n+1}^{+\infty}\hat\varphi_c(\ell)c_\ell(u)
$
and
$$
\varphi_n(u):=\sum_{\ell=0}^{n}\hat\varphi_c(\ell)c_\ell(u)+ C_{n}\chi_1(u) + C_{n+1}\chi_2(u),
$$
where
\[
  \begin{split}
    0=\sum_{\ell=0}^{n}\hat\varphi_c(\ell)c_\ell(0)+ C_{n},\qquad
    0=\sum_{\ell=0}^{n}\hat\varphi_c(\ell) c_\ell(1) +C_{n+1}.
\end{split}
\]
We have
$
\varphi_n(0)=0$, $\varphi_n(1)=0
$
and $\varphi_n\in C^\infty[0,1]\subset H^1_0[0,1]$. Therefore
$(\varphi_n)\subset H^{\al}_0[0,1]$. 
Note that also
\[
  \begin{split}
    &|C_{n}|\le \sqrt{2}\sum_{\ell=n+1}^{+\infty}|\hat\varphi_c(\ell)|\\
    &
    \le \sqrt{2}\left(\sum_{\ell=n+1}^{+\infty}\frac{1}{\ell^{3/2}}\right)^{1/2}\left(\sum_{\ell=n+1}^{+\infty}\ell^{3/2}|\hat\varphi_c(\ell)|^2\right)^{1/2}\to0,
\end{split}
\]
as $n\to+\infty$. Analogously, $C_{n+1}\to0$. We can write then
\begin{align*}
  \|\varphi-\varphi_n\|_{\al}\le \|r_n\|_{\al}+C_n\|\chi_1\|_{\al}+C_{n+1}\|\chi_2\|_{\al}\to0,
  \end{align*}
  as $n\to+\infty$, and this ends the proof of \eqref{010404-25}.

\qed

\medskip

As an immediate consequence of the above result we can formulate the
following.
\begin{corollary}
  \label{cor032009-24}
  $H_0^{\al}[0,1]$ is a closed subspace of $H^{\al}[0,1]$ of
  co-dimension $2$.
  \end{corollary}

 
 We let
 \begin{align}
   \label{101710-26}
     &
     {\cal b}^{(v)}(\varphi,\varrho):=\int_0^1 V_{\ga^2\varrho^4}(u,v)
    [\varphi(v)-\varphi(u) ] \; \dd u,\quad \varphi\in
    P[0,1].
\end{align}
Define $\Phi_{v}(\varrho;\varphi)$   by letting
\begin{align}
  \label{021610-26}
 &\Phi_{v}(\varrho;\varphi):=\Phi_{v}^{\rm dir}(\varrho;\varphi)
   -\Phi_{v}^{\rm cr} (\varrho;\varphi) ,\quad \mbox{where}\\
  &
  \Phi_{v}^{\rm dir} (\varrho;\varphi):=  F(\tilde\ga, \ga\varrho^2) {\cal b}^{(v)}(\varphi,\varrho),
 \notag \\
  &
    \Phi_{v}^{\rm cr} (\varrho;\varphi):=  G(\tilde\ga, \ga\varrho^2) {\cal b}^{(1-v)}(\varphi,\varrho)
   ,\quad \varphi\in
    P[0,1].
\end{align}
Note that
\begin{align}
  \label{051610-26}
  &
     \Phi_{v}^{\rm dir} (\varrho;\varphi)=\sum_{\ell=1}^{+\infty} \frac{c_\ell(v)\hat{\varphi}_c(\ell) (\pi\ell)^2 }
    { (\ell\pi )^2 +\ga^2 \varrho^4 }F(\tilde\ga,
    \ga\varrho^2) \\
  &
    \Phi_{v}^{\rm cr} (\varrho;\varphi)=\sum_{\ell=1}^{+\infty} \frac{c_\ell(1-v)\hat{\varphi}_c(\ell) (\pi\ell)^2 }
    { (\ell\pi )^2 +\ga^2 \varrho^4 }G(\tilde\ga,
    \ga\varrho^2).  \notag
\end{align}

The definition of $\Phi_{v}(\varrho;\varphi)$ extends to  $\varphi\in
L^2[0,1]$ for a.e. $\rho\in[0,+\infty)$. We also have the following.
\begin{lemma}
   \label{lm021509-24}
   There exists $C>0$ such that
  \begin{equation}
    \label{081509-24}
    \int_0^{+\infty} \Big(\Phi_{v}(\varrho;\varphi)\Big)^2\dd \varrho\le
    C\|\varphi\|_{3/4,0}^2,\quad \varphi\in H^{3/4}[0,1],\,v=0,1
  \end{equation}
  and
  \begin{equation}
    \label{081509-24a}
    \int_0^{+\infty} \Big( {\cal b}^{(v)}(\varphi,\varrho)\Big)^2\dd \varrho\le
    C\|\varphi\|_{3/4,0}^2,\quad \varphi\in H^{3/4}[0,1],\,v=0,1.
  \end{equation}
  \end{lemma}
\proof We only prove \eqref{081509-24}. The proof of
\eqref{081509-24a} goes along the same lines.

Suppose that $v=0$, the argument for $v=1$ is analogous.  We have
\begin{align}
  \label{012001-25}
 & \int_0^{+\infty}\Big(\Phi_{0}^{\rm dir}(\varrho;\varphi)\Big)^2\dd \varrho\notag
  \\
  &
    =\sum_{\ell,\ell'=1}^{+\infty}  \hat{\varphi  }_c(\ell) \hat{\varphi
   }_c(\ell') 
 \int_{0}^{+\infty}\frac{ (\pi\ell)^2c_\ell(0) F(\tilde\ga, \ga\varrho^2) } { (      \ell\pi )^2
   +\ga^2 \varrho^4 }\cdot \frac{ (\pi\ell')^2 c_{\ell'}(0) F(\tilde\ga, \ga\varrho^2)  (\tilde\ga, \ga\varrho^2)} { (      \ell'\pi )^2
   +\ga^2 \varrho^4 }\dd \varrho.
\end{align}
From the definition of $F$, see \eqref{F-G}, one can see that
for any $\tilde \ga>0$ there exists $C>0$ such that   $0\le F(\tilde\ga, k)\le C$.
    Using formula \eqref{031601-25} to integrate over $\varrho$,
   the right hand side of \eqref{081509-24}
  can be estimated by 
  \begin{align}
    \label{012001-25a}
    &
     C \sum_{\ell,\ell'=1}^{+\infty}  
      \frac{ (\pi\ell\pi\ell')^{1/2}[\pi\ell+\pi\ell'+(\pi\ell\pi\ell')^{1/2}]\hat{\varphi}_c(\ell)
      \hat{\varphi}_c(\ell')}{\ga^{1/2}[(\pi\ell)^{1/2}+(\pi\ell')^{1/2}](\pi\ell+\pi\ell')}\\
                 &
                   \le C'\sum_{\ell,\ell'=1}^{+\infty}  
 \frac{ (\pi\ell\pi\ell')^{1/2}|\hat{\varphi
   }(\ell) ||\hat{\varphi
   }(\ell')|}
                   {(\pi\ell)^{1/2}+(\pi\ell')^{1/2}}\le   C '' 
                   \|\varphi\|_{3/4,0}^2,\notag
   \end{align}
 for some constants $C,C',C''>0$ independent of $\varphi$,  by virtue of \eqref{032009-24},
 and \eqref{081509-24} follows for $v=0$. Analogously, we show that
 \begin{align}
  \label{041610-26}
 \int_0^{+\infty}\Big(\Phi_{0}^{\rm cr}(\varrho;\varphi)\Big)^2\dd
   \varrho 
  \le 
     C 
                   \|\varphi\|_{3/4,0}^2
\end{align} \qed

\medskip

\subsection{Proof of Theorem \ref{lm032209-24}}

 \label{secB.3}
\subsubsection{Reformulation of equation   \eqref{f-diff1}}
Fix $T_0,T_1>0$ as in Definition \ref{df1.5}. Define the functionals ${\cal b}^{(v)}_{\rm temp}:  [0,+\infty)\times L^2[0,1]\to\bbR$,
$v=0,\,1$ by
\begin{align}
  \label{021601-25}
 {\cal b}^{(v)}_{\rm temp} (\varrho,\varphi):=T_v- {\cal b}^{(v)}
  (\varrho,\varphi)=T_v- \sum_{\ell'=0}^{+\infty}
  \frac{(\ell' \pi)^2  c_{\ell'}(v)\hat \varphi_c(\ell')}{ (\ell' \pi)^2 +\ga^2 \varrho^4},\quad \varphi
   \in L^2[0,1],\,\varrho>0.
\end{align}
 For   $\varphi\in H^{3/4}[0,1]$ we can write
 \begin{align}
   \label{121710-26}
       {\cal b}^{(v)}_{\rm temp} (\varrho,\varphi) =T_v-\varphi(v)+ \sum_{\ell'=0}^{+\infty}
  \frac{\ga^2 \varrho^4 c_{\ell'}(v)\hat \varphi_c(\ell')}{ (\ell' \pi)^2 +\ga^2 \varrho^4}.
\end{align}
We also let
\begin{equation}
  \label{111710-26}
  \Phi_{v,{\rm temp}}(\varrho,\varphi):=F(\tilde\ga, \ga\varrho^2)  {\cal b}^{(v)}_{\rm temp} (\varrho,\varphi)-G(\tilde\ga, \ga\varrho^2)  {\cal b}^{(1-v)}_{\rm temp} (\varrho,\varphi).
\end{equation}
Thanks to \eqref{012111-24} we have
 \begin{align}
    \label{012212-24a}
  &\int_0^t\dd s\int_{0}^{+\infty}\big[{\cal b}^{(v)}_{\rm temp} (\varrho,T(s))\big]^2\dd
   \varrho<+\infty, \quad t>0\quad\mbox{and}\\
   &\int_0^t\dd s\int_{0}^{+\infty}\big[\Phi_{v,{\rm temp}}(\varrho,T(s)) \big]^2\dd
   \varrho<+\infty, \quad t>0,\,v=0,1.\notag
 \end{align}
 Equation \eqref{f-diff1} can be rewritten as
 \begin{equation}
    \label{011809-24a}
    \begin{split}
    & \langle \varphi,  T(t)-T_{\rm ini}\rangle_{L^2[0,1]}
 = -c_{\rm bulk}
 \sum_{\ell=1}^{+ \infty}
    (\pi \ell)^{3/2} \int_0^t\hat\varphi_c(\ell) \hat T_c(s,\ell)\dd s\\
     &
   +  4\ga^{1/2}c_{\rm bd}\sum_{v=0,1}  \int_0^t\dd s
   \int_0^{+\infty}\Phi_{v,{\rm temp}}(\varrho,T(s)) {\cal
     b}^{(v)}(\varrho,\varphi)\dd \varrho,\quad \varphi\in H_0^{3/4}[0,1].
    \end{split}
  \end{equation}
   for any $\varphi\in C_c^\infty(0,1)$. To show Theorem  \ref{lm032209-24} it is equivalent with  proving  
uniqueness of  solutions to \eqref{011809-24a} in the class of
functions for which
\eqref{021601-25} holds.

\begin{lemma}
  \label{lm012109-24}
Suppose that $T(\cdot)$,  is a solution to \eqref{011809-24a} in the
sense of Definition \ref{df1.5}. Then,
\begin{equation}
  \label{012109-24-0}
  \int_0^tT(s )\dd s\in H^{3/4}[0,1],\quad \mbox{for any }t>0.
\end{equation}
Furthermore, for any $t_*>0$ we have
  \begin{equation}
    \label{012109-24}
  \sup_{t\in[0,t_*]}  \sum_{\ell=1}^{+\infty}(\pi \ell)^{3/2}\big(\int_0^t\hat
    T_c(s,\ell)\dd s\big)^2<+\infty.
  \end{equation}
  In addition,  for  $t>0$, $v=0,1$
  \begin{equation}
    \label{032109-24}
     T_vt=\int_0^tT(s,v)\dd s .
   \end{equation}
  \end{lemma}
  \proof
  From equation \eqref{011809-24a} and Lemmas \ref{lm021509-24} and
  \ref{lm022009-24} we conclude that for any $t_*>0$ there exists
  $C>0$ such that
  \[
   \Big|\left\langle  \varphi,\int_0^t T(s )\dd
     s\right\rangle_{3/4,0}\Big|\le C \|\varphi\|_{3/4,0},\quad
   \varphi\in H_0^{3/4}[0,1],\,t\in[0,t_*].
 \]
 Define two bounded linear functionals
 $$
L_o[\varphi]:=\frac{1}{2}\Big(\varphi(0)+\varphi(1)\Big)\quad\mbox{and}\quad L_e[\varphi]:=\frac{1}{2^{3/2}}\Big(\varphi(0)-\varphi(1)\Big)
$$
for $\varphi\in H^{3/4}[0,1]$.
 Let
 $$
\tilde \varphi(u)= \varphi(u)-L_o[\varphi]c_0(u)-L_e[\varphi]c_1(u).
$$
We have
\begin{align*}
&\tilde
                 \varphi(0)=\varphi(0)-L_o[\varphi]c_0(0)-L_e[\varphi]c_1(0)=0,\\
  &
    \tilde
                 \varphi(1)=\varphi(1)-L_o[\varphi]c_0(1)-L_e[\varphi]c_1(1)=0.
\end{align*}
 According to \eqref{010404-25} we have $\tilde
 \varphi\in H^{3/4}_0[0,1]$. As a result 
 there exist constants $C,C'>0$ such that
 \begin{align*}
&
   \Big|\left\langle  \varphi,\int_0^t T(s )\dd
     s\right\rangle_{3/4,0}\Big|\le \Big|\left\langle  \tilde\varphi,\int_0^t T(s )\dd
     s\right\rangle_{3/4,0}\Big|+ \pi^{3/2}\Big|L_e[\varphi]\int_0^t \hat T_c(s,1 )\dd
                  s\Big|\\
   &
     \le C \|\tilde \varphi\|_{3/4,0}+ \pi^{3/2}|L_e[\varphi]|\Big|\int_0^t \hat T_c(s,1 )\dd
                  s\Big|\le C' \| \varphi\|_{3/4}
 \end{align*}
 for all $\varphi\in P[0,1]$.
 This proves \eqref{012109-24} (thus also \eqref{012109-24-0}).

  By virtue of \eqref{042009-24} we can write (see \eqref{121710-26})
  \begin{align}
  \label{021601-25a}
 \int_0^t {\cal b}^{(0)}_{\rm temp}(\varrho,T(s))\dd s= T_0t-\int_0^t {\hat T_c}(s,0)\dd s  +
     \sum_{\ell'=1}^{+\infty} \frac{(\ell' \pi)^2 c_{\ell'}(0)}{ (\ell' \pi)^2 
    +\ga^2 
   \varrho^4    }\int_0^t \hat T_c(s,\ell')\dd s  .
  \end{align}
  Thanks to \eqref{012109-24} the last term on the right hand side belongs to
  $L^2[0,+\infty)$ (in $\varrho$), and the left hand side does as well, cf
  \eqref{012212-24a}. Therefore,
  we conclude  the first formula of \eqref{032109-24} for $v=0$. The
  proof for $v=1$ is analogous. \qed

   \medskip

Define
\begin{equation}
    \label{eq:ELK}
    \begin{split}
&
{\cal E}_L( \varphi,\psi):= c_{\rm bulk}
 \left\langle  \varphi, \psi\right\rangle_{3/2,0}
 +{\cal E}_K( \varphi,\psi) \quad\mbox{and}\\
 &
{\cal E}_K( \varphi,\psi) := 4\ga^{1/2}c_{\rm bd}\sum_{v=0,1}
\int_0^{+\infty}\Phi_{v}(\varrho;\varphi){\cal b}_{v}(\varrho;\psi)\dd
\varrho,\quad \varrho,\varphi\in P[0,1].
    \end{split}
  \end{equation}
Note that
\begin{equation}
    \label{011610-26}
    \begin{split}
    & \sum_{v=0,1}   \Phi_{v}(\varrho;\varphi){\cal
      b}^{(v)}(\psi,\varrho) =F(\tilde\ga, \ga\varrho^2)  {\cal b}^{(0)}(\varrho,\varphi) {\cal b}^{(0)}(\psi,\varrho)- G(\tilde\ga, \ga\varrho^2) {\cal b}^{(1)}(\varrho,\varphi) {\cal b}^{(0)}(\psi,\varrho)
  \\
    & + F(\tilde\ga, \ga\varrho^2)  {\cal b}^{(1)}(\varrho,\varphi) {\cal b}^{(1)}(\psi,\varrho)
      - G(\tilde\ga, \ga\varrho^2) {\cal b}^{(0)}(\varrho,\varphi) {\cal b}^{(1)}(\psi,\varrho).
    \end{split}
  \end{equation}  
Therefore the form ${\cal E}_K( \cdot,\cdot) $ is symmetric.
\begin{proposition}
  \label{prop012111-24}
  There exists $C_K>0$ such that
  \begin{equation}
    \label{032111-24}
   0\le  {\cal
            E}_K(\varphi)\le C_K\|\varphi\|_{3/4,0}^2,\quad  \varphi\in P[0,1],
        \end{equation}
        where ${\cal
            E}_K(\varphi):={\cal
            E}_K(\varphi, \varphi)$. 
        \end{proposition}
        \proof 
  The expression in \eqref{011610-26} is a bilinear form in  ${\cal
    b}^{(v)}(\varrho,\varphi)$, $ {\cal b}^{(v')}(\psi,\varrho)$, with
 $M_{\widetilde\gamma}(\rho)$ - the matrix of coefficients, given by
 \eqref{eq:M-gamma-rho}, that is positive definite. This implies that  $0\le  {\cal
            E}_K(\varphi)$.

Recall that $P[0,1]$ is the space of all trigonometric polynomials in
 cosines. 
We can further write \begin{equation}
  \label{hK}
 \begin{split}         &
            {\cal
            E}_K(\varphi,\psi):=  \sum_{\ell,\ell'=1}^{+\infty}\widehat{\cal
            K}(\ell,\ell')   \hat{\varphi
            }_c(\ell)  \hat{\psi
            }_c(\ell'),\quad \varphi,\psi\in P[0,1],\,\mbox{where}\\
 & \widehat{\cal K}(\ell,\ell'):= 2\ga^{1/2}c_{\rm bd} \sum_{v=0,1}\int_{0}^{+\infty}\frac{ (\pi\ell)^2c_\ell(v) } { (      \ell\pi )^2
   +\ga^2 \varrho^4 }\cdot \frac{ (\pi\ell')^2 c_{\ell'}(v) } { (      \ell'\pi )^2
   +\ga^2 \varrho^4 }F(\tilde\ga, \ga\varrho^2)\dd \varrho\\
 &
 - \ga^{1/2}c_{\rm bd}  \int_{0}^{+\infty}\frac{ (\pi\ell)^2c_\ell(0) } { (      \ell\pi )^2
   +\ga^2 \varrho^4 }\cdot \frac{ (\pi\ell')^2 c_{\ell'}(1) } { (      \ell'\pi )^2
   +\ga^2 \varrho^4 }G(\tilde\ga, \ga\varrho^2)\dd \varrho\\
 &
 - \ga^{1/2}c_{\rm bd}  \int_{0}^{+\infty}\frac{ (\pi\ell)^2c_\ell(1) } { (      \ell\pi )^2
   +\ga^2 \varrho^4 }\cdot \frac{ (\pi\ell')^2 c_{\ell'}(0) } { (      \ell'\pi )^2
   +\ga^2 \varrho^4 }G(\tilde\ga, \ga\varrho^2)\dd \varrho\quad \ell,\ell'=1,2,\ldots. 
 \end{split}
\end{equation}

Using formula \eqref{031601-25} to integrate over $\varrho$ and estimate
 \eqref{032009-24} we conclude that there exist constants $C,C'>0$ such that
\begin{align*}
  &{\cal
  E}_K(\varphi)
    \le C\sum_{\ell,\ell'=1}^{+\infty}\frac{(\pi\ell)^{1/2}
    (\pi\ell')^{1/2}  |\hat{\varphi
            }_c(\ell) || \hat{\varphi
            }_c(\ell')|  }{(\pi\ell)^{1/2}+ (\pi\ell')^{1/2}
    }+C\sum_{\ell,\ell'=1}^{+\infty}\frac{  (\pi\ell)
   (\pi \ell') |\hat{\varphi
            }_c(\ell)||  \hat{\varphi
            }_c(\ell') |}{\big((\pi\ell)^{1/2}+ (\pi\ell')^{1/2}\big) (\pi\ell +\pi
   \ell') }\\
   &
    \le\frac{3}{2}C\sum_{\ell,\ell'=1}^{+\infty}\frac{(\pi\ell)^{1/2}
    (\pi\ell')^{1/2}  \hat{|\varphi
            }_c(\ell) || \hat{\varphi
            }_c(\ell')|  }{(\pi\ell)^{1/2}+ (\pi\ell')^{1/2}
    }
    \le C' \sum_{\ell =1}^{+\infty} (\pi\ell)^{3/2}   [\hat{\varphi
            }_c(\ell) ]^2,
\end{align*}
for all $\varphi\in P[0,1]$  and  \eqref{032111-24} follows.
\qed

\medskip

         Thanks to
 Proposition \ref{prop012111-24} the form  ${\cal
             E}_K(\cdot,\cdot)$ extends to a closed symmetric positive
          definite form on   $H^{3/4}[0,1]\times H^{3/4}[0,1]$.

\subsection{The end of the proof of Theorem \ref{lm032209-24}}
\label{secC33}

  Suppose that $T_m(\cdot)$,
  $m=1,2$ are two solutions of \eqref{011809-24a} in the class of
  functions described in Definition \ref{df1.5}. Let $\delta
  T=T_2-T_1$. Thanks to Lemma \ref{lm012109-24} we have $I(t):=\int_0^t\delta T(s )\dd
      s\in H_0^{3/4}[0,1]$ and   it   satisfies equation
    \begin{equation}
     \label{021309-24}
     \begin{split}
      & \langle \varphi,  \delta T(t) \rangle_{L^2[0,1]} 
  = -c_{\rm bulk}
  \left\langle  \varphi, I(t)\right\rangle_{3/2,0}
    -  c_{\rm bd} {\cal E}_K\Big(\varphi, I(t)\Big)  ,\quad \varphi\in H^{3/4}_0[0,1] .
     \end{split}
   \end{equation}
 Integrating both sides of \eqref{021309-24} in $t$ we conclude
 that $I(t)$   satisfies
 \begin{equation}
     \label{021309-24a}
     \begin{split}
      & \langle \varphi,  I(t) \rangle_{L^2[0,1]} 
  = -c_{\rm bulk}
  \left\langle  \varphi, \int_0^tI(s)\dd s\right\rangle_{3/2,0}
    -  c_{\rm bd} {\cal E}_K\Big(\varphi, \int_0^tI(s)\dd s\Big)  ,\quad \varphi\in H^{3/4}_0[0,1] .
     \end{split}
   \end{equation}
Since the function $[0,+\infty)\ni t\mapsto I(t)\in H^{3/4}_0[0,1] $  is
locally bounded (in $H^{3/4}_0[0,1]$) and
continuous  in the weak topology
of $L^2[0,1]$,
it is also  continuous in the strong topology.  Substituting $\int_0^tI(s)\dd s$ for a
test function $\varphi$ in   \eqref{021309-24a} we obtain
\begin{equation}
    \label{021309-24b}
    \begin{split}
   & \frac{\dd}{\dd t}\Big\|\int_0^t I(s)\dd s \Big\|_{L^2[0,1]}^2
 = -c_{\rm bulk}
 \left\|\int_0^t I(s )\dd
     s\right\|^2_{3/4,0} 
   -  c_{\rm bd} {\cal E}_K\Big( \int_0^t
   I(s )\dd s\Big) \le 0.
    \end{split}
  \end{equation}
  This proves that $\int_0^t I(s )\dd s\equiv 0$, for all
  $t\ge0$, which in turn implies that $\int_0^t \delta T(s )\dd
s\equiv0$, $t\ge0$, that ends
  the proof of Theorem \ref{lm032209-24}.\qed


\subsection{Regularity of solutions. Proof of  Theorem
  \ref{thm012611-24}}

\subsubsection{The case of a homogeneous boundary condition}

\label{sec41}

Consider first the case when $T_{\rm ini}\in H_0^{3/4}[0,1]$. Denote
by ${\cal E}_L^o(\cdot,\cdot)$ the form ${\cal E}_L(\cdot,\cdot)$,  defined
\eqref{eq:ELK},
restricted to the space $H_0^{3/4}[0,1] \times H_0^{3/4}[0,1]$. Thanks
to Proposition \eqref{prop012111-24}
we conclude that the respective quadratic form is equivalent with
$\|\cdot\|_{3/4,0}^2$ on the   space $H^{3/4}_0[0,1]$. Its
generator $L:{\cal D}(L)\to L^2[0,1]$   is a self-adjoint operator
whose kernel ${\rm ker}(L)=\{0\}$. 
The inverse  $L^{-1}$ extends  therefore to a compact operator on $ L^2[0,1]$.
Denote by
$
\vartheta _m
$, $m=1,2,\ldots$, the orthonormal base of eigenvectors of $L $
 together with the respective eigenvalues $ 0<\lambda_1\le
 \lambda_2\le \ldots$. 
 In fact, due to the fact that
$$
{\rm c}^* \|\varphi\|_{3/4,0}^2\ge
{\cal
  E}_L^o (\varphi)\ge {\rm c}_{\rm
  bulk}\|\varphi\|_{3/4,0}^2,\quad
\varphi\in H^{3/4}_0 [0,1] 
$$
for some constant $C_{\rm bulk} $,
by the min-max principle, see \cite[Theorem X.4.8, p. 908]{DS}, there
exist $C^*,C_*>0$ such that
\begin{equation}
\label{042001-25}
 C_*m^{3/2}\le \lambda_m \le C^*m^{3/2},\quad m=1,2,\ldots.
\end{equation}
As a result we conclude that the trace of  $L^{-1}$ is finite.

With the form we can associate,  in a standard way, a strongly continuous semigroup
$\big(Q_t \big)$  of
non-negative definite, symmetric contractions on $L^2_\iota[0,1]$. By
the definition $Q_t\big(H^{3/4}_0[0,1]\big)\subset H^{3/4}_0[0,1]$  for all $t\ge0$. 
Define $T(t,u):=Q_tT_{\rm ini}(u)$. Since  $T_{\rm ini}\in
H^{3/4}_0[0,1]$ we have $T(t)\in H^{3/4}_0[0,1]$.
Combining this with the definition of  the form ${\cal E}_L^o(\cdot,\cdot)$
we conclude that $T(t,u)$ is a solution to \eqref{f-diff1} in the
sense of Definition \eqref{df1.5}.


\subsubsection{Construction of a  stationary solution}

\label{sec34}
 Denote the subspaces of $L^2[0,1]$
  \begin{align*}
    &L^2_e[0,1]:=\Big[\varphi(u):=\sum_{\ell=0}^{+\infty}\hat\varphi_{c}(2\ell)
      c_{2\ell}(u)\Big],\\
    &L^2_o[0,1]:=\Big[\varphi(u):=\sum_{\ell=1}^{+\infty}\hat\varphi_{c}(2\ell-1)
      c_{2\ell-1}(u)\Big]
    \end{align*}
    and their respective counterparts $H^{3/4}_\iota[0,1]$,
    $\iota=e,o$ - subspaces of $H^{3/4}[0,1]$.
Denote by $P^{(\iota)}$ the respective orthogonal projections onto $L^2_\iota[0,1]$.

Consider the symmetric bilinear forms ${\cal E}^{(\iota)}_L(\cdot,
\cdot)$  obtained by restricting ${\cal E}_L(\cdot,
\cdot)$ to $(\varphi, \psi)\in H_\iota^{3/4}[0,1]\times
H_\iota^{3/4}[0,1]$, $\iota=e,o$.  They are given   by the respective formulas:
\begin{align}
  \label{021704-25}
&{\cal E}^{(e)}_L(\varphi, \psi):=c_{\rm bulk}\sum_{\ell=1}^{+\infty}
 (2\pi\ell)^{3/2}\hat\varphi_c(2\ell) \hat\psi_c(2\ell)+2^{3/2}\pi c_{\rm bd}\sum_{\ell, \ell'=1}^{+\infty} \widehat{\cal
                 K}_e(\ell,\ell') \hat\varphi_c(2\ell) \hat\psi_c(2\ell'),\notag\\
  &
   {\cal E}^{(o)}_L(\varphi, \psi):=c_{\rm bulk}\sum_{\ell=1}^{+\infty}
 \big(\pi(2\ell-1)\big)^{3/2}\hat\varphi_c(2\ell-1)
    \hat\psi_c(2\ell-1)\\
  &
    +2^{3/2}\pi c_{\rm bd}\sum_{\ell, \ell'=1}^{+\infty} \widehat{\cal
                 K}_o(\ell,\ell') \hat\varphi_c(2\ell-1)
    \hat\psi_c(2\ell'-1),\notag
\end{align}
Here
$
\widehat{\cal K}_e(\ell,\ell'):=\widehat{\cal K}(2\ell,2\ell')
$ and $
\widehat{\cal K}_o(\ell,\ell'):=\widehat{\cal K}(2\ell-1,2\ell'-1).
$
Note that, see   \eqref{hK}, ${\cal K}(2\ell,2\ell'+1)=0$ for all
$\ell,\ell'=0,1,\ldots$. Therefore, for any $\varphi\in
H^{3/4}_\iota[0,1]$ and $\psi\in
H^{3/4}[0,1]$
$$
{\cal E}_L(\varphi, \psi)={\cal E}^{(\iota)}_L(\varphi,
P^{(\iota)}\psi),\quad \iota\in\{e,o\}.
$$


The quadratic forms ${\cal E}^{(\iota)}_L(\cdot)$ are equivalent with
$\|\cdot\|_{3/4,0}^2$ on the respective spaces $H_\iota^{3/4}[0,1]$,
$\iota=e,o$. Their corresponding  generators are
self-adjoint operators   $L^{(\iota)}:{\cal D}\big(L^{(\iota)}\big)\to
L^2_\iota[0,1]$, that are given by
\begin{align*}
&{\cal D}\big(L^{(\iota)}\big):=\Big[\varphi:\,   {\cal
  E}^{(\iota)}_L(\varphi,\cdot)\,\mbox{extends to a bounded lin. funct. on
                 }L^2_\iota[0,1]\Big],\\
  &
  {\cal
  E}^{(\iota)}_L(\varphi,\psi)=  -\langle L^{(\iota)}\varphi
,\psi\rangle_{L^2[0,1]},\quad \psi\in H^{3/4}_\iota[0,1] .
\end{align*}

The null space of   $L^{(e)}$ is ${\rm span}\,(1)$, while $L^{(o)}$
is 
$1-1$.  The inverses   $\big(L^{(e)}\big)^{-1}$, restricted to ${\rm
  span}\,(1)^\perp$, and $\big(L^{(o)}\big)^{-1}$    are well defined
trace class symmetric operators.
Denote by
$
\vartheta^{(\iota)}_m
$, $m=1,2,\ldots$, the orthonormal bases of eigenvectors of $L^{(\iota)}$
 together with the respective eigenvalues $0<\lambda^{(\iota)}_1\le
 \lambda^{(\iota)}_2\le \ldots$. 
We let $\vartheta_0^{(e)}(u)\equiv 1$ and $\la_0^{(e)}=0$ and by
convention $\vartheta_0^{(o)}(u)\equiv 0$ and $\la_0^{(o)}=0$.
In fact,  
by the min-max principle,   there
exist $C^*,C_*>0$ such that
\begin{equation}
\label{042001-25a}
 C_*m^{3/2}\le \lambda^{(\iota)}_m \le C^*m^{3/2},\quad m=1,2,\ldots.
\end{equation}
In addition, since $\vartheta^{(\iota)}_m \in{\cal
  D}\big(L^{(\iota)}\big)$ we have $\vartheta^{(\iota)}_m\in
H^{3/4}[0,1]\subset C[0,1]$. Furthermore, 
\begin{equation}
\label{042001-25aa}
|{\cal E}^{(\iota)}_L(\varphi,\psi)|\le
C\|\varphi\|_{3/2,0}\|\psi\|_{3/2,0},\quad \varphi,\psi\in H^{3/4}_\iota[0,1],\,\iota=e,o.
\end{equation}

Let
$
\delta_e=\frac12(\delta_0+\delta_1)$  and $\delta_o=\frac12(\delta_0-\delta_1).
$
The above distributions belong to  $H^{-3/4}[0,1]$ - the dual to
$H^{3/4}[0,1]$, as well as to  $H^{-3/4}_e[0,1]$ and
$H^{-3/4}_o[0,1]$ - the
 duals to $H^{3/4}_e[0,1]$ and $H^{3/4}_o[0,1]$, respectively. Here
 \begin{equation}
   \label{H-34}
H^{-3/4}_\iota[0,1]=\Big[\varphi=\sum_{m=1}^{+\infty}\tilde \varphi_m\vartheta_{m}^{(\iota)}:\,\|\varphi\|_{-3/4,\iota}^2:= \sum_{m=1}^{+\infty}\frac{\tilde \varphi_m^2}{\la_m^{(\iota)}}<+\infty\Big],\quad\iota\in\{o,e\}.
\end{equation}
We have
\begin{align*}
&\langle\delta _\iota,\varphi\rangle=\bar\varphi^{(\iota)},\quad
                 \mbox{where}\quad  \bar\varphi^{(e)}=\frac12\big(\varphi(0)+\varphi(1)\big), \\
  &
   \bar\varphi^{(o)}=\frac12\big(\varphi(0)-\varphi(1)\big),\quad
\varphi \in H^{3/4}[0,1].
\end{align*}
Then,
$\delta _\iota(u)=\sum_{m=0}^{+\infty}\bar\vartheta_m^{(\iota)}\vartheta_m^{(\iota)}(u) $.

Let
\begin{align}
  \label{010704-24}
    &  \bar T:=\frac12(T_0+T_1)\quad\mbox{and} \quad {\scriptstyle\Delta} T:= T_0-T_1.
  \end{align}
Let
\begin{equation}
  \label{vts}
\vartheta_s(u):=\sum_{m=1}^{+\infty}\frac{ \sD\vartheta_m^{(o)}\vartheta_{m}^{(o)}(u)}{2\la_m^{(o)}}.
\end{equation}
Note that at least for some $m_0\ge 1$  we have
  $\sD\vartheta_{m_0}^{(o)}\not=0$. Otherwise, we would have
  $\vartheta_{m}^{(o)}(0)=\vartheta_{m}^{(o)}(1)$ and also
$\vartheta_m^{(o)}(0)=-\vartheta_m^{(o)}(1)$,  for all $m=1,2,\ldots$. This
  would imply that  any  $\varphi\in H^{3/4}_o[0,1]$, thus 
$H^{3/4}_o[0,1]\subset  H^{3/4}_0[0,1]$,   which is obviously false.
We have therefore
\begin{align}
  \label{011810-26}
0<\sD \vartheta_s = \sum_{m=1}^{+\infty}\frac{ \big(\sD\vartheta_m^{(o)}\big)^2}{2\la_m^{(o)}}=\frac12\| \delta_o\|_{-3/4,o}^2<+\infty,
\end{align}
since, see \eqref{H-34},
$$
\delta_o=\sum_{m=1}^{+\infty}(\sD\vartheta_m^{(o)})\vartheta_m^{(o)}\in H^{-3/4}_o[0,1].
$$
In addition, $\vartheta_s (0)=-\vartheta_s (1)=\frac12\sD \vartheta_s
$. Note that $\vartheta_s\in H^{3/4}[0,1]$ as, thanks to \eqref{011810-26},
\begin{equation}
  \label{vtsh}
\|\vartheta_s\|^2_{3/4,0}=\sum_{m=1}^{+\infty}\frac{ (\sD\vartheta_m^{(o)})^2}{4\la_m^{(o)}}<+\infty.
\end{equation}

Let
\begin{align}
  \label{TsT}
  T_s(u):=\bar T+\frac{\sD T}{ \sD\vartheta_s} \vartheta_s(u).
\end{align}
Since
$\sD\vartheta_m^{(o)}=2\vartheta_m^{(o)}(0)=-2\vartheta_m^{(o)}(1)$,
we have  $T_s(v)=T_v$, $v=0,1$.
  Substitute $T_s $ to   the right hand side of
  \eqref{011809-24a}. Then,    for any $\varphi\in H^{3/4}_0[0,1]$ 
\begin{align*}
 &  \langle |\Delta|^{3/4}\varphi, T_s\rangle_{L^2[0,1]}+{\cal
   E}_K(\varphi,T_s)=
   \frac{\sD T}{  \sD\vartheta_s}\sum_{m=1}^{+\infty}\langle 
   \delta_o,\vartheta_{m}^{(o)}\rangle\langle
  \vartheta_{m}^{(o)},\varphi\rangle_{L^2[0,1]}
 =  \frac{\sD T\sD\varphi}{ 2\sD \vartheta_s} =0,
\end{align*}
which shows that $T_s(t,u)\equiv T_s(u)$ given by \eqref{TsT} is a stationary solution of
\eqref{011809-24a}. To see that condition \eqref{012111-24} is satisfied we can
use the uniqueness argument made in Section \ref{secC33}.

\subsubsection{The case of an inhomogeneous boundary condition}

\label{sec43}

Finally assume that $T_{\rm ini}\in H^{3/4}[0,1]$. Let $T_v:=T_{\rm
  ini}(v)$, $v=0,1$. Suppose that $T_s$ is the stationary solution
constructed in Section \ref{sec34}. Let $\tilde T(t)$ be the solution
of the homogeneous problem, corresponding to the initial data $\tilde T_{\rm ini}:=T_{\rm ini}-
T_s \in H^{3/4}_0[0,1]$.  Thanks to the argument of Section
\ref{sec41} we have  $\tilde T(t)\in H^{3/4}_0[0,1]$. Then,
$T(t)=T_s+\tilde T(t)\in H^{3/4}[0,1]$ is the unique solution of
\eqref{f-diff1} in the sense of Definition \eqref{df1.5}, thanks to the uniqueness argument made in Section \ref{secC33}.
\qed

    \subsection{Some technical results}

          \begin{lemma}\label{lm011601-25}
  The following formulas hold: for any $a_j,b_j>0$, $j=1,2$ we have
  \begin{align}
    \label{031601-25}
   \int_0^{+\infty}\frac{\dd \la} {a_1^2
    +b_1^2 \la^4 }&=\frac{\pi}{(2a_1)^{3/2}b_1^{1/2}},\ \\
      \int_0^{+\infty}\frac{\dd \la} {(a^2_1
                                 +b^2_1 \la^4) (a^2_2
                                 +b^2_2 \la^4)
      }&=\frac{\pi\big(a_1b_2+a_2b_1+(a_1b_1a_2b_2)^{1/2}\big)}{2^{3/2}
      (a_1a_2)^{3/2}[(a_1b_2)^{1/2}+(a_2b_1)^{1/2}](a_1b_2+a_2b_1)} 
     .\notag
    \end{align}
  \end{lemma}
  \proof
  See \cite[formula (3.112), p. 253]{GR}. 
  \qed

  \medskip

  \begin{lemma}
    \label{lm012009-24}
 Suppose that $\al,\beta>0$  are such that $\al+2\beta=1$ (then $1/2>\beta>0$). Then, there exists $C>0$ such that
  \begin{equation}
    \label{011909-24}
   0\le \int_0^{+\infty} \int_0^{+\infty}\frac{f(x)f(y)\dd x\dd
      y}{(x^{\al}+y^{\al})x^{\beta}y^{\beta}}\le C
    \int_0^{+\infty}f^2(x)\dd x
  \end{equation}
  for all $f\in L^2(0,+\infty)$.
  \end{lemma}
  \proof
  We have
  \begin{align*}
    \int_0^{+\infty} \int_0^{+\infty}\frac{f(x)f(y)\dd x\dd
      y}{(x^{\al}+y^{\al})x^{\beta}y^{\beta}}=\int_0^{+\infty}\dd
    \rho\Big(\int_0^{+\infty}\frac{e^{-\rho x^{\al}}f(x)\dd x}{x^{\beta}}\Big)^2.
  \end{align*}
  Changing variables $\rho'=1/\rho$ we can write the right hand side
  as
  \begin{align*}
   & \int_0^{+\infty}\frac{\dd
    \rho}{\rho^2}\Big(\int_0^{+\infty}\frac{e^{-x^{\al}/\rho}f(x)\dd
     x}{x^{\beta}}\Big)^2
      \le \int_0^{+\infty}\frac{\dd
    \rho}{\rho^2}\Big(\sum_{n=0}^{+\infty}e^{-n^{\al}}\int_{n\rho^{1/\al}}^{(n+1)\rho^{1/\al}}\frac{f(x)\dd
      x}{x^{\beta}}\Big)^2.
  \end{align*}
  Since
  $
C_\al:=\sum_{n=0}^{+\infty}e^{-n^{\al}}<+\infty
$
the utmost right hand side can be estimated by
$C_\al {\rm I}_n$, where
\begin{align}
  \label{022009-24}
      &
        {\rm I}_n=\int_0^{+\infty}\frac{\dd
    \rho}{\rho^2} \Big(\int_{n\rho^{1/\al}}^{(n+1)\rho^{1/\al}}\frac{f(x)\dd
        x}{x^{\beta}}\Big)^2=\int_0^{+\infty}\frac{\dd
    \rho}{\rho^2} \Big(\int_{0}^{ \rho^{1/\al}}\frac{f(x+n\rho^{1/\al})\dd
        x}{(x+n\rho^{1/\al})^{\beta}}\Big)^2\notag\\
      &
      \le \int_0^{+\infty}\frac{\dd
    \rho}{\rho^2}\Big(\int_{0}^{ \rho^{1/\al}}\frac{f(x+n\rho^{1/\al})\dd
        x}{x^{\beta}}\Big)^2.
  \end{align}
  To estimate the utmost right hand side we show that
      \begin{equation}
        \label{012009-24}
        \begin{split}
         & {\rm I}[g]\le C\|g\|_{L^2(0,+\infty)}^2,\quad g\in
         L^2(0,+\infty),\quad\mbox{where}\\
         &
         {\rm I}[g]=\int_0^{+\infty}\frac{\dd
    \rho}{\rho^2}\Big(\int_{0}^{ \rho^{1/\al}}\frac{g(x)\dd
        x}{x^{\beta}}\Big)^2.
         \end{split}
\end{equation}
This combined with \eqref{022009-24} yields an estimate
\begin{align}
  \label{032009-24a}
       \int_0^{+\infty}\frac{\dd
    \rho}{\rho^2}\Big(\int_0^{+\infty}\frac{e^{-x^{\al}/\rho}f(x)\dd
     x}{x^{\beta}}\Big)^2
\le CC_\al\|f\|_{L^2(0,+\infty)}^2,
  \end{align}
  which ends the proof of \eqref{011909-24}.
The only remaining part is to show  \eqref{012009-24}

\subsubsection*{Proof of \eqref{012009-24}}
We omit the notation for a function writing the functional ${\rm I}$.
      Changing variables $\rho:=(\rho')^{\al}$ we obtain
  \begin{align*}
   & {\rm I}=\al\int_0^{+\infty}\frac{\dd
    \rho}{\rho^{\al+1}}\Big(\int_0^{\rho}\frac{g(x)\dd
     x}{x^{\beta}}\Big)^2\\
    &
      = \al\int_0^{+\infty} \dd
    \rho\Big(\frac{1}{\rho^{(\al+1)/2}}\int_1^{\rho }\frac{g(x)\dd
     x}{x^{\beta}}\Big)^2=\al\int_0^{+\infty} \dd
    \rho\Big(\frac{1}{\rho^{1-\beta}}\int_1^{\rho }\frac{g(x)\dd
     x}{x^{\beta}}\Big)^2.
  \end{align*}
  Recall the Hardy inequality: if $\beta+\frac1p<1$ and $p>1$, $\beta>0$, then
  \begin{align*}
   & \int_0^{+\infty} \dd
    \rho\Big(\frac{1}{\rho^{1-\beta}}\int_1^{\rho }\frac{g(x)\dd
     x}{x^{\beta}}\Big)^p\le
     \frac{1}{1-\beta-1/p}\int_0^{+\infty}|g(x)|^p\dd x,
  \end{align*}
  see \cite[A.4, p. 272]{stein}.
  Applying it in our case we get
  \begin{align*}
    {\rm I}
      \le \frac{\al\|g\|_{L^2(0,+\infty)}^2 }{1/2-\beta}
  \end{align*}
  and  estimate \eqref{012009-24} follows.
  \qed

  \medskip
  Here is an obvious corollary of the lemma.

  \begin{corollary}
    \label{cor012009-24}
 Under the assumptions of Lemma \ref{lm012009-24}   there exists a constant $C>0$ such that
    \begin{equation}
      \label{032009-24}
     0\le  \sum_{\ell,\ell'=1}^{+\infty}\frac{a_\ell
       a_{\ell'}}{(\ell^{\al}+(\ell')^{\al})\ell^{\beta}(\ell')^{\beta}}\le C\sum_{\ell=1}^{+\infty} a^2_\ell
   \end{equation}
   for all $(a_\ell)\in \ell^2$.
 \end{corollary}

\unappendix


\pagebreak

\begin{center}
  \textbf{\Large Supplemental Materials}
\end{center}

 \setcounter{supsection}{0}
\setcounter{equation}{0}
\setcounter{figure}{0}
\setcounter{table}{0}
\makeatletter

\renewcommand{\theequation}{S\thesupsection.\arabic{equation}}

\renewcommand{\thefigure}{S\arabic{figure}}

The following material is considered to be a supplement for the
foregoing paper, presenting   the calculations (referred to as being
straighforward, but tedious) that have not been performed in the paper.

\setcounter{supsection}{1}
\section{\thesupsection.   Proof of Proposition \ref{prop011612-24}}

\label{Ssec4.3}

Let
\begin{align*} {\cal H}_n^{(2)}(t) 
&=\frac 1{2(n+1)} 
\sum_{x,x'=0}^n\bigg\{\left(\mathbb E_n\left[ p_x(t)
  p_{x'}(t)\right] \right)^2+\left(\mathbb E_n\left[ r_x(t)
  r_{x'} (t)\right] \right)^2 \notag \\ & \quad \qquad \qquad \qquad +2 \left (\mathbb E_n\left[ p_x(t)
  r_{x'} (t)\right] \right)^2\bigg\}.\notag
\end{align*}
We have
\begin{align*}
&  \frac{\dd}{\dd t} {\cal H}_n^{(2)}(t) =I(t)+{\rm II}(t)+{\rm III}(t),\quad
                         \mbox{with}\\
  &
  I(t): =\frac 1{n+1} \sum_{x,x'=0}^n\left(\mathbb E_n\left[ p_x(t)
  p_{x'}(t)\right] \right) \frac{\dd}{\dd t}\left(\mathbb E_n\left[ p_x(t)
                         p_{x'}(t)\right] \right)\\
  &
    {\rm II}(t):=\frac 1{n+1} \sum_{x,x'=1}^n\left(\mathbb E_n\left[
    r_x (t)
  r_{x'} (t)\right] \right) \frac{\dd}{\dd t}\left(\mathbb E_n\left[
    r_x (t)
    r_{x'} (t)\right] \right) \notag \\
  &  
   {\rm III}(t):=\frac 2{n+1} \sum_{x=0}^n \sum_{x'=1}^n   \left (\mathbb E_n\left[ p_x(t)
  r_{x'}(t)\right] \right) \frac{\dd}{\dd t}\left (\mathbb E_n\left[ p_x(t)
   r_{x'}(t)\right] \right).
\end{align*}

    Next, for $x,x'=1,\ldots,n-1$ we have
    \begin{align*}
&\frac{\dd}{\dd t}\mathbb E_n\left[  r_x(t)
  r_{x'}'(t)\right] =\mathbb E_n\left[ \dot{ r}_x(t)
  r_{x'}(t)\right] +\mathbb E_n\left[{  r}_x(t)
                     \dot{  r}_{x'}(t)\right] \\
      &=n^{3/2}\Big\{\mathbb E_n\left[ \nabla^\star p_x(t)
  r_{x'}(t)\right] +\mathbb E_n\left[{r}_x(t)
                     \nabla^\star p_{x'}(t)\right]\Big\},
    \end{align*}

    \begin{align*}
&\frac{\dd}{\dd t}\left (\mathbb E_n\left[ p_x(t)
  r_{x'}(t)\right] \right)=n^{3/2}\Big\{\mathbb E_n\left[ \nabla r_x 
  r_{x'}(t)\right] +\ga\mathbb E_n\left[ \Delta_{\rm N}
   p_x(t) 
      r_{x'}(t)\right] \\
      &+\mathbb E_n\left[ p_x(t)
  \nabla^\star p_{x'}(t)\right] \Big\}
\end{align*}
    and
\begin{align*}
&\dd\left(\mathbb E_n\left[ p_x(t)
                         p_{x'}(t)\right] \right)=\left(\mathbb
                 E_n\left[ \dd p_x(t)
                         p_{x'}(t)\right] \right)+\left(\mathbb E_n\left[ p_x(t)
                 \dd p_{x'}(t)\right] \right)\\
  &
 +  \left(\mathbb E_n\left[ \dd p_x(t)
                         \dd p_{x'}(t)\right] \right)
\end{align*}
\begin{align*}
&=n^{3/2}\Big\{\mathbb E_n\left[ \nabla r_x (t)
                 p_{x'}(t)\right] +\ga\mathbb E_n\left[ \Delta_{\rm N}
   p_x(t)
                 p_{x'}(t)\right] 
    +\mathbb E_n\left[ \nabla r_{x'} (t)
                 p_{x}(t)\right] \\
  &+\ga\mathbb E_n\left[ \Delta_{\rm N}
   p_{x'}(t)
                 p_{x}(t)\right]
   -\delta_{x'=x+1}\ga \mathbb E_n\left[ \nabla^\star  p_{x+1}(t)
    \nabla^\star  p_{x'}(t) \right]\\
  &
    -\delta_{x=x'+1}\ga \mathbb E_n\left[ \nabla^\star  p_{x'+1}(t)
    \nabla^\star  p_{x}(t) \right] +
    \delta_{x'=x}\ga \mathbb E_n\Big[\big( \nabla^\star  p_{x+1}(t)
    \big)^2+\big( \nabla^\star  p_{x}(t)
    \big)^2\Big]\Big\}\dd t
\end{align*}

    For $x=0$, $x'=1,\ldots,n-1$
     we have
    \begin{align*}
&\frac{\dd}{\dd t}\mathbb E_n\left[ r_0(t)
  r_{x'}(t)\right] =0,
    \end{align*}

    \begin{align*}
&\frac{\dd}{\dd t}\left (\mathbb E_n\left[ p_0(t)
  r_{x'}(t)\right] \right)=n^{3/2}\Big\{\mathbb E_n\left[ \nabla r_0 
  r_{x'}(t)\right] +\ga\mathbb E_n\left[ \Delta_{\rm N}
   p_0(t) 
      r_{x'}(t)\right] \\
      &+\mathbb E_n\left[ p_0(t)
  \nabla^\star p_{x'}(t)\right] -\tilde \ga \mathbb E_n\left[ p_0(t)
  r_{x'}(t)\right]\Big\}
\end{align*}
    and
\begin{align*}
&\dd\left(\mathbb E_n\left[ p_0(t)
                         p_{x'}(t)\right] \right)=\left(\mathbb E_n\left[ \dd p_0(t)
                         p_{x'}(t)\right] \right)+\left(\mathbb E_n\left[ p_0(t)
                 \dd p_{x'}(t)\right] \right)\\
  &
 +  \left(\mathbb E_n\left[ \dd p_0(t)
                         \dd p_{x'}(t)\right] \right)
\end{align*}
\begin{align*}
&=n^{3/2}\Big\{\mathbb E_n\left[ \nabla r_0 (t)
                 p_{x'}(t)\right] +\ga\mathbb E_n\left[ \Delta_{\rm N}
   p_0(t)
                 p_{x'}(t)\right] -\tilde\ga\mathbb E_n\left[ 
   p_0(t)
                 p_{x'}(t)\right] \\
  &
    +\mathbb E_n\left[ \nabla r_{x'} (t)
                 p_{0}(t)\right] +\ga\mathbb E_n\left[ \Delta_{\rm N}
   p_{x'}(t)
    p_{0}(t)\right]
   -\delta_{x'=1}\ga \mathbb E_n\left[ \Big(\nabla^\star  p_{1}(t)\Big)^2\right]\Big\}\dd t
\end{align*}

  For $x=0$, $x'=n$
     we have
    \begin{align*}
&\frac{\dd}{\dd t}\mathbb E_n\left[ r_0(t)
  r_{n}(t)\right] =0,
    \end{align*}

    \begin{align*}
&\frac{\dd}{\dd t}\left (\mathbb E_n\left[ p_0(t)
  r_{n}(t)\right] \right)=n^{3/2}\Big\{\mathbb E_n\left[ \nabla r_0 
  r_{n}(t)\right] +\ga\mathbb E_n\left[ \Delta_{\rm N}
   p_0(t) 
      r_{n}(t)\right] \\
      &+\mathbb E_n\left[ p_0(t)
  \nabla^\star p_{n}(t)\right] -\tilde \ga \mathbb E_n\left[ p_0(t)
  r_{n}(t)\right]\Big\}
\end{align*}
    and
\begin{align*}
&\dd\left(\mathbb E_n\left[ p_0(t)
                         p_{n}(t)\right] \right)=\left(\mathbb E_n\left[ \dd p_0(t)
                         p_{n}(t)\right] \right)+\left(\mathbb E_n\left[ p_0(t)
                 \dd p_{n}(t)\right] \right)
\end{align*}
\begin{align*}
&=n^{3/2}\Big\{\mathbb E_n\left[ \nabla r_0 (t)
                 p_{n}(t)\right] +\ga\mathbb E_n\left[ \Delta_{\rm N}
   p_0(t)
                 p_{n}(t)\right] -\tilde\ga\mathbb E_n\left[ 
   p_0(t)
                 p_{n}(t)\right] \\
  &
    +\mathbb E_n\left[ \nabla r_{n} (t)
                 p_{0}(t)\right] +\ga\mathbb E_n\left[ \Delta_{\rm N}
   p_{n}(t)    p_{0}(t)\right]
  \Big\}\dd t
\end{align*}

      For $x=x'=0$
     we have
    \begin{align*}
&\frac{\dd}{\dd t}\mathbb E_n\left[ \Big(r_0(t)
 \Big)^2\right] =0,
    \end{align*}

    \begin{align*}
&\frac{\dd}{\dd t}\left (\mathbb E_n\left[ p_0(t)
  r_{0}(t)\right] \right)=0
\end{align*}
 and
\begin{align*}
\dd\left(\mathbb E_n\left[ \Big(p_0(t)
                         \Big)^2\right] \right)=2\left(\mathbb E_n\left[ \dd p_0(t)
                         p_{0}(t)\right] \right)
 +  \left(\mathbb E_n\left[ \Big(\dd p_0(t)
                         \Big)^2\right] \right)
\end{align*}
\begin{align*}
&=n^{3/2}\Big\{\mathbb E_n\left[ 2\nabla r_0 (t)
                 p_{0}(t)\right] +2\ga\mathbb E_n\left[ \Delta_{\rm N}
   p_0(t)
                 p_{0}(t)\right] -2\tilde\ga\mathbb E_n\left[ 
   \Big(p_0(t)\Big)^2
                 \right] \\
  &
    +2\tilde \ga T_L
   +\ga \mathbb E_n\left[ \Big(\nabla^\star  p_{1}(t)\Big)^2\right]\Big\}\dd t
\end{align*}

    For $x=n$, $x'=1,\ldots,n-1$
     we have
     \begin{align*}
&\frac{\dd}{\dd t}\mathbb E_n\left[ r_n(t)
  r_{x'}(t)\right] =\mathbb E_n\left[ \dot{r}_n(t)
  r_{x'}(t)\right] +\mathbb E_n\left[{r}_n(t)
                     \dot{ r}_{x'}(t)\right] \\
      &=n^{3/2}\Big\{\mathbb E_n\left[ \nabla^\star p_n(t)
  r_{x'}(t)\right] +\mathbb E_n\left[{r}_n(t)
                     \nabla^\star p_{x'}(t)\right]\Big\},
    \end{align*}

    \begin{align*}
&\frac{\dd}{\dd t}\left (\mathbb E_n\left[ p_n(t)
  r_{x'}(t)\right] \right)=n^{3/2}\Big\{\mathbb E_n\left[ \nabla r_n 
  r_{x'}(t)\right] +\ga\mathbb E_n\left[ \Delta_{\rm N}
   p_n(t) 
      r_{x'}(t)\right] \\
      &+\mathbb E_n\left[ p_n(t)
  \nabla^\star p_{x'}(t)\right] -\tilde \ga \mathbb E_n\left[ p_n(t)
  r_{x'}(t)\right]\Big\}
\end{align*}
    and
\begin{align*}
&\dd\left(\mathbb E_n\left[ p_n(t)
                         p_{x'}(t)\right] \right)=\left(\mathbb E_n\left[ \dd p_n(t)
                         p_{x'}(t)\right] \right)+\left(\mathbb E_n\left[ p_n(t)
                 \dd p_{x'}(t)\right] \right)\\
  &
 +  \left(\mathbb E_n\left[ \dd p_n(t)
                         \dd p_{x'}(t)\right] \right)
\end{align*}
\begin{align*}
&=n^{3/2}\Big\{\mathbb E_n\left[ \nabla r_n (t)
                 p_{x'}(t)\right] +\ga\mathbb E_n\left[ \Delta_{\rm N}
   p_n(t)
                 p_{x'}(t)\right] -\tilde\ga\mathbb E_n\left[ 
   p_n(t)
                 p_{x'}(t)\right] \\
  &
    +\mathbb E_n\left[ \nabla r_{x'} (t)
                 p_{n}(t)\right] +\ga\mathbb E_n\left[ \Delta_{\rm N}
   p_{x'}(t)
    p_{n}(t)\right]
   -\delta_{x'=n-1}\ga \mathbb E_n\left[ \Big(\nabla^\star  p_{n}(t)\Big)^2\right]\Big\}\dd t
\end{align*}

  For $x=x'=n$
     we have
     \begin{align*}
&\frac{\dd}{\dd t}\mathbb E_n\left[ \Big(r_n(t)\Big)^2
  \right] =2\mathbb E_n\left[ \dot{r}_n(t)
  r_{n}(t)\right] =
      n^{3/2}\mathbb E_n\left[ \nabla^\star p_n(t)
  r_{n}(t)\right],
    \end{align*}

    \begin{align*}
&\frac{\dd}{\dd t}\left (\mathbb E_n\left[ p_n(t)
  r_{n}(t)\right] \right)=n^{3/2}\Big\{\mathbb E_n\left[ \nabla r_n 
  r_{n}(t)\right] +\ga\mathbb E_n\left[ \Delta_{\rm N}
   p_n(t) 
      r_{n}(t)\right] \\
      &+\mathbb E_n\left[ p_n(t)
  \nabla^\star p_{n}(t)\right] -\tilde \ga \mathbb E_n\left[ p_n(t)
  r_{n}(t)\right]\Big\}
\end{align*}

    and
\begin{align*}
&\dd\left(\mathbb E_n\left[ \Big(p_n(t)
                         \Big)^2\right] \right)=2\mathbb E_n\left[ \dd p_n(t)
                         p_{n}(t)\right]
 +  \mathbb E_n\left[ \Big(\dd p_n(t)
                         \Big)^2\right] 
\end{align*}
\begin{align*}
&=n^{3/2}\Big\{2\mathbb E_n\left[ \nabla r_n (t)
                 p_{n}(t)\right] +2\ga\mathbb E_n\left[ \Delta_{\rm N}
   p_n(t)
                 p_{n}(t)\right] -2\tilde\ga\mathbb E_n\left[ 
  \Big( p_n(t)\Big)^2\right] \\
  &
  +2\tilde \ga T_R +\ga \mathbb E_n\left[ \Big(\nabla^\star  p_{n}(t)\Big)^2\right]\Big\}\dd t
\end{align*}

\subsection*{Calculation for the $p-p$ covariance}

We have
\begin{align*}
I(t)=\frac{1}{n+1}\sum_{x,x'=0}^n\left(\mathbb E_n\left[ p_x(t)
  p_{x'}(t)\right] \right) \frac{\dd}{\dd t}\left(\mathbb E_n\left[ p_x(t)
                         p_{x'}(t)\right] \right)=\frac{1}{n+1}\sum_{j=1}^6I_j(t),
\end{align*}

with
\begin{align*}
&I_1(t)=\sum_{x,x'=1}^{n-1}\left(\mathbb E_n\left[ p_x(t)
  p_{x'}(t)\right] \right) \frac{\dd}{\dd t}\left(\mathbb E_n\left[ p_x(t)
                 p_{x'}(t)\right] \right)\\
  &
    =n^{3/2}\sum_{x,x'=1}^{n-1}\mathbb E_n\left[ p_x(t)
  p_{x'}(t)\right] \mathbb E_n\left[ \nabla r_x (t)
    p_{x'}(t)\right] \\
  &
  +  n^{3/2}\sum_{x,x'=1}^{n-1}\mathbb E_n\left[ p_x(t)
  p_{x'}(t)\right] \mathbb E_n\left[ \nabla r_{x'} (t)
                 p_{x}(t)\right] 
\end{align*}
\begin{align*}
  &
    +\ga n^{3/2}\sum_{x,x'=1}^{n-1}\mathbb E_n\left[ p_x(t)
  p_{x'}(t)\right] \mathbb E_n\left[ \Delta_{\rm N}
   p_x(t)
    p_{x'}(t)\right] \\
   &
    +\ga n^{3/2}\sum_{x,x'=1}^{n-1}\mathbb E_n\left[ p_x(t)
  p_{x'}(t)\right] \mathbb E_n\left[ \Delta_{\rm N}
   p_{x'}(t)
                 p_{x}(t)\right] 
\end{align*}
\begin{align*}
  &
    +\ga n^{3/2}\sum_{x=1}^{n-1}\mathbb E_n\left[ \left(p_x(t)\right)^2
   \right]  \mathbb E_n\Big[\big( \nabla^\star  p_{x+1}(t)
    \big)^2+\big( \nabla^\star  p_{x}(t)
    \big)^2\Big]\\
  &
  - 2\ga n^{3/2}\sum_{x=1}^{n-2}\mathbb E_n\left[ \left(p_x(t)\right) \left(p_{x+1}(t)\right)
   \right]   \mathbb E_n\left[ \big( \nabla^\star  p_{x+1}(t)
    \big)^2\right]   
\end{align*}

\begin{align*}
&I_2(t)=\sum_{x'=1}^{n-1}\left(\mathbb E_n\left[ p_0(t)
  p_{x'}(t)\right] \right) \frac{\dd}{\dd t}\left(\mathbb E_n\left[ p_0(t)
                 p_{x'}(t)\right] \right)\\
  &+\sum_{x=1}^{n-1}\left(\mathbb E_n\left[ p_0(t)
  p_{x}(t)\right] \right) \frac{\dd}{\dd t}\left(\mathbb E_n\left[ p_0(t)
                 p_{x}(t)\right] \right)
\end{align*}
\begin{align*}
&
=n^{3/2} \sum_{x'=1}^{n-1} \mathbb E_n\left[ p_0(t)
  p_{x'}(t)\right] \mathbb E_n\left[ \nabla r_0 (t)
                 p_{x'}(t)\right]   \\
  &+n^{3/2} \sum_{x=1}^{n-1}\left(\mathbb E_n\left[ p_0(t)
  p_{x}(t)\right] \right) \mathbb E_n\left[ \nabla r_0 (t)
                 p_{x}(t)\right] 
\end{align*}
\begin{align*}
&
+n^{3/2} \sum_{x'=1}^{n-1} \mathbb E_n\left[ p_0(t)
  p_{x'}(t)\right] \mathbb E_n\left[ p_0(t)
                 \nabla r_{x'}(t)\right]   \\
  &+n^{3/2} \sum_{x=1}^{n-1}\left(\mathbb E_n\left[ p_0(t)
  p_{x}(t)\right] \right) \mathbb E_n\left[ p_0 (t)
                \nabla r_{x}(t)\right] 
\end{align*}
\begin{align*}
&
+\ga  n^{3/2} \sum_{x'=1}^{n-1} \mathbb E_n\left[ p_0(t)
  p_{x'}(t)\right] \mathbb E_n\left[ \Delta_{\rm N}
   p_0(t)
                 p_{x'}(t)\right]   \\
  &+\ga  n^{3/2} \sum_{x=1}^{n-1}\left(\mathbb E_n\left[ p_0(t)
  p_{x}(t)\right] \right) \mathbb E_n\left[ \Delta_{\rm N}
   p_0(t)
                 p_{x}(t)\right] 
\\
&
-2\ga  n^{3/2}  \mathbb E_n\left[ p_0(t)
  p_{1}(t)\right] \mathbb E_n\left[ \Big(\nabla^\star
     p_{1}(t)\Big)^2\right]  \\
  &
-2\tilde \ga  n^{3/2}  \sum_{x'=1}^{n-1}\mathbb E_n\left[ p_0(t)
  p_{x'}(t)\right] \mathbb E_n\left[ p_0(t)
  p_{x'}(t)\right] 
\end{align*}

\begin{align*}
&I_3(t)=\mathbb E_n\left[ \left(p_0(t)
  \right)^2\right]\frac{\dd}{\dd t} \mathbb E_n\left[ \left(p_0(t)
  \right)^2\right]
    =2n^{3/2}\mathbb E_n\left[ \left(p_0(t)
  \right)^2\right]\mathbb E_n\left[ \nabla r_0 (t)
    p_{0}(t)\right]\\
  &
    +2\ga n^{3/2}\mathbb E_n\left[ \left(p_0(t)
  \right)^2\right]\mathbb E_n\left[ \Delta_{\rm N}
   p_0(t)
                 p_{0}(t)\right] 
\end{align*}
\begin{align*}
&
-2\tilde \ga n^{3/2}\mathbb E_n\left[ \left(p_0(t)
  \right)^2\right]\mathbb E_n\left[ 
   \Big(p_0(t)\Big)^2
                 \right]  
 +2\tilde \ga n^{3/2}T_L\mathbb E_n\left[ \left(p_0(t)
    \right)^2\right]   \\
  &
    +\ga n^{3/2} \mathbb E_n\left[ \left(p_0(t)
    \right)^2\right]\mathbb E_n\left[ \Big(\nabla^\star  p_{1}(t)\Big)^2\right]
\end{align*}

\begin{align*}
&I_4(t)=2\mathbb E_n\left[ \left(p_0(t)
  \right) \left(p_n(t)
  \right)\right]\frac{\dd}{\dd t} \mathbb E_n\left[ \left(p_0(t)
  \right) \left(p_n(t)
  \right)\right]\\
  &
    =2n^{3/2}\mathbb E_n\left[ \left(p_0(t)
  \right) \left(p_n(t)
  \right)\right]\mathbb E_n\left[ \nabla r_0 (t)
    p_{n}(t)\right] \\
  &
    + 2n^{3/2}\mathbb E_n\left[ \left(p_0(t)
  \right) \left(p_n(t)
  \right)\right]\mathbb E_n\left[ \nabla r_n (t)
    p_{0}(t)\right]
\end{align*}
\begin{align*}
  &
    +2\ga n^{3/2}\mathbb E_n\left[ \left(p_0(t)
  \right) \left(p_n(t)
  \right)\right]\mathbb E_n\left[ \Delta_{\rm N}
   p_0(t)
    p_{n}(t)\right]\\
    &
    +2\ga n^{3/2}\mathbb E_n\left[ \left(p_0(t)
  \right) \left(p_n(t)
  \right)\right]\mathbb E_n\left[ \Delta_{\rm N}
   p_n(t)
                 p_{0}(t)\right] 
\end{align*}
\begin{align*}
  &
    -2\tilde \ga n^{3/2}\mathbb E_n\left[ \left(p_0(t)
  \right) \left(p_n(t)
  \right)\right]\mathbb E_n\left[ 
   p_0(t)
                 p_{n}(t)\right] 
\end{align*}

\begin{align*}
&I_5(t)=\sum_{x'=1}^{n-1}\left(\mathbb E_n\left[ p_n(t)
  p_{x'}(t)\right] \right) \frac{\dd}{\dd t}\left(\mathbb E_n\left[ p_n(t)
                 p_{x'}(t)\right] \right)\\
  &
    +\sum_{x=1}^{n-1}\left(\mathbb E_n\left[ p_n(t)
  p_{x}(t)\right] \right) \frac{\dd}{\dd t}\left(\mathbb E_n\left[ p_n(t)
                 p_{x}(t)\right] \right)
\end{align*}
\begin{align*}
&=n^{3/2}\sum_{x'=1}^{n-1}\mathbb E_n\left[ p_n(t)
  p_{x'}(t)\right] \mathbb E_n\left[ \nabla r_n (t)
                 p_{x'}(t)\right] \\
  &
   + n^{3/2}\sum_{x'=1}^{n-1}\mathbb E_n\left[ p_n(t)
  p_{x'}(t)\right] \mathbb E_n\left[ \nabla r_{x'} (t)
                 p_{n}(t)\right] \\
  &
    +n^{3/2}\sum_{x=1}^{n-1}\mathbb E_n\left[ p_n(t)
  p_{x}(t)\right] \mathbb E_n\left[ \nabla r_n (t)
    p_{x}(t)\right]   \\
  &
   + n^{3/2}\sum_{x=1}^{n-1}\mathbb E_n\left[ p_n(t)
  p_{x}(t)\right] \mathbb E_n\left[ \nabla r_{x} (t)
                 p_{n}(t)\right] 
\end{align*}
\begin{align*}
  &+
    \ga n^{3/2}\sum_{x'=1}^{n-1}\mathbb E_n\left[ p_n(t)
  p_{x'}(t)\right] \mathbb E_n\left[ \Delta_{\rm N}
   p_n(t)
    p_{x'}(t)\right] \\
  &+
    \ga n^{3/2}\sum_{x'=1}^{n-1}\mathbb E_n\left[ p_n(t)
  p_{x'}(t)\right] \mathbb E_n\left[ \Delta_{\rm N}
   p_{x'}(t)
                 p_{n}(t)\right] \\
  &
    +\ga n^{3/2}\sum_{x=1}^{n-1}\mathbb E_n\left[ p_n(t)
  p_{x}(t)\right] \mathbb E_n\left[ \Delta_{\rm N}
   p_n(t)
    p_{x}(t)\right] \\
  &
    +\ga n^{3/2}\sum_{x=1}^{n-1}\mathbb E_n\left[ p_n(t)
  p_{x}(t)\right] \mathbb E_n\left[ \Delta_{\rm N}
   p_x(t)
                 p_{n}(t)\right] 
\end{align*}
\begin{align*}
  &-
\tilde    \ga n^{3/2}\sum_{x'=1}^{n-1}\mathbb E_n\left[ p_n(t)
  p_{x'}(t)\right] \mathbb E_n\left[ p_n(t)
    p_{x'}(t)\right] \\
   &-
\tilde    \ga n^{3/2}\sum_{x=1}^{n-1}\mathbb E_n\left[ p_n(t)
  p_{x}(t)\right] \mathbb E_n\left[ p_n(t)
  p_{x}(t)\right] 
\end{align*}
\begin{align*}
  &-
   2\ga n^{3/2} \mathbb E_n\left[ p_n(t)
  p_{n-1}(t)\right] \mathbb E_n\left[ \Big(\nabla^\star  p_{n}(t)\Big)^2\right]
\end{align*}


\begin{align*}
&I_6(t)= \mathbb E_n\left[ \left(p_n(t)
  \right)^2\right] \frac{\dd}{\dd t} \mathbb E_n\left[ \left(p_n(t)
                 \right)^2\right]
    =2n^{3/2}\mathbb E_n\left[ \left(p_n(t)
  \right)^2\right] \mathbb E_n\left[ \nabla r_n (t)
    p_{n}(t)\right] \\
  &
   + 2n^{3/2}\ga \mathbb E_n\left[ \left(p_n(t)
  \right)^2\right] \mathbb E_n\left[ \Delta_{\rm N}
   p_n(t)
                 p_{n}(t)\right] 
\end{align*}
\begin{align*}
&-2n^{3/2} \tilde\ga\mathbb E_n\left[ \left(p_n(t)
  \right)^2\right] \mathbb E_n\left[ 
                 \Big( p_n(t)\Big)^2\right]  
    +2n^{3/2} \tilde\ga T_R\mathbb E_n\left[ \left(p_n(t)
    \right)^2\right] \\
  &
   + n^{3/2}  \ga  \mathbb E_n\left[ \left(p_n(t)
  \right)^2\right] \mathbb E_n\left[ \Big(\nabla^\star  p_{n}(t)\Big)^2\right]
  \end{align*}

Recall that $p_{-1}(t)=p_0(t)$ and  $p_{n+1}(t)=p_n(t)$.
By summing up we get
\begin{align*}
& (n+1)I(t)=2n^{3/2}\sum_{x,x'=0}^{n}\mathbb E_n\left[ p_x(t)
  p_{x'}(t)\right] \mathbb E_n\left[ \nabla r_x (t)
  p_{x'}(t)\right] 
\end{align*}
\begin{align*}
  &
    +\ga n^{3/2}\sum_{x,x'=0}^{n}\mathbb E_n\left[ p_x(t)
  p_{x'}(t)\right] \mathbb E_n\left[ \Delta_{\rm N}
   p_x(t)
    p_{x'}(t)\right] \\
   &
    +\ga n^{3/2}\sum_{x,x'=0}^{n}\mathbb E_n\left[ p_x(t)
  p_{x'}(t)\right] \mathbb E_n\left[ \Delta_{\rm N}
   p_{x'}(t)
                 p_{x}(t)\right] 
\end{align*}
\begin{align*}
  &
    +\ga n^{3/2}\sum_{x=0}^{n}\mathbb E_n\left[ \left(p_x(t)\right)^2
   \right]  \mathbb E_n\Big[\big( \nabla^\star  p_{x+1}(t)
    \big)^2+\big( \nabla^\star  p_{x}(t)
    \big)^2\Big]\\
  &
  - 2\ga n^{3/2}\sum_{x=0}^{n-1}\mathbb E_n\left[ \left(p_x(t)\right) \left(p_{x+1}(t)\right)
   \right]   \mathbb E_n\left[ \big( \nabla^\star  p_{x+1}(t)
    \big)^2\right] 
\end{align*}
\begin{align*}
&
 +2\tilde \ga n^{3/2}\Big(T_L-\mathbb E_n\left[ \left(p_0(t)
                 \right)^2\right] \Big) \mathbb E_n\left[ \left(p_0(t)
                 \right)^2\right]  \\
  &
    +2n^{3/2} \tilde\ga\Big( T_R-\mathbb E_n\left[ \left(p_n(t)
    \right)^2\right] \Big)\mathbb E_n\left[ \left(p_n(t)
    \right)^2\right]  \\
  &-2n^{3/2} \tilde\ga\sum_{x=1}^n \left(\mathbb E_n\left[ p_0(t)
  p_{x}(t)\right] \right)^2 -
2\tilde    \ga n^{3/2}\sum_{x=0}^{n-1}\left(\mathbb E_n\left[ p_n(t)
  p_{x}(t)\right] \right)^2\\
\end{align*}

\begin{align*}
& =2n^{3/2}\sum_{x,x'=0}^{n}\mathbb E_n\left[ p_x(t)
  p_{x'}(t)\right] \mathbb E_n\left[ \nabla r_x (t)
  p_{x'}(t)\right]  
\end{align*}
\begin{align*}
  &
    -2\ga n^{3/2}\sum_{x=1}^{n}\sum_{x'=0}^{n}\left(\mathbb E_n\left[ \nabla^\star p_x(t)
  p_{x'}(t)\right] \right)^2 \\
  &
    +\ga n^{3/2}\sum_{x=0}^{n-1}\mathbb E_n\left[ \left(p_x(t)\right)^2
   \right]  \mathbb E_n\Big[\big( \nabla^\star  p_{x+1}(t)
    \big)^2 \Big]\\
  &
    +\ga n^{3/2}\sum_{x=0}^{n-1}\mathbb E_n\left[ \left(p_{x+1}(t)\right)^2
   \right]  \mathbb E_n\Big[\big( \nabla^\star  p_{x+1}(t)
    \big)^2\Big]\\
  &
  - 2\ga n^{3/2}\sum_{x=0}^{n-1}\mathbb E_n\left[ \left(p_x(t)\right) \left(p_{x+1}(t)\right)
   \right]   \mathbb E_n\left[ \big( \nabla^\star  p_{x+1}(t)
    \big)^2\right] 
\end{align*}
\begin{align*}
&
 +2\tilde \ga n^{3/2}\Big(T_L-\mathbb E_n\left[ \left(p_0(t)
                 \right)^2\right] \Big) \mathbb E_n\left[ \left(p_0(t)
                 \right)^2\right]  \\
  &
    +2n^{3/2} \tilde\ga\Big( T_R-\mathbb E_n\left[ \left(p_n(t)
    \right)^2\right] \Big)\mathbb E_n\left[ \left(p_n(t)
    \right)^2\right]  \\
  &-2n^{3/2} \tilde\ga\sum_{x=1}^n \left(\mathbb E_n\left[ p_0(t)
  p_{x}(t)\right] \right)^2 -
2\tilde    \ga n^{3/2}\sum_{x=0}^{n-1}\left(\mathbb E_n\left[ p_n(t)
  p_{x}(t)\right] \right)^2\\
\end{align*}

\begin{align*}
& =2n^{3/2}\sum_{x,x'=0}^{n}\mathbb E_n\left[ p_x(t)
  p_{x'}(t)\right] \mathbb E_n\left[ \nabla r_x (t)
  p_{x'}(t)\right]  
\end{align*}
\begin{align*}
  &
    -2\ga n^{3/2}\sum_{x=1}^{n}\sum_{x'=0}^{n}\left(\mathbb E_n\left[ \nabla^\star p_x(t)
  p_{x'}(t)\right] \right)^2  
    +\ga n^{3/2}\sum_{x=0}^{n-1}  \left(\mathbb E_n\Big[\big( \nabla^\star  p_{x+1}(t)
    \big)^2 \Big]\right)^2
\end{align*}
\begin{align*}
&
 +2\tilde \ga n^{3/2}\Big(T_L-\mathbb E_n\left[ \left(p_0(t)
                 \right)^2\right] \Big) \mathbb E_n\left[ \left(p_0(t)
                 \right)^2\right]  \\
  &
    +2n^{3/2} \tilde\ga\Big( T_R-\mathbb E_n\left[ \left(p_n(t)
    \right)^2\right] \Big)\mathbb E_n\left[ \left(p_n(t)
    \right)^2\right]  \\
  &-2n^{3/2} \tilde\ga\sum_{x=1}^n \left(\mathbb E_n\left[ p_0(t)
  p_{x}(t)\right] \right)^2 -
2\tilde    \ga n^{3/2}\sum_{x=0}^{n-1}\left(\mathbb E_n\left[ p_n(t)
  p_{x}(t)\right] \right)^2\\
\end{align*}
\begin{align*}
& =n^{3/2}\sum_{x,x'=0}^{n}\mathbb E_n\left[ p_x(t)
  p_{x'}(t)\right] \mathbb E_n\left[ \nabla r_x (t)
  p_{x'}(t)\right] \\
  &
    +  n^{3/2}\sum_{x,x'=0}^{n-1}\mathbb E_n\left[ p_x(t)
  p_{x'}(t)\right] \mathbb E_n\left[ \nabla r_{x'} (t)
                 p_{x}(t)\right] 
\end{align*}
\begin{align*}
  &
    -2\ga n^{3/2}\sum_{x=1}^{n}\mathop{\sum_{x'=0}^{n}}_{x'\not\in\{x-1,x\}}\left(\mathbb E_n\left[ \nabla^\star p_x(t)
  p_{x'}(t)\right] \right)^2  +
    P_n(t)
\end{align*}
\begin{align*}
&
 +2\tilde \ga n^{3/2}\Big(T_L-\mathbb E_n\left[ \left(p_0(t)
                 \right)^2\right] \Big) \mathbb E_n\left[ \left(p_0(t)
                 \right)^2\right]  \\
  &
    +2n^{3/2} \tilde\ga\Big( T_R-\mathbb E_n\left[ \left(p_n(t)
    \right)^2\right] \Big)\mathbb E_n\left[ \left(p_n(t)
    \right)^2\right]  \\
  &-2n^{3/2} \tilde\ga\sum_{x=1}^n \left(\mathbb E_n\left[ p_0(t)
  p_{x}(t)\right] \right)^2 -
2\tilde    \ga n^{3/2}\sum_{x=0}^{n-1}\left(\mathbb E_n\left[ p_n(t)
  p_{x}(t)\right] \right)^2,
\end{align*}
with
\begin{align*}
 & P_n(t):=-2\ga n^{3/2}\sum_{x=1}^{n} \left(\mathbb E_n\left[ \nabla^\star p_x(t)
   p_{x}(t)\right] \right)^2  -2\ga n^{3/2}\sum_{x=1}^{n} \left(\mathbb E_n\left[ \nabla^\star p_x(t)
   p_{x-1}(t)\right] \right)^2  \\
  &
    +\ga n^{3/2}\sum_{x=0}^{n-1}  \left(\mathbb E_n\Big[\big( \nabla^\star  p_{x+1}(t)
    \big)^2 \Big]\right)^2
\end{align*}
\begin{align*}
 &  =-2\ga n^{3/2}\sum_{x=1}^{n} \left(\mathbb E_n\left[ (p_x)^2(t)
    \right] -\mathbb E_n\left[ p_{x-1}(t)
   p_{x}(t)\right] \right)^2  \\
  &-2\ga n^{3/2}\sum_{x=1}^{n} \left(\mathbb E_n\left[ (p_{x-1})^2(t)
    \right] -\mathbb E_n\left[ p_{x-1}(t)
   p_{x}(t)\right] \right)^2  \\
  &
    +\ga n^{3/2}\sum_{x=0}^{n-1}  \left(\mathbb E_n\Big[\big( p_{x+1}(t)-  p_{x}(t)
    \big)^2 \Big]\right)^2
\end{align*}
\begin{align*}
 &  =-2\ga n^{3/2}\sum_{x=1}^{n} \left(\mathbb E_n\left[ (p_x)^2(t)
    \right] -\mathbb E_n\left[ p_{x-1}(t)
   p_{x}(t)\right] \right)^2  \\
  &-2\ga n^{3/2}\sum_{x=1}^{n} \left(\mathbb E_n\left[ (p_{x-1})^2(t)
    \right] -\mathbb E_n\left[ p_{x-1}(t)
   p_{x}(t)\right] \right)^2  \\
  &
    +\ga n^{3/2}\sum_{x=0}^{n-1}  \left(\mathbb E_n\Big[
    (p_{x+1})^2(t)+    (p_{x})^2(t)- 2   p_{x+1}'(t)   p_{x}(t)
      \Big]\right)^2
\end{align*}
\begin{align*}
 &  =-2\ga n^{3/2}\sum_{x=1}^{n} \left\{\left(\mathbb E_n\left[ (p_x)^2(t)
    \right]\right)^2 +\left(\mathbb E_n\left[ p_{x-1}(t)
   p_{x}(t)\right]\right)^2 -2 \mathbb E_n\left[ (p_x)^2(t)
    \right]\mathbb E_n\left[ p_{x-1}(t)
   p_{x}(t)\right]\right\}  \\
  &-2\ga n^{3/2}\sum_{x=1}^{n} \left\{\left(\mathbb E_n\left[ (p_{x-1})^2(t)
    \right] \right)^2 +\left(\mathbb E_n\left[ p_{x-1}(t)
   p_{x}(t)\right]\right)^2  -2 \mathbb E_n\left[ (p_{x-1})^2(t)
    \right]\mathbb E_n\left[ p_{x-1}(t)
   p_{x}(t)\right]\right\} \\
  &
    +\ga n^{3/2}\sum_{x=0}^{n-1}  \Big\{\left(\mathbb E_n\Big[
    (p_{x+1})^2(t)\Big]\right)^2+\left(\mathbb E_n\Big[
    (p_{x})^2(t)\Big]\right)^2+ 4 \left(\mathbb E_n\Big[
    p_{x+1}(t)    p_{x}(t) \Big]\right)^2
  \\
  &
   +2 \mathbb E_n\Big[ \tilde  p^2_{x+1}(t)\Big]\mathbb E_n\Big[ \tilde  p^2_{x}(t)\Big]-4 \mathbb E_n\Big[ \tilde  p^2_{x+1}(t)\Big]\mathbb E_n\Big[ \tilde  p_{x+1}(t) \tilde  p_{x}(t) \Big]-4 \mathbb E_n\Big[ \tilde  p^2_{x}(t)\Big]\mathbb E_n\Big[ \tilde  p_{x+1}(t) \tilde  p_{x}(t) \Big]\Big\}
\end{align*}
\begin{align*}
 &  =-\ga n^{3/2}\sum_{x=0}^{n-1} \left\{\cancel{2}\left(\mathbb E_n\left[ (p_{x+1})^2(t)
    \right]\right)^2 +\cancel{2\left(\mathbb E_n\left[ p_{x}(t)
   p_{x+1}(t)\right]\right)^2} -\cancel{4 \mathbb E_n\left[ (p_{x+1})^2(t)
    \right]\mathbb E_n\left[ p_{x}(t)
   p_{x+1}(t)\right]}\right\}  \\
  &-\ga n^{3/2}\sum_{x=0}^{n-1} \left\{\cancel{2}\left(\mathbb E_n\left[ (p_{x})^2(t)
    \right] \right)^2 +\cancel{2\left(\mathbb E_n\left[ p_{x}(t)
   p_{x+1}(t)\right]\right)^2}  -\cancel{4 \mathbb E_n\left[ (p_{x})^2(t)
    \right]\mathbb E_n\left[ p_{x}(t)
   p_{x+1}(t)\right]}\right\} \\
  &
    +\ga n^{3/2}\sum_{x=0}^{n-1}  \Big\{\cancel{\left(\mathbb E_n\Big[ \tilde  p^2_{x+1}(t)\Big]\right)^2}+\cancel{\left(\mathbb E_n\Big[ \tilde  p^2_{x}(t)\Big]\right)^2}+ \cancel{4 \left(\mathbb E_n\Big[ \tilde  p_{x+1}(t) \tilde  p_{x}(t) \Big]\right)^2}
  \\
  &
   +2 \mathbb E_n\Big[ \tilde  p^2_{x+1}(t)\Big]\mathbb E_n\Big[ \tilde  p^2_{x}(t)\Big]-\cancel{4 \mathbb E_n\Big[ \tilde  p^2_{x+1}(t)\Big]\mathbb E_n\Big[ \tilde  p_{x+1}(t) \tilde  p_{x}(t) \Big]}-\cancel{4 \mathbb E_n\Big[ \tilde  p^2_{x}(t)\Big]\mathbb E_n\Big[ \tilde  p_{x+1}(t) \tilde  p_{x}(t) \Big]}\Big\}
\end{align*}
\begin{align*}
  &
    =-\ga n^{3/2}\sum_{x=0}^{n-1}\Big[\nabla\left(\mathbb E_n\left[ (p_{x})^2(t)
    \right]\right) \Big]^2
\end{align*}
Summarizing, we have shown that
\begin{align*}
& I(t)=\frac{2n^{3/2}}{n+1}\sum_{x,x'=0}^{n}\mathbb E_n\left[ p_x(t)
  p_{x'}(t)\right] \mathbb E_n\left[ \nabla r_x (t)
                 p_{x'}(t)\right]\\
  &
    -2\ga \frac{n^{3/2}}{n+1}\sum_{x=1}^{n}\mathop{\sum_{x'=0}^{n}}_{x'\not\in\{x-1,x\}}\left(\mathbb E_n\left[ \nabla^\star p_x(t)
  p_{x'}(t)\right] \right)^2  -\ga \frac{n^{3/2}}{n+1}\sum_{x=0}^{n-1}\Big[\nabla\left(\mathbb E_n\left[ (p_{x})^2(t)
    \right]\right)^2 \Big]^2
\end{align*}
\begin{align*}
&
 +2\tilde \ga \frac{n^{3/2}}{n+1}\Big(T_L-\mathbb E_n\left[ \left(p_0(t)
                 \right)^2\right] \Big) \mathbb E_n\left[ \left(p_0(t)
                 \right)^2\right]  \\
  &
    +2 \frac{n^{3/2}}{n+1} \tilde\ga\Big( T_R-\mathbb E_n\left[ \left(p_n(t)
    \right)^2\right] \Big)\mathbb E_n\left[ \left(p_n(t)
    \right)^2\right]  \\
  &-2 \frac{n^{3/2}}{n+1}\tilde\ga\sum_{x=1}^n \left(\mathbb E_n\left[ p_0(t)
  p_{x}(t)\right] \right)^2 -
2\tilde    \ga \frac{n^{3/2}}{n+1}\sum_{x=0}^{n-1}\left(\mathbb E_n\left[ p_n(t)
  p_{x}(t)\right] \right)^2,
\end{align*}

\subsection*{Calculation for the $r-r$ covariance}

We have
\begin{align*}
  &
    {\rm II}(t)=\frac {1}{n+1} \sum_{x,x'=1}^n\left(\mathbb E_n\left[ r_x(t)
  r_{x'}(t)\right] \right) \frac{\dd}{\dd t}\left(\mathbb E_n\left[ r_x(t)
    r_{x'}(t)\right] \right) \\
  &
    = \frac{n^{3/2}}{n+1}\sum_{x,x'=1}^n\mathbb E_n\left[ r_x(t)
  r_{x'}(t)\right] \mathbb E_n\left[ \nabla^\star p_x(t)
    r_{x'}(t)\right] \\
  &
    +\frac{n^{3/2}}{n+1}\sum_{x,x'=1}^n\mathbb E_n\left[ r_x(t)
  r_{x'}(t)\right]\mathbb E_n\left[{r}_x(t)
                     \nabla^\star p_{x'}(t)\right]
\end{align*}
\begin{align*}
  &
    =\frac {2n^{3/2}}{n+1}  \sum_{x,x'=1}^n\mathbb E_n\left[ r_x(t)
  r_{x'}(t)\right] \mathbb E_n\left[ \nabla^\star p_x(t)
    r_{x'}(t)\right] 
\end{align*}

\subsection*{Calculation for the $r-p$ covariance}

We have
\begin{align*}
   {\rm III}(t)=\frac 2{n+1} \sum_{x=0}^n \sum_{x'=1}^n   \left (\mathbb E_n\left[ p_x(t)
  r_{x'}(t)\right] \right) \frac{\dd}{\dd t}\left (\mathbb E_n\left[ p_x(t)
  r_{x'}(t)\right] \right)=\frac 2{n+1} \sum_{j=1}^3 {\rm III}_j(t),
\end{align*}
with
\begin{align*}
 & {\rm III}_1(t)= n^{3/2}\sum_{x=1}^{n-1} \sum_{x'=1}^n   \mathbb E_n\left[ p_x(t)
  r_{x'}(t)\right]\mathbb E_n\left[ \nabla r_x 
  r_{x'}(t)\right] \\
  &
   + \ga n^{3/2}\sum_{x=1}^{n-1} \sum_{x'=1}^n    \mathbb E_n\left[ p_x(t)
  r_{x'}(t)\right]\mathbb E_n\left[ \Delta_{\rm N}
   p_x(t) 
    r_{x'}(t)\right]\\
  &
  +  n^{3/2}\sum_{x=1}^{n-1} \sum_{x'=1}^n    \mathbb E_n\left[ p_x(t)
  r_{x'}(t)\right]\mathbb E_n\left[ p_x(t)
  \nabla^\star p_{x'}(t)\right],
\end{align*}

\begin{align*}
 & {\rm III}_2(t)= n^{3/2}  \sum_{x'=1}^n   \mathbb E_n\left[ p_0(t)
  r_{x'}(t)\right]\mathbb E_n\left[ \nabla r_0 
  r_{x'}(t)\right] \\
  &
   + \ga n^{3/2}  \sum_{x'=1}^n    \mathbb E_n\left[ p_0(t)
  r_{x'}(t)\right]\mathbb E_n\left[ \Delta_{\rm N}
   p_0(t) 
      r_{x'}(t)\right] \\
  &
  +  n^{3/2}  \sum_{x'=1}^n    \mathbb E_n\left[ p_0(t)
  r_{x'}(t)\right]\mathbb E_n\left[ p_0(t)
    \nabla^\star p_{x'}(t)\right]\\
  &
    -\tilde \ga  n^{3/2}  \sum_{x'=1}^n    \mathbb E_n\left[ p_0(t)
  r_{x'}(t)\right]\mathbb E_n\left[ p_0(t)
  r_{x'}(t)\right]
\end{align*}

\begin{align*}
 & {\rm III}_3(t)= n^{3/2}   \sum_{x'=1}^n  \mathbb E_n\left[ p_n(t)
  r_{x'}(t)\right]\mathbb E_n\left[ \nabla r_n 
   r_{x'}(t)\right] \\
  &
   + \ga n^{3/2} \sum_{x'=1}^n    \mathbb E_n\left[ p_n(t)
  r_{x'}(t)\right]\mathbb E_n\left[ \Delta_{\rm N}
   p_n(t) 
      r_{x'}(t)\right] \\
  &
  +  n^{3/2}  \sum_{x'=1}^n    \mathbb E_n\left[ p_n(t)
  r_{x'}(t)\right]\mathbb E_n\left[ p_{n}(t)
    \nabla^\star p_{x'}(t)\right]\\
  &
    -\tilde \ga  n^{3/2} \sum_{x'=1}^n    \mathbb E_n\left[ p_n(t)
  r_{x'}(t)\right]\mathbb E_n\left[ p_n(t)
  r_{x'}(t)\right]
\end{align*}

Hence
\begin{align*}
&   {\rm III}(t)=\frac {2n^{3/2}}{n+1} \sum_{x=0}^{n} \sum_{x'=1}^n   \mathbb E_n\left[ p_x(t)
  r_{x'}(t)\right]\mathbb E_n\left[ \nabla r_x (t)
                 r_{x'}(t)\right]\\
 &
   + \ga \frac {2n^{3/2}}{n+1}\sum_{x=0}^{n} \sum_{x'=1}^n    \mathbb E_n\left[ p_x(t)
  r_{x'}(t)\right]\mathbb E_n\left[ \Delta_{\rm N}
   p_x(t) 
    r_{x'}(t)\right]\\
  &
  +  \frac {2n^{3/2}}{n+1}\sum_{x=0}^{n} \sum_{x'=1}^n    \mathbb E_n\left[ p_x(t)
  r_{x'}(t)\right]\mathbb E_n\left[ p_x(t)
  \nabla^\star p_{x'}(t)\right]
\end{align*}
\begin{align*}
& -\tilde \ga \frac {2n^{3/2}}{n+1} \sum_{x'=1}^n    \mathbb E_n\left[ p_0(t)
  r_{x'}(t)\right]\mathbb E_n\left[ p_0(t)
                 r_{x'}(t)\right]\\
  & -\tilde \ga \frac {2n^{3/2}}{n+1} \sum_{x'=1}^n    \mathbb E_n\left[ p_n(t)
  r_{x'}(t)\right]\mathbb E_n\left[ p_n(t)
                 r_{x'}(t)\right]
\end{align*}

\begin{align*}
&   =-\frac {2n^{3/2}}{n+1} \sum_{x=1}^{n} \sum_{x'=1}^n   \mathbb E_n\left[ \nabla^\star p_x(t)
  r_{x'}(t)\right]\mathbb E_n\left[ r_x (t)
                 r_{x'}(t)\right]\\
 &
   + \ga \frac {2n^{3/2}}{n+1}\sum_{x=0}^{n} \sum_{x'=1}^n    \mathbb E_n\left[ p_x(t)
  r_{x'}(t)\right]\mathbb E_n\left[ \Delta_{\rm N}
   p_x(t) 
    r_{x'}(t)\right]\\
  &
  +  \frac {2n^{3/2}}{n+1}\sum_{x=0}^{n} \sum_{x'=1}^n    \mathbb E_n\left[ p_x(t)
  r_{x'}(t)\right]\mathbb E_n\left[ p_x(t)
  \nabla^\star p_{x'}(t)\right]
\end{align*}
\begin{align*}
& -\tilde \ga \frac {2n^{3/2}}{n+1} \sum_{x'=1}^n    \mathbb E_n\left[ p_0(t)
  r_{x'}(t)\right]\mathbb E_n\left[ p_0(t)
                 r_{x'}(t)\right]\\
  & -\tilde \ga \frac {2n^{3/2}}{n+1} \sum_{x'=1}^n    \mathbb E_n\left[ p_n(t)
  r_{x'}(t)\right]\mathbb E_n\left[ p_n(t)
                 r_{x'}(t)\right]
\end{align*}
\begin{align*}
&   =-\frac {2n^{3/2}}{n+1} \sum_{x=1}^{n} \sum_{x'=1}^n   \mathbb E_n\left[ \nabla^\star p_x(t)
  r_{x'}(t)\right]\mathbb E_n\left[ r_x (t)
                 r_{x'}(t)\right]\\
 &
   - \ga \frac {2n^{3/2}}{n+1}\sum_{x=1}^{n} \sum_{x'=1}^n    \left(\mathbb E_n\left[ \nabla^\star p_x(t)
  r_{x'}(t)\right] \right)^2\\
  &
  - \frac {2n^{3/2}}{n+1}\sum_{x=0}^{n} \sum_{x'=0}^n    \mathbb E_n\left[ p_x(t)
 \nabla r_{x'}(t)\right]\mathbb E_n\left[ p_x(t)
  p_{x'}(t)\right]
\end{align*}
\begin{align*}
 -\tilde \ga \frac {2n^{3/2}}{n+1} \sum_{x'=1}^n   \left( \mathbb E_n\left[ p_0(t)
  r_{x'}(t)\right] \right)^2  -\tilde \ga \frac {2n^{3/2}}{n+1} \sum_{x'=1}^n  \left(  \mathbb E_n\left[ p_n(t)
  r_{x'}(t)\right] \right)^2
\end{align*}

\subsection*{The   equation for ${\cal H}_n^{(2)}(t)$}
We have the following   equation
 \begin{align*}
& \frac{\dd}{\dd t}{\cal H}_n^{(2)}(t) =\cancel{\frac{2n^{3/2}}{n+1}\sum_{x,x'=0}^{n}\mathbb E_n\left[ p_x(t)
  p_{x'}(t)\right] \mathbb E_n\left[ \nabla r_x (t)
                 p_{x'}(t)\right]}\\
  &
    -2\ga \frac{n^{3/2}}{n+1}\sum_{x=1}^{n}\mathop{\sum_{x'=0}^{n}}_{x'\not\in\{x-1,x\}}\left(\mathbb E_n\left[ \nabla^\star p_x(t)
  p_{x'}(t)\right] \right)^2  -\ga \frac{n^{3/2}}{n+1}\sum_{x=0}^{n-1}\Big[\nabla\left(\mathbb E_n\left[ (p_{x})^2(t)
    \right]\right) \Big]^2
\end{align*}
\begin{align*}
&
 +2\tilde \ga \frac{n^{3/2}}{n+1}\Big(T_L-\mathbb E_n\left[ \left(p_0(t)
                 \right)^2\right] \Big) \mathbb E_n\left[ \left(p_0(t)
                 \right)^2\right]  \\
  &
    +2 \frac{n^{3/2}}{n+1} \tilde\ga\Big( T_R-\mathbb E_n\left[ \left(p_n(t)
    \right)^2\right] \Big)\mathbb E_n\left[ \left(p_n(t)
    \right)^2\right]  \\
  &-2 \frac{n^{3/2}}{n+1}\tilde\ga\sum_{x=1}^n \left(\mathbb E_n\left[ p_0(t)
  p_{x}(t)\right] \right)^2 -
2\tilde    \ga \frac{n^{3/2}}{n+1}\sum_{x=0}^{n-1}\left(\mathbb E_n\left[ p_n(t)
  p_{x}(t)\right] \right)^2,
\end{align*}

\begin{align*}
  &
    +\cancel{\frac {2n^{3/2}}{n+1}  \sum_{x,x'=1}^n\mathbb E_n\left[ r_x(t)
  r_{x'}(t)\right] \mathbb E_n\left[ \nabla^\star p_x(t)
    r_{x'}(t)\right] }
\end{align*}

\begin{align*}
&   -\cancel{\frac {2n^{3/2}}{n+1} \sum_{x=1}^{n} \sum_{x'=1}^n   \mathbb E_n\left[ \nabla^\star p_x(t)
  r_{x'}(t)\right]\mathbb E_n\left[ r_x (t)
                 r_{x'}(t)\right]}\\
 &
   - \ga \frac {2n^{3/2}}{n+1}\sum_{x=1}^{n} \sum_{x'=1}^n    \left(\mathbb E_n\left[ \nabla^\star p_x(t)
  r_{x'}(t)\right] \right)^2\\
  &
  \cancel{- \frac {2n^{3/2}}{n+1}\sum_{x=0}^{n} \sum_{x'=0}^n    \mathbb E_n\left[ p_x(t)
 \nabla r_{x'}(t)\right]\mathbb E_n\left[ p_x(t)
  p_{x'}(t)\right]}
\end{align*}
\begin{align*}
 -\tilde \ga \frac {2n^{3/2}}{n+1} \sum_{x'=1}^n   \left( \mathbb E_n\left[ p_0(t)
  r_{x'}(t)\right] \right)^2  -\tilde \ga \frac {2n^{3/2}}{n+1} \sum_{x'=1}^n  \left(  \mathbb E_n\left[ p_n(t)
  r_{x'}(t)\right] \right)^2
\end{align*}

Summarizing, we have shown the following
\begin{align*}
& \frac{\dd}{\dd t}{\cal H}_n^{(2)}(t) = 
    - \frac{2\ga n^{3/2}}{n+1}\sum_{x=1}^{n}\mathop{\sum_{x'=0}^{n}}_{x'\not\in\{x-1,x\}}\left(\mathbb E_n\left[ \nabla^\star p_x(t)
  p_{x'}(t)\right] \right)^2  -\frac{\ga  n^{3/2}}{n+1}\sum_{x=0}^{n-1}\Big[\nabla\left(\mathbb E_n\left[ (p_{x})^2(t)
                 \right]\right) \Big]^2\\
    &
   - \frac {2 \ga n^{3/2}}{n+1}\sum_{x=1}^{n} \sum_{x'=1}^n    \left(\mathbb E_n\left[ \nabla^\star p_x(t)
  r_{x'}(t)\right] \right)^2\\
\end{align*}
\begin{align*}
&
 +\frac{2\tilde \ga n^{3/2}}{n+1}T_L\Big(T_L-\mathbb E_n\left[ \left(p_0(t)
                 \right)^2\right] \Big) 
    + \frac{2\tilde\ga n^{3/2}}{n+1} T_R\Big( T_R-\mathbb E_n\left[ \left(p_n(t)
    \right)^2\right] \Big)
    \end{align*}
\begin{align*}
&
 -\frac{2\tilde \ga n^{3/2}}{n+1}\Big(T_L-\mathbb E_n\left[ \left(p_0(t)
                 \right)^2\right] \Big)^2
    - \frac{2\tilde\ga n^{3/2}}{n+1} \Big( T_R-\mathbb E_n\left[ \left(p_n(t)
    \right)^2\right] \Big)^2  \\
  &- \frac{2\tilde \ga n^{3/2}}{n+1} \sum_{x=1}^n \left(\mathbb E_n\left[ p_0(t)
  p_{x}(t)\right] \right)^2 -
  \frac{2\tilde \ga n^{3/2}}{n+1}\sum_{x=0}^{n-1}\left(\mathbb E_n\left[ p_n(t)
  p_{x}(t)\right] \right)^2.
\end{align*}
\begin{align*}
 -\tilde \ga \frac {2n^{3/2}}{n+1} \sum_{x'=1}^n   \left( \mathbb E_n\left[ p_0(t)
  r_{x'}(t)\right] \right)^2  -\tilde \ga \frac {2n^{3/2}}{n+1} \sum_{x'=1}^n  \left(  \mathbb E_n\left[ p_n(t)
  r_{x'}(t)\right] \right)^2
\end{align*}

In the integral form we can write
\begin{align}
  \label{Senergy-bal}
& {\cal H}_n^{(2)}(t) +
     \frac{2\ga n^{3/2}}{n+1}\sum_{x=1}^{n}\mathop{\sum_{x'=0}^{n}}_{x'\not\in\{x-1,x\}}\int_0^t\left(\mathbb E_n\left[ \nabla^\star p_x(s)
                 p_{x'}(s)\right] \right)^2 \dd s \\
  &
    +\frac{\ga  n^{3/2}}{n+1}\sum_{x=0}^{n-1}\int_0^t\Big[\nabla\left(\mathbb E_n\left[ (p_{x})^2(s)
                 \right]\right)\Big]^2\dd s 
   + \frac {2 \ga n^{3/2}}{n+1}\sum_{x=1}^{n} \sum_{x'=1}^n   \int_0^t \left(\mathbb E_n\left[ \nabla^\star p_x(s)
  r_{x'}(s)\right] \right)^2\dd s\notag
\end{align}

\begin{align*}
&
 +\frac{2\tilde \ga n^{3/2}}{n+1}\int_0^t\Big(T_L-\mathbb E_n\left[ \left(p_0(s)
                 \right)^2\right] \Big)^2\dd s 
    + \frac{2\tilde\ga n^{3/2}}{n+1}\int_0^t \Big( T_R-\mathbb E_n\left[ \left(p_n(s)
    \right)^2\right] \Big)^2\dd s  \\
  &+ \frac{2\tilde \ga n^{3/2}}{n+1} \sum_{x=1}^n \int_0^t\left(\mathbb E_n\left[ p_0(s)
  p_{x}(s)\right] \right)^2 \dd s+
  \frac{2\tilde \ga n^{3/2}}{n+1}\sum_{x=0}^{n-1}\int_0^t\left(\mathbb E_n\left[ p_n(s)
  p_{x}(s)\right] \right)^2\dd s
\end{align*}

\begin{align*}
 &
 + \frac {2 \tilde \ga n^{3/2}}{n+1} \sum_{x'=1}^n   \int_0^t\left( \mathbb E_n\left[ p_0(s)
  r_{x'}(t)\right] \right)^2  \dd s+  \frac {2 \tilde \ga n^{3/2}}{n+1} \sum_{x'=1}^n \int_0^t \left(  \mathbb E_n\left[ p_n(s)
   r_{x'}(s)\right] \right)^2\dd s
\end{align*}
\begin{align*}
&
 ={\cal H}_n^{(2)}(0) +\frac{2\tilde \ga n^{3/2}}{n+1}\int_0^t\left[T_L\Big(T_L-\mathbb E_n\left[ \left(p_0(s)
                 \right)^2\right] \Big) 
    +  T_R\Big( T_R-\mathbb E_n\left[ \left(\tilde
                 p_n(s)\right)^2\right]\Big)\right]\dd s
    \end{align*}

\setcounter{supsection}{2}
\section{\thesupsection.  Proof of formula \eqref{tFjj}}

Recall that
\begin{align}
  \label{S012409-25}
  &
    \psi_j(x)=\left(\frac{2-\delta_{0,j}}{n+1}\right)^{1/2}\cos\left(\frac{\pi
    j(2x+1)}{2(n+1)}\right),\\
  &\phi_j(x)=\left(\frac{2}{n+1}\right)^{1/2}\sin\left(\frac{jx\pi}{n+1}\right),\notag\\
&\nabla^\star \psi_j(x)=-\ga_j \phi_j(x),\quad x,j=0,\ldots,n, \notag\\
&
                                                                   \nabla \phi_j(x)=\ga_j \psi_j(x),\quad j=1,\ldots,n,\,x=0,\ldots,n, \notag\\
  &
\la_j=\ga_j^2,\quad  \ga_j=   2\sin\left(\frac{j\pi}{2(n\!+\!1)}\right),\quad j=1,\ldots,n. \notag
\end{align}

We have
            \begin{align*}
                 &\tilde F_{j,j'}:=
                 \ga\sum_{y=0}^n\psi_j(y)\psi_{j'}(y)\Big[\lang
                 (\nabla^\star p_y)^2\rang_t+\lang (\nabla^\star p_{y+1})^2\rang_t\Big]     \\          
              &
                -\ga\sum_{y=1}^n\big[\psi_j(y-1)\psi_{j'}(y)+ \psi_j(y)\psi_{j'}(y-1)\big]\lang
                 (\nabla^\star p_y)^2\rang_t   , \quad j,j'=0,\ldots,n.
                \end{align*}
Due to the convention $p_{-1}=p_0$ and $p_{n+1}=p_n$ we have $\nabla^\star p_0=\nabla^\star p_{n+1}=0$.
Using the relations \eqref{S012409-25} we can then write
\begin{align*}
                 &\tilde F_{j,j'}=
                 \ga\sum_{y=1}^n\psi_j(y)\psi_{j'}(y) \lang
                 (\nabla^\star p_y)^2\rang_t+\ga\sum_{y=1}^n \psi_j(y-1)\psi_{j'}(y-1) \lang (\nabla^\star
                   p_{y})^2\rang_t \\                
              &
                -\ga\sum_{y=1}^n\big[\psi_j(y-1)\psi_{j'}(y)+ \psi_j(y)\psi_{j'}(y-1)\big]\lang
                 (\nabla^\star
                p_y)^2\rang_t           \\
  &
=\ga\ga_j\ga_{j'}\sum_{y=1}^n\phi_j(y) \phi_{j'}(y) \lang
                   (\nabla^\star p_y)^2\rang_t.
 \end{align*}

 \setcounter{supsection}{3}
\section{\thesupsection.   Solution of the system \eqref{163011-21czx}}

  With the  notation introduced in \eqref{S012409-25}
we can rewrite the system (\ref{163011-21czx})  as follows
\begin{align}
  \label{S163011-21azx}
  &\ga_{j'}\tilde S^{(r,p)}_{j,j'}
   =-\ga_{j}\tilde S^{(p,r)}_{j,j'}+ \frac{1}{n^{3/2}t}\Big[{\tilde
    S}_{j,j'}^{(r)}(0)-{\tilde S}_{j,j'}^{(r)}(n^{3/2}t)\Big],\quad j,j'=1,\ldots,n,\notag\\
 &-\ga_{j} \tilde S^{(r)}_{j,j'}+\ga \la_j \tilde
   S^{(p,r)}_{j,j'}+\tilde \ga \sum_{\ell=0}^n[\psi_{\ell}(0)\psi_{j}(0)+\psi_{\ell}(n)\psi_{j}(n)]\tilde S^{(p,r)}_{\ell,j'}
   + \ga_{j'}\tilde S^{(p)}_{j,j'}\notag\\
  &
    =\frac{1}{n^{3/2}t}\Big[{\tilde S}_{j,j'}^{(p,r)}(0)-{\tilde
    S}_{j,j'}^{(p,r)}(n^{3/2}t)\Big], \quad j=0,\ldots,n,\,j'=1,\ldots,n,
   \notag\\
  &-\ga_{j'} \tilde S^{(r)}_{j,j'}+ \gamma \la_{j'}\tilde S^{(r,p)}_{j,j'}
    +\tilde \ga
    \sum_{\ell=0}^n[\psi_{\ell}(0)\psi_{j'}(0)+\psi_{\ell}(n)\psi_{j'}(n)]\tilde
    S^{(r,p)}_{j,\ell}+\ga_j \tilde S^{(p)}_{j,j'}\notag\\
  &
    =\frac{1}{n^{3/2}t}\Big[{\tilde S}_{j,j'}^{(r,p)}(0)-{\tilde S}_{j,j'}^{(r,p)}(n^{3/2}t)\Big],  \quad j=1,\ldots,n,\,j'=0,\ldots,n,\notag\\
  & -\ga_{j}\tilde S^{(r,p)}_{j,j'}-\ga_{j'}\tilde S^{(p,r)}_{j,j'}=  \tilde F_{j,j'}
    -\ga(\la_j+\la_{j'})\tilde  S^{(p)}_{j,j'}\\
  &
                   +2\tilde \ga \Big(T_L\psi_j(0)\psi_{j'}(0)+T_R\psi_j(n)\psi_{j'}(n)\Big)  -\tilde \ga
    \sum_{\ell=0}^n[\psi_{\ell}(0)\psi_{j}(0)+\psi_{\ell}(n)\psi_{j}(n)]\tilde
    S^{(p)}_{\ell,j'} \notag\\
  &
    -\tilde \ga
    \sum_{\ell=0}^n[\psi_{\ell}(0)\psi_{j'}(0)+\psi_{\ell}(n)\psi_{j'}(n)]S^{(p)}_{j,\ell} 
    +\frac{1}{n^{3/2}t}\Big[{\tilde S}_{j,j'}^{(p)}(0)-{\tilde
    S}_{j,j'}^{(p)}(n^{3/2}t)\Big],  \quad j,j'=0,\ldots,n .\notag
\end{align}

Let
\begin{align*}
 & \tilde s^{(p,\tilde r)}_{x,j}=\tilde s^{(\tilde r,p)}_{j,x}=\sum_{\ell=0}^n\psi_{\ell}(x)\tilde
  S^{(r,p)}_{j,\ell}=\lang \tilde r_j p_x\rang_t,\quad
   x=0,\ldots,n,\,j=1,\ldots,n,\\
  &
  \tilde s^{(p)}_{x,j}=  \tilde s^{(p)}_{j,x}=\sum_{\ell=0}^n\psi_{\ell}(x)\tilde
  S^{(p)}_{j,\ell}=\lang p'_j p_x\rang_t
\end{align*}
and
\begin{align*}
  &
  \delta_{0,t}{\tilde S}_{j,j'}^{(p)}  :={\tilde S}_{j,j'}^{(p)}(0)-{\tilde
    S}_{j,j'}^{(p)}(t),\quad  \delta_{0,t}{\tilde S}_{j,j'}^{(r)}  :={\tilde S}_{j,j'}^{(r)}(0)-{\tilde
    S}_{j,j'}^{(r)}(n^{3/2}t)
  \\
  &
    \delta_{0,t}{\tilde S}_{j,j'}^{(p,r)}  :={\tilde S}_{j,j'}^{(p,r)}(0)-{\tilde
    S}_{j,j'}^{(p,r)}(n^{3/2}t),\quad  \delta_{0,t}{\tilde S}_{j,j'}^{(r,p)}  :={\tilde S}_{j,j'}^{(r,p)}(0)-{\tilde
    S}_{j,j'}^{(r,p)}(n^{3/2}t)
    \end{align*}
We can rewrite the above system in the form
\begin{align}
  \label{S163011-21czxx}
  &\ga_{j'}\tilde S^{(r,p)}_{j,j'}
   =-\ga_{j}\tilde S^{(p,r)}_{j,j'}+\frac{1}{n^{3/2}t}\delta_{0,t}{\tilde S}_{j,j'}^{(r)} ,\quad j,j'=1,\ldots,n,\notag\\
 &-\ga_j \tilde S^{(r)}_{j,j'}+\ga \la_j \tilde
   S^{(p,r)}_{j,j'}+\tilde \ga \big(\psi_{j}(0) \tilde s^{(r,p)}_{j',0} +\psi_{j}(n) \tilde s^{(r,p)}_{j',n}\big) 
   + \ga_{j'}\tilde S^{(p)}_{j,j'}\notag\\
  &
    =\frac{1}{n^{3/2}t}\delta_{0,t}{\tilde S}_{j,j'}^{(p,r)} 
,\quad j=0,\ldots,n,\, j'=1,\ldots,n,   \notag\\
  &-\ga_{j'} \tilde S^{(r)}_{j,j'}+ \gamma \la_{j'}\tilde S^{(r,p)}_{j,j'}
    +\tilde \ga \big(\psi_{j'}(0) \tilde s^{(r,p)}_{j,0} +\psi_{j'}(n)
    \tilde s^{(r,p)}_{j,n}\big) +\ga_j \tilde S^{(p)}_{j,j'}\notag\\
  &
    =\frac{1}{n^{3/2}t}\delta_{0,t}{\tilde S}_{j,j'}^{(r,p)} ,\quad j=1,\ldots,n,\, j'=0,\ldots,n, \notag\\
  & -\ga_{j}\tilde S^{(r,p)}_{j,j'}-\ga_{j'}\tilde S^{(p,r)}_{j,j'}=
    \tilde F_{j,j'}\\
  &+2\tilde \ga t \Big(T_L\psi_j(0)\psi_{j'}(0)+T_R\psi_j(n)\psi_{j'}(n)\Big) 
    -\ga(\la_j+\la_{j'})\tilde  S^{(p)}_{j,j'}\notag\\
  &
    -\tilde \ga\big(\psi_{j}(0) \tilde s^{(p)}_{j',0} +\psi_{j}(n)
    \tilde s^{(p)}_{j',n} \big) -\tilde \ga\big(\psi_{j'}(0) \tilde
    s^{(p)}_{j,0} +\psi_{j'}(n) \tilde s^{(p)}_{j,n} \big)\notag
  \\
  &
    +\frac{1}{n^{3/2}}\delta_{0,t}{\tilde S}_{j,j'}^{(p)} ,\quad j,j'=0,\ldots,n  .\notag
\end{align}

Suppose now that $j,j'=1,\ldots,n.$
Adding the second and third equations of   \eqref{S163011-21czxx}, then
subtracting the third one from the second one we obtain
\begin{align}
\label{011306-22bo}
 &-(\ga_j +\ga_{j'} )\tilde S^{(r)}_{j,j'}+ \gamma \Big(\la_j \tilde S^{(p,r)}_{j,j'}+\la_{j'}\tilde S^{(r,p)}_{j,j'}\Big)
   + (\ga_j +\ga_{j'} )\tilde S^{(p)}_{j,j'} \\
  &
    =-\tilde \ga
    \big(\psi_{j}(0) \tilde s^{(p,r)}_{0,j'}+\psi_{j}(n) \tilde s^{(p,r)}_{n,j'}\big)
    -\tilde \ga \big(\psi_{j'}(0) \tilde s^{(p,r)}_{0,j}+\psi_{j'}(n)
    \tilde s^{(p,r)}_{n,j}\big) \notag\\
  &
    +\frac{1}{n^{3/2}t}\Big(\delta_{0,t}{\tilde S}_{j,j'}^{(p,r)} + \delta_{0,t}{\tilde S}_{j,j'}^{(r,p)} \Big),
  \notag\\
&
-(\ga_j -\ga_{j'} )\tilde S^{(r)}_{j,j'}+ \gamma \Big(\la_j \tilde S^{(p,r)}_{j,j'}-\la_{j'}\tilde S^{(r,p)}_{j,j'}\Big)
              - (\ga_j -\ga_{j'} )\tilde S^{(p)}_{j,j'}\notag\\
  &
    =-\tilde \ga
    \big(\psi_{j}(0) \tilde s^{(p,r)}_{0,j'}+\psi_{j}(n) \tilde s^{(p,r)}_{n,j'}\big)
    +\tilde \ga \big(\psi_{j'}(0) \tilde s^{(p,r)}_{0,j}+\psi_{j'}(n)
    \tilde s^{(p,r)}_{n,j}\big) \notag\\
  &
    +\frac{1}{n^{3/2}t}\Big(\delta_{0,t}{\tilde S}_{j,j'}^{(p,r)} - \delta_{0,t}{\tilde S}_{j,j'}^{(r,p)} \Big),
  \notag
\end{align}

From the first equation of \eqref{S163011-21azx} we get
\begin{equation}
  \label{S012210-22bo}
\tilde S^{(r,p)}_{j,j'}
   =-\frac{\ga_{j}}{\ga_{j'}}\tilde S^{(p,r)}_{j,j'}+\frac{1}{\ga_{j'} n^{3/2}}\delta_{0,t}{\tilde S}_{j,j'}^{(r)}.
   \end{equation}

   Since $\la_j=\ga_j^2$, when $j,j'=1,\ldots,n$ we use  \eqref{S012210-22bo} and get
\begin{align}
\label{S011306-22bo}
 &-(\ga_j +\ga_{j'} )\tilde S^{(r)}_{j,j'}+ \gamma \ga_j \Big(\ga_j -\ga_{j'}\Big) \tilde S^{(p,r)}_{j,j'}
   + (\ga_j +\ga_{j'} )\tilde S^{(p)}_{j,j'} 
    =G^{(1)}_{j,j'}\\
&
-(\ga_j -\ga_{j'} )\tilde S^{(r)}_{j,j'}+ \gamma \ga_j \Big(\ga_j +\ga_{j'}\Big) \tilde S^{(p,r)}_{j,j'}
              - (\ga_j -\ga_{j'} )\tilde S^{(p)}_{j,j'}
    =G^{(2)}_{j,j'}.\notag
\end{align}
Here
\begin{align*}
  &
  G^{(1)}_{j,j'}:=  -\tilde \ga
    \big(\psi_{j}(0) \tilde s^{(p,\tilde r)}_{0,j'}+\psi_{j}(n) \tilde s^{(p,\tilde r)}_{n,j'}\big)
    -\tilde \ga \big(\psi_{j'}(0) \tilde s^{(p,\tilde r)}_{0,j}+\psi_{j'}(n)
    \tilde s^{(p,\tilde r)}_{n,j}\big) \notag\\
  &
    +\frac{1}{n^{3/2}t}\Big(\delta_{0,t}{\tilde S}_{j,j'}^{(p,r)} +
    \delta_{0,t}{\tilde S}_{j,j'}^{(r,p)} \Big)
    -\frac{\ga\ga_{j'}}{  n^{3/2}t}\delta_{0,t}{\tilde S}_{j,j'}^{(r)}
\end{align*}
\begin{align*}
  &
  G^{(2)}_{j,j'}:=  -\tilde \ga
    \big(\psi_{j}(0) \tilde s^{(p,\tilde r)}_{0,j'}+\psi_{j}(n) \tilde s^{(p,\tilde r)}_{n,j'}\big)
    +\tilde \ga \big(\psi_{j'}(0) \tilde s^{(p,\tilde r)}_{0,j}+\psi_{j'}(n)
    \tilde s^{(p,\tilde r)}_{n,j}\big) \notag\\
  &
    +\frac{1}{n^{3/2}t}\Big(\delta_{0,t}{\tilde S}_{j,j'}^{(p,r)} -
    \delta_{0,t}{\tilde S}_{j,j'}^{(r,p)} \Big)
    +\frac{\ga\ga_{j'}}{  n^{3/2}t}\delta_{0,t}{\tilde S}_{j,j'}^{(r)}
  \end{align*}

The fourth equation of \eqref{S163011-21czxx} reads
\begin{align}
  \label{012702-25}
  \Big(\frac{\ga^2_{j}}{\ga_{j'}}-\ga_{j'}\Big)\tilde
    S^{(p,r)}_{j,j'}+\ga(\la_j+\la_{j'})\tilde  S^{(p)}_{j,j'}=
  \tilde   G_{j,j'}^{(3)}.
\end{align}
Here
\begin{align*}
&\tilde G_{j,j'}^{(3)}  := \tilde F_{j,j'}+ 2\tilde \ga 
                 \Big(T_L\psi_j(0)\psi_{j'}(0)+T_R\psi_j(n)\psi_{j'}(n)\Big)  
  \\
  &
    -\tilde \ga
   \big(\psi_{j}(0) \tilde s^{(p)}_{0,j'}+\psi_{j}(n) \tilde s^{(p)}_{n,j'}\big)
   -\tilde \ga
   \big(\psi_{j'}(0) \tilde s^{(p)}_{0,j}+\psi_{j'}(n) \tilde
    s^{(p)}_{n,j}\big)\\
  &
    +\frac{1}{n^{3/2}t}\delta_{0,t}{\tilde S}_{j,j'}^{(p)}+\frac{ \ga_{j}}{  \ga_{j'}n^{3/2}t}\delta_{0,t}{\tilde S}_{j,j'}^{(r)}.
\end{align*}

\subsection*{Solution of the system of equations for the covariances}

Consider the following
    system of equations
\begin{align}
\label{S011912-23}
 &-(\la^{1/2}_j +\la^{1/2}_{j'} )\tilde S^{(r)}_{j,j'}+ \gamma \la^{1/2}_j \Big(\la^{1/2}_j -\la^{1/2}_{j'}\Big) \tilde S^{(p,r)}_{j,j'}
   + (\la^{1/2}_j +\la^{1/2}_{j'} )\tilde S^{(p)}_{j,j'}
    =G_{j,j'}^{(1)},
   \notag\\
&
-(\la^{1/2}_j -\la^{1/2}_{j'} )\tilde S^{(r)}_{j,j'}+ \gamma \la^{1/2}_j \Big(\la^{1/2}_j +\la^{1/2}_{j'}\Big) \tilde S^{(p,r)}_{j,j'}
              - (\la^{1/2}_j -\la^{1/2}_{j'} )\tilde
              S^{(p)}_{j,j'}=G_{j,j'}^{(2)},\notag\\
&\Big(\la_{j}-\la_{j'}\Big)\tilde
    S^{(p,r)}_{j,j'}+\ga \la^{1/2}_{j'} (\la_j+\la_{j'})\tilde  S^{(p)}_{j,j'}=   G^{(3)}_{j,j'}.
\end{align}
By computing $\tilde S^{(p,r)}_{j,j'} $ by adding  the first two
equations of \eqref{S011912-23} 
we get
\begin{align}
  \label{S021912-23}
\tilde S^{(p,r)}_{j,j'}=\frac{\la^{1/2}_j}{\ga\la_j}  \tilde
                 S^{(r)}_{j,j'}-\frac{\la^{1/2}_{j'}}{\ga\la_j}\tilde
                 S^{(p)}_{j,j'}+\frac{1}{2\ga\la_j}
                 \Big(G_{j,j'}^{(1)}+G_{j,j'}^{(2)}\Big).
\end{align}
Substituting into the first   equation of \eqref{S011912-23} we get
\begin{align*}
&-(\la^{1/2}_j +\la^{1/2}_{j'} )\tilde S^{(r)}_{j,j'}+ \gamma \la^{1/2}_j \Big(\la^{1/2}_j -\la^{1/2}_{j'}\Big) \Big[\frac{1}{\ga \la^{1/2}_j}  \tilde
                 S^{(r)}_{j,j'}-\frac{\la^{1/2}_{j'}}{\ga\la_j}\tilde
                 S^{(p)}_{j,j'}+\frac{1}{2\ga\la_j}\Big(
                     G_{j,j'}^{(1)}+G_{j,j'}^{(2)}\Big)\Big]
  \\
  &
   + (\la^{1/2}_j +\la^{1/2}_{j'} )\tilde S^{(p)}_{j,j'}
    =G_{j,j'}^{(1)},
\end{align*}
Hence
\begin{align*}
-2 \la^{1/2}_{j'} \tilde S^{(r)}_{j,j'}
   + \frac{\la_j +\la_{j'}}{\la^{1/2}_j}\tilde S^{(p)}_{j,j'}
    =\frac{\la^{1/2}_j +\la^{1/2}_{j'} }{2 \la^{1/2}_j}
                     G_{j,j'}^{(1)} +\frac{\la^{1/2}_{j'} -\la^{1/2}_j }{2 \la^{1/2}_j}G_{j,j'}^{(2)}
\end{align*}
and
\begin{align}
  \label{S021903-23}
\tilde S^{(r)}_{j,j'}
   = \frac{\la_j +\la_{j'}}{2 \la^{1/2}_j \la^{1/2}_{j'} }\tilde S^{(p)}_{j,j'}
    -\frac{\la^{1/2}_j +\la^{1/2}_{j'} }{4 \la^{1/2}_j \la^{1/2}_{j'} }
                     G_{j,j'}^{(1)} +\frac{\la^{1/2}_j -\la^{1/2}_{j'}
  }{4 \la^{1/2}_j \la^{1/2}_{j'} }G_{j,j'}^{(2)}.
\end{align}
Furthermore, from \eqref{S021912-23} and \eqref{S021903-23}
\begin{align*}
&\tilde S^{(p,r)}_{j,j'}
    =\frac{1}{\ga \la^{1/2}_j}  \Big(\frac{\la_j +\la_{j'}}{2 \la^{1/2}_j \la^{1/2}_{j'} }\tilde S^{(p)}_{j,j'}
    -\frac{\la^{1/2}_j +\la^{1/2}_{j'} }{4 \la^{1/2}_j \la^{1/2}_{j'} }
                     G_{j,j'}^{(1)} +\frac{\la^{1/2}_j -\la^{1/2}_{j'}
  }{4 \la^{1/2}_j \la^{1/2}_{j'} }G_{j,j'}^{(2)}\Big)-\frac{\la^{1/2}_{j'}}{\ga\la_j}\tilde
                 S^{(p)}_{j,j'}\\
  &
    +\frac{1}{2\ga\la_j}
                 \Big( G_{j,j'}^{(1)} + G_{j,j'}^{(2)} \Big),
\end{align*}
hence
\begin{align}
  \label{S010702-24}
\tilde S^{(p,r)}_{j,j'}
    =  \frac{\la_j-\la_{j'} }{2\ga\la_j \la^{1/2}_{j'} }\tilde
                 S^{(p)}_{j,j'}
  +\frac{\la^{1/2}_{j'}-\la^{1/2}_j }{4\ga\la_j \la^{1/2}_{j'} }
    G_{j,j'}^{(1)}    +\frac{\la^{1/2}_j+\la^{1/2}_{j'} }{4\ga\la_j \la^{1/2}_{j'} }
    G_{j,j'}^{(2)}  
\end{align}

and 

$$
\Big(\la_{j}-\la_{j'}\Big)\tilde
    S^{(p,r)}_{j,j'}+\ga \la^{1/2}_{j'} (\la_j+\la_{j'})\tilde  S^{(p)}_{j,j'}=   G^{(3)}_{j,j'}.
$$

Substituting into the last equation of \eqref{S011912-23}  we get
\begin{align*}
  &\Big(\la_{j}-\la_{j'}\Big) \frac{\la_j-\la_{j'} }{2\ga\la_j \la^{1/2}_{j'} }\tilde
    S^{(p)}_{j,j'} 
  +  \Big(\la_{j}-\la_{j'}\Big) \frac{\la^{1/2}_{j'}-\la^{1/2}_j }{4\ga\la_j \la^{1/2}_{j'} }
    G_{j,j'}^{(1)}
  \\
  &
   \Big(\la_{j}-\la_{j'}\Big)\frac{\la^{1/2}_j+\la^{1/2}_{j'} }{4\ga\la_j \la^{1/2}_{j'} }
    G_{j,j'}^{(2)}  +\ga \la^{1/2}_{j'}  (\la_j+\la_{j'})\tilde  S^{(p)}_{j,j'}=
     G_{j,j'}^{(3)}.
\end{align*}
In consequence

\begin{align*}
  &\Big( \la_{j}-\la_{j'}\Big)  \frac{\la_j-\la_{j'} }{2\ga\la_j \la_{j'} }\tilde
                 S^{(p)}_{j,j'}
  +\Big( \la_{j}-\la_{j'}\Big)  \frac{\la^{1/2}_{j'}-\la^{1/2}_j }{4\ga\la_j \la_{j'} }
    G_{j,j'}^{(1)}     \\
  &
    +\Big( \la_{j}-\la_{j'}\Big)  \frac{\la^{1/2}_j+\la^{1/2}_{j'} }{4\la_j \la_{j'} }
    G_{j,j'}^{(2)}  
    +\ga  (\la_j+\la_{j'})\tilde  S^{(p)}_{j,j'}=   \frac{1}{\la^{1/2}_{j'}}  G^{(3)}_{j,j'}
\end{align*}


Hence,
\begin{align*}
  &\Theta_p^{-1}(\la_j,\la_{j'})\tilde S^{(p)}_{j,j'}
  = (\la_j-\la_{j'})\frac{\la^{1/2}_{j}-\la^{1/2}_{j'} }{4\ga\la_j \la_{j'} }
    G_{j,j'}^{(1)} \\
  &
    -
     (\la_j-\la_{j'})\frac{\la^{1/2}_{j'}+\la^{1/2}_{j}  }{4\ga\la_j \la_{j'} }
    G_{j,j'}^{(2)}  + \frac{1}{\la^{1/2}_{j'}}   G^{(3)}_{j,j'},
\end{align*}
where
\begin{align*}
&\Theta_p(c,c'):=\frac{(c-c')^2}{2\ga
  cc'}+\ga(c+c')=\frac {2\ga cc'}{\theta(c,c')},\\
  &
 \theta(c,c')= (c-c')^2+ 2\ga^2cc'(c+c') .
\end{align*}

We can write
\begin{align}
  \label{S011803-23}
 & \tilde S^{(p)}_{j,j'}
  =  \sum_{\ell=1}^3\Psi^{(p)}_\ell(\la_j,\la_{j'})
    G_{j,j'}^{(\ell)} 
\end{align}

Here
\begin{align*}
  &\Psi^{(p)}_1(c,c')
    =\frac{(\sqrt{c}+\sqrt{c'}) (\sqrt{c}-\sqrt{c'})^2
    }{2\theta(c,c')},
\\
  &\Psi^{(p)}_2(c,c')=-\frac{(\sqrt{c}-\sqrt{c'}) (\sqrt{c}+\sqrt{c'})^2
    }{2\theta(c,c')}, \\
  &\Psi^{(p)}_3(c,c')
    =\frac{2\ga c c'
    }{(c-c')^2+2\ga^2 cc'  (c+c')} .
  \end{align*}

Furthermore, from \eqref{S021903-23}, for $j,j'=1,\ldots,n$ we have
\begin{align*}
&\tilde S^{(r)}_{j,j'}
 =\sum_{\ell=1}^{3} \Psi^{(r)}_\ell(\la_j,\la_{j'})G_{j,j'}^{(\ell)},
\end{align*}
with
\begin{align*}
   &\Psi^{(r)}_1(c,c')
     =-\frac{(\sqrt{c}+\sqrt{c'})  [(\sqrt{c}-\sqrt{c'})^2+\ga^2 \sqrt{cc'}  (c+c')]
  }{2\theta(c,c')  } ,
\\
  &
   \Psi^{(r)}_2(c,c')
    =\frac{(\sqrt{c}-\sqrt{c'})[\ga^2 \sqrt{cc'}  (c+c') -   (\sqrt{c}+\sqrt{c'})^2]}
    {2\theta(c,c') } 
\\
  &
    \Psi^{(r)}_3(c,c') 
    =\frac{\ga(c +c')\sqrt{c}}{(c-c')^2
    +2\ga^2c\sqrt{c'}(c+c')}=\frac{\ga(c +c')\sqrt{c}}{\theta(c,c')}.
\end{align*}

Finally, from \eqref{S010702-24} we get for $j,j'=1,\ldots,n$
\begin{align*}
&\tilde S^{(p,r)}_{j,j'}
                  =\sum_{   \ell=1}^3\Psi^{(p,r)}_\ell(\la_j,\la_{j'}) G_{j,j'}^{(\ell)}  , 
\end{align*}
with
\begin{align*}
  &\Psi^{(p,r)}_1(c,c')
    =\frac{\ga\sqrt{c'}( c+c') (\sqrt{c'}-\sqrt{c})
    }{2\theta(c.c')},\\
  &\Psi^{(p,r)}_2(c,c')
    =\frac{ \ga\sqrt{c'}(\sqrt{c}+\sqrt{c'})(c+c')
    }{2\theta(c,c')},\\ 
  &\Psi^{(p,r)}_3(c,c')
    =\frac{c-c' 
    }{(c-c')^2+2\ga^2 cc'  (c+c')} .
\end{align*}

\subsection*{Formulas for the covariances}

\label{sec4.3}

Denote
\begin{align*}
 &\hat F_{j,j'} =\ga\ga_j\ga_{j'}\sum_{y=1}^n\phi_j(y) \phi_{j'}(y) \lang
                   (\nabla^\star p_y)^2\rang_t,\\
  &
    B_{j,j'}^{(p, r)}=\psi_{j}(0) \tilde s^{(p,\tilde
    r)}_{0,j'}+\psi_{j}(n) \tilde s^{(p,\tilde r)}_{n,j'},\quad B_{j,j'}^{(r,p)}=B_{j',j}^{(p, r)},\\
    &
     B_{j,j'}^{(p)}=2 
 \Big(T_L\psi_j(0)\psi_{j'}(0)+T_R\psi_j(n)\psi_{j'}(n)\Big) 
  \\
  &
    -\Big(\psi_j(0)\tilde s^{(p)}_{0,j'}+\psi_{j}(n)
      \tilde s^{(p)}_{n,j'}+\psi_{j'}(0) \tilde s^{(p)}_{0,j}+\psi_{j'}(n)
      \tilde s^{(p)}_{n,j}\Big),\\
  &
    R_{j,j'}^{(\iota)}=\frac{1}{n^{3/2}t} \delta_{0,t}{\tilde
    S}_{j,j'}^{(\iota)} ,\quad \iota\in I.
\end{align*}
Recall that $I=\{p,pr,rp,r\}$.
Then
\begin{align}
  \label{G1aa}
  &
  G^{(1)}_{j,j'}:=  -\tilde \ga
     B_{j,j'}^{(p, r)}
    -\tilde \ga  B_{j,j'}^{(r,p)}
    + R_{j,j'}^{(p,r)}+
    R_{j,j'}^{(r,p)} 
    - \ga\la^{1/2}_{j'} R_{j,j'}^{(r)},\\
  &
  G^{(2)}_{j,j'}:=  -\tilde \ga
     B_{j,j'}^{(p, r)}
    +\tilde \ga  B_{j,j'}^{(r,p)}
    +R_{j,j'}^{(p,r)}-
    R_{j,j'}^{(r,p)} 
    + \ga\la^{1/2}_{j'}R_{j,j'}^{(r)}\notag
  \end{align}
and
\begin{align*}
& G_{j,j'}^{(3)}  := \la^{1/2}_{j'}\tilde F_{j,j'}+ 2\tilde \ga \la^{1/2}_{j'}
                 \Big(T_L\psi_j(0)\psi_{j'}(0)+T_R\psi_j(n)\psi_{j'}(n)\Big)  
  \\
  &
    -\tilde \ga\la^{1/2}_{j'}
   \big(\psi_{j}(0) \tilde s^{(p)}_{0,j'}+\psi_{j}(n) \tilde s^{(p)}_{n,j'}\big)
   -\tilde \ga\la^{1/2}_{j'}
   \big(\psi_{j'}(0) \tilde s^{(p)}_{0,j}+\psi_{j'}(n) \tilde
    s^{(p)}_{n,j}\big)\\
  &
    +\frac{\la^{1/2}_{j'}}{n^{3/2}t}\delta_{0,t}{\tilde S}_{j,j'}^{(p)}+\frac{ \ga_{j}}{  n^{3/2}t}\delta_{0,t}{\tilde S}_{j,j'}^{(r)}.
\end{align*}


Rearranging the formulas for the covariances we obtain
\begin{align}
  \label{011803-23}
 & \tilde S^{(p)}_{j,j'} =  \Theta_p(\la_j,\la_{j'})\hat F_{j,j'}+\sum_{\iota\in
  I}\Xi^{(p)}_\iota(\la_j,\la_{j'})
  R_{j,j'}^{(\iota)} 
    +\sum_{\iota\in I}\Pi^{(p)}_\iota(\la_j,\la_{j'})
    B_{j,j'}^{(\iota)} .
\end{align}

  Here
  \begin{align*}
   &\Theta_p(c,c')=\frac {2\ga
     cc'}{\theta(c,c')},\quad\mbox{where}\quad 
     \theta(c,c')= (c-c')^2+ 2\ga^2cc'(c+c') ,\\
    &
      \Xi^{(p)}_p(c,c')= \Theta_p(c,c'),\quad
      \Pi^{(p)}_p(c,c')=\tilde \ga\Theta_p(c,c'),\\
    &
        \Xi^{(p)}_{p,r}(c,c')=\frac{ \sqrt{c'} (c'-
      c  )
      }{\theta(c,c')},\quad   \Xi^{(p)}_{r,p}(c,c')=\frac{ \sqrt{c} (c-
      c'  )
      }{\theta(c,c')},\\
  &    \Pi^{(p)}_{p,r}(c,c')=-\tilde \ga  \Xi^{(p)}_{p,r}(c,c'),\quad
    \Pi^{(p)}_{r,p}(c,c') =-\tilde\ga  \Xi^{(p)}_{r,p}(c,c')\\
       &
        \Xi^{(p)}_{r}(c,c')=\frac{ \ga\sqrt{cc'} (c  +c' 
     )
      }{\theta(c,c')},\quad 
        \Pi^{(p)}_{r}(c,c')=0.
\end{align*}

From \eqref{S010702-24} we obtain
\begin{align*}
&\tilde S^{(p,r)}_{j,j'}    
    =\Theta_{p,r}(\la_j,\la_{j'})\hat
                 F_{j,j'}+\sum_{\iota\in I}\Xi^{(p,r)}_\iota(\la_j,\la_{j'})
                 R_{j,j'}^{(\iota)}+\sum_{\iota\in I}\Pi^{(p,r)}_\iota(\la_j,\la_{j'})
                 B_{j,j'}^{(\iota)},
\end{align*}

where
\begin{align*}
&
\Theta_{p,r}(c,c')=\frac{c-c' }{2\ga c\sqrt{c'} } \Theta_p(c,c')=\frac
                 { (c-c' )\sqrt{c'}}{\theta(c,c')}\quad\mbox{where}\quad 
     \theta(c,c')= (c-c')^2+ 2\ga^2cc'(c+c')\\
  &
    \Xi^{(p,r)}_p(c,c')= \Theta_{p,r}(c,c'),\quad
    \Pi^{(p,r)}_p(c,c')=\tilde \ga \Theta_{p,r}(c,c'), \\
  &
    \Xi^{(p,r)}_{p,r}(c,c') =\frac
    {  \ga c'(c+c')}{\theta(c,c')},\quad \Pi^{(p,r)}_{p,r}(c,c')
    =-\tilde \ga  \Xi^{(p,r)}_{p,r}(c,c') ,\\
  &
    \Xi^{(p,r)}_{r,p}(c,c') =-\frac
                 { \ga
    \sqrt{cc'}(c+c')}{\theta(c,c')},\quad \Pi^{(p,r)}_{r,p}(c,c') =-\tilde \ga \Xi^{(p,r)}_{r,p}(c,c') ,  \\
       &
    \Xi^{(p,r)}_{r}(c,c') =\frac
                 { 1}{2 \sqrt{c} }\Big[1+\frac{(c-c'
         )(c+c') }{ \theta(c,c')}\Big]=\frac
                 { 1}{2 \sqrt{c} }\Big[1+\frac{(c^2-(c')^2)
         }{ \theta(c,c')}\Big]
  \\
  &
    =\frac
                 { 1}{2 \theta(c,c')\sqrt{c} }\Big[2c^2-2cc'+2\ga^2cc'(c+c')\Big]
    =\frac{ \sqrt{c}\Big[c-c'+\ga^2c'(c+c')\Big]}{ \theta(c,c') },\\
  &
    \Pi^{(p,r)}_r(c,c')=0
\end{align*}

Finally,
\begin{align*}
&\tilde S^{(r)}_{j,j'}
                 =\Theta_r(\la_j,\la_{j'}) \hat F_{j,j'}
                 +\sum_{\iota\in I}\Xi^{(r)}_{\iota}(\la_j,\la_{j'})
                 R_{j,j'}^{(\iota)}+\sum_{\iota\in
                 I}\Pi^{(r)}_{\iota}(\la_j,\la_{j'})
                 B_{j,j'}^{(\iota)} ,
\end{align*}
where
\begin{align*}
 & \Theta_r(c,c') =\frac{\ga (c+c')
   \sqrt{c}\sqrt{c'}}{\theta(c,c')}=\Xi^{(r)}_{p}(c,c'), \quad
   \Pi^{(r)}_{p}(c,c')=\tilde \ga \Theta_r(c,c'),\\
  &
    \Xi^{(r)}_{p,r}(c,c')=\frac{\sqrt{c}[c'-c-\ga^2
    c'(c+c')]}{\theta(c,c')},\quad 
    \Pi^{(r)}_{p,r}(c,c')=-\tilde\ga \Xi^{(r)}_{p,r}(c,c')\\
  &
    \Xi^{(r)}_{r,p}(c,c')=\frac{\sqrt{c'}[c-c'-\ga^2
    c(c+c')]}{\theta(c,c')}=\Xi^{(r)}_{p,r}(c',c)\\
  &
    \Pi^{(r)}_{r,p}(c,c')=-\tilde \ga\Xi^{(r)}_{r,p}(c,c') \\
  &
    \Xi^{(r)}_{r}(c,c')=\ga\frac{
    c^2+(c')^2+\ga^2cc'(c+c')}{\theta(c,c')},\quad \Pi^{(r)}_{r}(c,c')=0.
  \end{align*}

\setcounter{supsection}{4}
\section{\thesupsection.   Proof of Lemma \ref{lmW}}

  For a given function $f:[0,1]\to\bbR$ define its even and odd
extensions by letting
$$
 f _e(u)=\left\{
  \begin{array}{ll}
    f (u),& u\in[0,1],\\
    &\\
    f (-u),&u\in[-1,0]
  \end{array}
\right.
$$
and $f _o(0)=f_o(1):=0$,
$$
 f _o(u)=\left\{
  \begin{array}{ll}
    f (u),& u\in(0,1),\\
    &\\
    -f (-u),&u\in(-1,0).
  \end{array}
\right.
$$
For a given $2$-periodic $f:\bbR\to\mathbb C$  denote its discrete
Fourier transform
$$
\hat f(j;n):=\frac{1}{2^{1/2}(n+1)}\sum_{x=-n-1}^{n}\exp
\{-i\pi j u_x \}f(u_x),\quad j\in\bbZ.
$$
Here
\begin{equation}
  \label{u-x}
  u_x=\frac{x}{n+1}.
\end{equation}
Observe that
\begin{align*}
 & i\hat f_o(j;n)=\frac{\sqrt 2}{n+1} \sum_{x=1}^n \sin(\pi j u_x)
   f(u_x),\\
  &
    \hat f_e(j;n)=\frac{\sqrt 2}{n+1} \sum_{x=1}^n \cos(\pi j u_x)
   f(u_x)
  \end{align*}
Define
\begin{align*}
 & {\cal W}_{j,j'}:=  n^{1/2}\sum_{x=0}^{n} 
   \phi_{j'}(x+1) \psi_{j}(x)  \varphi'(u_{x+1}).
\end{align*}
Recall that ${\rm supp}\,\varphi'\subset(0,1)$. We prove that
  \begin{equation}
    \label{SW}
    \begin{split}
 &  {\cal W}_{j,j'} =- i  \Big(\frac{n}{2}\Big)^{1/2} \left(1-\frac{\delta_{0,j}}{2}\right)^{1/2}\cos
   \Big(\frac{\pi k_j}{2}\Big) \Big[ \widehat{(\varphi ')_o}(j-j') -\widehat{(\varphi ')_o}(j+j')\Big]
  \\
  &
    -\Big(\frac{n}{2}\Big)^{1/2}\left(1-\frac{\delta_{0,j}}{2}\right)^{1/2}
\sin
   \Big(\frac{\pi k_j}{2}\Big)    \Big[ \widehat{(\varphi
     ')_e}(j+j')-\widehat{(\varphi ')_e}(j-j')  \Big].
\end{split}
\end{equation}
Here
\begin{equation}
  \label{Sk-j}
  k_j:=\frac{j}{n+1}.
\end{equation}
 
We have
\begin{align}
  \label{S061801-24a}
  &
  {\cal W}_{j,j'}:=  \frac{ 2 n^{1/2}  }{n+1} \left(1-\frac{\delta_{0,j}}{2}\right)^{1/2}\sum_{x=0}^{n}
  \sin\Big(\frac{j'(x+1)\pi}{n+1}\Big)
  \cos\Big(\frac{j(2x+1)\pi}{2(n+1)}\Big) \varphi '\left(\frac{x+1}{n+1}\right).
\end{align}

We have
\begin{align*}
 & 
  {\cal W}_{j,j'}=  \frac{ n^{1/2}}{2i(n+1)}\left(1-\frac{\delta_{0,j}}{2}\right)^{1/2}\ \sum_{x=0}^{n}
\left[ \exp \Big\{\frac{j'(x+1)i\pi}{n+1}\Big\}-\exp
   \Big\{-\frac{j'(x+1)i\pi}{n+1}\Big\}\right]\\
  &
    \times
  \left[ \exp \Big\{\frac{j(2x+1)i\pi}{2(n+1)}\Big\} +\exp
    \Big\{-\frac{j(2x+1)i\pi}{2(n+1)}\Big\} \right]\varphi '\left(\frac{x+1}{n+1}\right)
\end{align*}

\begin{align*}
 & 
  =  \frac{ n^{1/2}}{2i(n+1)} \left(1-\frac{\delta_{0,j}}{2}\right)^{1/2}  \sum_{x=0}^{n}
\left[ \exp \Big\{\frac{j'(2x+2)i\pi}{2(n+1)}\Big\}-\exp
   \Big\{-\frac{j'(2x+2)i\pi}{2(n+1)}\Big\}\right]\\
  &
    \times
  \left[ \exp \Big\{\frac{j(2x+1)i\pi}{2(n+1)}\Big\} +\exp
    \Big\{-\frac{j(2x+1)i\pi}{2(n+1)}\Big\} \right] \varphi '\left(\frac{x+1}{n+1}\right)
\end{align*}

\begin{align*}
 & 
  =  \frac{ n^{1/2}}{2i(n+1)} \left(1-\frac{\delta_{0,j}}{2}\right)^{1/2}\sum_{x=0}^{n}
  \exp \Big\{\frac{(j'+j)xi\pi}{n+1}\Big\}\exp
   \Big\{\frac{(2j'+j)i\pi}{2(n+1)}\Big\}  \varphi '\left(\frac{x+1}{n+1}\right)  
\end{align*}
\begin{align*}
 & 
 +\frac{ n^{1/2}}{2i(n+1)} \left(1-\frac{\delta_{0,j}}{2}\right)^{1/2}\sum_{x=0}^{n}
\exp \Big\{\frac{(j'-j)ix\pi}{n+1}\Big\}\exp
   \Big\{\frac{(2j'-j)i\pi}{2(n+1)}\Big\}  \varphi '\left(\frac{x+1}{n+1}\right)
\end{align*}

\begin{align*}
 & 
 - \frac{ n^{1/2}}{2i(n+1)} \left(1-\frac{\delta_{0,j}}{2}\right)^{1/2}\sum_{x=0}^{n}
\exp
   \Big\{\frac{(j-j')xi\pi}{n+1}\Big\} \exp
   \Big\{\frac{(j-2j')i\pi}{2(n+1)}\Big\}   \varphi '\left(\frac{x+1}{n+1}\right)
   \end{align*}
\begin{align*}
 & 
  -  \frac{ n^{1/2}}{2i(n+1)}\left(1-\frac{\delta_{0,j}}{2}\right)^{1/2}\sum_{x=0}^{n}
 \exp
   \Big\{-\frac{(j+j')xi\pi}{n+1}\Big\} \exp
   \Big\{-\frac{(2j'+j)i\pi}{2(n+1)}\Big\}   \varphi '\left(\frac{x+1}{n+1}\right)
    \end{align*}

\begin{align*}
 & 
   =  \frac{ n^{1/2}}{2i(n+1)} \left(1-\frac{\delta_{0,j}}{2}\right)^{1/2}\sum_{x=0}^{n}
   \exp \Big\{\frac{(j'+j)(x+1)i\pi}{n+1}\Big\}\exp
   \Big\{-\frac{ji\pi}{2(n+1)}\Big\}  \varphi '\left(\frac{x+1}{n+1}\right)
   \end{align*}
\begin{align*}
 & 
  +  \frac{ n^{1/2}}{2i(n+1)} \left(1-\frac{\delta_{0,j}}{2}\right)^{1/2}\sum_{x=0}^{n}
\exp \Big\{\frac{(j'-j)(x+1)\pi}{n+1}\Big\}\exp
   \Big\{\frac{ji\pi}{2(n+1)}\Big\}  \varphi '\left(\frac{x+1}{n+1}\right)
   \end{align*}

\begin{align*}
 & 
 - \frac{ n^{1/2}}{2i(n+1)} \left(1-\frac{\delta_{0,j}}{2}\right)^{1/2}\sum_{x=0}^{n}
\exp
   \Big\{\frac{(j-j')(x+1)i\pi}{n+1}\Big\} \exp
   \Big\{-\frac{ji\pi}{2(n+1)}\Big\}   \varphi
   '\left(\frac{x+1}{n+1}\right)
\end{align*}
\begin{align*}
 & 
  -  \frac{ n^{1/2}}{2i(n+1)} \left(1-\frac{\delta_{0,j}}{2}\right)^{1/2}\sum_{x=0}^{n}
 \exp
   \Big\{-\frac{(j+j')(x+1)i\pi}{n+1}\Big\} \exp
   \Big\{\frac{ji\pi}{2(n+1)}\Big\}   
  \varphi '\left(\frac{x+1}{n+1}\right)
\end{align*}

\begin{align*}
 & 
   =  \frac{ n^{1/2}}{2i(n+1)} \left(1-\frac{\delta_{0,j}}{2}\right)^{1/2}\cos
   \Big(\frac{j\pi}{2(n+1)}\Big) \sum_{x=0}^{n}
   \exp \Big\{\frac{(j'+j)(x+1)i\pi}{n+1}\Big\} \varphi '\left(\frac{x+1}{n+1}\right)
\end{align*}

\begin{align*}
 & 
   -  \frac{ n^{1/2}}{2(n+1)} \left(1-\frac{\delta_{0,j}}{2}\right)^{1/2}\sin
   \Big(\frac{j\pi}{2(n+1)}\Big) \sum_{x=0}^{n}
   \exp \Big\{\frac{(j'+j)(x+1)i\pi}{n+1}\Big\} \varphi '\left(\frac{x+1}{n+1}\right)
\end{align*}

\begin{align*}
 & 
  +  \frac{ n^{1/2}}{2i(n+1)} \left(1-\frac{\delta_{0,j}}{2}\right)^{1/2}\cos
   \Big(\frac{j\pi}{2(n+1)}\Big) \sum_{x=0}^{n}
\exp \Big\{\frac{(j'-j)(x+1)\pi}{n+1}\Big\} \varphi '\left(\frac{x+1}{n+1}\right)
\end{align*}
\begin{align*}
 & 
  +  \frac{ n^{1/2}}{2(n+1)} \left(1-\frac{\delta_{0,j}}{2}\right)^{1/2}\sin
   \Big(\frac{j\pi}{2(n+1)}\Big) \sum_{x=0}^{n}
\exp \Big\{\frac{(j'-j)(x+1)\pi}{n+1}\Big\}\varphi '\left(\frac{x+1}{n+1}\right)
   \end{align*}

\begin{align*}
 & 
 - \frac{ n^{1/2}}{2i(n+1)} \left(1-\frac{\delta_{0,j}}{2}\right)^{1/2}\cos
   \Big(\frac{j\pi}{2(n+1)}\Big)   \sum_{x=0}^{n}
\exp
   \Big\{\frac{(j-j')(x+1)i\pi}{n+1}\Big\} \varphi
   '\left(\frac{x+1}{n+1}\right)
\end{align*}
\begin{align*}
 & 
 +\frac{ n^{1/2}}{2(n+1)} \left(1-\frac{\delta_{0,j}}{2}\right)^{1/2}\sin
   \Big(\frac{j\pi}{2(n+1)}\Big)   \sum_{x=0}^{n}
\exp
   \Big\{\frac{(j-j')(x+1)i\pi}{n+1}\Big\} \varphi
   '\left(\frac{x+1}{n+1}\right)
\end{align*}

\begin{align*}
 & 
  -  \frac{ n^{1/2}}{2i(n+1)} \left(1-\frac{\delta_{0,j}}{2}\right)^{1/2}\cos
   \Big(\frac{j\pi}{2(n+1)}\Big)  \sum_{x=0}^{n}
 \exp
   \Big\{-\frac{(j+j')(x+1)i\pi}{n+1}\Big\} 
  \varphi '\left(\frac{x+1}{n+1}\right)
\end{align*}
\begin{align*}
 & 
  -  \frac{ n^{1/2}}{2(n+1)} \left(1-\frac{\delta_{0,j}}{2}\right)^{1/2}\sin
   \Big(\frac{j\pi}{2(n+1)}\Big) \sum_{x=0}^{n}
 \exp
   \Big\{-\frac{(j+j')(x+1)i\pi}{n+1}\Big\} 
  \varphi '\left(\frac{x+1}{n+1}\right)
\end{align*}

\begin{align*}
 & 
   =  -  n^{1/2}i  \left(1-\frac{\delta_{0,j}}{2}\right)^{1/2}\cos
   \Big(\frac{j\pi}{2(n+1)}\Big)  \left(\frac{1}{2(n+1)} \sum_{x=-n-1}^{n}
   \exp \Big\{\frac{(j'+j)xi\pi}{n+1}\Big\} (\varphi
   ')_o\left(\frac{x}{n+1}\right)\right)\\
   & 
   - n^{1/2}  \left(1-\frac{\delta_{0,j}}{2}\right)^{1/2}\sin
   \Big(\frac{j\pi}{2(n+1)}\Big)  \left(\frac{1}{2(n+1)} \sum_{x=-n-1}^{n}
   \exp \Big\{\frac{(j'+j)xi\pi}{n+1}\Big\} (\varphi
     ')_e\left(\frac{x}{n+1}\right)\right)\\
   &
    +  n^{1/2}  \left(1-\frac{\delta_{0,j}}{2}\right)^{1/2}\sin
   \Big(\frac{j\pi}{2(n+1)}\Big)  \left(\frac{1}{2(n+1)} \Big[
  \varphi
     '\left(0\right) -(-1)^{j+j'}\varphi
     '\left(1\right)\right)\Big]\\
 & 
 +  n^{1/2}i \left(1-\frac{\delta_{0,j}}{2}\right)^{1/2}
\cos
   \Big(\frac{j\pi}{2(n+1)}\Big)   \left(\frac{1}{2(n+1)} \sum_{x=-n-1}^{n}
   \exp \Big\{\frac{(j-j')xi\pi}{n+1}\Big\} (\varphi
   ')_o\left(\frac{x}{n+1}\right)\right)\\
  & 
 +  n^{1/2} \left(1-\frac{\delta_{0,j}}{2}\right)^{1/2}
\sin
   \Big(\frac{j\pi}{2(n+1)}\Big)  \left(\frac{1}{2(n+1)} \sum_{x=-n-1}^{n}
   \exp \Big\{\frac{(j'-j)xi\pi}{n+1}\Big\} (\varphi
    ')_e\left(\frac{x}{n+1}\right)\right)\\
    &
    -  n^{1/2} \left(1-\frac{\delta_{0,j}}{2}\right)^{1/2}\sin
   \Big(\frac{j\pi}{2(n+1)}\Big)  \left(\frac{1}{2(n+1)} \Big[
  \varphi
     '\left(0\right) -(-1)^{j-j'}\varphi
     '\left(1\right)\right)\Big]\\
\end{align*}

\begin{align*}
 & 
   =  -  n^{1/2}i \left(1-\frac{\delta_{0,j}}{2}\right)^{1/2}\cos
   \Big(\frac{j\pi}{2(n+1)}\Big)  \left(\frac{1}{2(n+1)} \sum_{x=-n-1}^{n}
   \exp \Big\{\frac{(j'+j)xi\pi}{n+1}\Big\} (\varphi
   ')_o\left(\frac{x}{n+1}\right)\right)\\
   & 
   -  n^{1/2}  \left(1-\frac{\delta_{0,j}}{2}\right)^{1/2}\sin
   \Big(\frac{j\pi}{2(n+1)}\Big)  \left(\frac{1}{2(n+1)} \sum_{x=-n-1}^{n}
   \exp \Big\{\frac{(j'+j)xi\pi}{n+1}\Big\} (\varphi
     ')_e\left(\frac{x}{n+1}\right)\right)\\
 & 
 + n^{1/2}i \left(1-\frac{\delta_{0,j}}{2}\right)^{1/2}
\cos
   \Big(\frac{j\pi}{2(n+1)}\Big)   \left(\frac{1}{2(n+1)} \sum_{x=-n-1}^{n}
   \exp \Big\{\frac{(j-j')xi\pi}{n+1}\Big\} (\varphi
   ')_o\left(\frac{x}{n+1}\right)\right)\\
  & 
 +  n^{1/2} \left(1-\frac{\delta_{0,j}}{2}\right)^{1/2}
\sin
   \Big(\frac{j\pi}{2(n+1)}\Big)  \left(\frac{1}{2(n+1)} \sum_{x=-n-1}^{n}
   \exp \Big\{\frac{(j'-j)xi\pi}{n+1}\Big\} (\varphi
    ')_e\left(\frac{x}{n+1}\right)\right)
\end{align*}

Here
$$
 \varphi_e(u)=\left\{
  \begin{array}{ll}
    \varphi(u),& u\in[0,1],\\
    &\\
    \varphi(-u),&u\in[-1,0]
  \end{array}
\right.
$$
and
$$
 \varphi_o(u)=\left\{
  \begin{array}{ll}
    \varphi(u),& u\in[0,1],\\
    &\\
    -\varphi(-u),&u\in[-1,0).
  \end{array}
\right.
$$ 
We have concluded therefore formula \eqref{SW}.

\setcounter{supsection}{5}
\section{\thesupsection.   Proof of formula \eqref{eq:27}}

Using  formula (\ref{eq:theta})  we arrive at
\begin{align}
  \label{S061801-24}
 & \frac{1}{n+1}\theta_{pr}(\varphi';n)=-\frac{1}{n+1}  \sum_{j,j'=1}^{n} {\cal W}_{j,j'}
   \Theta_{p,r}^{(1)} (\la_j,\la_{j'})    F_{j,j'}\quad\mbox{with}\\
    &
    \Theta_{p,r}^{(1)}(c,c')=\sqrt{cc'} \Theta_{p,r}(c,c')=\frac
                 { (c-c' )c'\sqrt{c}}{\theta(c,c')},\notag\\
    &
  F_{j,j'} =\ga\sum_{y=1}^n\phi_j(y) \phi_{j'}(y) \lang
      (\nabla^\star p_y)^2\rang_t,\notag\\
  &
  {\cal W}_{j,j'}:=  n^{1/2}\sum_{x=0}^{n} 
    \phi_{j'}(x+1) \psi_{j}(x)  \varphi'(u_{x+1}) ,\notag\\
            &\Pi^{(pr)}_p(c,c')=\tilde \ga\Theta_{pr}(c,c'),\quad \Theta_{pr}(c,c')=\frac
              { (c-c' )\sqrt{c'}}{\theta(c,c')},\notag\\
      &  {\cal W}_{j,j'} =- i \Big(\frac{n}{2}\Big)^{1/2} \left(1-\frac{\delta_{0,j}}{2}\right)^{1/2}\cos
   \Big(\frac{\pi k_j}{2}\Big) \Big[ \widehat{(\varphi ')_o}(j-j') -\widehat{(\varphi ')_o}(j+j')\Big]
\notag  \\
  &
    -\Big(\frac{n}{2}\Big)^{1/2}\left(1-\frac{\delta_{0,j}}{2}\right)^{1/2}
\sin
   \Big(\frac{\pi k_j}{2}\Big)    \Big[ \widehat{(\varphi
     ')_e}(j+j')-\widehat{(\varphi ')_e}(j-j')  \Big] \notag 
\end{align}
and
\begin{align*}
&\la_j=4\sin^2\left(\frac{\pi j}{2(n+1)}\right),\quad
\psi_j(x)=\left(\frac{2-\delta_{0,j}}{n+1}\right)^{1/2}\cos\left(\frac{\pi
    j(2x+1)}{2(n+1)}\right),\\
&\phi_j(x)=\left(\frac{2}{n+1}\right)^{1/2}\sin\left(\frac{jx\pi}{n+1}\right),\quad
                 \la_j 
                                 ,\quad x,j=1,\ldots,n.
\end{align*}

  Subsituting for ${\cal W}_{j,j'}$ we obtain
 \begin{align*}
 & 
    \frac{1}{n+1}\theta_{pr}(\varphi';n)  =\frac{1}{n+1}\theta_{pr}^{(o)}(\varphi';n)
   +\frac{1}{n+1}\theta_{pr}^{(e)}(\varphi';n) ,\quad\mbox{where}\\
  &\frac{1}{n+1}\theta_{pr}^{(o)}(\varphi';n)= \frac{ i n^{1/2}  }{2^{1/2}(n+1)}\sum_{j,j'=1}^{n}  
       \Big[ \widehat{(\varphi ')_o}(j-j') -\widehat{(\varphi ')_o}(j+j')\Big]
    \Theta_{pr}^{(o)} (j,j') F_{j,j'}\\
        &\frac{1}{n+1}\theta_{pr}^{(e)}(\varphi';n)= \frac{ n^{1/2}  }{2^{1/2}(n+1)}\sum_{j,j'=1}^{n}  
       \Big[ \widehat{(\varphi ')_e}(j+j') -\widehat{(\varphi ')_e}(j-j')\Big]
    \Theta_{pr}^{(e)} (j,j') F_{j,j'}
\end{align*}
where
\begin{align*}
  &
   \Theta_{pr}^{(o)}(j,j') =\frac
                 { \cos
   \Big(\frac{\pi k_j}{2}\Big)  (\la_j-\la_{j'}) \la_{j'}\ga_j}{\theta (\la_j,\la_{j'})}= \frac
                 { \sin
   (\pi k_j)  (\la_j-\la_{j'}) \la_{j'} }{\theta (\la_j,\la_{j'})}  ,\\
  &
   \Theta_{pr}^{(e)}(j,j')  = \frac
                 { \sin
   \Big(\frac{\pi k_j}{2 }\Big)  (\la_j-\la_{j'}) \la_{j'}\ga_j}{\theta (\la_j,\la_{j'})}= \frac
                 {   (\la_j-\la_{j'}) \la_{j'}\la_j}{2\theta (\la_j,\la_{j'})}.
   \end{align*}

Taking into accound parity $F_{-j,j'} =F_{j,-j'}=-F_{j,j'}$
and  the fact that
\begin{align*}
 \cos
   \Big(\frac{-(n+1)\pi}{2(n +1)}\Big) =\phi_{0}(n)=\phi_{-n-1}(n+1)
   =\phi_{n+1}(n)=0
\end{align*}
we conclude that
\begin{align*}
  &\frac{1}{n+1}\theta_{pr}^{(o)}(\varphi';n)= \frac{ i n^{1/2}  }{2^{1/2}(n+1)}\sum_{j,j'=1}^{n}  
       \Big[ \widehat{(\varphi ')_o}(j-j') -\widehat{(\varphi ')_o}(j+j')\Big]
    \Theta_{pr}^{(o)} (j,j') F_{j,j'}\\
        &= \frac{ i n^{1/2}  }{2^{1/2}(n+1)}\sum_{j=1}^{n}  \sum_{j'=0}^{n}  
       \widehat{(\varphi ')_o}(j-j') 
          \Theta_{pr}^{(o)} (j,j') F_{j,j'}\\
  &
   - \frac{ i n^{1/2}  }{2^{1/2}(n+1)}\sum_{j=1}^{n}  \sum_{j'=-n-1}^{-1}  
       \widehat{(\varphi ')_o}(j-j')\Big]
    \Theta_{pr}^{(o)} (j,-j') F_{j,-j'}
\end{align*}
\begin{align*}
  &= \frac{ i n^{1/2}  }{2^{1/2}(n+1)}\sum_{j=1}^{n}  \sum_{j'=0}^{n}  
       \widehat{(\varphi ')_o}(j-j') 
          \Theta_{pr}^{(o)} (j,j') F_{j,j'}\\
  &
   + \frac{ i n^{1/2}  }{2^{1/2}(n+1)}\sum_{j=1}^{n}  \sum_{j'=-n-1}^{-1}  
       \widehat{(\varphi ')_o}(j-j')\Big]
    \Theta_{pr}^{(o)} (j,j') F_{j,j'}
\end{align*}
\begin{align*}
  &= \frac{ i n^{1/2}  }{2^{1/2}(n+1)}\sum_{j=1}^{n}  \sum_{j'=-n-1}^{n}  
       \widehat{(\varphi ')_o}(j-j') 
    \Theta_{pr}^{(o)} (j,j') F_{j,j'}\\
  &
    =\frac{ i n^{1/2}  }{2^{3/2}(n+1)}\sum_{j=1}^{n}  \sum_{j'=-n-1}^{n}  
       \widehat{(\varphi ')_o}(j-j') 
    \Theta_{pr}^{(o)} (j,j') F_{j,j'}\\
  &+\frac{ i n^{1/2}  }{2^{3/2}(n+1)}\sum_{j=-n-1}^{-1}  \sum_{j'=-n-1}^{n}  
       \widehat{(\varphi ')_o}(-j-j') 
    \Theta_{pr}^{(o)} (-j,j') F_{-j,j'}
\end{align*}
\begin{align*}
  &
    =\frac{ i n^{1/2}  }{2^{3/2}(n+1)}\sum_{j=1}^{n}  \sum_{j'=-n-1}^{n}  
       \widehat{(\varphi ')_o}(j-j') 
    \Theta_{pr}^{(o)} (j,j') F_{j,j'}\\
  &+\frac{ i n^{1/2}  }{2^{3/2}(n+1)}\sum_{j=-n-1}^{-1}  \sum_{j'=-n-1}^{n}  
       \widehat{(\varphi ')_o}(-j+j') 
    \Theta_{pr}^{(o)} (-j,-j') F_{-j,-j'}
\end{align*}
\begin{align*}
  &
    =\frac{ i n^{1/2}  }{2^{3/2}(n+1)}\sum_{j=1}^{n}  \sum_{j'=-n-1}^{n}  
       \widehat{(\varphi ')_o}(j-j') 
    \Theta_{pr}^{(o)} (j,j') F_{j,j'}\\
  &+\frac{ i n^{1/2}  }{2^{3/2}(n+1)}\sum_{j=-n-1}^{-1}  \sum_{j'=-n-1}^{n}  
       \widehat{(\varphi ')_o}(j-j') 
    \Theta_{pr}^{(o)} (j,j') F_{j,j'}\\
  &
    =\frac{ i n^{1/2}  }{2^{3/2}(n+1)}  \sum_{j,j'=-n-1}^{n}  
       \widehat{(\varphi ')_o}(j-j') 
    \Theta_{pr}^{(o)} (j,j') F_{j,j'}.
\end{align*}
Likewise
\begin{align*}
 &\frac{1}{n+1}\theta_{pr}^{(e)}(\varphi';n)= \frac{ n^{1/2}  }{2^{1/2}(n+1)}\sum_{j,j'=1}^{n}  
       \Big[ \widehat{(\varphi ')_e}(j+j') -\widehat{(\varphi ')_e}(j-j')\Big]
   \Theta_{pr}^{(e)} (j,j') F_{j,j'}\\
  &
    = \frac{ n^{1/2}  }{2^{1/2}(n+1)}\sum_{j,j'=1}^{n}  
        \widehat{(\varphi ')_e}(j+j')  
    \Theta_{pr}^{(e)} (j,j') F_{j,j'}\\
  &- \frac{ n^{1/2}  }{2^{1/2}(n+1)}\sum_{j,j'=1}^{n}  
       \widehat{(\varphi ')_e}(j-j') 
    \Theta_{pr}^{(e)} (j,j') F_{j,j'}
\end{align*}
\begin{align*}
  &
    = \frac{ n^{1/2}  }{2^{1/2}(n+1)}\sum_{j=1}^{n}  \sum_{j'=0}^{n}  
        \widehat{(\varphi ')_e}(j+j')  
    \Theta_{pr}^{(e)} (j,j') F_{j,j'}\\
  &- \frac{ n^{1/2}  }{2^{1/2}(n+1)}\sum_{j=1}^{n}  \sum_{j'=-n-1}^{-1}  
       \widehat{(\varphi ')_e}(j+j') 
    \Theta_{pr}^{(e)} (j,-j') F_{j,-j'}
\end{align*}
\begin{align*}
  &
    = \frac{ n^{1/2}  }{2^{1/2}(n+1)}\sum_{j=1}^{n}  \sum_{j'=-n-1}^{n}  
        \widehat{(\varphi ')_e}(j+j')  
    \Theta_{pr}^{(e)} (j,j') F_{j,j'}\\
  &
    = \frac{ n^{1/2}  }{2^{3/2}(n+1)}\sum_{j=0}^{n}  \sum_{j'=-n-1}^{n}  
        \widehat{(\varphi ')_e}(j+j')  
    \Theta_{pr}^{(e)} (j,j') F_{j,j'}\\
  &
    + \frac{ n^{1/2}  }{2^{3/2}(n+1)}\sum_{j=-n-1}^{-1}  \sum_{j'=-n-1}^{n}  
        \widehat{(\varphi ')_e}(-j+j')  
    \Theta_{pr}^{(e)} (-j,j') F_{-j,j'}
\end{align*}
\begin{align*}
  &
    = \frac{ n^{1/2}  }{2^{3/2}(n+1)}\sum_{j=0}^{n}  \sum_{j'=-n-1}^{n}  
        \widehat{(\varphi ')_e}(j+j')  
    \Theta_{pr}^{(e)} (j,j') F_{j,j'}\\
  &
    + \frac{ n^{1/2}  }{2^{3/2}(n+1)}\sum_{j=-n-1}^{-1}  \sum_{j'=-n-1}^{n}  
        \widehat{(\varphi ')_e}(-j-j')  
    \Theta_{pr}^{(e)} (-j,-j') F_{-j,-j'}
\end{align*}
\begin{align*}
  &
    = \frac{ n^{1/2}  }{2^{3/2}(n+1)}\sum_{j=0}^{n}  \sum_{j'=-n-1}^{n}  
        \widehat{(\varphi ')_e}(j+j')  
    \Theta_{pr}^{(e)} (j,j') F_{j,j'}\\
  &
    + \frac{ n^{1/2}  }{2^{3/2}(n+1)}\sum_{j=-n-1}^{-1}  \sum_{j'=-n-1}^{n}  
        \widehat{(\varphi ')_e}(j+j')  
    \Theta_{pr}^{(e)} (j,j') F_{j,j'}\\
  &
    = \frac{ n^{1/2}  }{2^{3/2}(n+1)}   \sum_{j,j'=-n-1}^{n}  
        \widehat{(\varphi ')_e}(j+j')  
    \Theta_{pr}^{(e)} (j,j') F_{j,j'}\\
  &
    =\frac{ n^{1/2}  }{2^{3/2}(n+1)}   \sum_{j,j'=-n-1}^{n}  
        \widehat{(\varphi ')_e}(j+j')  
    \Theta_{pr}^{(e)} (j',j) F_{j,j'}\\
  &
    =-\frac{ n^{1/2}  }{2^{3/2}(n+1)}   \sum_{j,j'=-n-1}^{n}  
        \widehat{(\varphi ')_e}(j+j')  
    \Theta_{pr}^{(e)} (j,j') F_{j,j'}.
\end{align*}
It proves that
\begin{align*}
    \theta_{pr}^{(e)}(\varphi';n)=0.
   \end{align*}

\end{document}